\documentclass[12pt,a4paper]{article}
 
\usepackage{jheppub}
\usepackage[usenames,dvipsnames,table]{xcolor}
\usepackage{amssymb} 
\usepackage{amsmath}
\usepackage{mathtools}
\usepackage{amsfonts}    
\usepackage{dsfont}
\usepackage{pdfpages}
\usepackage[table]{xcolor}
\arrayrulecolor{gray} 
\renewcommand{\arraystretch}{1.3}
\usepackage{verbatim}
\usepackage{stmaryrd}
\usepackage{graphicx}
\usepackage{float}         

\usepackage{tensor}
\usepackage{mathrsfs}
\usepackage{textgreek} 
\usepackage{euscript}
\usepackage[smalltableaux]{ytableau}
\ytableausetup{centertableaux}
\usepackage{tikz}
\usetikzlibrary{decorations.pathreplacing, decorations.markings,calc,shapes.misc,decorations.pathmorphing,patterns.meta, math}
\usetikzlibrary{decorations.pathmorphing}
\usetikzlibrary{decorations.markings}
\usepackage{slashed}
\usepackage{datetime}
\usepackage{hyperref}
\usepackage{braket}
\hypersetup{
    pdfencoding=unicode,
	colorlinks=true,
	urlcolor=Maroon,
	linkcolor=RoyalBlue,
	citecolor=Maroon,
	pdftitle={On the three-point functions in timelike $\mathcal{N}=1$ Liouville CFT},
	pdfauthor={Beatrix M\"uhlmann, Vladimir Narovlansky and Ioannis Tsiares},
	pdfdisplaydoctitle=true,
	pdfstartview=FitH,
	linktocpage=true
}

\usetikzlibrary{shapes}
\usetikzlibrary{arrows}
\usetikzlibrary{decorations.pathmorphing}
\usetikzlibrary{decorations.markings}
\usetikzlibrary{shapes.misc}
\tikzset{snake it/.style={decorate, decoration=snake}}
\tikzset{cross/.style={cross out, draw=black, minimum size=2*(#1-\pgflinewidth), inner sep=0pt, outer sep=0pt},
cross/.default={1pt}}
\usetikzlibrary{shapes.geometric}
\tikzset{
    partial ellipse/.style args={#1:#2:#3}{
        insert path={+ (#1:#3) arc (#1:#2:#3)}
    }
}

\usepackage{subcaption}

\newcommand{\ba}{\begin{align}}
\newcommand{\bea}{\begin{aligned}}

\newcommand{\be}{\begin{equation}}
\newcommand{\ee}{\end{equation}}
\def\bd{\begin{tikzpicture}}
\def\ed{\end{tikzpicture}}

\allowdisplaybreaks

\def\XXint#1#2#3{{\setbox0=\hbox{$#1{#2#3}{\int}$}
     \vcenter{\hbox{$#2#3$}}\kern-.5\wd0}}

\definecolor{light-gray}{gray}{0.75}
\definecolor{mygreen}{RGB}{220,230,220}

\renewcommand\d{\text{d}}

\newcommand{\e}{\mathrm{e}}

\renewcommand{\leq}{\leqslant}
\renewcommand{\geq}{\geqslant}

\newcommand{\bb}{\hat{b}}
\newcommand{\pp}{\hat{p}}

\definecolor{bleudefrance}{rgb}{0.19, 0.55, 0.91}

\definecolor{vert}{rgb}{0.1367 0.543 0.1367}

\definecolor{pink}{rgb}{1.0, 0.13, 0.32}

\definecolor{myforestgreen}{RGB}{24, 150, 144}

\title{On the three-point functions in timelike $\mathcal{N}=1$ Liouville CFT}

\author[1]{\!\! Beatrix M\"uhlmann}\emailAdd{beatrix@ias.edu}
\author[1]{\!\!, Vladimir Narovlansky}\emailAdd{vladi@ias.edu}
\author[2]{\!\!, Ioannis Tsiares}\emailAdd{ioannis.tsiares@ipht.fr}
\affiliation[1]{School of Natural Sciences, Institute for Advanced Study, Princeton, NJ 08540, USA}
\affiliation[2]{Universit\'e Paris-Saclay, CNRS, CEA, Institut de Physique Th\'eorique, \\91191, Gif-sur-Yvette, France}

\abstract{
We use analytic (super-)conformal bootstrap methods to derive explicit expressions for the structure constants of $\mathcal{N}=1$ Liouville CFT in the `timelike' regime of the superconformal central charge. The obtained expressions take the form of inverses of the appropriate spacelike counterparts, which we explain concretely by elucidating the analytic properties of the corresponding shift relations in the NS- and R-sectors for the normalization-independent bootstrap data on the sphere. In a particular normalization, the timelike structure constants are shown to agree with the OPE coefficients of $\mathcal{N}=1$ Minimal Models when specified at degenerate values of the momenta, exactly as in the non-supersymmetric case. We discuss possible applications of our results, with emphasis on the construction of the $\mathcal{N}=1$ supersymmetric analog of the Virasoro Minimal String.
}

\begin{document}

\maketitle

\makeatletter
\g@addto@macro\bfseries{\boldmath}
\makeatother

\section{Introduction}
Liouville field theory was first introduced by Polyakov in 1981 in the study of bosonic string theory and two-dimensional quantum gravity \cite{Polyakov:1981rd}. At its core it is a nontrivial two-dimensional conformal field theory that is fully solvable: the spectrum of its primary operators, along with the associated structure constants, are known in closed analytic form \cite{Dorn:1994xn,Zamolodchikov:1995aa,Teschner:1995yf, Ponsot:1999uf,  Zamolodchikov:2005fy,Kostov:2005kk,Schomerus:2003vv,Ribault:2015sxa}. As a result, all correlation functions on a given Riemann surface are, in principle, exactly computable. This level of exact control, combined with its deep origins, has established Liouville CFT as an outstanding example of a quantum field theory.
\par After its discovery, Liouville theory has been proven remarkably versatile with a wide range of applications and various natural extensions (for some reviews see \cite{Seiberg:1990eb,Ginsparg:1993is, Nakayama:2004vk,Schomerus:2005aq, Anninos:2020ccj}). To mention just a few notable connections, in the early days the seminal work \cite{Knizhnik:1988ak} (see also \cite{David:1988hj,Distler:1988jt}) opened up an exciting new relation with 2d quantum gravity and matrix models, while a few years later Liouville correlation functions were shown to
have a deep relationship to four dimensional Yang Mills theories via the AGT correspondence\cite{Alday:2009aq}. Connections with probability theory were very recently put forward in \cite{David:2014aha,Kupiainen:2017eaw,Guillarmou:2020wbo} (see also \cite{Chatterjee:2024phq} for a nice introduction to these methods) where the authors proved in a rigorous set-up the celebrated DOZZ formula \cite{Dorn:1994xn,Zamolodchikov:1995aa} (along with its later bootstrap incarnations by Teschner in \cite{Teschner:1995yf,Teschner:2001rv}, as well as by Ponsot, Teschner in \cite{Ponsot:1999uf,Ponsot:2000mt}). On the string theory side, Liouville theory has been a pivotal building block of the worldsheet theory in various guises, particularly in theories with a low number of target spacetime dimensions—such as the Minimal String \cite{Seiberg:2003nm,Seiberg:2004at}, the $c=1$ \cite{Moore:1991zv, Balthazar:2017mxh} or type 0A/0B models \cite{Douglas:2003up, Klebanov:2003wg, Takayanagi:2003sm, Balthazar:2022atu}, and more recently, the Virasoro Minimal String \cite{Collier:2023cyw} and the Complex Liouville String \cite{Collier:2024kmo}. In all these cases, the associated string amplitudes are described by a simpler `holographic' dual theory and exhibit remarkably simple yet rich expressions, stemming from the exact solvability of Liouville theory. The Hilbert space spanned by Liouville conformal blocks also naturally describes the Hilbert space of 3d gravity with negative cosmological constant via (the holomorphic half of) the so-called Virasoro TQFT \cite{Collier:2023fwi, Collier:2024mgv}. This connection is closely related with the fact that Liouville theory can be shown to capture a universal chaotic sector of irrational unitary 2d CFTs characterized by a twist gap above the vacuum \cite{Jackson:2014nla,Turiaci:2016cvo, Collier:2018exn, Collier:2019weq}\footnote{See also \cite{Sonner:2024tqd} on a recent connection of Liouville theory with chaos.}. Finally, natural generalizations of Liouville theory include the presence of boundaries \cite{Fateev:2000ik, Teschner:2000md, Zamolodchikov:2001ah,Hosomichi:2001xc,Ponsot:2001ng}, higher spin symmetry algebras known as Toda conformal field theories \cite{Fateev:2007ab}, or various levels of supersymmetry \cite{Polyakov:1981re, Poghossian:1996agj,Rashkov:1996np,  Belavin:2007gz, Belavin:2007eq, Belavin:2006pv, Fukuda:2002bv,Hori:2001ax,Hosomichi:2004ph}. The latter, in particular, will play a central role in the present work.
\vspace{0.5cm}
\begin{table}[H]
\begin{center}
\begin{tabular}{>{\columncolor{mygreen}}l | >{\columncolor{gray!5}}c >{\columncolor{gray!5}}c}
\rowcolor{green!20}
 & \textbf{Spacelike Liouville CFT} & \textbf{Timelike Liouville CFT} \\ \hline
$\mathcal{N}=0$ & \cite{Dorn:1994xn,Zamolodchikov:1995aa,Teschner:1995yf,Ponsot:1999uf} & \cite{Zamolodchikov:2005fy,Kostov:2005kk,Schomerus:2003vv,Ribault:2015sxa} \\
$\mathcal{N}=1$ & \cite{Poghossian:1996agj,Rashkov:1996np,Belavin:2007gz} & this paper \& \cite{Rangamani:2025wfa} \\
$\mathcal{N}=2$ & ?  & ? 
\end{tabular}
\caption{\footnotesize{Summary of the main references that have established the structure constants (from the conformal bootstrap p.o.v.) of the quantum spacelike and timelike Liouville theory with various levels of supersymmetry.}}
\end{center}
\end{table}
\par There is yet another important distinction of Liouville theory, which came to be known as `timelike' Liouville theory. This terminology refers to the usual bosonic Liouville theory realized as a distinct solution to the conformal bootstrap equations with central charge $c \leq 1$, as opposed to the `spacelike'  theory that usually refers to Liouville theory at central charge $c\geq 25$.\footnote{We will occasionally make use of this terminology throughout this work, suitably adjusted in the supersymmetric setup.} Timelike Liouville theory first appeared as a model for time-dependent string theory in \cite{Strominger:2003fn}, but has since come to be understood as a fully consistent conformal field theory in its own right, characterized by a bounded (continuum) spectrum of primary operators and, crucially, by structure constants that are \textit{different} (but closely related) to those of its spacelike counterpart. Notably, timelike Liouville theory defines a \textit{non-unitary} CFT, in that its spectrum includes primary operators with negative conformal dimensions that violate the unitarity bound. Also, a remarkable feature of the theory is that, even though a conformal weight zero operator is part of its spectrum, this does \textit{not} correspond to the identity operator with the usual fusion rule $\mathds{1}\times \Phi\sim \Phi$. This was already noticed in the original references \cite{Strominger:2003fn, Zamolodchikov:2005fy,Schomerus:2003vv}. Despite these features, the theory has been developed from various viewpoints throughout the years \cite{Kostov:2005kk,Ribault:2015sxa, Harlow:2011ny, Bautista:2019jau, Martinec:2003ka, Anninos:2021ene, Muhlmann:2022duj, Anninos:2024iwf, McElgin:2007ak,Giribet:2011zx, Chatterjee:2025yzo}. 

From the perspective of \cite{Bautista:2019jau, Martinec:2003ka, Anninos:2021ene, Muhlmann:2022duj, Anninos:2024iwf} timelike Liouville theory constitutes a rigorous model of two-dimensional de Sitter quantum gravity. In that setup, the theory is coupled to a conformal field theory of large and positive central charge and the corresponding path integral admits a round two-sphere saddle, which is the geometry of Euclidean dS$_2$, while its disk path integral leads to a Hartle-Hawking type wavefunction \cite{Anninos:2024iwf}.
Furthermore, the negative sign of the kinetic term for the Weyl factor mimics the conformal mode problem of higher dimensional Euclidean quantum gravity \cite{Polchinski:1988ua, Gibbons:1978ac}. Timelike Liouville theory thus captures key features of higher-dimensional de Sitter quantum gravity while remaining more tractable than the higher-dimensional models. In a different direction, the structure constants of timelike Liouville theory have recently been shown to capture the connectivity probabilities in 2D percolation \cite{Delfino:2010xm} as well as the correlation functions of the so-called conformal loop models \cite{Ikhlef:2015eua}\footnote{See \cite{Ang:2021tjp} for the proof of these connections.}. 
\newline
\par In light of these wide-ranging applications, one would expect that supersymmetric extensions of timelike Liouville theory are not only natural but arguably essential—yet, with some exceptions such as \cite{Distler:1989nt, Antoniadis:1990mx, Anninos:2022ujl, Anninos:2023exn} that discuss the semiclassical supersymmetric (timelike) Liouville path integral, to date, no rigorous constructions coming from the (super-)conformal bootstrap exist in the literature.
\par We fill this gap in the present work by introducing the exact structure constants of quantum $\mathcal{N} = 1$ Liouville CFT in the `timelike' regime of the superconformal central charge. We do so purely from the perspective of the conformal bootstrap, without relying on specific details related to the Liouville action. In other words, we show that $\mathcal{N}=1$ timelike Liouville theory can be realized as a \textit{novel} solution to the superconformal bootstrap equations in a particular range of the central charge. As we will discover, many of the bizarre yet intriguing features of the usual bosonic timelike Liouville theory find natural analogues in the $\mathcal{N}=1$ setting. In particular, and in parallel to their bosonic counterparts, the structure constants in the $\mathcal{N}=1$ timelike theory admit an elegant expression, which we summarize in the table below.
\begin{table}[H]
\centering
\renewcommand{\arraystretch}{1.9}
\rowcolors{2}{gray!5}{white}
\begin{tabular}{>{\columncolor{green!20}}c | c | c}
\rowcolor{mygreen}
 & \footnotesize{\textbf{Spacelike} $\mathcal{N}=1$~ $(b\in\mathbb{R}_{(0,1]})$} & \footnotesize{\textbf{Timelike} $\mathcal{N}=1$ ~ $(\hat{b}\in\mathbb{R}_{(0,1]})$} \\
\hline
\footnotesize{\textbf{NS-sector}} & \footnotesize{$C^{(b)}_{\mathrm{NS}}(p_1,p_2,p_3)\ , \ \widetilde{C}^{(b)}_{\mathrm{NS}}(p_1,p_2,p_3)$} & \footnotesize{$\left(\frac{1}{i}\widetilde{C}^{(\hat{b})}_{\mathrm{NS}}(i{p}_1,i{p}_2,i{p}_3)\right)^{-1} , \ \left(\frac{1}{i}C^{(\hat{b})}_{\mathrm{NS}}(i{p}_1,i{p}_2,i{p}_3)\right)^{-1}$} \\
\footnotesize{\textbf{R-sector}} & \footnotesize{$C^{(b)}_{\mathrm{even}}(p_1,p_2;p_3)\ , \ C^{(b)}_{\mathrm{odd}}(p_1,p_2;p_3)$} & \footnotesize{$\left(C^{(\hat{b})}_{\mathrm{odd}}(ip_1,ip_2;ip_3)\right)^{-1}  , \ \left(C^{(\hat{b})}_{\mathrm{even}}(i{p}_1,i{p}_2;i{p}_3)\right)^{-1}$} \\
\hline
\end{tabular}
\caption{\footnotesize{Summary of the structure constants in $\mathcal{N}=1$ Liouville theory.}}
\label{tab:summaryintro}
\end{table}
To get a first grasp of what these results actually mean it is useful to introduce some minimal notation. We will henceforth refer to the ``spacelike'' $\mathcal{N}=1$ Liouville CFT at (superconformal) central charge $c$ for the values $c\geq9$. Analogously, we refer to the ``timelike'' $\mathcal{N}=1$ Liouville CFT at (superconformal) central charge $\hat{c}$ for the values $\hat{c}\leq 1$. We will accordingly adopt the following parametrization
\begin{align}
    \text{spacelike $\mathcal{N}=1$ Liouville CFT:}& \quad  \quad \quad  \ \ c =1 + 2Q^2~,\quad Q = b+b^{-1}~,\quad b \in\mathbb{R}_{(0,1]}~, \nonumber \\
    \text{timelike $\mathcal{N}=1$ Liouville CFT:}& \quad  \quad \quad  \ \ \hat{c} =1 - 2\widehat{Q}^2~,\quad \widehat{Q} = \hat{b}^{-1} - \hat{b}~,\quad \hat{b} \in\mathbb{R}_{(0,1]}~.\nonumber 
\end{align} 
For each version of the theory (spacelike or timelike), there are two types of structure constants, belonging either in the NS-sector or in the R-sector. In the NS-sector, there are \textit{two independent} structure constants which for the spacelike theory we denote as $C_{\mathrm{NS}},\widetilde{C}_{\mathrm{NS}}$. These depend on the central charge $b$ as well as the conformal dimensions of three NS primary operators via the momentum variable $p$, which is roughly the square root of the conformal dimension (c.f. (\ref{eq:hNS})). In the timelike theory, the relevant structure constants are \textit{not} obtained by analytic continuation of the spacelike expressions—mirroring the subtle and well-known distinction already present in bosonic Liouville theory\cite{Zamolodchikov:2005fy}. We will show that (in a particular normalization) the timelike NS-sector structure constants take the form that is written in table \ref{tab:summaryintro}, namely they are simply the inverses of the corresponding spacelike expressions, evaluated at appropriately rotated momentum variables, with the additional replacement $b\rightarrow\hat{b}$. Similarly in the R-sector, there are two structure constants that depend on the conformal dimensions of two R primaries via ($p_1,p_2$) (c.f. (\ref{eq:hR})) and the conformal dimension of one NS primary via $p_3$. For the spacelike case we denoted them as $C_{\mathrm{even}},C_{\mathrm{odd}}$, even though these are not really independent (as we will explain, they are related by $C_{\mathrm{odd}}(p_1,-p_2;p_3)=C_{\mathrm{odd}}(-p_1,p_2;p_3)=C_{\mathrm{even}}(p_1,p_2;p_3)$). In the timelike theory, we will show that, again, the two Ramond structure constants can be written as the inverses of the spacelike expressions exactly in the same fashion as in the NS case. The explicit expressions for all the structure constants are summarized in the beginning of section \ref{sec:discussion}.

\par Our derivation of these results proceeds by carefully studying the \textit{shift relations} that determine the structure constants, in parallel to the standard techniques employed in usual bosonic Liouville theory \cite{Teschner:1995yf, Zamolodchikov:2005fy}. More specifically, we will study the shift relations in the $p$ variables obeyed by the `bootstrap ratios'
\begin{equation}
    \frac{C^{(b)}_{\mathrm{NS}}(p_1,p_2,p_3)^2}{B^{(b)}_{\mathrm{NS}}(p_1)}, \ \frac{\widetilde{C}^{(b)}_{\mathrm{NS}}(p_1,p_2,p_3)^2}{B^{(b)}_{\mathrm{NS}}(p_1)} \quad \text{and} \quad \frac{C^{(b)}_{\mathrm{even}}(p_1,p_2;p_3)^2}{B^{(b)}_{\mathrm{R}}(p_1)},\frac{C^{(b)}_{\mathrm{odd}}(p_1,p_2;p_3)^2}{B^{(b)}_{\mathrm{R}}(p_1)}
\end{equation}
for the NS- and R-sectors respectively. Here $B^{(b)}_{\mathrm{NS}}, B^{(b)}_{\mathrm{R}}$ are the corresponding two-point function normalizations. Those shift relations `couple' non-trivially the above ratios in a way that does not depend on the specific choice of two-point function normalization. More importantly, these shift relations arise universally from the superconformal symmetry algebra and the existence of degenerate representations thereof; hence they do not distinguish between spacelike and timelike theories and are, consequently, analytic functions of the parameter $b$. It is precisely this feature that we harness in a systematic way (using the so-called `Virasoro-Wick Rotation'\cite{Ribault:2023vqs}) in order to extract the structure constants presented in table \ref{tab:summaryintro}.
\par This paper is organized as follows. In section \ref{sec: spacelike} we start by reviewing the spacelike $\mathcal{N}=1$ Liouville theory. In particular, we present the NS-sector and R-sector structure constants in a natural normalization that does not depend on the cosmological constant of the Liouville action. A detailed derivation of the shift relations obeyed by the normalization-independent bootstrap data on the sphere is presented in appendices \ref{app:NS} and \ref{app:R} without the use of the superspace formalism. In section \ref{sec:Timelike} we introduce the $\mathcal{N}=1$ timelike Liouville CFT. We discuss in detail the analytic properties of the shift relations when continued to the timelike central charge regime, and implement the so-called Virasoro-Wick Rotation to derive explicit expressions for the relevant structure constants. We then show explicitly that the newly-derived structure constants match, in a particular normalization of the two-point functions, to those of the $\mathcal{N}=1$ Minimal Models, analogously to the situation in the non-supersymmetric case. In section \ref{sec:discussion} we summarize the various structure constants and discuss some interesting future directions, placing a particular emphasis on the $\mathcal{N}=1$ analog of the Virasoro minimal string. In appendix \ref{app:SpecialFunctions} we review the relevant special functions for the construction of the Liouville structure constants and, for completeness, in appendix \ref{app:N0Liouville} we provide a pedagogical review of the derivation of the spacelike and timelike structure constants in standard bosonic Liouville theory. Finally, in Appendix \ref{app:fusionkernel} we review the $\mathcal{N} = 1$ fusion kernel in Liouville theory and discuss its connection with the structure constants.
\newline
\paragraph{Note added.}
While this work was near its completion, we became aware of \cite{Rangamani:2025wfa}, who also investigate the $\mathcal{N}=1$ timelike Liouville structure constants, and with whom we have coordinated publications.

\section{Spacelike $\mathcal{N}=1$ Liouville CFT}\label{sec: spacelike}
\par The $\mathcal{N}=1$ supersymmetric extension of Liouville field theory was first introduced in \cite{Polyakov:1981re,Distler:1989nt} in relation to the quantization of two-dimensional supergravity and non-critical superstring theory. 
On a two-dimensional surface, the theory is described by the following action of a real bosonic field $\phi$, and a Majorana spinor $\psi$
\begin{equation}\label{eq:N1action}
    S^{\mathcal{N}=1}_{\mathrm{sL}}=\frac{1}{4\pi}\int \d^2 x\,\tilde{\e}\left(\frac{1}{2}\tilde{g}^{\mu\nu}\partial_\mu\phi\partial_\nu\phi-\frac{i}{2}\overline{\psi}\slashed{D}\psi+\frac{1}{2}Q\widetilde{R}\phi + \frac{1}{2}\mu^2\e^{2b\phi} -\frac{i}{2}\mu b \e^{b\phi}\overline{\psi}\psi\right)~.
\end{equation}
The scale parameter $\mu$ is the (super) cosmological constant, and $b$ is the standard parameter of the background/reference charge $Q=b+b^{-1}$;  $\tilde{\e}_\mu^a$ is the zweibein of the reference metric $\tilde{g}_{\mu\nu}$ (with $\widetilde{R}$ the corresponding Ricci scalar). For the fermion, $\overline{\psi} = \psi^T \mathcal{C}$ with the charge conjugation matrix $\mathcal{C}$ (see e.g. \cite{Anninos:2023exn}). 
The quantum field theory described by the action (\ref{eq:N1action}) is characterized by $\mathcal{N}=1$ superconformal symmetry, where the central charge of the superconformal algebra is given by\footnote{We distinguish this from the central charge of the Virasoro subalgebra which is given by $\frac{3}{2}{c}$.}
\begin{equation}\label{eq:ccN1}
    c =1 + 2Q^2.
\end{equation}
The $\mathcal{N}=1$ superconformal algebra is generated by two copies (left-moving and right-moving) of the usual Virasoro stress tensor $(T(z), \widetilde{T}(\bar{z}))$ and, in addition, two copies of the fermionic supercurrent $(G(z), \widetilde{G}(\bar{z}))$. Their respective operator products read
\begin{subequations}\label{eq:algN1ope}
\begin{align}
    G(z)G(z') &\sim \frac{{c}}{(z-z')^3} + \frac{2T(z')}{z-z'} + \mathrm{reg.}~\\
    T(z)G(z') &\sim \frac{\frac{3}{2}G(z')}{(z-z')^2} + \frac{\partial G(z')}{z-z'}+ \mathrm{reg.}~\\
    T(z)T(z') &\sim \frac{\frac{3{c}}{4}}{(z-z')^4} + \frac{2T(z')}{(z-z')^2} + \frac{\partial T(z')}{z-z'} + \mathrm{reg.}
\end{align}   
\end{subequations}
and similarly for the right-moving ones. The algebra (\ref{eq:algN1ope}) has a $\mathbb{Z}_2$ automorphism, $G(z)\rightarrow -G(z)$, which captures the two possible moddings for the Laurent modes of the supercurrent $G(z)$, defining the usual NS (Neveu-Schwarz) and R (Ramond) algebras:
\begin{align}\label{eq:currents}
G(z)&=\sum_{r\in\mathbb{Z}+\nu}\frac{G_r}{z^{r+3/2}}~ , \quad G_r=\oint_0\frac{\d z}{2\pi i }z^{r+1/2}G(z) ~, ~ r\in\mathbb{Z}+\nu ~ , \\
T(z)&=\sum_{n\in\mathbb{Z}}\frac{L_n}{z^{n+2}} ~, \quad  \quad~~ L_n=\oint_0\frac{\d z}{2\pi i }z^{n+1}T(z) ~ , \quad  n~\in~\mathbb{Z} \ 
\end{align}
where $\nu =1/2$ in the NS- and $\nu=0$ in the R-sector. In terms of the modes $L_n, G_r$, the operator product expansions (\ref{eq:algN1ope}) are equivalent to the (anti-)commutation relations
\begin{align}\label{eq:N1algebra}
    \{G_k, G_l\} &= 2L_{k+l} + \frac{{c}}{2}\left(k^2 - \frac{1}{4}\right)\delta_{k+l}~,\cr
    [L_n,G_k] &= \left(\frac{n}{2} - k\right) G_{n+k}~,\cr
    [L_n,L_m] &= (n-m)L_{n+m} + \frac{{c}}{8}n(n^2-1)\delta_{m+n}~,
\end{align}
and similarly for the right-moving generators\footnote{We also implicitly choose the hermiticity conditions
$ G_k^\dagger = G_{-k}, L_n^\dagger = L_{-n}$.}. In addition, 
\begin{align}\label{eq:leftrightmodes}
\left\{G_k,\widetilde{G}_l\right\}&=0 \ , \nonumber \\
\left[L_n,\widetilde{G}_k\right]&=0=\left[\widetilde{L}_n,G_k\right] \ , \nonumber \\ 
\left[L_m,\widetilde{L}_n\right]&=0~.
\end{align}
We will usually refer to (\ref{eq:N1algebra}) as the chiral $\mathcal{N}=1$ \textit{super-Virasoro algebra} (either in the NS- or R-sector). 
The implications of this symmetry algebra in the quantum theory have been extensively analyzed for decades, and from many viewpoints; for a partial list of references see \cite{Bershadsky:1985dq, Friedan:1985ge,Nam:1985qe,Friedan:1984rv}. \par For the purposes of the present work, we will view $\mathcal{N}=1$ Liouville theory not as arising from the action (\ref{eq:N1action}), but rather as a concrete example of a `non-compact' two-dimensional quantum field theory whose correlation functions satisfy the 
$\mathcal{N}=1$ super-Virasoro Ward identities. To that end, the basic CFT data governing its correlation functions — namely, the theory's spectrum of primary operators and structure constants — can be determined in a systematic way as a function of the central charge using the methods of the analytic (super)conformal bootstrap together with some crucial assumptions of analyticity that we will make clear as we go along.

\subsection{Spectrum}
Spacelike $\mathcal{N}=1$ Liouville theory with $c\geq 9$ has been extensively studied (see \cite{Poghossian:1996agj,Rashkov:1996np,Belavin:2006pv,Belavin:2007gz, Belavin:2007eq,Fukuda:2002bv, Hadasz:2007wi, Chorazkiewicz:2008es, Suchanek:2010kq, Chorazkiewicz:2011zd, Hadasz:2008dt}) and, as we will review later on, many aspects of the theory have been well understood. It is a \textit{unitary} theory whose spectrum consists of a continuum of scalar NS- and R-sector primary operators which we will denote as $V_p^{\mathrm{NS}}$ and $\mathrm{V}_p^{\mathrm{R},\pm}$, labelled by the continuous momentum variable $p$. The $\pm$ superscript refers to the double degeneracy of the R-primaries, which we explain below. The corresponding conformal dimensions are given by the usual reflection-symmetric (i.e. $p\leftrightarrow-p$) parametrization
\begin{subequations}\label{eq: conformal dimensions}
\begin{align}\label{eq:hNS}
    V_p^{\mathrm{NS}}:~\quad &h^{\mathrm{NS}}_p = \widetilde{h}^{\mathrm{NS}}_p =  \frac{Q^2}{8} - \frac{p^2}{2} ~,\\ \label{eq:hR}
    V^{\mathrm{R},\pm}_p :\quad &h^{\mathrm{R}}_{p} = \widetilde{h}^{\mathrm{R}}_{p} = \frac{Q^2}{8} - \frac{p^2}{2}+\frac{1}{16}~.
    \end{align}
\end{subequations}
Physical (normalizable) operators have $p\in i\mathbb{R}_{\geq0}$, which means that the spectrum is bounded from below as $h^{\mathrm{NS}}\geq \frac{c-1}{16}$ and $h^{\mathrm{R}}\geq \frac{c}{16}$. 
\par Degenerate representations of the algebra (\ref{eq:N1algebra}) occur at the following discrete values of the momenta (modulo reflections)
\begin{equation}\label{eq: degenerate prs}
    p_{\langle r, s\rangle } = \frac{rb^{-1}+sb}{2}~,\quad r,s \in \mathbb{Z}_{\geq 1}~.
\end{equation}
The corresponding primary field $V_{p_{\langle r,s\rangle}}$ possesses a null vector at level $\frac{rs}{2}$ \cite{Friedan:1984rv, Nam:1985qe} associated with a null vector equation of the form
\begin{equation}
    D_{r,s}V_{p_{\langle r,s\rangle}}\equiv \bigg((G_{-1/2})^{rs}+\cdots\bigg)V_{p_{\langle r,s\rangle}}=0.
\end{equation}
When $r-s \in 2\mathbb{Z}$ it is a null vector over an NS field, whereas when $r-s \in 2\mathbb{Z}+1$ it is a null vector over a R-sector field. In particular, the identity operator corresponds to a NS degenerate field with momentum $p_{\mathds{1}}\equiv p_{\langle1,1\rangle}=\frac{Q}{2}$. The corresponding vacuum state $\ket{0}_{\text{NS}}$ is invariant under the global superconformal algebra $\mathfrak{osp}(1|2)$, i.e. it is annihilated by the five generators $L_0,L_{\pm1},G_{\pm\frac{1}{2}}$. 
\par
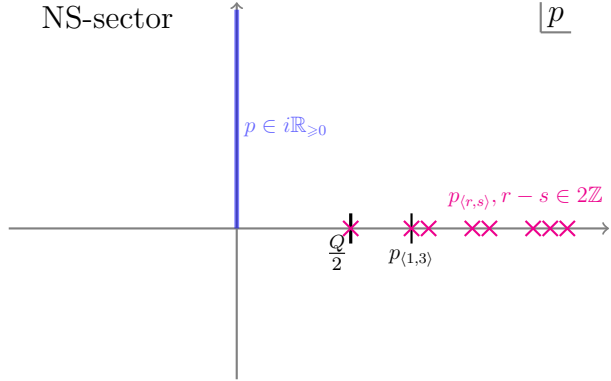
\begin{figure}[h]
\centering
\begin{minipage}{0.38\textwidth}
    \begin{tikzpicture}
        \draw[thick, gray, ->] (2,-2) -- (2,3);
        \draw[thick, gray, ->] (-1,0) -- (6.9,0); 
        \draw[very thick, blue,  line width=1.8, opacity = .5] (2,0) -- (2,2.9);
        \node[scale=.9,thick] at (3.32,-.32) {$\frac{Q}{2}$};   
        \node[scale=.7,thick] at (4.3,-.39) {${p_{\langle1,3\rangle}}$};  
        \node[scale=.7,thick, blue, opacity =.6] at (2.6,1.35) {~$p \in i\mathbb{R}_{\geq0}$}; 
        \node[scale=.7,thick, magenta] at (5.8,.4) {$p_{\langle r, s\rangle}, r-s\in2\mathbb{Z}$}; 
        \draw[very thick, black] (3.5,-.2) -- (3.5,.2);
        \draw[thick, black] (4.3,-.2) -- (4.3,.2);
        \node[cross out, thick, draw=magenta, scale=.79/1.2] at (2.475+1.025,0) {};    
        \node[cross out, thick, draw=magenta, scale=.79/1.2] at (2.475+1.825,0) {};   
        \node[cross out, thick, draw=magenta, scale=.79/1.2] at (2.475+2.05,0) {};      
        \node[cross out, thick, draw=magenta, scale=.79/1.2] at (2.475+2.625,0) {};   
        \node[cross out, thick, draw=magenta, scale=.79/1.2] at (2.475+2.85,0) {};      
        \node[cross out, thick, draw=magenta, scale=.79/1.2] at (2.475+3.425,0) {};      
        \node[cross out, thick, draw=magenta, scale=.79/1.2] at (2.475+3.65,0) {};   
        \node[cross out, thick, draw=magenta, scale=.79/1.2] at (2.475+3.875,0) {};  
        \draw[thick, gray] (6,2.6) -- (6,3);
        \draw[thick, gray] (6,2.6) -- (6.4,2.6);
        \node[scale=1.,thick] at (6.2,2.8) {$p$};  
        \node[scale=1.,thick] at (0.3,2.8) {NS-sector};  
    \end{tikzpicture}
\end{minipage}
\caption{\footnotesize{The physical NS-spectrum in the $p-$plane in spacelike $\mathcal{N}=1$ Liouville theory (blue) and the degenerate representations of the NS-sector algebra (magenta) for central charge values $c\geq 9$ (or $b\in\mathbb{R}_{(0,1]}$). The corresponding expressions are given in (\ref{eq:hNS}) and (\ref{eq: degenerate prs}). 
}}
\label{fig:NSspectra}
\end{figure}

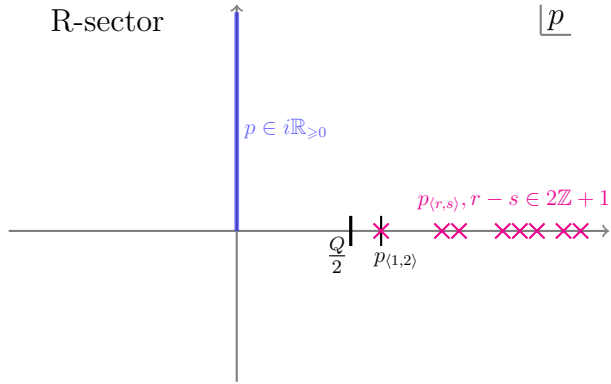
\begin{figure}[h]
\centering
\begin{minipage}{0.38\textwidth}
    \begin{tikzpicture}
        \draw[thick, gray, ->] (2,-2) -- (2,3);
        \draw[thick, gray, ->] (-1,0) -- (6.9,0); 
        \draw[very thick, blue,  line width=1.8, opacity = .5] (2,0) -- (2,2.9);
        \node[scale=.9,thick] at (3.32,-.32) {$\frac{Q}{2}$};   
        \node[scale=.7,thick] at (2.475+1.425+0.2,-.39) {${p_{\langle1,2\rangle}}$};  
        \node[scale=.7,thick, blue, opacity =.6] at (2.6,1.35) {~$p \in i\mathbb{R}_{\geq0}$}; 
        \node[scale=.7,thick, magenta] at (5.65,.4) {$p_{\langle r, s\rangle}, r-s\in2\mathbb{Z}+1$}; 
        \draw[very thick, black] (3.5,-.2) -- (3.5,.2);
        \draw[thick, black] (2.475+1.425,-.2) -- (2.475+1.425,.2);
        \node[cross out, thick, draw=magenta, scale=.79/1.2] at (2.475+1.425,0) {};    
        \node[cross out, thick, draw=magenta, scale=.79/1.2] at (2.475+2.225,0) {};   
        \node[cross out, thick, draw=magenta, scale=.79/1.2] at (2.475+2.45,0) {};      
        \node[cross out, thick, draw=magenta, scale=.79/1.2] at (2.475+3.025,0) {};   
        \node[cross out, thick, draw=magenta, scale=.79/1.2] at (2.475+3.25,0) {};      
        \node[cross out, thick, draw=magenta, scale=.79/1.2] at (2.475+3.475,0) {};      
        \node[cross out, thick, draw=magenta, scale=.79/1.2] at (2.475+3.825,0) {};   
        \node[cross out, thick, draw=magenta, scale=.79/1.2] at (2.475+4.05,0) {};  
        \draw[thick, gray] (6,2.6) -- (6,3);
        \draw[thick, gray] (6,2.6) -- (6.4,2.6);
        \node[scale=1.,thick] at (6.2,2.8) {$p$};  
        \node[scale=1.,thick] at (0.3,2.8) {R-sector}; 
    \end{tikzpicture}
\caption{\footnotesize{The physical R-spectrum in the $p-$plane in spacelike $\mathcal{N}=1$ Liouville theory (blue) and the degenerate representations of the R-sector algebra (magenta) for central charge values $c\geq 9$ (or $b\in\mathbb{R}_{(0,1]}$). The corresponding expressions are given in (\ref{eq:hR}) and (\ref{eq: degenerate prs}).}} 
\label{fig:Rspectra}
\end{minipage}
\end{figure}

 In the spacelike $\mathcal{N}=1$ Liouville theory the parameter $b$ takes values in $\mathbb{R}_{(0,1]}$,  and hence these degenerate values of the momenta are in principle outside the physical spectrum. However, as it is common in Liouville theory, these values can be reached via analytic continuation as we explain more below. 
The following low-lying null vectors are going to be of special importance for us (see e.g. \cite{Belavin:2006pv}):
\begin{itemize}
    \item R-sector, level 1. \begin{equation}\label{eq:nullvectorR}
    \left(L_{-1} - \frac{2b^2}{1+2b^2} G_{-1}G_0\right)V^{\mathrm{R},\pm}_{p}=0 \ , ~~~~ ~~ \ p=p_{\langle1,2\rangle}.
     \end{equation}
    \item NS-sector, level $\frac{3}{2}$.
    \begin{equation}\label{eq:nullvectorNS}
        \bigg(L_{-1}G_{-1/2} +b^{2}G_{-3/2}\bigg)V^{\mathrm{NS}}_p=0 \ , ~~~~ ~~~~~~ \ p=p_{\langle1,3\rangle}.
    \end{equation}
\end{itemize}
We summarize the various spectra in figures \ref{fig:NSspectra} and \ref{fig:Rspectra}.

We now proceed to a more detailed review of the operator content of the spacelike theory in the NS and R-sectors before moving on to discussing the two- and three-point functions in the following section.
\paragraph{NS-sector.}
The NS fields $V^{\mathrm{NS}}_{p}(z)$ are superconformal primary fields, namely they belong to the highest weight representation of the $\mathcal{N}=1$ super-Virasoro algebra (\ref{eq:N1algebra}) with half-integer modes of $G$. They satisfy
\begin{align}
    &L_0 V^{\mathrm{NS}}_p=\widetilde{L}_0V^{\mathrm{NS}}_p = h^{\mathrm{NS}}_pV^{\mathrm{NS}}_p \ , \nonumber \\
    &L_{n}V^{\mathrm{NS}}_p=0 \ , \ \ \widetilde{L}_{n}V^{\mathrm{NS}}_p=0 \ \ \text{for} \ n>0 \ , \nonumber \\
    &G_{k}V^{\mathrm{NS}}_p=0 \ , \ \ \widetilde{G}_{k}V^{\mathrm{NS}}_p=0 \ \ \text{for} \ k>0.
\end{align}
In terms of the Lagrangian (\ref{eq:N1action}) it is useful to think of these superconformal primaries as the properly normal ordered exponentials of the bosonic field $\phi$\footnote{This exponential expression can be given an exact sense in the region $\phi\rightarrow-\infty$ where the interaction terms can be dropped and the effective Lagrangian of the theory becomes that of a free boson and a free Majorana fermion.}
\begin{equation}
    V^{\mathrm{NS}}_p=:\e^{\left(Q/2-p\right)\phi}: \ .
\end{equation}
The rest of the operators in the NS-sector are superconformal descendants of these basic fields and can have either integer or half-integer level (at each holomorphic sector). Among those, it is important to distinguish three operators (which are primaries under the Virasoro subalgebra)
\begin{equation}\label{eq:supermultipletNS}
    \Lambda^{\mathrm{NS}}_p = G_{-1/2}V^{\mathrm{NS}}_p~,\quad \widetilde{\Lambda}^{\mathrm{NS}}_p = \widetilde{G}_{-1/2}V^{\mathrm{NS}}_p~,\quad  W^{\mathrm{NS}}_p = G_{-1/2}\widetilde{G}_{-1/2}V^{\mathrm{NS}}_p~.
\end{equation}
These fields are generated by the super Poincar\'e subalgebra of the full $\mathcal{N}=1$ symmetry algebra\footnote{In the terms of the fields and parameters of the Lagrangian (\ref{eq:N1action}), a common normalization for these operators in the literature reads for example (see e.g. \cite{Belavin:2007gz}) 
\begin{align}
    & W^{\mathrm{NS}}_p =\left\{G_{-1/2},\left[G_{-1/2},\phi\right]\right\} =  \mu \e^{(\alpha+b)\phi} -\frac{i}{2} \e^{\alpha \phi} \alpha\bar{\psi}\psi~,
\end{align}
where $\alpha = Q/2-p$.}. 
Some basic OPE of these fields with the stress tensor and the supercurrent read \cite{Fukuda:2002bv, Belavin:2007gz}
\begin{subequations}\label{eq: OPE NS}
\begin{align}
    T(z)V^{\mathrm{NS}}_{p}(0)&=\frac{h^{\mathrm{NS}}_p}{z^2}V^{\mathrm{NS}}_{p}(0)+\frac{1}{z}\partial V^{\mathrm{NS}}_{p}(0)+\text{reg}. \\
T(z)\Lambda^{\mathrm{NS}}_{p}(0)&=\frac{h^{\mathrm{NS}}_p+\frac{1}{2}}{z^2}\Lambda^{\mathrm{NS}}_{p}(0)+\frac{1}{z}\partial \Lambda^{\mathrm{NS}}_{p}(0)+\text{reg}. \\
G(z)V^{\mathrm{NS}}_{p}(0)&=\frac{1}{z}\Lambda^{\mathrm{NS}}_{p}(0)+\text{reg}. \\
G(z)\Lambda^{\mathrm{NS}}_{p}(0)&=\frac{2h^{\mathrm{NS}}_p}{z^2}V^{\mathrm{NS}}_{p}(0)+\frac{1}{z}\partial V^{\mathrm{NS}}_{p}(0)+\text{reg}.
\end{align}
\end{subequations}

\paragraph{R-sector.} 
Highest weight representations of the Ramond algebra are captured by the primary fields $V_p^{\mathrm{R},\pm}$. The $\pm$ superscript indicates the double degeneracy of these highest weight representations, which stems from the presence of zero modes $G_0$, $\widetilde{G}_0$. 
\par Let us explain briefly the origin of this degeneracy. From (\ref{eq:N1algebra}) we infer
\begin{equation}\label{eq:commutations R sector ground state}
    G_0^2=\widetilde{G
    }_0^2=L_0-\frac{c}{16} \ , \quad [L_0,G_0] = [\widetilde{L}_0,\widetilde{G}_0]=0~.
\end{equation}
This implies that, at each holomorphic sector, there are two degenerate highest weight states w.r.t $L_0$($\widetilde{L}_0$), ending up seemingly in a total four-fold degeneracy for the combined sectors. Following the convention of \cite{Fukuda:2002bv}, let us denote the corresponding four primary operators (acting on the combined holomorphic + anti-holomorphic Hilbert space) as $\Theta^{\pm\pm}$ and $\Theta^{\pm\mp}$. There is, however, still an important relation that we have not yet taken into account—namely, the fact that 
\begin{equation}\label{eq:Cartan2}
    \left\{G_0,\tilde{G}_0\right\}=0~.
\end{equation}
This will eventually bring the degeneracy (for the combined Hilbert space) down to two, giving us the two desired Ramond states which, following \cite{Hadasz:2008dt}, we choose as 
\begin{equation}\label{eq:Ramondfields}
    V_p^{\mathrm{R},+} = \frac{1}{\sqrt{2}} (\Theta_p^{++} - i \Theta_p^{--})~,\quad V_p^{\mathrm{R},-} = \frac{1}{\sqrt{2}} (\Theta^{+-}_p + \Theta^{-+}_p)~.
\end{equation}
A nice way to talk about the \textit{combined} algebra (\ref{eq:commutations R sector ground state}), (\ref{eq:Cartan2}) is by introducing the linear combinations
\begin{equation}\label{eq: linear combination G0}
    \mathcal{G}_0 \equiv  G_0 + i \widetilde{G}_0~,\quad \widetilde{\mathcal{G}}_0 \equiv  G_0 - i \widetilde{G}_0~,
\end{equation}
such that $\mathcal{G}_0^\dagger = \widetilde{\mathcal{G}}_0$. When acting on the Liouville physical states these operators obey the standard fermionic harmonic oscillator anticommutation relations
\begin{equation}\label{eq: fermionic harmonic oscillator}
    \{\mathcal{G}_0 ,\mathcal{G}_0\} = \{\widetilde{\mathcal{G}}_0 ,\widetilde{\mathcal{G}}_0\} = 0~,\quad \{\mathcal{G}_0, \widetilde{\mathcal{G}}_0\} = 4h_p^{\mathrm{R}} - \frac{c}{4}~.
\end{equation} 
It is now straightforward to see that there are two-dimensional irreducible representations of (\ref{eq: fermionic harmonic oscillator}), which are realized by the highest weight vectors $V_p^{\mathrm{R},\pm}$ in terms of $G_0,\widetilde{G}_0$ as
\begin{align}
    &G_0\left(\begin{array}{cc}
        V_p^{\mathrm{R},+}  \\
        V_p^{\mathrm{R},-} 
    \end{array}\right)=\begin{pmatrix}
0 & c_+p \\
c_-p & o
\end{pmatrix}\left(\begin{array}{cc}
       V_p^{\mathrm{R},+}  \\
        V_p^{\mathrm{R},-} 
    \end{array}\right) \ , \nonumber \\
    \nonumber \\
    &\widetilde{G}_0\left(\begin{array}{cc}
        V_p^{\mathrm{R},+}  \\
        V_p^{\mathrm{R},-} 
    \end{array}\right)=\begin{pmatrix}
0 & \tilde{c}_+p \\
\tilde{c}_-p & o
\end{pmatrix}\left(\begin{array}{cc}
       V_p^{\mathrm{R},+}  \\
        V_p^{\mathrm{R},-} 
    \end{array}\right).
\end{align}
The constants $c_+,c_-,\tilde{c}_+,\tilde{c}_-$ are only specified up to the requirement that 
\begin{equation}
   p^2( c_+\times c_- )= p^2(\tilde{c}_+\times \tilde{c}_-)= h^{\mathrm{R}}_p-\frac{c}{16}=-\frac{p^2}{2}~.
\end{equation}
Here we will choose $c_{\pm}=\frac{i\e^{\mp\frac{i\pi}{4}}}{\sqrt{2}}$ and $\tilde{c}_{\pm}=c_{\pm}^*$ (where star denotes complex conjugation).
Using (\ref{eq:commutations R sector ground state}) one can easily infer that $V_p^{\mathrm{R},\pm}$ have the same $L_0$ eigenvalue 
given by (\ref{eq:hR}).
Furthermore, they are annihilated by $L_n$ and $\widetilde{L}_n$ for $n> 0$ as well as $G_k$ and $\widetilde{G}_k$, for $k>0$.\footnote{In terms of the linear combinations $\mathcal{G}_0$ and $\widetilde{\mathcal{G}}_0$ the above relations translate to \cite{Klebanov:2003wg, Seiberg:2003nm}
\begin{subequations}\label{eq:mathcalG0 on VR}
    \begin{align}
        \mathcal{G}_0 V_p^{\mathrm{R},+} &= \e^{-\frac{i\pi}{4}}\sqrt{4h_p^{\mathrm{R}}-\frac{c}{4}}V_p^{\mathrm{R},-}~,\quad \mathcal{G}_0 V_p^{\mathrm{R},-} =0~,\\ 
        \widetilde{\mathcal{G}}_0 V_p^{\mathrm{R},-} &= \e^{\frac{i\pi}{4}}\sqrt{4h_p^{\mathrm{R}}-\frac{c}{4}}V_p^{\mathrm{R},+}~,\quad \ \ \widetilde{\mathcal{G}}_0 V_p^{\mathrm{R},+} =0.
    \end{align}
\end{subequations}
} 
\par In terms of the free field language, it is useful to think of these Ramond primaries as
\begin{equation}
    V_p^{\mathrm{R},\pm} = \sigma^{\pm} :\e^{(\frac{Q}{2}-p)\phi}:
\end{equation}
where $\sigma^{\pm}$ are the standard order and disorder spin fields of conformal dimension $1/16$ with respect to the free fermion (see e.g. \cite{Ginsparg:1988ui}).
\par Finally, the OPE of $V_p^{\mathrm{R},\pm}$ with the stress tensor and the supercurrent read \cite{Hadasz:2008dt} 
\begin{align}
    T(z) V_p^{\mathrm{R},\pm}(0) &=\frac{h_p^{\mathrm{R}}}{z^2} V_p^{\mathrm{R},\pm}(0) +\frac{1}{z}\partial V_p^{\mathrm{R},\pm}(0) +\mathrm{reg}~,\\
    G(z)V_p^{\mathrm{R},\pm} (0) &= \frac{i}{z^{3/2}}\,\frac{p}{\sqrt{2}}\e^{\mp\frac{i\pi}{4}} V_p^{\mathrm{R},\mp}(0) + \frac{1}{z^{1/2}}G_{-1}V_p^{\mathrm{R},\pm} (0)+ \mathrm{reg}~.
\end{align}

\subsection{Three-point functions}\label{sec:three-ptspacelike}
In this section we review the derivation of the structure constants for $\mathcal{N}=1$ spacelike Liouville theory. The results of this section are not new but, as we will see, we will adopt a slightly different normalization than the one considered so far in the literature. Also the expressions for the spacelike structure constants are going to be important later on in the construction of the timelike structure constants for $\hat{c}\leq 1$. We keep the technical calculations to a minimum in this section and refer to appendix \ref{app:NS} for more details (see also \cite{Belavin:2007gz}).

\subsubsection{NS-sector} The OPE between two NS fields closes upon itself, i.e. it takes the general form
\begin{equation}
    [\text{NS}] [\text{NS}]\sim  [\text{NS}]~. 
\end{equation}
Therefore we can treat this sector completely separately. In Liouville theory we have a continuous OPE that involves integration over the momentum variable $p$\footnote{The OPE contour is along the imaginary axis when $p_1,p_2\in i\mathbb{R}_{+}$, however it should be deformed appropriately for general complex values of $p_1,p_2$ due to poles coming from the structure constants that could cross the vertical contour.}. Crucially, in the case of $\mathcal{N}=1$ superconformal symmetry the contributions of integer and half-integer level descendants enter independently, therefore leading to two independent structure constants as follows
\begin{multline}
    V^{\mathrm{NS}}_{p_1}(z)V^{\mathrm{NS}}_{p_2}(0)\sim \int_{i\mathbb{R}_+}\!\! \frac{\d p}{i} \frac{(z\bar{z})^{h_p-h_{p_1}-h_{p_2}}}{B_{\text{NS}}^{(b)}(p)} \Big(C_{\mathrm{NS}}^{(b)}(p_1,p_2,p)\left[V^{\mathrm{NS}}_p(0)\right]_{\mathrm{ee}}\cr
    -\widetilde{C}_{\mathrm{NS}}^{(b)}(p_1,p_2,p)\left[V^{\mathrm{NS}}_p(0)\right]_{\mathrm{oo}}\Big)~.\label{eq:NSOPE}
\end{multline}
Following \cite{Belavin:2007gz}, we have normalized the chain operators as
\begin{align}
    \left[V^{\mathrm{NS}}_p(0)\right]_{\mathrm{ee}}&=\mathcal{C}_e^{p_1,p_2}(p,z)\overline{\mathcal{C}}_e^{p_1,p_2}(p,\overline{z})V^{\mathrm{NS}}_p(0), \nonumber \\
    \left[V^{\mathrm{NS}}_p(0)\right]_{\mathrm{oo}}&=\mathcal{C}_o^{p_1,p_2}(p,z)\overline{\mathcal{C}}_o^{p_1,p_2}(p,\overline{z})V^{\mathrm{NS}}_p(0),\label{eq:chains0}
\end{align}
where
\begin{align}
    \mathcal{C}_{\mathrm{e}}^{p_1,p_2}(p,z)&=1+z\frac{h_p+h_1-h_2}{2h_p}L_{-1}+O(z^2) \ , \nonumber \\
    \mathcal{C}_{\mathrm{o}}^{p_1,p_2}(p,z)&=\frac{z^{1/2}}{2h_p}G_{-1/2}+O(z^{3/2}) \label{eq:chains} \ ,
\end{align}
and similarly for the anti-holomorphic factors $\overline{\mathcal{C}}_{\mathrm{e}},\overline{\mathcal{C}}_{\mathrm{o}}$ which include the corresponding anti-holomorphic modes $\widetilde{L}_k,\widetilde{G}_k$. 
The subscripts $\mathrm{e}$ and $\mathrm{o}$ stand for `even' and `odd' and refer to the integer and half-integer level descendant contributions respectively. These factors are completely fixed at each order in $z(\bar{z})$ by superconformal symmetry. Note also that the minus sign of the second term in the OPE originates from the anticommutativity of $G_k$ and $\widetilde{G}_k$ modes leading, in turn, to the anticommutativity of the two odd chain operators. 
\par The CFT data $\left\{B_{\text{NS}}^{(b)},C_{\mathrm{NS}}^{(b)},\widetilde{C}_{\mathrm{NS}}^{(b)}\right\}$
comprises the basic NS two- and three-point functions of the theory\footnote{The corresponding NS \textit{structure constants} $\mathbb{C}_{p_1p_2}^p,\tilde{\mathbb{C}}_{p_1p_2}^p$ are related with the two- and three-point functions as $\mathbb{C}_{p_1p_2}^p=C_{\mathrm{NS}}^{(b)}(p_1,p_2,p)/B_{\text{NS}}^{(b)}(p)$ and $\widetilde{\mathbb{C}}_{p_1p_2}^p=-\widetilde{C}_{\mathrm{NS}}^{(b)}(p_1,p_2,p)/B_{\text{NS}}^{(b)}(p)$.}. They are defined as
\begin{subequations}\label{eq: spacelikeN1 structure constantsNS}
\begin{align}
 \langle V^{\mathrm{NS}}_{p_1}(0)V^{\mathrm{NS}}_{p_2}(1)\rangle &= B^{(b)}_{\text{NS}}(p_1)[\delta(p_1-p_2) + \delta(p_1+p_2)]~,\\
    \langle V^{\mathrm{NS}}_{p_1}(z_1)V^{\mathrm{NS}}_{p_2}(z_2)V^{\mathrm{NS}}_{p_3}(z_3)\rangle &= \frac{C_{\mathrm{NS}}^{(b)}(p_1,p_2,p_3)}{|z_{12}|^{2h_{1+2-3}}|z_{23}|^{2h_{2+3-1}}|z_{31}|^{2h_{3+1-2}}}~,\\
    \langle W^{\mathrm{NS}}_{p_1}(z_1)V^{\mathrm{NS}}_{p_2}(z_2)V^{\mathrm{NS}}_{p_3}(z_3)\rangle &= \frac{\widetilde{C}_{\mathrm{NS}}^{(b)}(p_1,p_2,p_3)}{|z_{12}|^{2h_{1+2-3}+1}|z_{23}|^{2h_{2+3-1}-1}|z_{31}|^{2h_{3+1-2}+1}}~,
\end{align}    
\end{subequations}
where $h_{1+2+3} \equiv h_{p_1}^{\mathrm{NS}} +h_{p_2}^{\mathrm{NS}} +h_{p_3}^{\mathrm{NS}}$ and similarly for the other combinations.
All other three-point functions involving the components (\ref{eq:supermultipletNS}) can be written in terms of the basic $C^{(b)}_{\mathrm{NS}},\widetilde{C}^{(b)}_{\mathrm{NS}}$ via superprojective Ward identities \cite{Belavin:2007gz}.
\par
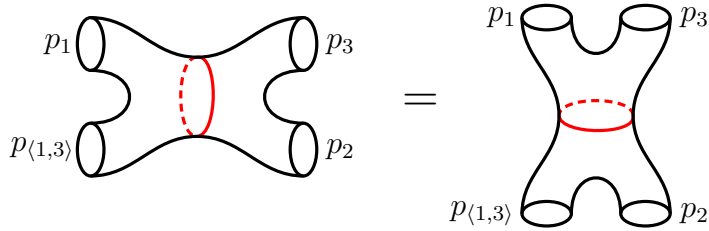
\begin{figure}[H]
\centering 
\begin{tikzpicture}
\begin{scope}[shift={(1.1,0)}, scale=.7]
\draw[very thick, red, looseness=.7] (2,1.75) to[out=10,in=5] (2,.25);
\draw[very thick, densely dashed, red, looseness=.7] (2,1.75) to[out=-180,in=-180] (2,.25);    
\draw[very thick] (0,0) ellipse (0.25 and .5);
\draw[very thick] (0,2) ellipse (0.25 and .5);
\draw[very thick] (4,0) ellipse (0.25 and .5);
\draw[very thick] (4,2) ellipse (0.25 and .5);
\node[scale=1., thick] at (-0.9,0) {$p_{\langle1,3\rangle}$};
\node[scale=1., thick] at (-0.6,2) {$p_1$};  
\node[scale=1., thick] at (4.7,2) {$p_3$};
\node[scale=1., thick] at (4.7,0) {$p_2$};
\draw[very thick] (0,.5) to[out=0, in=0, looseness=2.5] (0,1.5);
\draw[very thick] (4,.5) to[out=180, in=180, looseness=2.5] (4,1.5);
\draw[very thick] (0,2.5) to[out=0, in=180] (2,1.75) to[out=0, in=180] (4,2.5);
\draw[very thick] (0,-.5) to[out=0, in=180] (2,.25) to[out=0, in=180] (4,-.5); 
\node[scale=1.5] at (6.2,.9) {$=$};
\end{scope}
\begin{scope}[shift={(7.1,1.75)}, scale=.65, rotate=270]
\draw[very thick, red, looseness=.7] (2,1.75) to[out=10,in=5] (2,.25);
\draw[very thick, densely dashed, red, looseness=.7] (2,1.75) to[out=-180,in=-180] (2,.25);    
\draw[very thick] (0,0) ellipse (0.25 and .5);
\draw[very thick] (0,2) ellipse (0.25 and .5);
\draw[very thick] (4,0) ellipse (0.25 and .5);
\draw[very thick] (4,2) ellipse (0.25 and .5);
\node[scale=1., thick] at (0,-0.9) {$p_1$};
\node[scale=1., thick] at (0,3) {$p_3$};  
\node[scale=1., thick] at (4,3) {$p_2$};
\node[scale=1., thick] at (4,-1.3) {$p_{\langle1,3\rangle}$};
\draw[very thick] (0,.5) to[out=0, in=0, looseness=2.5] (0,1.5);
\draw[very thick] (4,.5) to[out=180, in=180, looseness=2.5] (4,1.5);
\draw[very thick] (0,2.5) to[out=0, in=180] (2,1.75) to[out=0, in=180] (4,2.5);
\draw[very thick] (0,-.5) to[out=0, in=180] (2,.25) to[out=0, in=180] (4,-.5);    
\end{scope}
\end{tikzpicture}
\caption{\footnotesize{Teschner's trick in the NS-sector of $\mathcal{N}=1$ Liouville theory: the analytic bootstrap problem involving crossing of the sphere four-point function between a NS degenerate field $V_{p_{\langle1,3\rangle}}$ and three general NS fields $V_{p_1},V_{p_2},V_{p_3}$. The analysis (done in App.\ref{app:NS}) leads to the shift relations (\ref{eq: shift equations NS sector1}), (\ref{eq: shift equations NS sector1binv}). 
}
}
\label{fig:N1NSbootstrap}
\end{figure}

 The explicit expressions for the two- and three-point functions as a function of the momenta and the central charge were computed a while ago in \cite{Poghossian:1996agj, Rashkov:1996np} (see also \cite{Belavin:2007gz, Belavin:2007eq}) through various methods that include the free field method \cite{Dotsenko:1984nm} or (a supersymmetric extension of) Teschner's trick which mixes both the NS- and R-sectors. Here we will essentially follow \cite{Belavin:2007gz} but, instead of the free field method, we will use a suitbale version of Teschner's trick adapted only for the NS-sector (without mixing the Ramond sector). More specifically, we will be studying the \textit{shift relations} of the normalization-independent bootstrap data on the sphere that arise from analyzing the crossing equation of a particular four-point function depicted in figure \ref{fig:N1NSbootstrap}, which we now explain.
\noindent
\paragraph{Shift relations.}
As we show explicitly in Appendix \ref{app:NS}, using the null vector equation (\ref{eq:nullvectorNS}) we can obtain a third order differential equation for the four point function of three physical operators and the degenerate field $V^{\mathrm{NS}}_{p_{\langle 1,3\rangle}}$,
\begin{equation}
    \langle V^{\mathrm{NS}}_{p_{\langle 1,3\rangle}}(z_0) V_{p_1}^{\mathrm{NS}}(z_1) V_{p_2}^{\mathrm{NS}}(z_2) V_{p_3}^{\mathrm{NS}}(z_3)\rangle~.
\end{equation}
Imposing crossing symmetry to the solutions of this differential equation naturally
leads to two independent sets of constraint equations for the two-and three-point functions in (\ref{eq: spacelikeN1 structure constantsNS}). In appendix \ref{app:NS} we provide an extensive analysis of this problem. When the dust settles, we obtain the following set of shift relations
\begin{align}\label{eq: shift equations NS sector1}
    &\frac{C_{\text{NS}}^{(b)}(p_1+ b,p_2,p_3)^2/B^{(b)}_{\text{NS}}(p_1+b)}{C_{\text{NS}}^{(b)}(p_1- b,p_2,p_3)^2/B^{(b)}_{\text{NS}}(p_1-b)}= \kappa_{\mathrm{NS}}^{(b)}(p_1|p_2,p_3)~, \nonumber \\
    &\frac{\widetilde{C}_{\text{NS}}^{(b)}(p_1,p_2,p_3)^2/B^{(b)}_{\text{NS}}(p_1)}{C_{\text{NS}}^{(b)}(p_1- b,p_2,p_3)^2/B^{(b)}_{\text{NS}}(p_1-b)}= \lambda_{\mathrm{NS}}^{(b)}(p_1|p_2,p_3)~.
\end{align}
The explicit expressions for the functions $\kappa_{\mathrm{NS}}^{(b)}$ and $\lambda_{\mathrm{NS}}^{(b)}$ are given in (\ref{eq:kappaNS}), (\ref{eq:lambdaNS}). They are meromorphic functions of the momenta $p_1,p_2,p_3$ as well as of the parameter $b$. We will return to the properties of these `bootstrap functions' shortly. 
\par For $b^2\notin \mathbb{Q}$, the same analysis leads also to another set of incommensurable shift relations with $b\leftrightarrow b^{-1}$, namely
\begin{align}\label{eq: shift equations NS sector1binv}
    &\frac{C_{\text{NS}}^{(b)}(p_1+ b^{-1},p_2,p_3)^2/B^{(b)}_{\text{NS}}(p_1+b^{-1})}{C_{\text{NS}}^{(b)}(p_1- b^{-1},p_2,p_3)^2/B^{(b)}_{\text{NS}}(p_1-b^{-1})}= \kappa_{\mathrm{NS}}^{(b^{-1})}(p_1|p_2,p_3)~, \nonumber \\
    &\frac{\widetilde{C}_{\text{NS}}^{(b)}(p_1,p_2,p_3)^2/B^{(b)}_{\text{NS}}(p_1)}{C_{\text{NS}}^{(b)}(p_1- b^{-1},p_2,p_3)^2/B^{(b)}_{\text{NS}}(p_1-b^{-1})}= \lambda_{\mathrm{NS}}^{(b^{-1})}(p_1|p_2,p_3)~.
\end{align}
 A crucial, but natural assumption that goes into (\ref{eq: shift equations NS sector1binv}) is the invariance of the two- and three-point functions under $b\leftrightarrow b^{-1}$, or in other words their dependency solely on the central charge. Furthermore, since the four-point function that we started with (c.f. fig. \ref{fig:N1NSbootstrap}) does not distinguish between the operators $V^{\mathrm{NS}}_{p_1},V^{\mathrm{NS}}_{p_2},V^{\mathrm{NS}}_{p_3}$, we are naturally seeking a solution for $C^{(b)}_{\text{NS}}(p_1,p_2,p_3),\widetilde{C}^{(b)}_{\text{NS}}(p_1,p_2,p_3)$ that is invariant under permutations of the three momenta.
\par The fact that the functions $\kappa_{\mathrm{NS}}^{(b)}$ and $\lambda_{\mathrm{NS}}^{(b)}$ are meromorphic in the parameter $b$ suggests that the shift relations can be analytically continued without worry to any central charge regime. It is precisely this feature that we leverage later in section \ref{sec:Timelike} to determine the timelike $\hat{c} \leq 1$ structure constants. As functions of the momenta, $\kappa^{(b)}_{\text{NS}},\lambda^{(b)}_{\text{NS}}$ are symmetric under the exchange of $p_2,p_3$,
\begin{equation}
    \kappa_{\mathrm{NS}}^{(b)}(p_1|p_3,p_2)=\kappa_{\mathrm{NS}}^{(b)}(p_1|p_2,p_3) \ , \ \ \lambda_{\mathrm{NS}}^{(b)}(p_1|p_3,p_2)=\lambda_{\mathrm{NS}}^{(b)}(p_1|p_2,p_3)~,
\end{equation}
and under reflections,
\begin{align}
    \kappa_{\mathrm{NS}}^{(-b)}(-p_1|p_2,p_3)=\kappa_{\mathrm{NS}}^{(b)}(p_1|-p_2,-p_3)=\kappa_{\mathrm{NS}}^{(b)}(p_1|p_2,p_3) \ , \nonumber \\
    \lambda_{\mathrm{NS}}^{(-b)}(-p_1|p_2,p_3)=\lambda_{\mathrm{NS}}^{(b)}(p_1|-p_2,-p_3)=\lambda_{\mathrm{NS}}^{(b)}(p_1|p_2,p_3) ~.
\end{align}
\par As we explain in detail in appendix \ref{app:NS}, it is crucial that there are two independent shift relations given by (\ref{eq: shift equations NS sector1}) (or (\ref{eq: shift equations NS sector1binv}) for shifts in $b^{-1}$) that couple the two independent three-point functions $C_{\mathrm{NS}}^{(b)},\widetilde{C}_{\mathrm{NS}}^{(b)}$. All other shift relations that arise from the same bootstrap problem of interest will eventually boil down to these two shift relations. For example, instead of (\ref{eq: shift equations NS sector1}), we could have equivalently derived the following system of independent shift relations
\begin{align}\label{eq: shift equations NS sector2}
\frac{\widetilde{C}^{(b)}_{\text{NS}}(p_1+b,p_2,p_3)^2/B^{(b)}_{\text{NS}}(p_1+b)}{\tilde{C}^{(b)}_{\text{NS}}(p_1-b,p_2,p_3)^2/B^{(b)}_{\text{NS}}(p_1-b)}&\equiv\widetilde{\kappa}_{\mathrm{NS}}^{(b)}(p_1|p_2,p_3) \ \nonumber , \\ 
\frac{C^{(b)}_{\text{NS}}(p_1,p_2,p_3)^2/B^{(b)}_{\text{NS}}(p_1)}{\widetilde{C}^{(b)}_{\text{NS}}(p_1-b,p_2,p_3)^2/B^{(b)}_{\text{NS}}(p_1-b)}&\equiv\widetilde{\lambda}_{\mathrm{NS}}^{(b)}(p_1|p_2,p_3)~,
\end{align}
together with their counterparts with $b\leftrightarrow b^{-1}$. It is straightforward to see that the functions $\widetilde{\kappa}_{\mathrm{NS}}^{(b)}, \widetilde{\lambda}_{\mathrm{NS}}^{(b)}$ are again analytic functions of $p_i$'s and $b$, and simply given in terms of $\kappa_{\mathrm{NS}}^{(b)},  \lambda_{\mathrm{NS}}^{(b)}$ as follows
\begin{align}\label{eq:tildekappatildelambda}
\widetilde{\kappa}_{\mathrm{NS}}^{(b)}(p_1|p_2,p_3)&=\frac{\kappa_{\mathrm{NS}}^{(b)}(p_1-b|p_2,p_3)\lambda^{(\mathrm{NS})}_{b}(p_1+b|p_2,p_3)}{\lambda^{(\mathrm{NS})}_{b}(p_1-b|p_2,p_3)}, \nonumber \\ \widetilde{\lambda}_{\mathrm{NS}}^{(b)}(p_1|p_2,p_3)&=\frac{\kappa_{\mathrm{NS}}^{(b)}(p_1-b|p_2,p_3)}{\lambda_{\mathrm{NS}}^{(b)}(p_1-b|p_2,p_3)}~.
\end{align}
It may seem redundant to mention these shift relations here. However, as we will see in section \ref{sec:VWR}, there is a non-trivial relation between the function $\widetilde{\kappa}_{\mathrm{NS}}^{(b)}$ evaluated at central charge $c\geq9$ and the function $\kappa_{\mathrm{NS}}^{(b)}$ evaluated at central charge $\hat{c}\leq 1$ (or vice versa, in terms of the central charge). A similar relation holds between $\lambda_{\mathrm{NS}}^{(b)}$ and $\widetilde{\lambda}_{\mathrm{NS}}^{(b)}$ as well. 
It will be exactly this interesting `coupling' between these functions that will play a crucial role in the derivation of the timelike structure constants as we will discuss in section \ref{sbsbsec:tNS}.

\paragraph{Natural normalization.} 
As we emphasized above, the shift relations (\ref{eq: shift equations NS sector1}) are constraint relations for the normalization-independent bootstrap data on the sphere, and hence they are true for any consistent choice of the two-point function and three-point functions. For example, one can explicitly check that the expressions derived in the original papers \cite{Poghossian:1996agj, Rashkov:1996np} for the NS-sector satisfy (\ref{eq: shift equations NS sector1}). Those expressions -- just like the familiar DOZZ formula in ordinary Liouville \cite{Dorn:1994xn, Zamolodchikov:1995aa, Teschner:1995yf} -- depend explicitly on the cosmological constant of the action (\ref{eq:N1action}). Here we will find it instructive to make a different choice for the two-point function, namely:
\begin{equation}\label{eq: BNS}
    B_{\mathrm{NS}}^{(b)}(p) \equiv \left(\rho^{(b)}_{\mathrm{NS}}(p)\right)^{-1} = -\frac{1}{4 \sin(\pi b p)\sin(\pi b^{-1}p)}~.
\end{equation}
The motivation for this choice comes from a similar convenient choice in usual bosonic Liouville (as e.g. in \cite{Collier:2019weq,Collier:2023cyw,Eberhardt:2023mrq}) and is closely related to the modular transformation of NS characters on a torus with modulus $\tau$. For $c>1$ non-degenerate representations of the NS algebra, there are two distinct torus characters  given by\footnote{The $\pm$ indicates the absence or presence of $(-1)^F$ in the definition of the trace.} (see for example \cite{Bae:2018qym,Benjamin:2020zbs})
\begin{align}
 \chi^{\text{NS},+}_p(\tau) := q^{h_p-\frac{c-1}{16}}\e^{-\frac{\pi i}{24}}\frac{\eta(\frac{\tau+1}{2})}{\eta(\tau)^2}~,\quad 
    \chi^{\text{NS},-}_p(\tau) :=  q^{h_p-\frac{c-1}{16}}\frac{\eta(\frac{\tau}{2})}{\eta(\tau)^2}~, ~~~~~~~ q\equiv \e^{2\pi i \tau},
\end{align}
where $\eta(\tau)$ is the Dedekind function. There are two corresponding vacuum characters given by
\begin{align}
\label{eq:idNScharacters}\chi_{\mathds{1}}^{\text{NS},\pm }(\tau) = \chi^{\text{NS},\pm}_{p=Q/2}(\tau) \mp \chi^{\text{NS},\pm}_{p = (b^{-1}-b)/2}(\tau) ~.
\end{align}
Under a modular S transform $\tau\rightarrow-1/\tau$, it is straightforward to verify that the characters $\chi_{p}^{\text{NS},+},\chi_\mathds{1}^{\text{NS},+}$ transform as\footnote{The other two characters $\chi_{p}^{\text{NS},-},\chi_\mathds{1}^{\text{NS},-}$ branch into Ramond torus characters under modular transformations, so they won't be relevant for us at this stage. We will mention them later on when we discuss the Ramond sector.}
\begin{align}
     \chi_{p}^{\text{NS},+}\left(-\frac{1}{\tau}\right) &= \int_{i\mathbb{R}_{+}} \frac{\d p'}{i}\, \mathbb{S}_{pp'}^{(\text{NS})} \ \chi^{\text{NS},+}_{p'}(\tau) \ , \quad ~~~~~ \mathbb{S}_{pp'}^{(\text{NS})}\equiv 2\cos(2\pi p' p) \ , \nonumber \\
     \chi_\mathds{1}^{\text{NS},+}\left(-\frac{1}{\tau}\right) &= \int_{i\mathbb{R}_{+}} \frac{\d p}{i}\, \mathbb{S}_{p\mathds{1}}^{(\text{NS})} \ \chi_p^{\text{NS},+}(\tau)\ , \quad ~~~~~~\mathbb{S}_{p\mathds{1}}^{(\text{NS})}\equiv\rho^{(b)}_{\mathrm{NS}}(p)
     ~.
\end{align}
The identity modular kernel $\rho^{(b)}_{\mathrm{NS}}$ is exactly the inverse of our choice (\ref{eq: BNS}). It is the natural analog of the identity modular kernel in the usual Virasoro case, which has been used as a convenient normalization for the two-point function in ordinary Liouville theory \cite{Collier:2019weq,Collier:2023cyw,Eberhardt:2023mrq}. With this choice the vertex operators are identified according to
\begin{equation}
    V^{\mathrm{NS}}_{p}=V^{\mathrm{NS}}_{-p}~.
\end{equation}
In other words, there is no reflection coefficient that relates vertex operators with reflected momenta. Furthermore, $\rho^{(b)}_{\mathrm{NS}}(p)$ can be identified with the Plancherel measure in a particular series of representations of the quantum group $\mathcal{U}_q\left(osp(1|2)\right)$ studied in \cite{Hadasz:2013bwa,Pawelkiewicz:2013wga,Fan:2021bwt}. This, again, mimics the situation in the Virasoro case where the identity modular kernel plays a similar role for the quantum group $\mathcal{U}_q\left(sl(2)\right)$\cite{Ponsot:2000mt}.

Going back to the shift relations (\ref{eq: shift equations NS sector1}), the choice (\ref{eq: BNS}) leads to an essential simplification. It is straightforward to check that the following ratios form a complete square
\begin{align}\label{eq: shift equations NS sector1v2}
    & \frac{\rho^{(b)}_{\mathrm{NS}}(p_1-b)}{\rho^{(b)}_{\mathrm{NS}}(p_1+b)}\kappa_{\mathrm{NS}}^{(b)}(p_1|p_2,p_3) = \left(~\cdot~\right)^2~,  \ \ \  \frac{\rho^{(b)}_{\mathrm{NS}}(p_1-b)}{\rho^{(b)}_{\mathrm{NS}}(p_1)} \lambda_{\mathrm{NS}}^{(b)}(p_1|p_2,p_3) = \left(~\cdot~\right)^2~,
\end{align}
and thereby leading, via (\ref{eq: shift equations NS sector1}), to the following relations for the three-point functions:
\begin{align}\label{eq:1stshift}
    &\frac{C_{\text{NS}}^{(b)}(p_1+ b,p_2,p_3)}{C_{\text{NS}}^{(b)}(p_1- b,p_2,p_3)}=\frac{\Gamma (1+ b p_1)\Gamma\left(1+b^2+bp_1\right)}{\Gamma (1- b p_1)\Gamma\left(1+b^2-bp_1\right)} \frac{\gamma \left(\frac{1-b^2}{2}+ b p_1\right)\gamma \left(\frac{1}{4} \left(1-b^2-2  b p_{1-2-3}\right)\right)}{\gamma \left(\frac{1-b^2}{2}- b p_1\right)\gamma \left(\frac{1}{4} (1-b^2+2  b p_{1-2-3})\right)} \  \nonumber \\
    & ~~~~~~~~~~~~~\times \frac{ \gamma \left(\frac{1}{4} \left(1-b^2-2  b p_{1+2-3}\right)\right) \gamma \left(\frac{1}{4} \left(1-b^2-2  b p_{1-2+3}\right)\right) \gamma \left(\frac{1}{4} \left(1-b^2-2  b p_{1+2+3}\right)\right)}{ \gamma \left(\frac{1}{4} (1-b^2+2  b p_{1+2-3})\right) \gamma \left(\frac{1}{4} (1-b^2+2  b p_{1-2+3})\right) \gamma \left(\frac{1}{4} (1-b^2+2  b p_{1+2+3})\right)}
\end{align}
and 
\begin{align}
    \frac{\widetilde{C}_{\text{NS}}^{(b)}(p_1,p_2,p_3)}{C_{\text{NS}}^{(b)}(p_1- b,p_2,p_3)}&=\frac{2 i \Gamma (1+b p_1) \Gamma \left(\frac{1-b^2}{2}+b p_1\right)}{b^2 \Gamma \left(1+b^2-b p_1 \right) \Gamma \left(\frac{1+b^2}{2}-b p_1\right)} \nonumber \\ 
    &\times\frac{\gamma\left(\frac{1}{4}\left(3+b^2-2bp_{1-2-3}\right)\right)\gamma\left(\frac{1}{4}\left(3+b^2-2bp_{1+2-3}\right)\right)}{\gamma\left(\frac{1}{4}\left(1-b^2+2bp_{1-2+3}\right)\right)\gamma\left(\frac{1}{4}\left(1-b^2+2bp_{1+2+3}\right)\right)}~ \label{eq:2ndshift}~.
\end{align}
Here $\gamma(x)=\frac{\Gamma(x)}{\Gamma(1-x)}$ and we used the abbreviations e.g. $p_{1+2-3}=p_1+p_2-p_3$ etc. These shift relations can be solved by introducing two common special functions in $\mathcal{N}=1$ Liouville theory built out of the usual Barnes double gamma function $\Gamma_b$, namely
\begin{equation}
    \Gamma_b^{\mathrm{NS}}(x)\equiv \Gamma_b\left(\frac{x}{2}\right)\Gamma_b\left(\frac{x+b+b^{-1}}{2}\right) \ , \ \ \
\Gamma_b^{\mathrm{R}}(x)\equiv\Gamma_b\left(\frac{x+b}{2}\right)\Gamma_b\left(\frac{x+b^{-1}}{2}\right)~,
\end{equation}
which themselves obey the shift relations\footnote{We review some of the properties of these functions (and their ``descendants'') in appendix \ref{app:SpecialFunctions}.}
\begin{equation}\label{eq:shiftsG01}
\frac{\Gamma_b^{\mathrm{NS}}(x+b)}{\Gamma_b^{\mathrm{R}}(x)}=\frac{\sqrt{2\pi} \ b^{ \frac{xb}{2}}}{\Gamma\left(\frac{1+bx}{2}\right)} \ , \ \ \ \ \frac{\Gamma_b^{\mathrm{R}}(x+b)}{\Gamma_b^{\mathrm{NS}}(x)}=\frac{\sqrt{2\pi} \ b^{\frac{1}{2}(bx-1)}}{\Gamma\left(\frac{xb}{2}\right)} ,
\end{equation} 
and similarly for $b\rightarrow b^{-1}$. It is now straightforward to verify that the shift relations (\ref{eq:1stshift}),(\ref{eq:2ndshift}) are solved by the following ansätze for the three-point functions
\begin{subequations}\label{eq: NS spacelike 1 and 2}
\begin{align} \label{eq: NS spacelike 1}
     C_{\rm NS}^{(b)} (p_1,p_2,p_3) &=  \frac{\Gamma_b^{\rm NS}(2Q)}{2\Gamma_b^{\rm NS}(Q)^3}\frac{\Gamma^{\rm NS}_b\left(\frac{Q}{2} \pm p_1 \pm p_2 \pm p_3\right)}{\prod_{j=1}^3\Gamma_b^{\rm NS}\left(Q \pm 2 p_j\right)}~,\\ \label{eq: NS spacelike 2}
    \widetilde{C}^{(b)}_{\rm NS} (p_1,p_2,p_3) &=i    \frac{\Gamma_b^{\rm NS}(2Q)}{\Gamma_b^{\rm NS}(Q)^3}\frac{\Gamma^{\rm R}_b\left(\frac{Q}{2} \pm p_1 \pm p_2 \pm p_3\right)}{\prod_{j=1}^3\Gamma_b^{\rm NS}\left(Q \pm 2 p_j\right)}~.
\end{align}    
\end{subequations}
In the above the $\pm$ signs denote that we take a product over all possible combinations. For example each numerator is a product of eight terms. The expressions (\ref{eq: BNS}), (\ref{eq: NS spacelike 1 and 2}) together with the OPE (\ref{eq:NSOPE}) provide the basic data that define the spacelike $\mathcal{N}=1$ Liouville CFT in the NS-sector, meaning that any correlation function of NS fields on any genus-$g$ Riemann surface (with a spin structure) can be computed in terms of these quantities.
We now turn to a more detailed discussion of the expressions (\ref{eq: NS spacelike 1 and 2}).
\paragraph{Properties of the three-point functions.} The shift relations (\ref{eq:1stshift}) determine the three-point function $C^{(b)}_{\rm NS}$ up to a momentum-independent constant that in principle depends on the central charge. In (\ref{eq: NS spacelike 1}) we chose succinctly this constant in a way that we will justify shortly. Once this is fixed, the associated constant in front of $\widetilde{C}^{(b)}_{\rm NS}$ is $2i$ times that constant, as can be seen from (\ref{eq: NS spacelike 2}). This follows directly from the second shift relation (\ref{eq:2ndshift}). 
\par In the same way that in ordinary bosonic Liouville theory the DOZZ formula for the structure constants is unique \cite{Teschner:1995yf}, it can be shown that both expressions (\ref{eq: NS spacelike 1 and 2}) are the unique solutions to the shift relations (\ref{eq:1stshift}), (\ref{eq:2ndshift}) for $b\in\mathbb{R}_{(0,1]}$ with the following features:
\begin{itemize}
    \item Continuous in $b$, and invariant under $b\leftrightarrow b^{-1}$,
    \item Meromorphic in the momenta $p_i$,
    \item Permutation symmetric under the exchange of any two momenta,
    \item Reflection symmetric under any $p_i\rightarrow -p_i$,
    \item In the limit where one of the operator momentum approaches the value $Q/2$, we get
    \begin{align}
        \lim_{p_3\rightarrow\frac{Q}{2}}C_{\rm NS}^{(b)} (p_1,p_2,p_3)&= \left(\rho^{(b)}_{\mathrm{NS}}(p_1)\right)^{-1}\delta(p_1-p_2)~,
         \nonumber \\
         \lim_{p_3\rightarrow\frac{Q}{2}}\widetilde{C}_{\rm NS}^{(b)} (p_1,p_2,p_3)&=0~.\label{eq:IdLimitNSspacelike}
    \end{align}    
\end{itemize}
Notice that the \textit{diagonal structure} (i.e., the delta function) appearing in the first limit\footnote{The best way to study this limit is by writing $p_3= Q/2-\epsilon, p_1=p_2+\epsilon$, and then taking the limit $\epsilon\rightarrow0$. We will need the residue of the double gamma function at $x=0$, which is given by $\text{Res}_{x=0}\Gamma_b(x)=\Gamma_b(Q)/2\pi$. Once this is done carefully, the appearance of the delta function relies on the compensation of a double pole (coming from the numerator) and a simple zero (coming from the denominator). The second limit in (\ref{eq:IdLimitNSspacelike}) is studied in exactly the same way, except this time there is only a simple zero (coming from the denominator).} serves as an essential check for the consistency of spacelike $\mathcal{N}=1$ Liouville theory. The judicious choice of the momentum-independent constant in (\ref{eq: NS spacelike 1}) is such that, in the limit, the coefficient of the delta function reproduces exactly our choice of two-point normalization. The second limit is also an important check and reflects the fact that the NS vacuum module possesses only integer-level descendants, in the sense that we described in (\ref{eq:NSOPE}) for the OPE.
\par For $b\in(0,1]$ and $p_i\in i\mathbb{R}$, it is easy to check that (\ref{eq: BNS}) as well as (\ref{eq: NS spacelike 1}) are positive definite functions. For the same reasons, the three-point function (\ref{eq: NS spacelike 2}) is purely imaginary with positive imaginary part. We emphasize that the latter does not violate unitarity, e.g. in the four-point function of $V_p$'s. Given our conventions (\ref{eq:NSOPE}), (\ref{eq:chains}) for the OPE and the chain operators, when expanding the four-point function of $V_p$'s into conformal blocks the three-point function squared $\widetilde{C}_{\mathrm{NS}}^2$ multiplies the odd-odd conformal blocks with an overall minus sign (see App.\ref{app:NS}). Therefore, the net result comes with a positive sign for these contributions \cite{Belavin:2007gz}. 
\par Finally it is instructive to record the poles and zeroes of the two three-point functions as dictated by the analytic structure of $\Gamma_b^{\mathrm{NS}},\Gamma_b^{\mathrm{R}}$ (c.f. Appendix \ref{app:SpecialFunctions}). In particular, both $C^{(b)}_{\mathrm{NS}}(p_1,p_2,p_3)$ and $\widetilde{C}^{(b)}_{\mathrm{NS}}(p_1,p_2,p_3)$ have\footnote{For the zeroes and poles to be simple, we have implicitly assumed $b^2\notin \mathbb{Q}$.}
\begin{equation}\label{eq:NSzeroes}
\text{\textit{simple zeroes} when $ p_j = \frac{r b+s b^{-1}}{2}$, $\quad r,s\in \mathbb{Z}_{\geq 1}$, $\quad r-s\in 2\mathbb{Z} $},
\end{equation}
for $j=1,2,3$ and all reflections $p_j\rightarrow-p_j$ thereof. These are exactly the values corresponding to the degenerate representations of the NS algebra. A similar phenomenon occurs in the analogous normalization of the three-point function in ordinary bosonic Liouville theory \cite{Collier:2019weq,Collier:2023cyw}, where the analogous function $C_0$ vanishes at the degenerate representations of the Virasoro algebra. On the other hand, the singularities are slightly different in the two expressions: 
\begin{align}
&\text{$C^{(b)}_{\mathrm{NS}}(p_1,p_2,p_3)$: ~ \textit{simple poles} when $p_1=p_2+p_3+\frac{Q}{2}+kb+lb^{-1}$}, \nonumber\\
&\text{$\widetilde{C}^{(b)}_{\mathrm{NS}}(p_1,p_2,p_3)$: ~ \textit{simple poles} when $p_1=p_2+p_3+\frac{Q}{2}+k'b+l'b^{-1}$},\label{eq:NSpoles} 
\end{align}
and all reflections $p_j\rightarrow-p_j$ and permutations of $(p_1,p_2,p_3)$ thereof. Here all $k,l,k',l'\in\mathbb{Z}_{\geq 0}$ but, crucially, $k-l\in2\mathbb{Z}$ whereas $k'-l'\in2\mathbb{Z}+1$. The location of these poles are reminiscent of the analogous case in ordinary bosonic Liouville, where the $C_0$ function possesses simple poles at the values of the so-called \textit{Virasoro double twist operators} \cite{Collier:2018exn, Kusuki:2018wpa}. Those are momenta of a similar form\footnote{To be precise, the Virasoro double twist spectrum is given in our notation by $p_1=p_2+p_3+\frac{Q}{2}+mb+nb^{-1}$ with $m,n$ \textit{any} set of non-negative integers.} and appear as universal discrete contributions in the spectrum of Virasoro primaries at large spin in any unitary 2D CFT with just Virasoro symmetry and a twist gap above the vacuum \cite{Collier:2018exn, Kusuki:2018wpa}. By analogy, we are tempted to refer to the states in (\ref{eq:NSpoles}) as the two families of \textit{NS Virasoro double-twist operators}, anticipating that they similarly provide universal discrete contributions at large spin in unitary 2D CFTs with $\mathcal{N}=1$ super-Virasoro symmetry and a twist gap in the spectrum of NS primaries above the vacuum. To the best of our knowledge no such large spin universality has been properly examined to date for the $\mathcal{N}=1$ case\footnote{This would require the proper analysis of the fusion kernel of NS four-point conformal blocks developed in \cite{Hadasz:2007wi,Chorazkiewicz:2008es}, in the limit where the vacuum dominates in the T-channel for pairwise identical external operators. We initiate to some extent this discussion in Appendix \ref{app:fusionkernel}.}.

\subsubsection{R-sector}The general structure of the OPE in the R-sector reads
\begin{equation}
   [\text{R}] [\text{NS}]\sim  [\text{R}] \ , \ \ \ \  [\text{R}] [\text{R}]\sim  [\text{NS}]~.
\end{equation}
From this it is evident that we cannot consider this sector separately, since both R and NS fields are mixed with each other. 
\par In Liouville theory, we explicitly have ($\epsilon = \pm$) 
\begin{align}\label{eq:ROPE}
     &V^{\mathrm{R},\epsilon}_{p_1}(z)V^{\mathrm{NS}}_{p_2}(0)\sim \int_{i\mathbb{R}_+} \frac{\d p}{i} \frac{(z\bar{z})^{h_{[p]}-h_{[p_1]}-h_{p_2}}}{B_{\text{R}}^{(b)}(p)} C^{(b)}_{\mathrm{R},\epsilon} (p_1,p;p_2)\left[V^{\mathrm{R},\epsilon}_{p}(0)\right] , \nonumber \\
    &V^{\mathrm{R},\epsilon}_{p_1}(z)V^{\mathrm{R},\epsilon}_{p_2}(0)\sim \int_{i\mathbb{R}_+} \frac{\d p}{i} \frac{(z\bar{z})^{h_{p}-h_{[p_1]}-h_{[p_2]}}}{B_{\text{NS}}^{(b)}(p)} \left(C_{\mathrm{NS}}^{(b)}(p_1,p_2,p)\left[V^{\mathrm{NS}}_p(0)\right]_{\mathrm{ee}}-\widetilde{C}_{\mathrm{NS}}^{(b)}(p_1,p_2,p)\left[V^{\mathrm{NS}}_p(0)\right]_{\mathrm{oo}}\right) \ , \nonumber \\
    &V^{\mathrm{R},\epsilon}_{p_1}(z)V^{\mathrm{R},-\epsilon}_{p_2}(0)\sim \int_{i\mathbb{R}_+} \frac{\d p}{i} \frac{(z\bar{z})^{h_{p}-h_{[p_1]}-h_{[p_2]}}}{B_{\text{NS}}^{(b)}(p)} \widetilde{C}_{\mathrm{NS}}^{(b)}(p_1,p_2,p)\left(\left[V^{\mathrm{NS}}_p(0)\right]_{\mathrm{oe}}+\left[V^{\mathrm{NS}}_p(0)\right]_{\mathrm{eo}}\right)~.
\end{align}
We denoted the conformal dimensions corresponding to R primaries with a square bracket $h_{[p]}$ to differentiate them from the ones corresponding to the NS-sector\footnote{We will adopt this notation in various places throughout the text whenever we find it appropriate.}. In the first line we also denoted the R-sector module by square brackets indicating the contributions from all the holomorphic+anti-holomorphic descendants built out of $L_{n\in\mathbb{Z}_{\leq-1}}$ and $G_{k\in\mathbb{Z}_{\leq-1}}$ (see e.g. \cite{Hadasz:2008dt} for a more rigorous definition). The rest of the chain operators in the second and third lines correspond to the NS-sector and are defined according to (\ref{eq:chains}).
\par
\begin{figure}[H]
\centering 
\begin{tikzpicture}
\begin{scope}[shift={(1.1,0)}, scale=.7]
\draw[very thick, red, looseness=.7] (2,1.75) to[out=10,in=5] (2,.25);
\draw[very thick, densely dashed, red, looseness=.7] (2,1.75) to[out=-180,in=-180] (2,.25);    
\draw[very thick] (0,0) ellipse (0.25 and .5);
\draw[very thick] (0,2) ellipse (0.25 and .5);
\draw[very thick] (4,0) ellipse (0.25 and .5);
\draw[very thick] (4,2) ellipse (0.25 and .5);
\node[scale=1., thick] at (-0.9,0) {$p^{\mathrm{R}}_{\langle1,2\rangle}$};
\node[scale=1., thick] at (-0.8,2) {$p^{\mathrm{R}}_1$};  
\node[scale=1., thick] at (4.9,2) {$p^{\mathrm{NS}}_3$};
\node[scale=1., thick] at (4.9,0) {$p^{\mathrm{NS}}_2$};
\draw[very thick] (0,.5) to[out=0, in=0, looseness=2.5] (0,1.5);
\draw[very thick] (4,.5) to[out=180, in=180, looseness=2.5] (4,1.5);
\draw[very thick] (0,2.5) to[out=0, in=180] (2,1.75) to[out=0, in=180] (4,2.5);
\draw[very thick] (0,-.5) to[out=0, in=180] (2,.25) to[out=0, in=180] (4,-.5); 
\node[scale=1.5] at (6.2,.9) {$=$};
\end{scope}
\begin{scope}[shift={(7.1,1.75)}, scale=.65, rotate=270]
\draw[very thick, red, looseness=.7] (2,1.75) to[out=10,in=5] (2,.25);
\draw[very thick, densely dashed, red, looseness=.7] (2,1.75) to[out=-180,in=-180] (2,.25);    
\draw[very thick] (0,0) ellipse (0.25 and .5);
\draw[very thick] (0,2) ellipse (0.25 and .5);
\draw[very thick] (4,0) ellipse (0.25 and .5);
\draw[very thick] (4,2) ellipse (0.25 and .5);
\node[scale=1., thick] at (-0.2,-0.9) {$p^{\mathrm{R}}_1$};
\node[scale=1., thick] at (0,3.2) {$p^{\mathrm{NS}}_3$};  
\node[scale=1., thick] at (4.3,3) {$p^{\mathrm{NS}}_2$};
\node[scale=1., thick] at (4,-1.3) {$p^{\mathrm{R}}_{\langle1,2\rangle}$};
\draw[very thick] (0,.5) to[out=0, in=0, looseness=2.5] (0,1.5);
\draw[very thick] (4,.5) to[out=180, in=180, looseness=2.5] (4,1.5);
\draw[very thick] (0,2.5) to[out=0, in=180] (2,1.75) to[out=0, in=180] (4,2.5);
\draw[very thick] (0,-.5) to[out=0, in=180] (2,.25) to[out=0, in=180] (4,-.5);    
\end{scope}
\end{tikzpicture}
\caption{\small{Teschner's trick in the R-sector of $\mathcal{N}=1$ Liouville theory: the analytic bootstrap problem involving crossing of the sphere four-point function between two NS fields with momenta $p_2,p_3$ and two R fields, one with momentum $p_{\langle1,2\rangle}$ and one with momentum $p_1$. The analysis (done in App.\ref{app:R}) leads to the shift relations (\ref{eq: shiftequations R sector}), (\ref{eq: shift equations R sectorbinv}). 
}
}
\label{fig:N1Rbootstrap}
\end{figure}
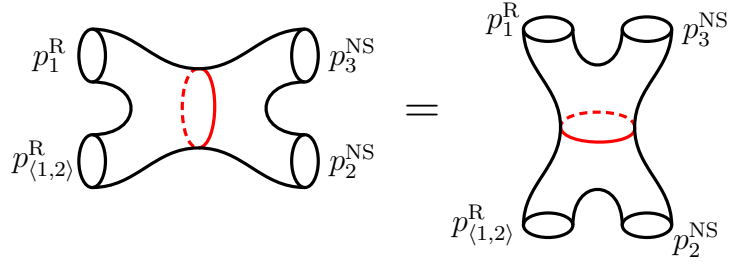

 The CFT data $\left\{B_{\text{R}}^{(b)},C_{\mathrm{R},\epsilon}^{(b)}\right\}$
now comprises the basic R-sector two- and three-point functions of the theory
\begin{subequations}\label{eq: spacelikeN1 structure constants}
\begin{align}
 \langle V^{\mathrm{R},\epsilon}_{p_1}(0)V^{\mathrm{R},\epsilon}_{p_2}(1)\rangle &= B_{\text{R}}^{(b)}(p_1)[\delta(p_1-p_2) + \epsilon\,\delta(p_1+p_2)]~, \\
    \langle V^{\mathrm{R},\epsilon}_{p_1}(z_1)V^{\mathrm{R},\epsilon}_{p_2}(z_2) V^{\mathrm{NS}}_{p_3}(z_3) \rangle &= \frac{C_{\mathrm{R},\epsilon}^{(b)}(p_1,p_2;p_3)}{|z_{12}|^{2h_{[1]+[2]-3}}|z_{23}|^{2h_{[2]+3-[1]}}|z_{31}|^{2h_{3+[1]-[2]}}}~.
\end{align}    
\end{subequations}
The explicit expressions for these data were obtained originally in \cite{Poghossian:1996agj, Rashkov:1996np}. Here we will mostly follow the logic of \cite{Poghossian:1996agj} and apply Teschner's trick in a particular four-point function depicted in figure \ref{fig:N1Rbootstrap} which we now discuss. 
\noindent
\paragraph{Shift relations.}
Using the null vector (\ref{eq:nullvectorR}) one can obtain a second order hypergeometric differential equation for the four-point function involving one degenerate R field, a general R field, and two general NS fields,
\begin{equation}
    \langle V^{\mathrm{R},\epsilon}_{p_{\langle 1,2\rangle}}(z_0) V_{p_1}^{\mathrm{R},\epsilon}(z_1) V_{p_2}^{\mathrm{NS}}(z_2) V_{p_3}^{\mathrm{NS}}(z_3)\rangle~. 
\end{equation}
We present the details in appendix \ref{app:R} (following \cite{Poghossian:1996agj,Fukuda:2002bv}). 
Imposing crossing symmetry in the $s$- and $t$-channel expansions then leads to shift relations for the following combinations of three-point functions which, using standard terminology, we refer to as $C_{\mathrm{even}},C_{\mathrm{odd}}$, 
\begin{subequations}
\begin{align}
 \hspace{-2cm} C_{\mathrm{even}}^{(b)}(p_1,p_2,p_3) &\equiv \frac{1}{2} \left(\langle V_{p_1}^{\mathrm{R,+}} V_{p_2}^{\mathrm{R,+}} V_{p_3}^{\mathrm{NS}}\rangle + \langle V_{p_1}^{\mathrm{R,-}} V_{p_2}^{\mathrm{R,-}}V_{p_3}^{\mathrm{NS}}\rangle \right)~,\\
 C_{\mathrm{odd}}^{(b)}(p_1,p_2,p_3) &\equiv \frac{1}{2} \left(\langle V_{p_1}^{\mathrm{R,+}} V_{p_2}^{\mathrm{R,+}} V_{p_3}^{\mathrm{NS}}\rangle - \langle V_{p_1}^{\mathrm{R,-}} V_{p_2}^{\mathrm{R,-}} V_{p_3}^{\mathrm{NS}}\rangle\right)~.
 \end{align}
\end{subequations}
In particular, we find 
\begin{align}\label{eq: shiftequations R sector}
\frac{C_{\mathrm{even}}^{(b)}(p_1+\frac{b}{2},p_2; p_3)^2/B_{\mathrm{R}}^{(b)}(p_1+\frac{b}{2})}{C_{\mathrm{odd}}^{(b)}(p_1-\frac{b}{2}, p_2; p_3)^2/B_{\mathrm{R}}^{(b)}(p_1-\frac{b}{2})} &=\kappa^{(b)}_{\mathrm{R}}\left(p_1|p_2;p_3\right) ~ ,  \cr
\frac{C_{\mathrm{odd}}^{(b)}(p_1+\frac{b}{2},p_2; p_3)^2/B_{\mathrm{R}}^{(b)}(p_1+\frac{b}{2})}{C_{\mathrm{even}}^{(b)}(p_1-\frac{b}{2}, p_2; p_3)^2/B_{\mathrm{R}}^{(b)}(p_1-\frac{b}{2})} &=\kappa^{(b)}_{\mathrm{R}}\left(p_1|-p_2;p_3\right)
\end{align}
where
\begin{align}\label{eq:kappaR}
\kappa^{(b)}_{\mathrm{R}}\left(p_1|p_2;p_3\right):=-\frac{\gamma\left(\frac{1-b^2}{2}+bp_1\right)}{\gamma\left(\frac{1-b^2}{2}-bp_1\right)} \left[\frac{\Gamma\left(bp_1\right)\gamma\left(\frac{3}{4}-\frac{b}{2}p_{1+2-3}\right)\gamma\left(\frac{3}{4}-\frac{b}{2}p_{1+2+3}\right)}{\Gamma\left(-bp_1\right)\gamma\left(\frac{3}{4}+\frac{b}{2}p_{1-2-3}\right)\gamma\left(\frac{3}{4}+\frac{b}{2}p_{1-2+3}\right)}\right]^2.
\end{align}
For $b^2\notin \mathbb{Q}$, the same analysis leads also to another set of incommensurable shift relations with $b\leftrightarrow b^{-1}$, namely
\begin{align}
\frac{C_{\mathrm{even}}^{(b)}(p_1+\frac{b^{-1}}{2},p_2; p_3)^2/B_{\mathrm{R}}^{(b)}(p_1+\frac{b^{-1}}{2})}{C_{\mathrm{odd}}^{(b)}(p_1-\frac{b^{-1}}{2}, p_2; p_3)^2/B_{\mathrm{R}}^{(b)}(p_1-\frac{b^{-1}}{2})} &=\kappa^{(b^{-1})}_{\mathrm{R}}\left(p_1|p_2;p_3\right) \ , \nonumber \\
     \frac{C_{\mathrm{odd}}^{(b)}(p_1+\frac{b^{-1}}{2},p_2; p_3)^2/B_{\mathrm{R}}^{(b)}(p_1+\frac{b^{-1}}{2})}{C_{\mathrm{even}}^{(b)}(p_1-\frac{b^{-1}}{2}, p_2; p_3)^2/B_{\mathrm{R}}^{(b)}(p_1-\frac{b^{-1}}{2})} &=\kappa^{(b^{-1})}_{\mathrm{R}}\left(p_1|-p_2;p_3\right)~.\label{eq: shift equations R sectorbinv}
\end{align}
As in the NS-sector, we are searching for solutions of $B_{\mathrm{R}}^{(b)},C_{\mathrm{even}}^{(b)},C_{\mathrm{odd}}^{(b)}$ that are invariant under $b\leftrightarrow b^{-1}$. However, unlike the NS-sector case, the structure constants are \textit{not} permutation symmetric for the obvious reason that the momentum $p_3$ (of the NS field) is not on the same grounds as the momenta $p_1,p_2$ (of the R fields). However, we do require that our solutions are permutation symmetric in $p_1,p_2$. From the structure of the shift relations (\ref{eq: shiftequations R sector}), (\ref{eq: shift equations R sectorbinv}) it is further evident that the structure constants should obey the relationship
\begin{equation}
   C_{\mathrm{even}}^{(b)}(p_1,-p_2; p_3)=C_{\mathrm{odd}}^{(b)}(p_1,p_2; p_3)=C_{\mathrm{even}}^{(b)}(-p_1,p_2; p_3).
\end{equation}
The second equation comes from the aforementioned permutation symmetry in $p_1,p_2$.
\par As in the case of the NS-sector, the explicit expression (\ref{eq:kappaR}) for the `bootstrap function' $\kappa_{\mathrm{R}}$ shows that the shift relations are manifestly meromorphic in the momenta as well as in $b$. Thus they can be analytically continued to the timelike regime $\hat{c}\leq1$ without problems, as we will see in detail in section \ref{sec:Timelike}. As a function of the momenta, $\kappa_{\mathrm{R}}$ does not possess any obvious properties under permutations (for the reasons we discussed). However under reflections we get
\begin{align}  \kappa^{(-b)}_{\mathrm{R}}\left(-p_1|p_2;p_3\right)=\kappa^{(b)}_{\mathrm{R}}\left(p_1|-p_2;p_3\right) \ , \ \kappa^{(b)}_{\mathrm{R}}\left(p_1|p_2;-p_3\right)=\kappa^{(b)}_{\mathrm{R}}\left(p_1|p_2;p_3\right).
\end{align}
\paragraph{Natural normalization.} In accordance with the NS-sector case, we will proceed by making a specific natural choice for the two-point function (\ref{eq: spacelikeN1 structure constants}) that does not depend on the cosmological constant of the action. In particular, we will choose
\begin{equation}\label{eq: osp12 normalization 2pt function spacelike R sector}
    B_{\mathrm{R}}^{(b)}(p) \equiv \left(\rho_\mathrm{R}^{(b)}(p)\right)^{-1}= \frac{1}{2\sqrt{2}\cos(\pi b p)\cos(\pi b^{-1}p)}~.
\end{equation}
The denominator in the expression is related with the modular transformation of the second identity character $\chi_{\mathds{1}}^{\mathrm{NS},-}(\tau)$ (c.f. (\ref{eq:idNScharacters}))
which can be now branched into an integral over the non-trivial Ramond characters\cite{Benjamin:2020zbs}
\begin{equation}
    \chi^{\text{R},+}_p(\tau) := 2q^{h_{[p]}-\frac{c}{16}}\frac{\eta(2\tau)}{\eta(\tau)^2} ~
\end{equation}
evaluated on the modular transformed channel. Specifically,
\begin{align}
\chi_\mathds{1}^{\text{NS},-}\left(-\frac{1}{\tau}\right) = \int_{i\mathbb{R}_{+}} \frac{\d p}{i}\, \mathbb{S}^{(\mathrm{R})}_{p\mathds{1}}\ \chi_p^{\text{R},+}(\tau)\ , \quad ~~~~~~\mathbb{S}_{p\mathds{1}}^{(\text{R})}\equiv\rho^{(b)}_{\mathrm{R}}(p)
     ~.
\end{align}
With this convention, the vertex operators are identified with a trivial reflection coefficient as
\begin{equation}
   V_{p}^{\mathrm{R},\epsilon}=\epsilon V_{-p}^{\mathrm{R},\epsilon}  \ , \ \ \ \ \ \ \epsilon=\pm.
\end{equation}
\par Going back to the shift relations (\ref{eq: shiftequations R sector}), the choice (\ref{eq: osp12 normalization 2pt function spacelike R sector}) leads to the same essential simplification as in the NS-sector, namely the ratio $\frac{\rho^{(b)}_{\mathrm{R}}(p_1-\frac{b}{2})}{\rho^{(b)}_{\mathrm{R}}(p_1+\frac{b}{2})}\kappa_{\mathrm{R}}^{(b)}(p_1|\pm p_2;p_3)$ (for either signs of $p_2$) forms a complete square. We then obtain the following simplified shift relations for the structure constants 
\begin{equation}
    \frac{ C_{\mathrm{even}}^{(b)}(p_1+\frac{b}{2},p_2;p_3)}{ C_{\mathrm{odd}}^{(b)}(p_1-\frac{b}{2},p_2;p_3)} = \frac{\Gamma(1+bp_1)\Gamma(\frac{1}{2}(1+b^2+2bp_1))}{\Gamma(1-bp_1)\Gamma(\frac{1}{2}(1+b^2- 2bp_1))} \frac{\gamma(\frac{3}{4} - \frac{b}{2}p_{1+2-3})\gamma(\frac{3}{4} -\frac{b}{2}p_{1+2+3})}{\gamma(\frac{3}{4} + \frac{b}{2}p_{1-2-3})\gamma(\frac{3}{4} +\frac{b}{2}p_{1-2+3})}~
\end{equation}
and 
\begin{equation}
    \frac{ C_{\mathrm{odd}}^{(b)}(p_1+\frac{b}{2},p_2;p_3)}{ C_{\mathrm{even}}^{(b)}(p_1-\frac{b}{2},p_2;p_3)} = \frac{\Gamma(1+bp_1)\Gamma(\frac{1}{2}(1+b^2+2bp_1))}{\Gamma(1-bp_1)\Gamma(\frac{1}{2}(1+b^2- 2bp_1))} \frac{\gamma(\frac{3}{4} -\frac{b}{2}p_{1-2-3})\gamma(\frac{3}{4} - \frac{b}{2}p_{1-2+3})}{\gamma(\frac{3}{4} + \frac{b}{2}p_{1+2-3})\gamma(\frac{3}{4} +\frac{b}{2}p_{1+2+3})}~.
\end{equation}
Harnessing the same special functions $\Gamma_b^{\mathrm{NS}},\Gamma_b^{\mathrm{R}}$, it is straightforward to show that the solutions take the following form:
\begin{subequations}\label{eq: R spacelike}
\begin{align}
   C_{\mathrm{even}}^{(b)}(p_1,p_2;p_3)
    &=\frac{\Gamma_b^{\rm NS}(2Q)}{\sqrt{2}\Gamma_b^{\rm NS}(Q)^3} \frac{\Gamma_b^{\mathrm{R}}\left(\frac{Q}{2}\pm (p_1+p_2)\pm p_3\right)\Gamma_b^{\mathrm{NS}}\left(\frac{Q}{2}\pm (p_1-p_2)\pm p_3\right)}{\Gamma_b^{\mathrm{R}}(Q\pm 2p_1)\Gamma_b^{\mathrm{R}}(Q\pm 2p_2)\Gamma_b^{\mathrm{NS}}(Q\pm 2p_3)},\\
    C_{\mathrm{odd}}^{(b)}(p_1,p_2;p_3)&= \frac{\Gamma_b^{\rm NS}(2Q)}{\sqrt{2}\Gamma_b^{\rm NS}(Q)^3} \frac{\Gamma_b^{\mathrm{NS}}\left(\frac{Q}{2}\pm (p_1+p_2)\pm p_3\right)\Gamma_b^{\mathrm{R}}\left(\frac{Q}{2}\pm (p_1-p_2)\pm p_3\right)}{\Gamma_b^{\mathrm{R}}(Q\pm 2p_1)\Gamma_b^{\mathrm{R}}(Q\pm 2p_2)\Gamma_b^{\mathrm{NS}}(Q\pm 2p_3)}~.
\end{align}    
\end{subequations}
The expressions (\ref{eq: osp12 normalization 2pt function spacelike R sector}), (\ref{eq: R spacelike}) together with the OPE (\ref{eq:ROPE}) provide the basic data that define the spacelike $\mathcal{N}=1$ Liouville CFT in the R-sector, meaning that any correlation function involving R and NS fields on a genus-$g$ Riemann surface (with a spin structure) can be computed in principle in terms of these quantities.
We now turn to a more detailed discussion of the expressions (\ref{eq: R spacelike}).
\paragraph{Properties of the three-point functions}
As in the NS-sector, the shift relations (\ref{eq: shiftequations R sector}), (\ref{eq: shift equations R sectorbinv}) determine \textit{uniquely} the three-point functions up to a momentum-independent constant that can in principle depend on the central charge. In particular, (\ref{eq: R spacelike}) are the unique solutions with the following features:
\begin{itemize}
    \item Continuous in $b$, and invariant under $b\leftrightarrow b^{-1}$,
    \item Meromorphic in the momenta $p_i$,
    \item Permutation symmetric under the exchange of $p_1 \leftrightarrow p_2$,
    \item Reflection symmetric under the NS-sector momentum $p_3\rightarrow -p_3$, while
  \begin{equation}\label{eq: properties R sector constants}
\frac{C^{(b)}_{\mathrm{even}}(p_1,-p_2;p_3)}{C^{(b)}_{\mathrm{odd}}(p_1,p_2;p_3)}=1~,\quad \frac{C^{(b)}_{\mathrm{even}}(-p_1,p_2;p_3)}{C^{(b)}_{\mathrm{odd}}(p_1,p_2;p_3)}=1~,
\end{equation}  
    \item In the limit where the NS-operator momentum approaches the value $Q/2$, we get
    \begin{align}
        \lim_{p_3\rightarrow\frac{Q}{2}}C_{\rm even}^{(b)} (p_1,p_2;p_3)&= \left(\rho^{(b)}_{\mathrm{R}}(p_1)\right)^{-1}\delta(p_1-p_2)~,\cr
         \lim_{p_3\rightarrow\frac{Q}{2}}{C}_{\rm odd}^{(b)} (p_1,p_2;p_3)&=\left(\rho^{(b)}_{\mathrm{R}}(p_1)\right)^{-1}\delta(p_1+p_2)~.
    \end{align}  
\end{itemize}
The overall $b-$dependent constants in (\ref{eq: R spacelike}) are chosen so that the coefficient of the delta functions in the above limits is exactly reproduced by the two-point function (\ref{eq: osp12 normalization 2pt function spacelike R sector}). Just as in the NS-sector, we observe that one can consistently recover the diagonal structure of the two-point function of Ramond fields by analytic continuation of the corresponding three point structure constants. We note also that for $b\in\mathbb{R}$ and $p_i\in i \mathbb{R}$ the structure constants (\ref{eq: R spacelike}) are both positive definite.
\par Finally, it is again instructive to record the poles and zeroes. In particular, both $C^{(b)}_{\mathrm{even}}(p_1,p_2;p_3)$ and $C^{(b)}_{\mathrm{odd}}(p_1,p_2;p_3)$ have
\begin{equation}\label{eq:Rzeroes p3}
\text{\textit{simple zeros} when $ p_3 = \frac{r b+s b^{-1}}{2}$, $\quad r,s\in \mathbb{Z}_{\geq 1}$, $\quad r-s\in 2\mathbb{Z} $},
\end{equation}
and reflections $p_3\rightarrow-p_3$ thereof. These are exactly the values corresponding to the degenerate representations of the NS algebra, which is consistent with the fact that $V_{p_3}^{\mathrm{NS}}$ is indeed an NS-primary. In addition, we also get
\begin{equation}\label{eq:Rzeroes pj}
\text{\textit{simple zeros} when $ p_j = \frac{r b+s b^{-1}}{2}$, $\quad r,s\in \mathbb{Z}_{\geq 1}$, $\quad r-s\in 2\mathbb{Z}+1 $},\quad j=1,2~.
\end{equation}
These are exactly the values corresponding to the degenerate representations of the R algebra. For the singularities, on the other hand, we obtain the following structure
\begin{align}
&\text{$C^{(b)}_{\mathrm{even}}(p_1,p_2;p_3)$: ~ \textit{simple poles} when $\pm (p_1-p_2)=\pm p_3+\frac{Q}{2}+kb+lb^{-1}$}, \nonumber\\
&\text{${C}^{(b)}_{\mathrm{odd}}(p_1,p_2;p_3)$: ~ \textit{simple poles} when $\pm (p_1+p_2)=\pm p_3+\frac{Q}{2}+kb+lb^{-1}$}~,\label{eq:Rpoles a} 
\end{align}
as well as
\begin{align}
&\text{$C^{(b)}_{\mathrm{even}}(p_1,p_2;p_3)$: ~ \textit{simple poles} when $\pm (p_1+p_2)=\pm p_3+\frac{Q}{2}+k'b+l'b^{-1}$}, \nonumber\\
&\text{${C}^{(b)}_{\mathrm{odd}}(p_1,p_2;p_3)$: ~ \textit{simple poles} when $\pm (p_1-p_2)=\pm p_3+\frac{Q}{2}+k'b+l'b^{-1}$}~.\label{eq:Rpoles b} 
\end{align}
 Here all $k,l,k',l'\in\mathbb{Z}_{\geq 0}$ with $k-l\in2\mathbb{Z}$, whereas $k'-l'\in2\mathbb{Z}+1$.
 \\
\par This concludes our discussion for the spacelike $\mathcal{N}=1$ Liouville theory. We will next move on to describing the timelike theory.

\section{Timelike $\mathcal{N}=1$ Liouville CFT}\label{sec:Timelike}
\par Ordinary timelike Liouville theory in the non-supersymmetric setting has been explored in various works, including \cite{Strominger:2003fn, Zamolodchikov:2005fy,Schomerus:2003vv,Kostov:2005kk,Ribault:2015sxa, Harlow:2011ny, Bautista:2019jau, Kostov:2005av,Kostov:2006zp,McElgin:2007ak,Giribet:2011zx,Delfino:2010xm,Ikhlef:2015eua,Ang:2021tjp,Chatterjee:2025yzo}.
Across these references, the theory has been studied in a range of different contexts and roles. While the works \cite{Zamolodchikov:2005fy,Kostov:2005kk,Schomerus:2003vv,Ribault:2015sxa} primarily focus on constructing the theory from the bootstrap point of view, \cite{Harlow:2011ny} examines its comparison with the semiclassical gravitational path integral, and \cite{Anninos:2021ene, Muhlmann:2022duj, Anninos:2024iwf, Bautista:2019jau, Martinec:2003ka} consider it as part of a quantum gravity theory exhibiting de Sitter vacua. Yet another interesting connection of the theory has been established in \cite{Delfino:2010xm,Ikhlef:2015eua,Ang:2021tjp} in relation with percolation and the conformal loop models. Recently, attempts to bring timelike Liouville theory on firmer mathematical footings were initiated in \cite{Chatterjee:2025yzo}.

Another application was proposed in \cite{Collier:2023cyw} for the 'Virasoro minimal string', in the context of solvable string theories with low-dimensional target space. The Virasoro minimal string is a two-dimensional critical string theory whose worldsheet CFT consists of timelike Liouville theory with central charge $\hat{c} \leq 1$, coupled to spacelike Liouville theory with central charge $26 - \hat{c} \geq 25$. The exact solvability of both the spacelike and timelike sectors played a pivotal role in formulating the Virasoro minimal string as a dual random matrix theory where the string amplitudes were shown to take surprisingly simple forms. It is natural to ponder about a supersymmetric extension of this setup\footnote{We will return to this question in Section~\ref{sec:discussion}.}, specifically, when the spacelike sector could be replaced by the known $\mathcal{N}=1$ Liouville theory that we described in section \ref{sec: spacelike}. A corresponding \textit{timelike} theory at the quantum level, however, has yet to be formulated on the same grounds.

Indeed, there has been notably less focus on supersymmetric timelike Liouville theory in the existing literature. In relation to dS gravity, the authors of \cite{Anninos:2022ujl, Anninos:2023exn} introduced the supersymmetric $\mathcal{N}=1$ (and $\mathcal{N}=2$) timelike Liouville theory from a path integral perspective with action
\begin{equation}\label{eq:N1actiontL}
S^{\mathcal{N}=1}_{\mathrm{tL}}=\frac{1}{4\pi}\int \d^2 x\,\tilde{\e}\left(-\frac{1}{2}\tilde{g}^{\mu\nu}\partial_\mu\phi\partial_\nu\phi+\frac{i}{2}\overline{\psi}\slashed{D}\psi-\frac{1}{2}\widehat{Q}\widetilde{R}\phi + \frac{1}{2}\mu^2\e^{2\hat{b}\phi} +\frac{1}{2}\mu\hat{b} \e^{\hat{b}\phi}\overline{\psi}\psi\right)~.
\end{equation}
Similarly to the non-supersymmetric case this theory exhibits a negative kinetic term for the conformal mode factor $\phi$.
(\ref{eq:N1actiontL}) can be viewed as a Weyl gauge-fixed supergravity theory that admits two-dimensional de Sitter vacua. \par In the present section, instead of a path integral analysis of (\ref{eq:N1actiontL}), we will follow the bootstrap approach and view $\mathcal{N}=1$ timelike Liouville theory as a two-dimensional conformal field theory that satisfies the $\mathcal{N}=1$ superconformal Ward identities and crossing equations for a specific range of the (superconformal) central charge.

\subsection{Spectrum}
As we stated in the introduction, the $\mathcal{N}=1$ timelike Liouville theory is defined for the (superconformal) central charge range $\hat{c}\leq1$, which we parametrize by\footnote{We will henceforth denote by hat all the quantities related to the timelike theory.}
\begin{equation}
    \hat{c} =1 - 2\widehat{Q}^2~,\quad \widehat{Q} = \hat{b}^{-1} - \hat{b}~,\quad \hat{b} \in\mathbb{R}_{(0,1]}~.
\end{equation}
Just as in the spacelike case, its spectrum consists of a \textit{continuum} of NS and R primary operators with conformal dimensions
\begin{equation}\label{eq:hTimelike}
    \hat{h}_{\hat{p}}^{\mathrm{NS}} =  -\frac{\widehat{Q}^2}{8} -\frac{\hat{p}^2}{2}~,\quad \hat{h}_{\hat{p}}^{\mathrm{R}} =  -\frac{\widehat{Q}^2}{8} -\frac{\hat{p}^2}{2} +\frac{1}{16}~,
\end{equation}
where the physical spectrum of the theory is expected for
\begin{equation}
    \hat{p}\in i\mathbb{R}_{\geq0}-\epsilon~.
\end{equation}
Similarly to the non-supersymmetric case \cite{Ribault:2015sxa}, an arbitrary $\epsilon>0$ shift is necessary in order to avoid the singularities of the conformal blocks coming from degenerate representations inside correlation functions. This means that the spectrum is again bounded from below, i.e. $\hat{h}_{\hat{p}}^{\mathrm{NS}}\geq -\frac{1-\hat{c}}{16}$ and $\hat{h}_{\hat{p}}^{\mathrm{R}}\geq\frac{\hat{c}}{16}$, except that now the conformal dimensions can take negative values, which renders the theory non-unitary.
\par Degenerate representations of the superconformal algebra for $\hat{c}\leq 1$ occur at momenta
\begin{equation}\label{eq:degtimelike}
    \hat{p}_{\langle r,s\rangle}=\frac{i\left(r\hat{b}^{-1}-s\hat{b}\right)}{2}, ~~~~ r,s\in\mathbb{Z}_{\geq 1},
\end{equation}
where, again, we have a null NS state when $r-s\in2\mathbb{Z}$, and a null R state when $r-s\in2\mathbb{Z}+1$. Notice that, in contrast to the spacelike case, the degenerate representations `overlap' with the support of the timelike Liouville spectrum (modulo the $\epsilon$ shift, as in the non-supersymmetric timelike case \cite{Ribault:2015sxa}). In particular the field with $\hat{p}_{\langle 1,1\rangle}$ has $\hat{h}^{\mathrm{NS}}_{\hat{p}_{\langle 1,1\rangle}}=0$ and it is tempting to identify it with the usual identity operator. However, this is subtle and we are going to return to this issue when we discuss the structure constants. 
\par
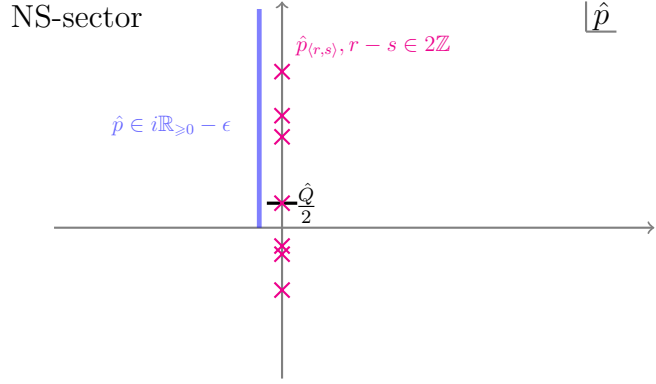
\begin{figure}[h]
\centering
    \begin{tikzpicture}
        \draw[thick, gray, ->] (2,-2) -- (2,3);
        \draw[thick, gray, ->] (-1,0) -- (6.9,0); 
        \draw[very thick, blue,  line width=1.8, opacity = .5] (1.7,-1.9) -- (1.7,2.9);   \node[scale=.9,thick] at (2.32,1.025-0.7) {$\frac{\hat{Q}}{2}$};
        \node[scale=.7,thick, blue, opacity =.6] at (0.5,1.35) {~$\hat{p} \in i\mathbb{R}-\epsilon$}; 
        \node[scale=.7,thick, magenta] at (3.2,2.4) {$\hat{p}_{\langle r, s\rangle},r-s\in2\mathbb{Z}$}; 
         \draw[very thick, black] (1.8,1.025-0.7) -- (2.2,1.025-0.7);
        \node[cross out, thick, draw=magenta, scale=.79/1.2] at (2,1.025-0.7) {};    
        \node[cross out, thick, draw=magenta, scale=.79/1.2] at (2,-0.351) {};   
        \node[cross out, thick, draw=magenta, scale=.79/1.2] at (2,1.205056) {};      
        \node[cross out, thick, draw=magenta, scale=.79/1.2] at (2,-0.827) {};   
        \node[cross out, thick, draw=magenta, scale=.79/1.2] at (2,-0.242889) {};      
        \node[cross out, thick, draw=magenta, scale=.79/1.2] at (2,1.48511) {};  
        \node[cross out, thick, draw=magenta, scale=.79/1.2] at (2,2.06922) {};
        \draw[thick, gray] (6,2.6) -- (6,3);
        \draw[thick, gray] (6,2.6) -- (6.4,2.6);
        \node[scale=1.,thick] at (6.2,2.8) {$\hat{p}$}; 
        \node[scale=1.,thick] at (-0.7,2.8) {NS-sector};
    \end{tikzpicture}
\caption{\footnotesize{The expected physical NS-spectrum in timelike $\mathcal{N}=1$ Liouville theory (blue) and the degenerate representations for the NS algebra (magenta) for central charge values $\hat{c}\leq 1$ (or $\hat{b}\in\mathbb{R}_{(0,1]}$). The corresponding expressions are given in (\ref{eq:hTimelike}), (\ref{eq:degtimelike}).}}
\label{fig:tNSspectra}
\end{figure}

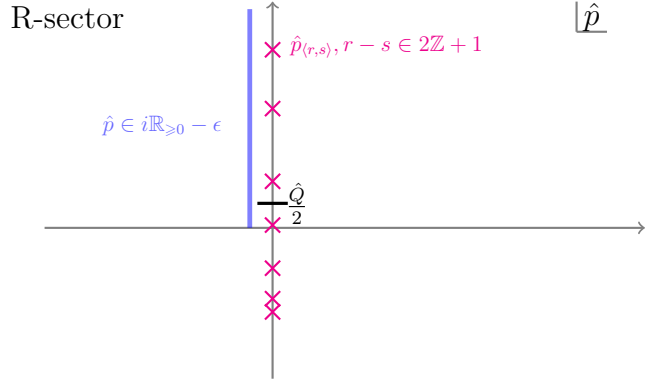
\begin{figure}[h]
\centering
    \begin{tikzpicture}
        \draw[thick, gray, ->] (2,-2) -- (2,3);
        \draw[thick, gray, ->] (-1,0) -- (6.9,0); 
        \draw[very thick, blue,  line width=1.8, opacity = .5] (1.7,-1.9) -- (1.7,2.9);  \node[scale=.9,thick] at (2.32,1.025-0.7) {$\frac{\hat{Q}}{2}$}; 
        \draw[very thick, black] (1.8,1.025-0.7) -- (2.2,1.025-0.7);
        \node[scale=.7,thick, blue, opacity =.6] at (0.5,1.35) {~$\hat{p} \in i\mathbb{R}-\epsilon$}; 
        \node[scale=.7,thick, magenta] at (3.5,2.4) {$\hat{p}_{\langle r, s\rangle},r-s\in2\mathbb{Z}+1$}; 
        \node[cross out, thick, draw=magenta, scale=.79/1.2] at (2,0.037) {};    
        \node[cross out, thick, draw=magenta, scale=.79/1.2] at (2,-0.939) {};   
        \node[cross out, thick, draw=magenta, scale=.79/1.2] at (2,0.617056) {};      
        \node[cross out, thick, draw=magenta, scale=.79/1.2] at (2,-1.115) {};   
        \node[cross out, thick, draw=magenta, scale=.79/1.2] at (2,-0.534944) {};      
        \node[cross out, thick, draw=magenta, scale=.79/1.2] at (2,1.58122) {};
        \node[cross out, thick, draw=magenta, scale=.79/1.2] at (2,2.36128) {};
        \draw[thick, gray] (6,2.6) -- (6,3);
        \draw[thick, gray] (6,2.6) -- (6.4,2.6);
        \node[scale=1.,thick] at (6.2,2.8) {$\hat{p}$}; 
        \node[scale=1.,thick] at (-0.7,2.8) {R-sector};
    \end{tikzpicture}
\caption{\footnotesize{The expected physical R-spectrum in timelike $\mathcal{N}=1$ Liouville theory (blue) and the degenerate representations of the R algebra (magenta) for central charge values $\hat{c}\leq 1$ (or $\hat{b}\in\mathbb{R}_{(0,1]}$). The corresponding expressions are given in (\ref{eq:hTimelike}), (\ref{eq:degtimelike}).}}
\label{fig:tRspectra}
\end{figure}

 We depict the various spectra in figures \ref{fig:tNSspectra} and \ref{fig:tRspectra}.

\paragraph{NS-sector.}
In the NS-sector of the theory, we denote the primaries with conformal dimension $\hat{h}_{\hat{p}}^{\mathrm{NS}}$ given in (\ref{eq:hTimelike}) as $\widehat{V}_{\hat{p}}^{\mathrm{NS}}$ \footnote{In terms of the Lagrangian (\ref{eq:N1actiontL}) we may want to think of them as the normal ordered exponentials
$\widehat{V}_{\hat{p}}^{\mathrm{NS}}= :\e^{(-\frac{\widehat{Q}}{{2}} -i\hat{p})\phi}: .$}.
We distinguish again the three operators generated by the super Poincar\'e subalgebra
\begin{equation}\label{eq:supermultipletNS timelike}
    \widehat{\Lambda}^{\mathrm{NS}}_{\hat{p}} = \widehat{G}_{-1/2}\widehat{V}^{\mathrm{NS}}_{\hat{p}}~,\quad \widehat{\widetilde{\Lambda}}^{\mathrm{NS}}_{\hat{p}} = \widehat{\widetilde{G}}_{-1/2}\widehat{V}^{\mathrm{NS}}_{\hat{p}}~,\quad  \widehat{W}^{\mathrm{NS}}_{\hat{p}} = \widehat{G}_{-1/2}\widehat{\widetilde{G}}_{-1/2}\widehat{V}^{\mathrm{NS}}_{\hat{p}}~.
\end{equation}
Their OPE with the super Virasoro stress tensor and the supercurrent are analogous to the spacelike case (\ref{eq: OPE NS}) with the obvious replacements.
\paragraph{R-sector.}
Similarly in the R-sector, we introduce again the combinations
\begin{equation}
    \widehat{\mathcal{G}}_0 = \widehat{G}_0 + i \widehat{\widetilde{G}}_0~, \quad \widehat{\widetilde{\mathcal{G}}}_0 = \widehat{G}_0 - i \widehat{\widetilde{G}}_0~,
\end{equation}
which satisfy the fermionic harmonic oscillator commutation relations (\ref{eq: fermionic harmonic oscillator}). On the doubly degenerate operator $\widehat{V}_{\hat{p}}^{\mathrm{R},\pm}$ they act as 
\begin{subequations}\label{eq: R sector timelike}
    \begin{align}
        \widehat{\mathcal{G}}_0 \widehat{V}_{\hat{p}}^{\mathrm{R},+} &= \e^{-\frac{i\pi}{4}}\sqrt{4\hat{h}_{\hat{p}}^{\mathrm{R}}-\frac{\hat{c}}{4}}\widehat{V}_{\hat{p}}^{\mathrm{R},-}~,\quad \widehat{\mathcal{G}}_0 \widehat{V}_{\hat{p}}^{\mathrm{R},-} =0~,\\
        \widehat{\widetilde{\mathcal{G}}}_0 \widehat{V}_{\hat{p}}^{\mathrm{R},-} &= \e^{\frac{i\pi}{4}}\sqrt{4\hat{h}_{\hat{p}}^{\mathrm{R}}-\frac{\hat{c}}{4}}\widehat{V}_{\hat{p}}^{\mathrm{R},+}~,\quad \ \ \widehat{\widetilde{\mathcal{G}}}_0 \widehat{V}_{\hat{p}}^{\mathrm{R},+} =0~,
    \end{align}
\end{subequations}
with $\hat{h}_{\hat{p}}^{\mathrm{R}}$ given in (\ref{eq:hTimelike}).

\subsection{Three-point functions}
In this section we describe our main new results, which are the explicit expressions for the $\hat{c} \leq 1$ structure constants of $\mathcal{N}=1$ timelike Liouville theory. We view these results as part of the fully quantum definition of the timelike theory, together with the spectra described in the previous section. Just as in ordinary timelike Liouville theory \cite{Zamolodchikov:2005fy}, the key idea relies on the observation that, even though the spacelike structure constants (\ref{eq: NS spacelike 1 and 2}), (\ref{eq: R spacelike}) \textit{cannot} be analytically continued as a function of the central charge, the basic shift relations that determine those data (\ref{eq: shift equations NS sector1}), (\ref{eq: shiftequations R sector}) can. The strategy then is to look for new solutions to those analytically continued shift relations at $\hat{c}\leq1$. We will systematize this analytic continuation by implementing the so-called `Virasoro-Wick Rotation' symmetry, first discussed at the level of the shift relations for the Virasoro modular and fusion kernels in \cite{Ribault:2023vqs}. This procedure will then clearly reveal how to construct new solutions for the structure constants valid at $\hat{c}\leq 1$ from the solutions valid at $c\geq9$.

\subsubsection{Virasoro--Wick Rotation}\label{sec:VWR}

\par Following \cite{Ribault:2023vqs}, we define the Virasoro–Wick Rotation (VWR) as the (anti-)involution map
\begin{equation}
    \left\{
\begin{aligned}
    p &\rightarrow \hat{p} = ip \\
    b &\rightarrow \hat{b} = -ib
\end{aligned}
\right\} 
\end{equation}
on the parameters $(p,b)$ characterizing the (holomorphic) conformal dimension $h$ and the central charge $c$ respectively. In usual bosonic Liouville theory, the central charge and conformal dimensions are parametrized as in (\ref{eq:hcN0Liouville}), which means that under VWR they map to
\begin{equation}\label{eq:VWRN0}
    \mathcal{N}=0: ~~~~ \left\{
\begin{aligned}
    c &\rightarrow  \hat{c}=26-c \\
    h_p &\rightarrow    \hat{h}_{\hat{p}}=1-h_p
\end{aligned}
\right\}  \ .
\end{equation}
Similarly, the $\mathcal{N}=1$ theory we parametrize as in (\ref{eq:ccN1}), (\ref{eq: conformal dimensions}) and therefore
\begin{equation}\label{eq:VWRN1}
    \mathcal{N}=1: ~~~~ \left\{
\begin{aligned}
    c &\rightarrow    \hat{c}=10-c \\
h^{\mathrm{NS}}_p &\rightarrow \hat{h}^{\mathrm{NS}}_{\hat{p}}=  \frac{1}{2}-h^{\mathrm{NS}}_p \\
h^{\mathrm{R}}_{p} &\rightarrow   \hat{h}^{\mathrm{R}}_{\hat{p}}=\frac{5}{8}-h^{\mathrm{R}}_{p}
\end{aligned}
\right\}  \ .
\end{equation} 
\par All CFT quantities of interest (structure constants, correlation functions etc.) depend in principle on $(h,c)$, which are \textit{even} functions w.r.t the parametrization $(p,b)$. Therefore, applying the VWR map \textit{twice} leaves all the corresponding expressions unaffected. It is interesting however to study the transformation of the CFT quantities under a \textit{single} VWR, assuming some analyticity in $p,b$. The structure constants of Liouville theory -- such as the DOZZ formula or the $\mathcal{N}=1$ structure constants (\ref{eq: NS spacelike 1 and 2}), (\ref{eq: R spacelike}) -- are famously non-analytic in $b$, and hence a single VWR is ill defined\footnote{Here we implicitly assume that we start with either a purely real or purely imaginary value of $b$, before performing the VWR.}. On the other hand, the basic shift relations that determine those data, such as eqns (\ref{eq: shift equations NS sector1}), (\ref{eq: shiftequations R sector}) in the $\mathcal{N}=1$ case, are analytic in $b$ and possess interesting properties under a single VWR which we now explain.
\par For the NS-sector, let us return back to the main shift relations (\ref{eq: shift equations NS sector1}) or (\ref{eq: shift equations NS sector2}) that determine the structure constants. The `bootstrap functions' $(\kappa_{\mathrm{NS}},\lambda_{\mathrm{NS}})$ and $(\widetilde{\kappa}_{\mathrm{NS}},\widetilde{\lambda}_{\mathrm{NS}})$ defined in (\ref{eq:kappaNS}), (\ref{eq:lambdaNS}) and (\ref{eq:tildekappatildelambda}) can be shown to obey the following non-trivial relations
\begin{equation}\label{eq:property kappa NS sector VWR}
\widetilde{\kappa}_{\mathrm{NS}}^{(b)}(p_1|p_2,p_3)= \frac{\left(\frac{p_1+ b}{p_1- b}\right)^2}{\kappa_{\mathrm{NS}}^{(-ib)}(-ip_1|-ip_2,-ip_3)} \ , \ \widetilde{\lambda}_{\mathrm{NS}}^{(b)}(p_1|p_2,p_3)= \frac{\left(\frac{p_1}{p_1- b}\right)^2}{\lambda_{\mathrm{NS}}^{(-ib)}(-ip_1|-ip_2,-ip_3)}~.
\end{equation}
Similarly for the R-sector, the bootstrap function 
 $\kappa_\mathrm{R}$ that determines the main shift relations (\ref{eq: shiftequations R sector}) and defined in (\ref{eq:kappaR}), satisfies
\begin{equation}\label{eq:property kappa R sector VWR}
    \kappa_{\mathrm{R}}^{(b)}(p_1|p_2;p_3) =\frac{1}{\kappa_{\mathrm{R}}^{(-ib)}(-ip_1|ip_2;-ip_3)}~.
\end{equation}
These equations demonstrate an almost inverse relation between the bootstrap functions evaluated at the original arguments and the bootstrap functions evaluated at the VWR-ed arguments (appropriately `coupled', in the NS-sector case). It is exactly this feature, as we will see, that will allow us to construct the timelike solutions for the structure constants as, roughly speaking, the inverse of the spacelike ones. This is reminiscent of the almost inverse relation between the timelike and spacelike structure constants in ordinary bosonic Liouville theory \cite{Kostov:2005kk}, which can be explained exactly along the same lines\footnote{For completeness, we present this derivation and the analogous equations for the bootstrap function in $\mathcal{N}=0$ Liouville in Appendix \ref{app:N0Liouville} (c.f. eqns (\ref{eq:N0cond2}),(\ref{eq:N0vwrcond})).}. We will describe concretely the construction in the $\mathcal{N}=1$ case in the next sections.
\par More generally, in \cite{Ribault:2023vqs} it was shown that a particular class of shift relations that determine the Virasoro modular and fusion kernels behave in a similar fashion under VWR. Therefore, one can in principle leverage this feature to construct analogous `timelike' solutions for the crossing kernels with (Virasoro) central charge $c\leq1$, starting from the kernels valid at $c\geq 25$. It can actually be shown that the structure constants of bosonic Liouville theory arise as special cases of the more general fusion and modular kernels \cite{Teschner:2001rv, Collier:2019weq}, which explains the applicability of the VWR from this broader framework to the specific quantities of Liouville theory. In the $\mathcal{N}=1$ case, it is expected that a similar connection between the fusion and modular kernels and Liouville's structure constant is also true, although it is not as concretely understood to date as in the non-supersymmetric case\footnote{In Appendix \ref{app:fusionkernel} we discuss the explicit relation of the NS fusion kernel with the spacelike NS structure constants of Liouville theory.}. 
\subsubsection{NS-sector}\label{sbsbsec:tNS} 
We start by discussing the operator product in the NS-sector, which we normalize in the same way as in the spacelike theory
\begin{multline}\label{eq:OPEtimelikeNS}
    \widehat{V}^{\mathrm{NS}}_{\hat{p}_1}(z)\widehat{V}^{\mathrm{NS}}_{\hat{p}_2}(0)\sim \int_{i\mathbb{R}_+-\epsilon} \frac{\d \hat{p}}{i} \frac{(z\bar{z})^{\hat{h}_{\hat{p}}-\hat{h}_{\hat{p}_1}-\hat{h}_{\hat{p}_2}}}{\widehat{B}_{\text{NS}}^{(\hat{b})}(\hat{p})}
  \Big(\widehat{C}_{\mathrm{NS}}^{(\hat{b})} (\hat{p}_1,\hat{p}_2,\hat{p}) \left[\widehat{V}^{\mathrm{NS}}_{\hat{p}}(0)\right]_{\mathrm{ee}}\cr-
     \widehat{\widetilde{C}}_{\mathrm{NS}}^{(\hat{b})}(\hat{p}_1,\hat{p}_2,\hat{p})\left[\widehat{V}^{\mathrm{NS}}_{\hat{p}}(0)\right]_{\mathrm{oo}}\Big).
\end{multline}
The corresponding two- and three-point functions that define $\mathcal{N}=1$ timelike Liouville theory in the NS-sector take the form
\begin{subequations}\label{eq: timelikeN1 structure constants}
\begin{align}
 \langle \widehat{V}^{\mathrm{NS}}_{\hat{p}_1}(0)\widehat{V}^{\mathrm{NS}}_{\hat{p}_2}(1)\rangle &= \widehat{B}^{(\hat{b})}_{\text{NS}}(\hat{p}_1)[\delta(\hat{p}_1-\hat{p}_2) + \delta(\hat{p}_1+\hat{p}_2)]~,\\
    \langle \widehat{V}^{\mathrm{NS}}_{\hat{p}_1}(z_1)\widehat{V}^{\mathrm{NS}}_{\hat{p}_2}(z_2)\widehat{V}^{\mathrm{NS}}_{\hat{p}_3}(z_3)\rangle &= \frac{\widehat{C}_{\mathrm{NS}}^{(\hat{b})}(\hat{p}_1,\hat{p}_2,\hat{p}_3)}{|z_{12}|^{2\hat{h}_{1+2-3}}|z_{23}|^{2\hat{h}_{2+3-1}}|z_{31}|^{2\hat{h}_{3+1-2}}}~,\\
    \langle \widehat{W}^{\mathrm{NS}}_{\hat{p}_1}(z_1)\widehat{V}^{\mathrm{NS}}_{\hat{p}_2}(z_2)\widehat{V}^{\mathrm{NS}}_{\hat{p}_3}(z_3)\rangle &= \frac{\widehat{\widetilde{C}}_{\mathrm{NS}}^{(\hat{b})}(\hat{p}_1,\hat{p}_2,\hat{p}_3)}{|z_{12}|^{2\hat{h}_{1+2-3}+1}|z_{23}|^{2\hat{h}_{2+3-1}-1}|z_{31}|^{2\hat{h}_{3+1-2}+1}}~.
\end{align}    
\end{subequations}

Our goal is to derive explcit expressions for these data starting from basic constraints coming from representation theory, exactly parallel to the spacelike case. More specifically, our strategy is to look for \textit{new} solutions to the main shift relations (\ref{eq: shift equations NS sector1}) in the timelike regime $b=-i\hat{b}$, $\hat{b}\in\mathbb{R}_{(0,1]}$. To achieve this, we will implement the properties of the shift relations under Virasoro Wick Rotation, as discussed in section \ref{sec:VWR}.

For $b\in (0,1]$ or $c\geq 9$, we have already found solutions to the main shift relations (i.e. the spacelike structure constants (\ref{eq: BNS}),(\ref{eq: NS spacelike 1 and 2})). In particular, let us express them as solutions to the alternative set of independent NS shift relations (\ref{eq: shift equations NS sector2}). We then have
\begin{align}\label{eq: shift equations NS sector2 again}
\frac{\widetilde{C}^{(b)}_{\text{NS}}(p_1+b,p_2,p_3)^2/B^{(b)}_{\text{NS}}(p_1+b)}{\tilde{C}^{(b)}_{\text{NS}}(p_1-b,p_2,p_3)^2/B^{(b)}_{\text{NS}}(p_1-b)}&=\widetilde{\kappa}_{\mathrm{NS}}^{(b)}(p_1|p_2,p_3) \ \nonumber , \\ 
\frac{C^{(b)}_{\text{NS}}(p_1,p_2,p_3)^2/B^{(b)}_{\text{NS}}(p_1)}{\widetilde{C}^{(b)}_{\text{NS}}(p_1-b,p_2,p_3)^2/B^{(b)}_{\text{NS}}(p_1-b)}&=\widetilde{\lambda}_{\mathrm{NS}}^{(b)}(p_1|p_2,p_3)~.
\end{align}
The functions $\widetilde{\kappa}_{\text{NS}}^{(b)}, \widetilde{\lambda}_{\text{NS}}^{(b)}$ are given in (\ref{eq:tildekappatildelambda}) and are analytic in both $b$ and the $p_i$'s. They further obey the `inverse' property (\ref{eq:property kappa NS sector VWR}) under VWR which we will shortly harness.

For $b= - i \hat{b}$, $\hat{b} \in \mathbb{R}_{(0,1]}$, or $\hat{c} \leq 1$, the CFT data (\ref{eq: timelikeN1 structure constants}) satisfy the \textit{same} shift relations in the corresponding central charge regime. Let us express those as \textit{unknown} solutions to, instead,  the original set of NS shift relations (\ref{eq: shift equations NS sector1}), namely
\begin{align}\label{eq: shift equations timelike NS 1}
    &\frac{\widehat{{C}}_{\text{NS}}^{(\hat{b})}(\hat{p}_1+(-i\hat{b}),\hat{p}_2,\hat{p}_3)^2/\widehat{B}^{(\hat{b})}_{\text{NS}}(\hat{p}_1+(-i\hat{b}))}{\widehat{{C}}_{\text{NS}}^{(\hat{b})}(\hat{p}_1-(-i\hat{b}),\hat{p}_2,\hat{p}_3)^2/\widehat{B}^{(\hat{b})}_{\text{NS}}(\hat{p}_1-(-i\hat{b}))}= {\kappa}_{\mathrm{NS}}^{(-i\hat{b})}(\hat{p}_1|\hat{p}_2,\hat{p}_3)~, \nonumber \\
    &\frac{\widehat{\widetilde{C}}_{\text{NS}}^{(\hat{b})}(\hat{p}_1,\hat{p}_2,\hat{p}_3)^2/\widehat{B}^{(\hat{b})}_{\text{NS}}(\hat{p}_1)}{\widehat{{C}}_{\text{NS}}^{(\hat{b})}(\hat{p}_1-(-i\hat{b}),\hat{p}_2,\hat{p}_3)^2/\widehat{B}^{(\hat{b})}_{\text{NS}}(\hat{p}_1-(-i\hat{b}))}= {\lambda}_{\mathrm{NS}}^{(-i\hat{b})}(\hat{p}_1|\hat{p}_2,\hat{p}_3)~.
\end{align}
There is an additional set of shift relations involving shifts by $b^{-1}=i\hat{b}^{-1}$, and hence identical to (\ref{eq: shift equations timelike NS 1}) with the substitution $\hat{b}\rightarrow-\hat{b}^{-1}$. The two sets of shift relations in $\hat{b},\hat{b}^{-1}$ are incommensurable for $\hat{b}^2\notin\mathbb{Q}$, which is what we assume henceforth.

We will now derive the new timelike solutions by massaging equation (\ref{eq: shift equations NS sector2 again}) into (\ref{eq: shift equations timelike NS 1}). We start by relabeling $b= \hat{b} \in \mathbb{R}$ in (\ref{eq: shift equations NS sector2 again}) and using the `inverse' property (\ref{eq:property kappa NS sector VWR}). This yields
\begin{align}\label{eq: shift equations NS sector2 VWR}
\frac{\widetilde{C}^{(\hat{b})}_{\text{NS}}(p_1+\hat{b},p_2,p_3)^2/B^{(\hat{b})}_{\text{NS}}(p_1+\hat{b})}{\tilde{C}^{(\hat{b})}_{\text{NS}}(p_1-\hat{b},p_2,p_3)^2/B^{(\hat{b})}_{\text{NS}}(p_1-\hat{b})}&=\widetilde{\kappa}_{\mathrm{NS}}^{(\hat{b})}(p_1|p_2,p_3) = \frac{\left(\frac{p_1+ \hat{b}}{p_1- \hat{b}}\right)^2}{\kappa_{\mathrm{NS}}^{(-i\hat{b})}(-ip_1|-ip_2,-ip_3)}\ \nonumber , \\ 
\frac{C^{(\hat{b})}_{\text{NS}}(p_1,p_2,p_3)^2/B^{(\hat{b})}_{\text{NS}}(p_1)}{\widetilde{C}^{(\hat{b})}_{\text{NS}}(p_1-\hat{b},p_2,p_3)^2/B^{(\hat{b})}_{\text{NS}}(p_1-\hat{b})}&=\widetilde{\lambda}_{\mathrm{NS}}^{(\hat{b})}(p_1|p_2,p_3)= \frac{\left(\frac{p_1}{p_1- \hat{b}}\right)^2}{\lambda_{\mathrm{NS}}^{(-i\hat{b})}(-ip_1|-ip_2,-ip_3)}~.
\end{align}
If we now rename $p_k \rightarrow i \hat{p}_k$, we get
\begin{align}\label{eq: shift equations NS sector2 VWR step2}
\frac{\widetilde{C}^{(\hat{b})}_{\text{NS}}(i(\hat{p}_1+(-i\hat{b})),i\hat{p}_2,i\hat{p}_3)^2/B^{(\hat{b})}_{\text{NS}}(i(\hat{p}_1+(-i\hat{b}))}{\tilde{C}^{(\hat{b})}_{\text{NS}}(i(\hat{p}_1-(-i\hat{b}),i\hat{p}_2,i\hat{p}_3)^2/B^{(\hat{b})}_{\text{NS}}(i(\hat{p}_1-(-i\hat{b}))} = \frac{\left(\frac{\hat{p}_1+ (-i\hat{b})}{\hat{p}_1-(-i \hat{b})}\right)^2}{\kappa_{\mathrm{NS}}^{(-i\hat{b})}(\hat{p}_1|\hat{p}_2,\hat{p}_3)}\ \nonumber , \\ 
\frac{C^{(\hat{b})}_{\text{NS}}(i\hat{p}_1,i\hat{p}_2,i\hat{p}_3)^2/B^{(\hat{b})}_{\text{NS}}(i\hat{p}_1)}{\widetilde{C}^{(\hat{b})}_{\text{NS}}(i(\hat{p}_1-(-i\hat{b}),i\hat{p}_2,i\hat{p}_3)^2/B^{(\hat{b})}_{\text{NS}}(i(\hat{p}_1-(-i\hat{b}))}= \frac{\left(\frac{\hat{p}_1}{\hat{p}_1-(-i\hat{b})}\right)^2}{\lambda_{\mathrm{NS}}^{(-i\hat{b})}(\hat{p}_1|\hat{p}_2,\hat{p}_3)}~,
\end{align}
which is equivalent to
\begin{align}\label{eq: shift equations NS sector2 VWR step3}
\left(\frac{\hat{p}_1+ (-i\hat{b})}{\hat{p}_1-(-i \hat{b})}\right)^2\frac{\widetilde{C}^{(\hat{b})}_{\text{NS}}(i(\hat{p}_1-(-i\hat{b})),i\hat{p}_2,i\hat{p}_3)^2/B^{(\hat{b})}_{\text{NS}}(i(\hat{p}_1-(-i\hat{b}))}{\tilde{C}^{(\hat{b})}_{\text{NS}}(i(\hat{p}_1+(-i\hat{b}),i\hat{p}_2,i\hat{p}_3)^2/B^{(\hat{b})}_{\text{NS}}(i(\hat{p}_1+(-i\hat{b}))} = {\kappa_{\mathrm{NS}}^{(-i\hat{b})}(\hat{p}_1|\hat{p}_2,\hat{p}_3)}\ \nonumber , \\ 
\left(\frac{\hat{p}_1}{\hat{p}_1-(-i\hat{b})}\right)^2\frac{\widetilde{C}^{(\hat{b})}_{\text{NS}}(i(\hat{p}_1-(-i\hat{b}),i\hat{p}_2,i\hat{p}_3)^2/B^{(\hat{b})}_{\text{NS}}(i(\hat{p}_1-(-i\hat{b}))}{ C^{(\hat{b})}_{\text{NS}}(i\hat{p}_1,i\hat{p}_2,i\hat{p}_3)^2/B^{(\hat{b})}_{\text{NS}}(i\hat{p}_1)}= {\lambda_{\mathrm{NS}}^{(-i\hat{b})}(\hat{p}_1|\hat{p}_2,\hat{p}_3)}~.
\end{align}
This system of shift relations, together with its corresponding one with $\hat{b}$ $\rightarrow -\hat{b}^{-1}$,
has exactly the form (\ref{eq: shift equations timelike NS 1}) that we initially set out to solve. A consistent choice for the new solutions then reads
\begin{subequations}\label{eq: NS timelike}
\begin{align}\label{eq: NS timelike 0}
\widehat{B}^{(\hat{b})}_{\text{NS}}(\hat{p}) &=\frac{1}{\hat{p}^2B^{(\hat{b})}_{\text{NS}}(i\hat{p})}   ~,\\ \label{eq: NS timelike 1}
     \widehat{C}_{\rm NS}^{(\hat{b})} (\hat{p}_1,\hat{p}_2,\hat{p}_3) &=\pm\frac{ i}{\widetilde{C}^{(\hat{b})}_{\text{NS}}(i\hat{p}_1,i\hat{p}_2,i\hat{p}_3)}  ~,\\ \label{eq: NS timelike 2}
    \widehat{\widetilde{C}}^{(\hat{b})}_{\rm NS} (\hat{p}_1,\hat{p}_2,\hat{p}_3) &=\pm \frac{ i}{C^{(\hat{b})}_{\text{NS}}(i\hat{p}_1,i\hat{p}_2,i\hat{p}_3)}  ~,
\end{align}    
\end{subequations}
or, explicitly
\begin{subequations}\label{eq: NS timelike explicit}
\begin{align}
\widehat{B}^{(\hat{b})}_{\text{NS}}(\hat{p}) &= \frac{4 \sinh(\pi \hat{b} \hat{p})\sinh(\pi \hat{b}^{-1}\hat{p})}{\hat{p}^2}\ ,  \\
     \widehat{C}_{\rm NS}^{(\bb)} (\pp_1,\pp_2,\pp_3) &=\pm \frac{ \Gamma_{\hat{b}}^{\rm NS}(\hat{b}^{-1}+ \hat{b})^3}{\Gamma_{\hat{b}}^{\rm NS}(2(\hat{b}^{-1} + \hat{b}))}\frac{\prod_{j=1}^3\Gamma_{\bb}^{\rm NS}\left(\bb+\bb^{-1} \pm 2 i\pp_j\right)}{\Gamma^{\rm R}_{\bb}\left(\frac{\bb+\bb^{-1}}{2} \pm i\pp_1 \pm i\pp_2 \pm i\pp_3\right)}~,\\
    \widehat{\widetilde{C}}^{(\bb)}_{\rm NS} (\pp_1,\pp_2,\pp_3)&= 
   \pm \frac{ 2i\Gamma_{\hat{b}}^{\rm NS}(\hat{b}^{-1}+ \hat{b})^3}{\Gamma_{\hat{b}}^{\rm NS}(2(\hat{b}^{-1} + \hat{b}))}
    \frac{\prod_{j=1}^3\Gamma_{\bb}^{\rm NS}\left(\bb+\bb^{-1} \pm 2 i\pp_j\right)}{\Gamma^{\rm NS}_{\bb}\left(\frac{\bb+\bb^{-1}}{2} \pm i\pp_1 \pm i\pp_2 \pm i\pp_3\right)}~.
\end{align} 
\end{subequations}
\par These are our main results for the timelike NS structure constants. It is evident from the above that, as in the non-supersymmetric setting, the $\mathcal{N}=1$ timelike structure constants are \textit{not} the analytic continuation of their spacelike counterparts. It is also interesting that the structure constant $\widehat{C}_{\rm NS}$, corresponding roughly to the integer-descendant module (c.f. the OPE (\ref{eq:OPEtimelikeNS})), is given by the inverse of $\widetilde{C}_{\mathrm{NS}}$ in the spacelike case (which corresponds to the half-integer descendant module), and similarly for the pair $\widehat{\widetilde{C}}_{\rm NS}\leftrightarrow C_{\mathrm{NS}}$. This fact traces back to the way the shift relations are coupled under VWR, as dictated by (\ref{eq:property kappa NS sector VWR}).
\par Notice that we have kept explicit a $\pm$ ambiguity in the definition of the structure constants, as the shift relations are insensitive to this choice\footnote{Of course, similar ambiguities come with the two-point normalization, though we will not discuss any particular choice at this stage. We will return to this issue in section \ref{subsec:MM} where we discuss the connection with the Minimal Models.}. We will return to this point in section \ref{sec:discussion}. On the other hand, the choice for the overall $b-$dependent factor (which the shift relations are also insensitive to) is essentially fixed by the spacelike expressions in our conventions \footnote{We simply chose an extra factor of $i$ in (\ref{eq: NS timelike 1}) in order to cancel the $i$ in the definition of $\widetilde{C}_{\mathrm{NS}}^{(b)}$, while in (\ref{eq: NS timelike 2}) we choose an $i$ to make the overall constant look similar to the spacelike case.}.
\paragraph{Properties of the three-point functions.} 
\par The expressions (\ref{eq: NS timelike explicit}) are the \textit{unique} solutions to the NS shift relations in the superconformal central charge regime $\hat{c}\leq 1$, exactly in the same sense that the bosonic timelike structure constants are \cite{Zamolodchikov:1995aa, Teschner:1995yf}. In particular, modulo a momentum-independent factor, they are the unique solutions with the following features:
\begin{itemize}
    \item Continuous in $\hat{b}$, and invariant under $\hat{b}\leftrightarrow -\hat{b}^{-1}$,
    \item Meromorphic in the momenta $\hat{p}_i$,
    \item Permutation symmetric under the exchange of any two momenta,
    \item Reflection symmetric under any $\hat{p}_i\rightarrow - \hat{p}_i$.
\end{itemize}
It is now interesting to ask about the limits $\hat{p}_3\rightarrow \hat{p}_{\langle1,1\rangle}$ in the expressions (\ref{eq: NS timelike explicit}). It is straightforward to check that
\begin{align}
    \lim_{\pp_3\rightarrow\hat{p}_{\langle1,1\rangle}}\widehat{C}_{\rm NS}^{(\bb)} (\pp_1,\pp_2,\pp_3)&\neq  \delta(\hat{p}_1\pm \hat{p}_2)~,
         \nonumber \\
         \lim_{\pp_3\rightarrow\hat{p}_{\langle1,1\rangle}}\widehat{\widetilde{C}}^{(\bb)}_{\rm NS} (\pp_1,\pp_2,\pp_3)&\neq 0.
\end{align}
In particular, both of these limits turn out to be finite (and non-zero). In other words, we observe that not only the `diagonal' structure of the two-point functions is not recovered from the limit $\pp_3\rightarrow\hat{p}_{\langle1,1\rangle}$ of $\widehat{C}_{\rm NS}$, but also the fact that the identity module is built purely out of integer-level descendants is not encoded in the expression for $\widehat{\widetilde{C}}_{\rm NS}$. This is in contrast to the analogous statements we discussed in the spacelike case (c.f. (\ref{eq:IdLimitNSspacelike})) and reflects the fact that the operator with momentum $\hat{p}_{\langle1,1\rangle}$ is \textit{not} to be identified with the usual identity operator. These apparent `problematic' features of the theory are attributed to its non-unitary nature, analogously to the non-supersymmetric setup \cite{Ribault:2015sxa,Harlow:2011ny}.   
\par Instead, the correct statement is that the two-point functions $\langle\widehat{V}^{\mathrm{NS}}\widehat{V}^{\mathrm{NS}}\rangle$, $\langle\widehat{W}^{\mathrm{NS}}\widehat{V}^{\mathrm{NS}}\rangle$ are \textit{not} obtained by analytic continuation from the corresponding three-point functions, and we define the theory by imposing
    \begin{align}
        \langle \widehat{V}^{\mathrm{NS}}_{\hat{p}_1}(0) \widehat{V}^{\mathrm{NS}}_{\hat{p}_2}(1)\rangle &= \frac{\rho^{(\hat{b})}_{\mathrm{NS}}(i\hat{p}_1)}{\hat{p}_1^2} \left(\delta(\hat{p_1}- \hat{p}_2) +\delta(\hat{p_1} +\hat{p}_2) \right), ~~ \text{and} ~~\\
        \langle \widehat{W}^{\mathrm{NS}}_{\hat{p}_1}(0) \widehat{V}^{\mathrm{NS}}_{\hat{p}_2}(1)\rangle &=0.
    \end{align}


\subsubsection{R-sector}\label{sbsbsec:tR}
In the R-sector we proceed similarly. We define the corresponding two- and three-point functions for $\hat{c}\leq1$ as
\begin{align}
    \langle \widehat{V}^{\mathrm{R},\epsilon}_{\hat{p}_1}(0)\widehat{V}^{\mathrm{R},\epsilon}_{\hat{p}_2}(1)\rangle &= \widehat{B}_{\text{R}}^{(\hat{b})}(\hat{p}_1)[\delta(\hat{p}_1-\hat{p}_2) + \epsilon\,\delta(\hat{p}_1+\hat{p}_2)]~, \nonumber\\
     \langle \widehat{V}^{\mathrm{R},\epsilon}_{\hat{p}_1}(z_1)\widehat{V}^{\mathrm{R},\epsilon}_{\hat{p}_2}(z_2) \widehat{V}^{\mathrm{NS}}_{\hat{p}_3}(z_3) \rangle &= \frac{\widehat{C}_{\mathrm{R},\epsilon}^{(\hat{b})}(\hat{p}_1,\hat{p}_2;\hat{p}_3)}{|z_{12}|^{2\hat{h}_{[1]+[2]-3}}|z_{23}|^{2\hat{h}_{[2]+3-[1]}}|z_{31}|^{2\hat{h}_{3+[1]-[2]}}}~,\label{eq:tscR}
\end{align}
and we define again the following combinations of three-point correlators
\begin{align}
 \hspace{-2cm} \widehat{C}_{\mathrm{even}}^{(\hat{b})}(\hat{p}_1,\hat{p}_2;\hat{p}_3) &\equiv \frac{1}{2} \left(\langle \widehat{V}_{\hat{p}_1}^{\mathrm{R,+}} \widehat{V}_{\hat{p}_2}^{\mathrm{R,+}} \widehat{V}_{\hat{p}_3}^{\mathrm{NS}}\rangle + \langle \widehat{V}_{\hat{p}_1}^{\mathrm{R,-}} \widehat{V}_{\hat{p}_2}^{\mathrm{R,-}}\widehat{V}_{\hat{p}_3}^{\mathrm{NS}}\rangle \right)~,\nonumber\\
 \widehat{C}_{\mathrm{odd}}^{(\hat{b})}(\hat{p}_1,\hat{p}_2;\hat{p}_3) &\equiv \frac{1}{2} \left(\langle \widehat{V}_{p_1}^{\mathrm{R,+}} \widehat{V}_{\hat{p}_2}^{\mathrm{R,+}} \widehat{V}_{\hat{p}_3}^{\mathrm{NS}}\rangle - \langle \widehat{V}_{\hat{p}_1}^{\mathrm{R,-}} \widehat{V}_{\hat{p}_2}^{\mathrm{R,-}} \widehat{V}_{\hat{p}_3}^{\mathrm{NS}}\rangle\right)~.\label{eq:CeCotl}
 \end{align}
\par We will determine those data starting from the basic Ramond sector shift relations (\ref{eq: shiftequations R sector}), (\ref{eq: shift equations R sectorbinv}) which are analytic in $b$, and use their properties under VWR. 
\par For $b\in\mathbb{R}_{(0,1]}$ or $c\geq9$, the spacelike solutions (\ref{eq: osp12 normalization 2pt function spacelike R sector}), (\ref{eq: R spacelike}) satisfy
\begin{align}
    \frac{C_{\mathrm{even}}^{(b)}(p_1+\frac{b}{2},p_2; p_3)^2/B_{\mathrm{R}}^{(b)}(p_1+\frac{b}{2})}{C_{\mathrm{odd}}^{(b)}(p_1-\frac{b}{2}, p_2; p_3)^2/B_{\mathrm{R}}^{(b)}(p_1-\frac{b}{2})} &=\kappa^{(b)}_{\mathrm{R}}\left(p_1|p_2;p_3\right) ~ ,  \cr
\frac{C_{\mathrm{odd}}^{(b)}(p_1+\frac{b}{2},p_2; p_3)^2/B_{\mathrm{R}}^{(b)}(p_1+\frac{b}{2})}{C_{\mathrm{even}}^{(b)}(p_1-\frac{b}{2}, p_2; p_3)^2/B_{\mathrm{R}}^{(b)}(p_1-\frac{b}{2})} &=\kappa^{(b)}_{\mathrm{R}}\left(p_1|-p_2;p_3\right).\label{eq:brealshiftsR}
\end{align}
The function $\kappa^{(b)}_{\mathrm{R}}$ is given in (\ref{eq:kappaR}) and is manifestly analytic in both $p_i$'s and $b$. In particular, as we discussed, it satisfies the `inverse' property (\ref{eq:property kappa R sector VWR}) which we will put to work shortly.
\par For $b=-i\hat{b}$, $\hat{b}\in\mathbb{R}_{(0,1]}$, or $\hat{c}\leq1$, the Ramond data (\ref{eq:tscR}), (\ref{eq:CeCotl}) satisfy the \textit{same} shift relations in the corresponding central charge regime, namely
\begin{align}
\frac{\widehat{C}_{\mathrm{even}}^{(\hat{b})}(\hat{p}_1+\frac{(-i\hat{b})}{2}, \hat{p}_2; \hat{p}_3)^2/\widehat{B}_{\mathrm{R}}^{(\hat{b})}(\hat{p}_1+\frac{(-i\hat{b})}{2})}{\widehat{C}_{\mathrm{odd}}^{(\hat{b})}(\hat{p}_1-\frac{(-i\hat{b})}{2}, \hat{p}_2; \hat{p}_3)^2/\widehat{B}_{\mathrm{R}}^{(\hat{b})}(\hat{p}_1-\frac{(-i\hat{b})}{2})} &= \kappa_{\mathrm{R}}^{(-i\hat{b})}(\hat{p}_1|\hat{p}_2;\hat{p}_3)~,\nonumber\\
\frac{\widehat{C}_{\mathrm{odd}}^{(\hat{b})}(\hat{p}_1+\frac{(-i\hat{b})}{2},\hat{p}_2; \hat{p}_3)^2/\widehat{B}_{\mathrm{R}}^{(\hat{b})}(\hat{p}_1+\frac{(-i\hat{b})}{2})}{\widehat{C}_{\mathrm{even}}^{(\hat{b})}(\hat{p}_1-\frac{(-i\hat{b})}{2}, \hat{p}_2; \hat{p}_3)^2/\widehat{B}_{\mathrm{R}}^{(\hat{b})}(\hat{p}_1-\frac{(-i\hat{b})}{2})} &=\kappa_{\mathrm{R}}^{(-i\hat{b})}(\hat{p}_1|-\hat{p}_2;\hat{p}_3)~,\label{eq:bhrealshiftsR}
\end{align}
together with the complement shift relations with $\hat{b}\rightarrow-\hat{b}^{-1}$.
\par In the same way as before, one can massage equation (\ref{eq:brealshiftsR}) into (\ref{eq:bhrealshiftsR}) by renaming $b\rightarrow\hat{b}$, $p_k\rightarrow i\hat{p}_k$, together with the `inverse' property (\ref{eq:property kappa R sector VWR}). This yields
\begin{align}
\frac{C_{\mathrm{odd}}^{(\hat{b})}(i(\hat{p}_1-\frac{(-i\hat{b})}{2}), i\hat{p}_2; i\hat{p}_3)^2/B_{\mathrm{R}}^{(\hat{b})}(i(\hat{p}_1-\frac{(-i\hat{b})}{2}))}{C_{\mathrm{even}}^{(\hat{b})}(i(\hat{p}_1+\frac{(-i\hat{b})}{2}, i\hat{p}_2; i\hat{p}_3)^2/B_{\mathrm{R}}^{(b)}(i(\hat{p}_1+\frac{(-i\hat{b})}{2}))} &= {\kappa_{\mathrm{R}}^{(-i\hat{b})}(\hat{p}_1|-\hat{p}_2;\hat{p}_3)}~,\nonumber\\
\frac{C_{\mathrm{even}}^{(\hat{b})}(i(\hat{p}_1-\frac{(-i\hat{b})}{2}), i\hat{p}_2; i\hat{p}_3)^2/B_{\mathrm{R}}^{(\hat{b})}(\hat{p}_1-\frac{(-i\hat{b})}{2}))}{C_{\mathrm{odd}}^{(\hat{b})}(i(\hat{p}_1+\frac{(-i\hat{b})}{2}), i\hat{p}_2; i\hat{p}_3)^2/B_{\mathrm{R}}^{(\hat{b})}(i(\hat{p}_1+\frac{(-i\hat{b})}{2}))} &={\kappa_{\mathrm{R}}^{(-i\hat{b})}(\hat{p}_1|\hat{p}_2;\hat{p}_3)}~.
\end{align}
Comparing this with (\ref{eq:bhrealshiftsR}), a consistent choice for the new solutions reads
\begin{align}
\widehat{B}_{\mathrm{R}}^{(\hat{b})}(\hat{p}) &= \frac{1}{B_{\mathrm{R}}^{(\hat{b})}(i\hat{p})}~,\nonumber\\
    \widehat{C}_{\mathrm{even}}^{(\hat{b})}(\hat{p}_1,\hat{p}_2;\hat{p}_3) &= \pm\frac{ 1}{C_{\mathrm{odd}}^{(\hat{b})}(i\hat{p}_1,i\hat{p}_2;i\hat{p}_3)} ~,\nonumber\\
    \widehat{C}_{\mathrm{odd}}^{(\hat{b})}(\hat{p}_1,\hat{p}_2;\hat{p}_3) & = \pm\frac{ 1}{C_{\mathrm{even}}^{(\hat{b})}(i\hat{p}_1,i\hat{p}_2;i\hat{p}_3)} ~.
\end{align}    
or, explicitly
\begin{align}
  \widehat{B}_{\text{R}}^{(\hat{b})}(\hat{p})&=2\sqrt{2}\cosh(\pi \hat{b} \pp)\cosh(\pi \hat{b}^{-1}\pp)~,\nonumber \\
  \widehat{C}_{\mathrm{R},\epsilon}^{(\hat{b})}(\hat{p}_1,\hat{p}_2;\hat{p}_3) &\equiv  \widehat{C}_{\mathrm{even}}^{(\hat{b})}(\hat{p}_1,\hat{p}_2;\hat{p}_3) +\epsilon\,  \widehat{C}_{\mathrm{odd}}^{(\hat{b})}(\hat{p}_1,\hat{p}_2;\hat{p}_3) ~,\nonumber \\
 \widehat{C}_{\mathrm{even}}^{(\hat{b})}(\hat{p}_1,\hat{p}_2;\hat{p}_3) &=   \frac{\pm\sqrt{2} \Gamma_{\hat{b}}^{\rm NS}(\hat{b}^{-1}+ \hat{b})^3\Gamma_{\hat{b}}^{\mathrm{NS}}(\hat{b}+\hat{b}^{-1}\pm 2i\hat{p}_3)\prod_{j=1}^2\Gamma_{\hat{b}}^{\mathrm{R}}(\hat{b}+\hat{b}^{-1}\pm 2i\hat{p}_j)}{\Gamma_{\hat{b}}^{\rm NS}(2(\hat{b}^{-1} + \hat{b}))\Gamma_{\hat{b}}^{\mathrm{NS}}\left(\frac{\hat{b}+\hat{b}^{-1}}{2}\pm i(\hat{p}_1+\hat{p}_2)\pm i\hat{p}_3\right)\Gamma_{\hat{b}}^{\mathrm{R}}\left(\frac{\hat{b}+\hat{b}^{-1}}{2}\pm i(\hat{p}_1-\hat{p}_2)\pm i\hat{p}_3\right)}~,\nonumber \\
 \widehat{C}_{\mathrm{odd}}^{(\hat{b})}(\hat{p}_1,\hat{p}_2;\hat{p}_3) &=  \frac{\pm \sqrt{2}\Gamma_{\hat{b}}^{\rm NS}(\hat{b}^{-1}+ \hat{b})^3\Gamma_{\hat{b}}^{\mathrm{NS}}(\hat{b}+\hat{b}^{-1}\pm 2i\hat{p}_3)\prod_{j=1}^2\Gamma_{\hat{b}}^{\mathrm{R}}(\hat{b}+\hat{b}^{-1}\pm 2i\hat{p}_j)}{\Gamma_{\hat{b}}^{\rm NS}(2(\hat{b}^{-1} + \hat{b}))\Gamma_{\hat{b}}^{\mathrm{R}}\left(\frac{\hat{b}+\hat{b}^{-1}}{2}\pm i(\hat{p}_1+\hat{p}_2)\pm i\hat{p}_3\right)\Gamma_{\hat{b}}^{\mathrm{NS}}\left(\frac{\hat{b}+\hat{b}^{-1}}{2}\pm i(\hat{p}_1-\hat{p}_2)\pm i\hat{p}_3\right)}~.\label{eq:collecttR}
\end{align}  
\par These are our main results for the $\mathcal{N}=1$ timelike Liouville structure constants in the Ramond sector. In particular, it is evident again that they are \textit{not} given by the analytic continuation of their spacelike counterparts. We have also kept explicit a $\pm$ ambiguity and, just as in the NS-sector, the choice for the overall $b-$dependent factor (which the shift relations are insensitive to) is essentially fixed by the spacelike expressions in our conventions.
\paragraph{Properties of the three-point functions.} The expressions (\ref{eq:collecttR}) are again the \textit{unique} solutions to the shift relations (\ref{eq:bhrealshiftsR}) valid for $\hat{c}\leq 1$ (modulo a momentum-independent factor) with the following features: 
\begin{itemize}
    \item Continuous in $\hat{b}$, and invariant under $\hat{b}\leftrightarrow -\hat{b}^{-1}$
    \item Meromorphic in the momenta $\hat{p}_i$
    \item Permutation symmetric under the exchange of $\hat{p}_1\leftrightarrow \hat{p}_2$
    \item Reflection symmetric under the NS-sector momentum $\hat{p}_3\rightarrow -\hat{p}_3$, while 
    \begin{equation}
        \frac{\widehat{C}_{\mathrm{even}}^{(\hat{b})}(\hat{p}_1,-\hat{p_2};\hat{p}_3)}{\widehat{C}_{\mathrm{odd}}^{(\hat{b})}(\hat{p}_1,\hat{p}_2;\hat{p}_3)}=1 \ , \ \  \frac{\widehat{C}_{\mathrm{even}}^{(\hat{b})}(-\hat{p}_1,\hat{p_2};\hat{p}_3)}{\widehat{C}_{\mathrm{odd}}^{(\hat{b})}(\hat{p}_1,\hat{p}_2;\hat{p}_3)}=1~.
    \end{equation}
\end{itemize}
It is straightforward to check finally that also in this case the limit $\hat{p}_3\rightarrow \hat{p}_{\langle 1,1\rangle }$ of the structure constants does \textit{not} reproduce the diagonal structure of the two-point function, and to define the theory we simply impose
    \begin{equation}
        \langle \widehat{V}^{\mathrm{R},\pm}_{\hat{p}_1}(0) \widehat{V}^{\mathrm{R},\pm}_{\hat{p}_2}(1)\rangle = {\rho^{(\hat{b})}_{\mathrm{R}}(i\hat{p}_1)} \left(\delta(\hat{p_1}- \hat{p}_2) \pm \delta(\hat{p_1} +\hat{p}_2) \right)~.
    \end{equation}


\subsection{Generalized $\mathcal{N}=1$ Minimal Models}\label{subsec:MM}
We now turn to the connection of $\mathcal{N}=1$ timelike Liouville theory with the $\mathcal{N}=1$ superconformal minimal models. In the non-supersymmetric case \cite{Zamolodchikov:2005fy}, a connection of the timelike theory was established with the so-called \textit{generalized minimal models} which have roughly the features of the usual minimal models except certain analyticity in the central charge and the spectrum is assumed. We will see that the same connection is true in the $\mathcal{N}=1$ case. We start with a small review of the ordinary $\mathcal{N}=1$ minimal models \cite{Friedan:1984rv}, \cite{Zamolodchikov:1988nm}.  
\par The $\mathcal{N}=1$ superconformal minimal models are described by a pair of positive integers $(p,p'>p)$ with $p,p'\geq 2$. The unitary series is characterized by $p'=p+2$, whereas more generally one can have either $(p,p')$ odd and coprime, or $(p,p')$ even, with $(p/2,p'/2)$ coprime and $(p-p')/2$ odd \cite{DiFrancesco:1988xz} \footnote{This last requirement comes from modular invariance; see also Appendix B of \cite{, Klebanov:2003wg} on this point.}. The central charge is characterized by rational $\hat{b}^2$, i.e.   
\begin{equation}
    \hat{b}^2=p'/p \quad \Longrightarrow \quad \hat{c}_{p,p'}=1-\frac{2(p-p')^2}{pp'}.
\end{equation}
The primary operators $\mathcal{O}_{r,s}$ are labelled by two positive integers $r,s$ subject to
\begin{equation}
    1\leq s\leq p'-1, \quad 1\leq r\leq p-1, \quad rp'\geq sp,
\end{equation}
and are identified as $\mathcal{O}_{r,s}\equiv \mathcal{O}_{p-r,p'-s}$. Their conformal dimensions read
\begin{equation}
    h_{(r,s)}=\frac{\left(rp'-sp\right)^2-\left(p-p'\right)^2}{8pp'}+\frac{1-(-1)^{r-s}}{32},
\end{equation}
and correspond to NS operators for even $s-r$, and to R operators for odd $s-r$. A prototypical example that belongs in the unitary class is the tri-critical Ising model with $p=3,p'=5$. It contains two operators in the NS-sector, namely $\mathcal{O}_{1,1}$ (i.e. the vacuum) and $\mathcal{O}_{1,3}$, and two operators in the R-sector, $\mathcal{O}_{1,2}$, $\mathcal{O}_{2,1}$\footnote{Of course it is known that the tri-critical Ising model also happens to be minimal w.r.t the Virasoro subalgebra.} \cite{Qiu:1986if}.
\par Due the existence of null states, the operators  of the $(p,p')$ minimal model are characterized by a truncated fusion algebra in the OPE of two fields, analogously to the usual Virasoro minimal models \cite{Belavin:1984vu, Friedan:1983xq}. In particular, it has been shown that the Coulomb gas formalism of the Virasoro minimal models, as prescribed by Dotsenko and Fateev \cite{Dotsenko:1984nm, Dotsenko:1984ad, Dotsenko:1985hi}, can be extended to the case of $\mathcal{N}=1$ superconformal minimal models, allowing in principle for the extraction of explicit expressions for the OPE coefficients \cite{Kitazawa:1987za, Alvarez-Gaume:1991nvs,Mussardo:1987eq,Jayaraman:1989as}.
\par Similar to what was observed in \cite{Zamolodchikov:2005fy}, we will see that the timelike structure constants of $\mathcal{N}=1$ Liouville theory, when certain momenta are evaluated at the values of the degenerate representations, agree with the OPE coefficients of the $\mathcal{N}=1$ minimal models. However, just as in \cite{Zamolodchikov:2005fy}, one \textit{cannot} recover the truncated fusion rules from the $\mathcal{N}=1$ timelike expressions for the structure constants. It is in this sense that the latter should be understood as describing a \textit{formal} series of the so-called \textit{generalized} $\mathcal{N}=1$ minimal models, where the central charge is treated as a continuous parameter (in particular, $\hat{b}^2\notin\mathbb{Q}$) and the (formal) spectrum of primary operators $\mathcal{O}_{r,s}$ is not characterized by a truncated fusion algebra \cite{Belavin:2008vc}.  
\par To establish this connection we will have to switch to the canonical normalization of operators in timelike Liouville theory, namely one where all the two-point functions are unit-normalized. This is different than the one we discussed in section \ref{sec:Timelike}. Doing that consistently, we obtain the following expressions for the structure constants for $\hat{b}\in\mathbb{R}_{(0,1]}$ (or $\hat{c}\leq1$):
\begin{align}
    &\hat{\mathcal{B}}_{\mathrm{NS}}(\hat{p})=1 \ , \nonumber \\
    &\widehat{\mathcal{C}}^{(\hat{b})}_{\mathrm{NS}}(\hat{p}_1,\hat{p}_2,\hat{p}_3)=A_{\mathrm{NS}}\frac{\left[\prod_{j=1}^3\Gamma_{\hat{b}}^{\mathrm{R}}\left(\pm2i\hat{p}_j+\hat{b}\right)\Gamma_{\hat{b}}^{\mathrm{R}}\left(\pm2i\hat{p}_j+\hat{b}^{-1}\right)\right]^{1/2}}{\Gamma_{\hat{b}}^{\mathrm{R}}\left(\frac{\hat{b}+\hat{b}^{-1}}{2}\pm i\hat{p}_1\pm i\hat{p}_2\pm i\hat{p}_3\right)}, \nonumber \\
    &\widehat{\widetilde{\mathcal{C}}}^{(\hat{b})}_{\mathrm{NS}}(\hat{p}_1,\hat{p}_2,\hat{p}_3)=2iA_{\mathrm{NS}}\frac{\left[\prod_{j=1}^3\Gamma_{\hat{b}}^{\mathrm{R}}\left(\pm2i\hat{p}_j+\hat{b}\right)\Gamma_{\hat{b}}^{\mathrm{R}}\left(\pm2i\hat{p}_j+\hat{b}^{-1}\right)\right]^{1/2}}{\Gamma_{\hat{b}}^{\mathrm{NS}}\left(\frac{\hat{b}+\hat{b}^{-1}}{2}\pm i\hat{p}_1\pm i\hat{p}_2\pm i\hat{p}_3\right)}\label{eq:MMnormNS}
\end{align}
for the NS-sector, and
\begin{align}
    &\hat{\mathcal{B}}_{\mathrm{R}}(\hat{p})= 1 \ , \nonumber \\
    &\widehat{\mathcal{C}}^{(\hat{b})}_{\mathrm{even}}(\hat{p}_1,\hat{p}_2;\hat{p}_3)=A_{\mathrm{R}}\times\nonumber \\ 
    &~~\frac{\left[\Gamma_{\hat{b}}^{\mathrm{R}}\left(\pm2i\hat{p}_3+\hat{b}\right)\Gamma_{\hat{b}}^{\mathrm{R}}\left(\pm2i\hat{p}_3+\hat{b}^{-1}\right)\prod_{j=1}^2\Gamma_{\hat{b}}^{\mathrm{NS}}\left(\pm2i\hat{p}_j+\hat{b}\right)\Gamma_{\hat{b}}^{\mathrm{NS}}\left(\pm2i\hat{p}_j+\hat{b}^{-1}\right)\right]^{1/2}}{\Gamma_{\hat{b}}^{\mathrm{NS}}\left(\frac{\hat{b}+\hat{b}^{-1}}{2}\pm i(\hat{p}_1+\hat{p}_2)\pm i\hat{p}_3\right)\Gamma_{\hat{b}}^{\mathrm{R}}\left(\frac{\hat{b}+\hat{b}^{-1}}{2}\pm i(\hat{p}_1-\hat{p}_2)\pm i\hat{p}_3\right)}, \nonumber \\
    &\widehat{\mathcal{C}}^{(\hat{b})}_{\mathrm{odd}}(\hat{p}_1,\hat{p}_2;\hat{p}_3)=A_{\mathrm{R}}\times\nonumber\\
    &~~\frac{\left[\Gamma_{\hat{b}}^{\mathrm{R}}\left(\pm2i\hat{p}_3+\hat{b}\right)\Gamma_{\hat{b}}^{\mathrm{R}}\left(\pm2i\hat{p}_3+\hat{b}^{-1}\right)\prod_{j=1}^2\Gamma_{\hat{b}}^{\mathrm{NS}}\left(\pm2i\hat{p}_j+\hat{b}\right)\Gamma_{\hat{b}}^{\mathrm{NS}}\left(\pm2i\hat{p}_j+\hat{b}^{-1}\right)\right]^{1/2}}{\Gamma_{\hat{b}}^{\mathrm{R}}\left(\frac{\hat{b}+\hat{b}^{-1}}{2}\pm i(\hat{p}_1+\hat{p}_2)\pm i\hat{p}_3\right)\Gamma_{\hat{b}}^{\mathrm{NS}}\left(\frac{\hat{b}+\hat{b}^{-1}}{2}\pm i(\hat{p}_1-\hat{p}_2)\pm i\hat{p}_3\right)}\label{eq:MMnormR}
\end{align}
for the R-sector. It is straightforward to verify that these expressions satisfy the main shift relations (\ref{eq: shift equations NS sector1}), (\ref{eq: shiftequations R sector}) respectively for $b=-i\hat{b}$, $\hat{b}\in\mathbb{R}_{(0,1]}$. \par Following \cite{Zamolodchikov:2005fy}, we will determine the overall constants $A_{\mathrm{NS}},A_{\mathrm{R}}$ from the requirement that the two-point function of identical operators yields unity. Denoting $\hat{p}_{\mathds{1}}\equiv\frac{i(\hat{b}^{-1}-\hat{b})}{2}$, somewhat surprisingly we find that the two constants are given by the same expression
\begin{align}\label{eq:MMnormalizationId}
   &\widehat{\mathcal{C}}^{(\hat{b})}_{\mathrm{NS}}\left(\hat{p},\hat{p},\hat{p}_{\mathds{1}}\right)=1 \ \ \ \ \ \ \ \ \ \Longrightarrow \ \ \quad \quad   A_{\mathrm{NS}}\equiv\frac{\hat{b}^{\frac{\hat{b}^{-2}-\hat{b}^{2}}{2}-1}\left[\gamma\left(\frac{\hat{b}^2+1}{2}\right)\gamma\left(\frac{\hat{b}^{-2}-1}{2}\right)\right]^{1/2}}{\Upsilon_{\hat{b}}^{\mathrm{R}}(\hat{b})} \ , \nonumber \\ &\widehat{\mathcal{C}}^{(\hat{b})}_{\mathrm{even}}\left(\hat{p},\hat{p};\hat{p}_{\mathds{1}}\right)=\widehat{\mathcal{C}}^{(\hat{b})}_{\mathrm{odd}}\left(\hat{p},-\hat{p};\hat{p}_{\mathds{1}}\right)=1 \ \ \ \ \ \ \ \ \ \Longrightarrow \ \ \ \ \ \ \    A_{\mathrm{R}}=A_{\mathrm{NS}}~.
\end{align} 
Note that we do not claim that we somehow recovered the \textit{diagonal} structure of the two-point function in this normalization. This is not possible in general, regardless of the chosen normalization of the timelike structure constants as we emphasized in section \ref{sec:Timelike}. Having said that, it is understood that the quantities $\widehat{\mathcal{C}}^{(\hat{b})}_{\mathrm{NS}}\left(\hat{p},\hat{p},\hat{p}_{\mathds{1}}\right),\widehat{\mathcal{C}}^{(\hat{b})}_{\mathrm{even}}\left(\hat{p},\hat{p};\hat{p}_{\mathds{1}}\right)$ in (\ref{eq:MMnormalizationId}) are well-defined and finite.
\par It might initially appear problematic that the structure constants in the normalization (\ref{eq:MMnormNS}), (\ref{eq:MMnormR}) involve square roots, as this comes at the cost of meromorphicity. However, the advantage is that these expressions match with the corresponding OPE coefficients of the $\mathcal{N}=1$ minimal models, when certain momenta are specified at degenerate values. Indeed, one can compare directly e.g. with the expressions in the NS-sector written down in \cite{Kitazawa:1987za,Alvarez-Gaume:1991nvs} (or, slightly more clearly, with the refined expressions given in \cite{Fredenhagen:2007tk}\footnote{See equations (111), (113) in Appendix A of the paper.}) and obtain an explicit match with our formulas (\ref{eq:MMnormNS}), (\ref{eq:MMnormalizationId}). 
\par A concrete example of this matching can be seen in the general form of the OPE between a degenerate NS field $\mathcal{O}_{1,3}$ with a general NS field $\mathcal{O}_{r,s}$ at a given minimal model with $\hat{b}^2=p'/p$, which obeys the general fusion rule (assuming all $\mathcal{O}$'s are unit-normalized) \cite{Friedan:1984rv}
\begin{equation}
    \mathcal{O}_{1,3}(z)\mathcal{O}_{r,s}(0)=\sum_{\varepsilon=\{1,0,-1\}}C_{(1,3)(r,s)}\text{}^{(r,s+2\varepsilon)}(z\bar{z})^{h_{(r,s+2\varepsilon)}-h_{(r,s)}-h_{(1,3)}}\left[\mathcal{O}_{r,s+2\varepsilon}\right](0)
\end{equation}
with (see e.g. \cite{Belavin:2008vc}) 
\begin{align}
    &C_{(1,3)(r,s)}\text{}^{(r,s+2\varepsilon)}=\left[\frac{\gamma\left(\frac{1+\hat{b}^2}{2}\right)\gamma\left(\varepsilon i\hat{b}(p_{\langle r,s\rangle}-\varepsilon i\hat{b})\right)}{\gamma\left(\frac{3\hat{b}^2-1}{2}\right)\gamma\left(\varepsilon i\hat{b}(p_{\langle r,s\rangle}-\varepsilon i\hat{b}^{-1})\right)}\right]^{1/2} \ , \quad \quad \varepsilon=\pm, \nonumber \\
    &\left.C_{(1,3)(r,s)}\text{}^{(r,s+2\varepsilon)}\right|_{\varepsilon=0}\equiv\widetilde{C}_{(1,3)(r,s)}\text{}^{(r,s)}\\
    &~~~~~~~~~~~~=\frac{i\hat{b}^{-2}\gamma\left(\frac{1+\hat{b}^2}{2}\right)\gamma\left(\frac{1+\hat{b}^2}{2}-i\hat{b}p_{\langle r,s\rangle}\right)}{\gamma\left(\frac{1-\hat{b}^2}{2}-i\hat{b}p_{\langle r,s\rangle}\right)}\left[\frac{\gamma\left(1-\hat{b}^2\right)\gamma\left(\frac{\hat{b}^2-1}{2}\right)}{\gamma\left(\hat{b}^2-1\right)\gamma\left(\frac{3\hat{b}^2-1}{2}\right)}\right]^{1/2},
\end{align}
and $p_{\langle r,s\rangle}$ is defined as in (\ref{eq:degtimelike}). It is straightforward to check that these expressions agree with (\ref{eq:MMnormNS}) under the normalization (\ref{eq:MMnormalizationId}). Nevertheless, it is manifest that the timelike Liouville formulas do \textit{not} return zero for operator dimensions different than the ones corresponding to the modules $\left[\mathcal{O}_{r,s+2\varepsilon}\right]$ with $\varepsilon=\pm,0$\footnote{Here we have suppressed the even/odd notation for the modules in the OPE. It should be understood that $\left[\mathcal{O}_{r,s\pm 2}\right]\equiv\left[\mathcal{O}_{r,s\pm 2}\right]_{\mathrm{ee}} $ whereas $\left.\left[\mathcal{O}_{r,s+2\varepsilon}\right]\right|_{\varepsilon\rightarrow0}\equiv \left[\mathcal{O}_{r,s}\right]_{\mathrm{oo}}$, according to the notation in (\ref{eq:NSOPE})-(\ref{eq:chains}).} and hence there is no observed truncation of the fusion algebra in $\mathcal{N}=1$ timelike Liouville theory, exactly as in the non-supersymmetric case.

\section{Summary \& Applications}\label{sec:discussion}
We now summarize the structure constants of $\mathcal{N}=1$ spacelike and timelike Liouville theory in the natural normalization that we have studied so far. We then proceed to a discussion of applications and possible future directions.

\paragraph{Spacelike structure constants}\
\\
In the NS-sector,
\begin{align}
 \langle V^{\mathrm{NS}}_{p_1}(0)V^{\mathrm{NS}}_{p_2}(1)\rangle &= \left(\rho_{\mathrm{NS}}^{(b)}(p_1)\right)^{-1}[\delta(p_1-p_2) + \delta(p_1+p_2)]~,\nonumber\\
    \langle V^{\mathrm{NS}}_{p_1}(0)V^{\mathrm{NS}}_{p_2}(1)V^{\mathrm{NS}}_{p_3}(\infty)\rangle &=C_{\mathrm{NS}}^{(b)}(p_1,p_2,p_3)~,\nonumber\\
    \langle W^{\mathrm{NS}}_{p_1}(0)V^{\mathrm{NS}}_{p_2}(1)V^{\mathrm{NS}}_{p_3}(\infty)\rangle &= \widetilde{C}_{\mathrm{NS}}^{(b)}(p_1,p_2,p_3)~,\label{eq: spacelikeN1 structure constantsNS summary}
\end{align}    
where
\begin{align}
\rho_{\mathrm{NS}}^{(b)}(p)&:=-4 \sin(\pi b p)\sin(\pi b^{-1}p)\ , \nonumber\\
     C_{\rm NS}^{(b)} (p_1,p_2,p_3) &:=  \frac{\Gamma_b^{\rm NS}(2Q)}{2\Gamma_b^{\rm NS}(Q)^3}\frac{\Gamma^{\rm NS}_b\left(\frac{Q}{2} \pm p_1 \pm p_2 \pm p_3\right)}{\prod_{j=1}^3\Gamma_b^{\rm NS}\left(Q \pm 2 p_j\right)}~,\nonumber\\
    \widetilde{C}^{(b)}_{\rm NS} (p_1,p_2,p_3) &:=i    \frac{\Gamma_b^{\rm NS}(2Q)}{\Gamma_b^{\rm NS}(Q)^3}\frac{\Gamma^{\rm R}_b\left(\frac{Q}{2} \pm p_1 \pm p_2 \pm p_3\right)}{\prod_{j=1}^3\Gamma_b^{\rm NS}\left(Q \pm 2 p_j\right)}~.\label{eq: NS spacelike 1 and 2 summary}
\end{align}    
In the R-sector,
\begin{align}
 \langle V^{\mathrm{R},\pm}_{p_1}(0)V^{\mathrm{R},\pm}_{p_2}(1)\rangle &= \left(\rho_{\mathrm{R}}^{(b)}(p_1)\right)^{-1}[\delta(p_1-p_2) \pm \delta(p_1+p_2)]~,\nonumber \\
    \langle V^{\mathrm{R},\pm}_{p_1}(0)V^{\mathrm{R},\pm}_{p_2}(1)V^{\mathrm{NS}}_{p_3}(\infty)\rangle &= C_{\mathrm{even}}^{(b)}(p_1,p_2;p_3) \pm C_{\mathrm{odd}}^{(b)}(p_1,p_2;p_3) ~,\label{eq: spacelikeN1 structure constantsR symmary}
\end{align}    
where
\begin{align}
\rho_{\mathrm{R}}^{(b)}(p)&:=2\sqrt{2}\cos(\pi b p)\cos(\pi b^{-1}p) \ , \nonumber \\
   C_{\mathrm{even}}^{(b)}(p_1,p_2;p_3)
    &:=\frac{\Gamma_b^{\rm NS}(2Q)}{\sqrt{2}\Gamma_b^{\rm NS}(Q)^3} \frac{\Gamma_b^{\mathrm{R}}\left(\frac{Q}{2}\pm (p_1+p_2)\pm p_3\right)\Gamma_b^{\mathrm{NS}}\left(\frac{Q}{2}\pm (p_1-p_2)\pm p_3\right)}{\Gamma_b^{\mathrm{R}}(Q\pm 2p_1)\Gamma_b^{\mathrm{R}}(Q\pm 2p_2)\Gamma_b^{\mathrm{NS}}(Q\pm 2p_3)},\nonumber \\
    C_{\mathrm{odd}}^{(b)}(p_1,p_2;p_3)&:= \frac{\Gamma_b^{\rm NS}(2Q)}{\sqrt{2}\Gamma_b^{\rm NS}(Q)^3} \frac{\Gamma_b^{\mathrm{NS}}\left(\frac{Q}{2}\pm (p_1+p_2)\pm p_3\right)\Gamma_b^{\mathrm{R}}\left(\frac{Q}{2}\pm (p_1-p_2)\pm p_3\right)}{\Gamma_b^{\mathrm{R}}(Q\pm 2p_1)\Gamma_b^{\mathrm{R}}(Q\pm 2p_2)\Gamma_b^{\mathrm{NS}}(Q\pm 2p_3)}~.\label{eq: R spacelike summary}
\end{align}    

\paragraph{Timelike structure constants}\
\\
In the NS-sector, 
\begin{align}
 \langle \widehat{V}^{\mathrm{NS}}_{\hat{p}_1}(0)\widehat{V}^{\mathrm{NS}}_{\hat{p}_2}(1)\rangle &= \frac{\rho_{\mathrm{NS}}^{(\hat{b})}(ip_1)}{\hat{p}_1^2}[\delta(\hat{p}_1-\hat{p}_2) + \delta(\hat{p}_1+\hat{p}_2)]~, \nonumber\\
    \langle \widehat{V}^{\mathrm{NS}}_{\hat{p}_1}(0)\widehat{V}^{\mathrm{NS}}_{\hat{p}_2}(1)\widehat{V}^{\mathrm{NS}}_{\hat{p}_3}(\infty)\rangle &= \pm \frac{i}{\widetilde{C}_{\mathrm{NS}}^{(\hat{b})}(i\hat{p}_1,i\hat{p}_2,i\hat{p}_3)}~,\nonumber\\
    \langle \widehat{W}^{\mathrm{NS}}_{\hat{p}_1}(0)\widehat{V}^{\mathrm{NS}}_{\hat{p}_2}(1)\widehat{V}^{\mathrm{NS}}_{\hat{p}_3}(\infty)\rangle &= \pm \frac{i}{C^{(\hat{b})}_{\text{NS}}(i\hat{p}_1,i\hat{p}_2,i\hat{p}_3)} ~.\label{eq: NS timelike symmary}
\end{align}    
In the R-sector,
\begin{align}
 \langle \widehat{V}^{\mathrm{R},\pm}_{\hat{p}_1}(0)\widehat{V}^{\mathrm{R},\pm}_{\hat{p}_2}(1)\rangle &= \rho_{\mathrm{R}}^{(\hat{b})}(i\hat{p}_1)[\delta(\hat{p}_1-\hat{p}_2) \pm \delta(\hat{p}_1+\hat{p}_2)]~,\nonumber\\
    \langle \widehat{V}^{\mathrm{R},\pm}_{\hat{p}_1}(0)\widehat{V}^{\mathrm{R},\pm}_{\hat{p}_2}(1)\widehat{V}^{\mathrm{NS}}_{\hat{p}_3}(\infty)\rangle &= \widehat{C}_{\mathrm{even}}^{(\hat{b})}(\hat{p}_1,\hat{p}_2;\hat{p}_3) \pm \,  \widehat{C}_{\mathrm{odd}}^{(\hat{b})}(\hat{p}_1,\hat{p}_2;\hat{p}_3) ~,
\end{align}    
with
\begin{align}
    \widehat{C}_{\mathrm{even}}^{(\hat{b})}(\hat{p}_1,\hat{p}_2;\hat{p}_3)  \equiv \pm\frac{ 1}{C_{\mathrm{odd}}^{(\hat{b})}(i\hat{p}_1,i\hat{p}_2;i\hat{p}_3)} ~,\nonumber\\
    \widehat{C}_{\mathrm{odd}}^{(\hat{b})}(\hat{p}_1,\hat{p}_2;\hat{p}_3)  \equiv  \pm \frac{1}{C_{\mathrm{even}}^{(\hat{b})}(i\hat{p}_1,i\hat{p}_2;i\hat{p}_3)} ~.\label{eq: R timelike symmary}
  \end{align}  
We have explicitly kept the sign ambiguity in the expressions (\ref{eq: NS timelike symmary}), (\ref{eq: R timelike symmary}) for the structure constants. We now highlight several promising directions to explore given our results.

\paragraph{$\mathcal{N}=1$ Virasoro minimal string.}
There is at least one possible application of $\mathcal{N}=1$ timelike and spacelike Liouville theory in constructing solvable models of string theory\footnote{We are especially grateful to Lorenz Eberhardt and Mukund Rangamani for discussions on this topic.}. In \cite{Collier:2023cyw} the Virasoro minimal string (VMS) was introduced as a stringy realization of JT gravity. The Virasoro minimal string is a two-dimensional critical string theory, which from a worldsheet perspective corresponds to a spacelike and timelike non-supersymmetric Liouville CFT coupled to each other. The main observables in the VMS, the string amplitudes, were denoted as $\mathsf{V}_{g,n}^{(b)}$ and are polynomials in the Liouville momenta and central charge. In the semiclassical limit (namely, at large spacelike Liouville central charge) the string amplitudes reduce to the standard Weil–Petersson volumes \cite{Saad:2019lba}, hence in \cite{Collier:2023cyw} $\mathsf{V}_{g,n}^{(b)}$ were interpreted as quantum volumes. The VMS furthermore exhibits a dual Hermitian double scaled matrix integral, whose topological recursion provides access to the quantum volumes for arbitrary genus $g$ and number of punctures $n$.

 It is tempting to imagine a $\mathcal{N}=1$ extension of the VMS that consists of a worldsheet theory that couples in the same way the spacelike and timelike $\mathcal{N}=1$ Liouville CFTs, as described in the present work. Here we refrain from giving a complete presentation of the $\mathcal{N}=1$ VMS but merely want to present an appetizer, and leave a more detailed discussion to future work. From a worldhseet perspective the $\mathcal{N}=1$ VMS is the following superstring theory (see also \cite{Johnson:2024fkm})
\begin{equation}
\begin{array}{c}
\text{$c\geq 9$ $\mathcal{N}=1$} \\ \text{Liouville CFT}
\end{array}
\ \oplus\  
\begin{array}{c} \text{$\hat c\leq 1$ $\mathcal{N}=1$} \\ \text{Liouville CFT} \end{array} \ \oplus\  \text{$\mathfrak{b}\mathfrak{c}$-ghosts}  \ \oplus \  \text{$\beta\gamma$-ghosts}~. 
\label{eq:superVMS}
\end{equation}
In the above, $c\geq9$ is the quantum spacelike $\mathcal{N}=1$ super Liouville CFT whereas $\hat{c}\leq 1$ is the quantum timelike $\mathcal{N}=1$ super Liouville CFT.
Imposing the vanishing of the conformal anomaly of the combined superstring theory  (\ref{eq:superVMS})
 we obtain the constraint 
 \begin{equation}\label{eq:conformal anomaly cancellation}
    c + \hat{c} -10=0~\Leftrightarrow~ \hat{b} = b~.
\end{equation}
We now introduce the standard picture number convention, i.e. $-1$ for the NS-sector and $-1/2$ for the R-sector. Explicitly, if we denote by $\e^{q\varphi}$ the ``bosonized" $\beta,\gamma$-system with conformal dimension $h_q= -q(q+2)/2$ \cite{deLacroix:2017lif}, we construct 
\begin{subequations}
    \begin{align}
       \mathcal{V}^{\mathrm{NS}}_p &= \mathsf{N}(p)\mathfrak{c}\tilde{\mathfrak{c}} \e^{-\varphi}\e^{- \tilde{\varphi}} V^{\mathrm{NS}}_p \widehat{V}^{\mathrm{NS}}_{ip}~, \\
       \mathcal{V}^{\mathrm{R}}_p &= \mathsf{R}(p)\mathfrak{c}\tilde{\mathfrak{c}} \e^{-\frac{\varphi}{2} }\e^{- \frac{\tilde{\varphi}}{2}} (V_p^{\mathrm{R},+} \widehat{V}_{ip}^{\mathrm{R},-} +iV_p^{\mathrm{R},-} \widehat{V}_{ip}^{\mathrm{R},+})~.\label{eq:ramondvertex}
    \end{align}
\end{subequations}
where $\mathsf{N}(p),\mathsf{R}(p)$ are some arbitrary normalizations. The above vertex operators satisfy the mass-shell conditions 
\begin{equation}
   h_p^{\mathrm{NS}} + \hat{h}^{\mathrm{NS}}_{\hat{p}} = \frac{1}{2} ~,\quad  h_p^{\mathrm{R}} + \hat{h}^{\mathrm{R}}_{\hat{p}} = \frac{5}{8}  ~,
\end{equation}
which imply $\hat{p} = ip$. 
Combining the mass-shell condition with (\ref{eq:conformal anomaly cancellation}) we obtain 
\begin{equation}\label{eq: relation square root}
    4h_p^{\mathrm{R}} - \frac{c}{4} = -\left(4\hat{h}_p^{\mathrm{R}} - \frac{\hat{c}}{4}\right)~.
\end{equation}
The choice of the Ramond state in (\ref{eq:ramondvertex}) guarantees that it is annihilated by the total holomorphic supercharge $\mathcal{G}_{\mathrm{tot}} \equiv \mathcal{G}_{0}+\widehat{\mathcal{G}}_0$ (and similarly for the total antiholomorphic one $\widetilde{\mathcal{G}}_{\mathrm{tot}} \equiv \widetilde{\mathcal{G}}_{0}+\widehat{\widetilde{\mathcal{G}}}_0$)\footnote{To see this we use (\ref{eq:mathcalG0 on VR}) and (\ref{eq: R sector timelike}) combined with (\ref{eq: relation square root}) and choose $\sqrt{-1} =i$.}. Alternatively, one could have chosen the Ramond state with a relative minus sign in (\ref{eq:ramondvertex}), in which case it would be annihilated by $\mathcal{G}_{\mathrm{tot}} \equiv \mathcal{G}_{0}-\widehat{\mathcal{G}}_0$ (and similarly for $\widetilde{\mathcal{G}}_{\mathrm{tot}}$), corresponding to a different but consistent way of gauging the worldsheet theory. These two different choices, however, seem to lead to inequivalent string theories, even with a given GSO projection.

Indeed, the above defines a type 0B superstring theory. In type 0A there is in fact no R-sector vertex operator \cite{Klebanov:2003wg, Seiberg:2003nm}.

We can now look at some first simple string amplitudes. 
Using the notation $\mathsf{V}^{(b)}_{g,n_{\mathrm{NS}}, n_\mathrm{R}}$, we consider the NS-NS-NS and R-R-NS sphere amplitudes
\begin{equation}\label{eq: string amplitudes}
    \mathsf{V}_{0,3,0}^{(b)}(p_1,p_2,p_3)~\quad \quad \mathrm{and} \quad \quad \mathsf{V}_{0,1,2}^{(b)}(p_1,p_2;p_3)~.
\end{equation}
The dimension of the corresponding super-moduli space $\mathcal{M}_{g,n_{\mathrm{NS}}, n_{\mathrm{R}}}^{\mathcal{N}=1}$ is given by (see e.g.\cite{Witten:2012ga})
\begin{equation}
    \mathrm{dim}(\mathcal{M}_{g,n_{\mathrm{NS}}, n_{\mathrm{R}}}^{\mathcal{N}=1}) = 3g-3+n_{\mathrm{NS}} + n_{\mathrm{R}}| 2g-2+ n_{\mathrm{NS}} + \frac{1}{2}n_{\mathrm{R}}~,
\end{equation}
and therefore
\begin{equation}
 \mathrm{dim}(\mathcal{M}_{0,3, 0}^{\mathcal{N}=1}) = 0|1~,\quad \mathrm{dim}(\mathcal{M}_{0,1, 2}^{\mathcal{N}=1}) = 0|0~.
\end{equation}
To evaluate the string amplitudes (\ref{eq: string amplitudes}) we will use the NSR formalism and replace the integration over the worldsheet gravitino by a picture changing operator (PCO) \cite{Friedan:1985ge}. On a genus $g$ surface the number of $\beta$ minus the number of $\gamma$ zero modes is $2g-2$, or equivalently the total $\varphi$ charge must be $2g-2$. Therefore, on a sphere we have net picture number $(-2,-2)$. To achieve this, on a given amplitude with $n_{\mathrm{NS}}$ number of $\mathrm{NS}$ punctures and $n_{\mathrm{R}}$ number of $\mathrm{R}$ `punctures', we need to introduce $2g-2+n_{\mathrm{NS}} + \frac{1}{2}n_{\mathrm{R}}$ number of (holomorphic) PCOs. Therefore, for the amplitudes (\ref{eq: string amplitudes}) we need one (+ one antiholomorphic) PCO for $\mathsf{V}_{0,3,0}^{(b)}$ and none for $\mathsf{V}_{0,1,2}^{(b)}$. Denoting the holomorphic PCO by $\chi$ \cite{Friedan:1985ge, Balthazar:2022atu} we hence consider
\begin{equation}
   \mathsf{V}_{0,3,0}^{(b)}(p_1,p_2,p_3) = \langle \chi \tilde{\chi} \mathcal{V}^{\mathrm{NS}}_{p_1} \mathcal{V}^{\mathrm{NS}}_{p_2}\mathcal{V}^{\mathrm{NS}}_{p_3} \rangle ~,\quad \mathsf{V}_{0,1,2}^{(b)}(p_1,p_2,p_3) = \langle \mathcal{V}^{\mathrm{R}}_{p_1} \mathcal{V}^{\mathrm{R}}_{p_2}\mathcal{V}^{\mathrm{NS}}_{p_3} \rangle  ~.
\end{equation}
Following standard methods (see e.g. \cite{Balthazar:2022atu, Belavin:2008vc}) we find
\begin{align}
   \mathsf{V}_{0,3,0}^{(b)}(p_1,p_2,p_3)  &= \mathsf{N}(p_1)\mathsf{N}(p_2)\mathsf{N}(p_3) \bigg[ \langle V_{p_1}^{\mathrm{NS}} V_{p_2}^{\mathrm{NS}} V_{p_3}^{\mathrm{NS}}\rangle  \langle \widehat{W}_{ip_1}^{\mathrm{NS}} \widehat{V}_{ip_2}^{\mathrm{NS}} \widehat{V}_{ip_3}^{\mathrm{NS}}\rangle\cr
   &\quad +\langle W_{p_1}^{\mathrm{NS}} V_{p_2}^{\mathrm{NS}} V_{p_3}^{\mathrm{NS}}\rangle  \langle \widehat{V}_{ip_1}^{\mathrm{NS}} \widehat{V}_{ip_2}^{\mathrm{NS}} \widehat{V}_{ip_3}^{\mathrm{NS}}\rangle \bigg]\cr
   &= \mathsf{N}(p_1)\mathsf{N}(p_2)\mathsf{N}(p_3) \bigg[C_{\mathrm{NS}}^{(b)}(p_1,p_2,p_3)\widehat{\widetilde{C}}_{\mathrm{NS}}^{(b)}(ip_1,ip_2,ip_3)\cr
   &\quad + \widetilde{C}_{\mathrm{NS}}^{(b)}(p_1,p_2,p_3)\widehat{{C}}_{\mathrm{NS}}^{(b)}(ip_1,ip_2,ip_3)\bigg]~.\label{eq:NSNSNScompute}
\end{align}
Similarly we obtain for the R-R-NS three point amplitude 
\begin{align}
  \mathsf{V}_{0,1,2}^{(b)}(p_1,p_2;p_3) & =   \mathsf{R}(p_1)\mathsf{R}(p_2)\mathsf{N}(p_3)\bigg[  \langle V_{p_1}^{\mathrm{R},+} V_{p_2}^{\mathrm{R},+} V_{p_3}^{\mathrm{NS}}\rangle  \langle \widehat{V}_{ip_1}^{\mathrm{R},-} \widehat{V}_{ip_2}^{\mathrm{R},-} \widehat{V}_{ip_3}^{\mathrm{NS}}\rangle\cr
  &\quad -   \langle V_{p_1}^{\mathrm{R},-} V_{p_2}^{\mathrm{R},-} V_{p_3}^{\mathrm{NS}}\rangle  \langle \widehat{V}_{ip_1}^{\mathrm{R},+} \widehat{V}_{ip_2}^{\mathrm{R},+} \widehat{V}_{ip_3}^{\mathrm{NS}}\rangle  \bigg]\cr
  &= 2\mathsf{R}(p_1)\mathsf{R}(p_2)\mathsf{N}(p_3) \bigg[C_{\mathrm{odd}}^{(b)}(p_1,p_2;p_3) \widehat{C}_{\mathrm{even}}^{(b)}(ip_1,ip_2;ip_3) \cr
  &\quad - C_{\mathrm{even}}^{(b)}(p_1,p_2;p_3) \widehat{C}_{\mathrm{odd}}^{(b)}(ip_1,ip_2;ip_3)\bigg]~.\label{eq:RRNScompute}
\end{align}
\par We are now facing an interesting puzzle. Notice that all the hatted quantities in (\ref{eq:NSNSNScompute}), (\ref{eq:RRNScompute}) -- 
 corresponding to the $\mathcal{N}=1$ timelike structure constants -- are given exactly by the inverses of the spacelike structure constants that they multiply, as in  (\ref{eq: NS timelike symmary}), (\ref{eq: R timelike symmary}). However, as we explained in sections \ref{sec: spacelike} and \ref{sec:Timelike}, the structure constants of $\mathcal{N}=1$ Liouville theory can be determined from the basic shift relations only up to an overall sign ambiguity and hence, from the point of view of a \textit{single} Liouville CFT, there is really no distinction for either sign. On the other hand, when we couple two $\mathcal{N}=1$ Liouville CFTs to define (\ref{eq:superVMS}), the corresponding string amplitudes clearly depend on the relative signs, as these can determine whether the net result in (\ref{eq:NSNSNScompute}), (\ref{eq:RRNScompute}) is non-zero or vanishes entirely\footnote{It is useful to contrast this with the ordinary VMS case, where a similar sign ambiguity arises in the timelike DOZZ formula but ultimately plays no significant role, as there is only a single structure constant in that case.}. 
\par Let us explore some expectations that might guide our interpretation. In the case of $\mathcal{N}=1$ JT supergravity, \cite{Stanford:2019vob, Turiaci:2024cad} argued that the presence of fermionic moduli forces the super Weil–Petersson volumes $V_{0,n_{\mathrm{NS}},0}=0$ to vanish. If we interpret $\mathcal{N}=1$ VMS as the superstringy realization of $\mathcal{N}=1$ JT gravity one may be inclined to believe that also $\mathsf{V}^{(b)}_{0,n_{\mathrm{NS}},0}=0$\footnote{Of course it is understood that this argument is a bit fast, since the `quantum super-volumes' of $\mathcal{N}=1$ VMS can in general be non-trivial and still yield something vanishing in the JT limit.}, and hence choose accordingly the relative signs in (\ref{eq:NSNSNScompute}). On the other hand, in \cite{Collier:2023cyw} the quantum volume corresponding to the sphere amplitude with three punctures was interpreted (through the connection with chiral three-dimensional gravity) as counting the number of Liouville sphere three-point blocks, which simply is unity. To back this even more, from a Random Matrix Theory perspective the fact that $\mathsf{V}_{0,3}^{(b)}=1$ is a universal consequence of a ‘square root edge’ spectral curve\footnote{A ‘square root edge’ double-scaled random matrix has a universal form for the leading three-point correlator of resolvents, usually denoted by $\omega_{0,3}(z_1, z_2, z_3)$, which after Laplace transforming in $z_{1,2,3}$ yields $\mathsf{V}_{0,3}(l_1,l_2,l_3) = 1$.} which is also true e.g. in JT gravity \cite{Saad:2019lba}.
For the $\mathcal{N}=1$ VMS case,  in the NS-sector of Liouville theory there are two independent structure constants and hence one might believe that the net result of (\ref{eq:NSNSNScompute}) should instead be non-trivial and equal to 2 (modulo obvious normalizations). This would consequently lead to yet a different choice of the relative signs in the above expression.  
Similarly, for the case of the R-R-NS amplitude, and following the intuition of the corresponding vanishing amplitude in $\mathcal{N}=1$ JT \cite{Stanford:2019vob}, it could be possible that $\mathsf{V}^{(b)}_{0,1,2}=0$. However this does not seem to be compatible with the existence of a single Ramond sector Liouville conformal block according to the previous logic, thereby leading to an expected non-trivial net result for (\ref{eq:RRNScompute}).
\par These arguments collectively seem to suggest that possibly either choice of relative signs in (\ref{eq:NSNSNScompute}) and (\ref{eq:RRNScompute}) may yield a viable string theory, reflecting eventually the two distinct ways of gauging the two $\mathcal{N}=1$ algebras present on the worldsheet (exactly in the sense that we described below (\ref{eq: relation square root})). We leave a clearer understanding and resolutions of these interesting puzzles to future work \cite{upcoming}.


\paragraph{Sphere partition function from timelike structure constants.}
In \cite{Anninos:2023exn} a systematic semiclassical expansion of the sphere path integral of $\mathcal{N}=1$ timelike Liouville theory was established. In particular, the sphere partition function of timelike Liouville theory reads
\begin{equation}
\label{eq: Ztl}
    \mathcal{Z}^{\mathcal{N}=1}_{\mathrm{tL}}[\mu] = \int \frac{[\mathcal{D}\phi][\mathcal{D}\psi]}{\mathrm{vol}_{OSP(1|2;\mathbb{C})}}\,\e^{-S_{\mathrm{tL}}^{\mathcal{N}=1}}~,
\end{equation}
where the action is given by (\ref{eq:N1actiontL}). The background metric $\tilde{\e}_\mu^a$, related to the physical metric by $\e_\mu^a= \e^{\hat{b}\phi}\tilde{\e}_\mu^a$, is now a round metric on the sphere with radius $r$ leading to $\widetilde{R} = 2/r^2$ in (\ref{eq:N1actiontL}). The path integral (\ref{eq: Ztl}) admits a real two-sphere saddle and a systematic non-vanishing semiclassical loop expansion.  We can ask how our explicit expressions for the $\mathcal{N}=1$ timelike structure constants compare with existing approaches in the literature that evaluate the path integral. In particular, we can ask whether the following equality holds
\begin{equation}\label{eq: sphere to Cbbb}
    \partial^3_\mu \mathcal{Z}^{\mathcal{N}=1}_{\mathrm{tL}}[\mu]  \approx \widehat{\widetilde{C}}_{\mathrm{NS}}^{(\hat{b})}(\hat{p}^*,\hat{p}^*,\hat{p}^*)~,\quad \quad \text{for}~~ \hat{p}^* = \frac{i}{2}(\hat{b}^{-1}+\hat{b})~,
\end{equation}
where the approximation indicates equality up to terms independent of $\hat{b}$. The RHS of (\ref{eq: sphere to Cbbb}) encodes the three-point correlator $\langle \widehat{W}^{\mathrm{NS}}_{p_1} \widehat{W}^{\mathrm{NS}}_{p_2}\widehat{W}^{\mathrm{NS}}_{p_3}\rangle $\footnote{In the timelike theory we have $\widehat{W}^{\mathrm{NS}}_p = \mu \e^{(\hat{\alpha}+ \hat{b})\phi}+ \frac{1}{2} \hat{\alpha}\overline{\psi}\psi\e^{\hat{\alpha}\phi}$ where $\hat{\alpha} = -\frac{\widehat{Q}}{2}- i \hat{p}$.} which, via superconformal Ward identities \cite{Belavin:2007gz}, is related to the correlator of $\langle \widehat{W}^{\mathrm{NS}}_{p_1} \widehat{V}^{\mathrm{NS}}_{p_2}\widehat{V}^{\mathrm{NS}}_{p_3}\rangle$ . 
The choice $\hat{p}= \hat{p}^*$ captures the operator
\begin{equation}
    \widehat{W}^{\mathrm{NS}}_{\hat{p}^*} = \mu \e^{2\hat{b} \phi} + \frac{1}{2}\hat{b} \e^{\hat{b}\phi}\overline{\psi}\psi~,
\end{equation}
which is the analog of the area operator in the non-supersymmetric case. 
The analogous problem in the  $\mathcal{N}=1$ spacelike case was examined in \cite{Anninos:2023exn} where it was shown that a non-vanishing prediction coming from the semiclassical expansion of the explicit expression for the structure constant (i.e. the analogous of the RHS of (\ref{eq: sphere to Cbbb})) can be matched with the semiclassical path integral expansion coming from the LHS to one-loop order. In the timelike case however, it is straightforward to compute from our expressions (\ref{eq: NS timelike symmary}) that\footnote{Note that, for the other structure constant, $\lim_{\hat{p}_2\rightarrow \hat{p}_1} \lim_{\hat{p}_3\rightarrow \hat{p}^*} \widehat{C}_{\mathrm{NS}}^{(\hat{b})}(\hat{p}_1,\hat{p}_2,\hat{p}_3)$ diverges.}
\begin{equation}
    \widehat{\widetilde{C}}^{(\bb)}_{\rm NS} (\pp^*,\pp^*,\pp^*) =0~.
\end{equation}
Just as in the non-supersymmetric case \cite{Anninos:2021ene}, this result contradicts the non-vanishing semiclassical gravity calculation. It is still an interesting open problem to reconcile these two pictures in the timelike case, both for the non-supersymmetric and $\mathcal{N}$=1 setups.

\paragraph{$\mathcal{N}=2$ Liouville theories.}
It is a natural step to include more supersymmetry. It has been shown that $\mathcal{N}=2$ Liouville theory is related by mirror symmetry to the $SL(2,\mathbb{R})/U(1)$ Kazama–Suzuki supercoset (the 2D fermionic “cigar” black hole) \cite{Hori:2001ax}. Such a duality does not exist for the timelike ($c<3$) theory \cite{Anninos:2023exn}. The general obstruction toward the structure constants of the $\mathcal{N}=2$ Liouville theory partly has its roots on the fact that the classical background charge $Q=1/b$ does not receive quantum corrections, and hence the quantum central charge does not end up having the $b\leftrightarrow b^{-1}$ symmetry\footnote{In particular, $c^{\mathcal{N}=2} = 3+6b^{-2}$.} that we are used to from the $\mathcal{N}=0,1$ cases. It would be interesting to study what kind of alternative methods could give us access to the theory, such as supersymmetric localization \cite{Anninos:2023exn}. In particular, it seems reasonable to expect that any potential analogue of the shift relations (which would originate purely from the symmetry algebra) would still be analytic as functions of $b$—even though there is no $b^{-1}$ counterpart. As such, one can analytically continue these relations and subsequently study the resulting solutions, subject to the usual uniqueness arguments (which might turn out to be tricky in this case).

\paragraph{Crossing symmetry in $\mathcal{N}=1$ timelike Liouville.} A crucial step toward fully establishing 
$\mathcal{N}=1$ timelike Liouville theory as a two-dimensional superconformal field theory is to demonstrate that the structure constants derived in this work give rise to crossing-symmetric four-point functions on the sphere, as well as modular-covariant one-point functions on the torus. As we discussed in Section \ref{sec:Timelike}, the expected primary spectrum over which crossing/modular covariance will be established should be
\begin{equation}
     \hat{p}\in i\mathbb{R}_{\geq0}-\epsilon
\end{equation}
for both the NS and the Ramond sectors (c.f. (\ref{eq:hTimelike})). This was checked numerically for some cases in \cite{Rangamani:2025wfa}, following similar analysis as in the non-supersymmetric case \cite{Ribault:2015sxa}. What is still missing, both in the bosonic and $\mathcal{N}=1$ case, is an analytic derivation of these statements. A key component in this direction is the construction of the fusion and modular kernels for $\hat{c}\leq 1$ (as continuous functions of $\hat{b}$), which have yet to be developed for either case\footnote{See however \cite{Ribault:2023vqs}, \cite{Roussillon:2024wmr} for the non-supersymmetric case.}.
In particular we emphasize that, at least in the spacelike case, the Liouville structure constants (in the natural normalization) are particular instances of these kernels (see Appendix \ref{app:fusionkernel}) \cite{Teschner:2001rv,Poghosyan:2016kvd}. The latter, in turn, are solutions to Moore-Seiberg consistency conditions which can be recast as shift relations that uniquely determine the kernels in the corresponding spacelike and timelike regimes of the central charge \cite{Eberhardt:2023mrq, Ribault:2023vqs}. It would be worthwhile to explore further the analogous statements for the timelike case in the future.

\section*{Acknowledgements}

We would like to thank Dionysios Anninos, Vladimir Belavin, Scott Collier, Lorenz Eberhardt, Raghu Mahajan, Juan Maldacena, Mukund Rangamani, Sylvain Ribault, Victor Rodriguez and Edward Witten
for stimulating discussions. We would in particular like to thank Mukund Rangamani and Jianming Zheng for sharing their draft with us and coordinating submission. We also thank Dionysios Anninos, Scott Collier, Lorenz Eberhardt and Mukund Rangamani for carefully reading our draft and valuable comments. BM gratefully acknowledges funding provided by the Sivian Fund at the Institute for Advanced Study and the National Science Foundation with grant number PHY-2207584.
VN gratefully acknowledges support from NSF Grant PHY-2207584.
IT is supported by the ERC Starting Grant 853507. BM and IT would also like to thank l'Institut Pascal at Universite Paris-Saclay, with the support of the program ``Investissements d’avenir" ANR-11-IDEX-0003-01, where this project started.

\appendix

\section{Special functions}\label{app:SpecialFunctions}

In this appendix we review the special functions necessary for the construction of the structure constants in $\mathcal{N}=0,1$ Liouville theory.

\subsection*{$\mathcal{N}=0$}\label{app.N0}
We start by briefly mentioning the special functions relevant for the structure constants of ordinary  bosonic Liouville theory. More specific details and further properties of these functions are well-documented and can be found, e.g., in \cite{Harlow:2011ny, Collier:2018exn}. 
\par The main special function is the Barnes double gamma function $\Gamma_b(x)$ with the integral representation
\begin{equation}
    \log\Gamma_b(x) = \int_0^\infty{\d t\over t}\left[{\e^{-xt}-\e^{-Qt/2}\over (1-\e^{-bt})(1-\e^{-b^{-1}t})}-{1\over 2}(Q/2-x)^2\e^{-t}-{Q/2-x\over t}\right]
\end{equation}
convergent for $x$ in the right half-plane. Away from this region it is defined via analytic continuation from its shift relations in $b,b^{-1}$
\begin{equation}\label{eq:shiftsG}
\frac{\Gamma_b(x+b^{\pm1})}{\Gamma_b(x)}=\frac{\sqrt{2\pi}b^{\pm b^{\pm 1}x\mp \frac{1}{2}}}{\Gamma(b^{\pm1}x)},
\end{equation}
where we implicitly assume that $b^2\notin \mathbb{Q}$\footnote{In this case the two shift relations in $b$ and $b^{-1}$ are incommensurable.}. As a function of $b$, it is analytic in the whole $b^2-$complex
plane 
except for the negative part of the real axis, where it meets with a natural bound of
analyticity. It also has the property $\Gamma_b=\Gamma_{b^{-1}}$. As a function of $x$, it is meromorphic with no zeroes, and simple poles at $x=-mb-nb^{-1}$, for $m,n\in\mathbb{N}$\footnote{$\mathbb{N}$ here (and for the rest of this appendix) stands for the natural numbers \textit{including} zero.}. 
\par There are two main special functions built out of $\Gamma_b(x)$:
\begin{equation}
  S_b(x):=\frac{\Gamma_b(x)}{\Gamma_b(Q-x)}, \quad \Upsilon_b (x):=\frac{1}{\Gamma_b(x)\Gamma_b(Q-x)} ~.
\end{equation}
From the properties of $\Gamma_b$ it is straightforward to deduce all the analytic properties of these functions. For example their corresponding shift relations read
\begin{equation}\label{eq:shiftsSandY}
    \frac{S_b(x+b^{\pm1})}{S_b(x)}=2\sin{(\pi b^{\pm1} x)} \ , ~~  \frac{\Upsilon_b(x+b^{\pm1})}{\Upsilon_b(x)}=b^{\pm 1\mp 2b^{\pm 1}x}\frac{\Gamma(b^{\pm 1}x)}{\Gamma(1-b^{\pm 1}x)}~.
\end{equation}

\subsection*{$\mathcal{N}=1$}\label{app.N1}

 The relevant special functions for $\mathcal{N}=1$ Liouville theory are again constructed from $\Gamma_b$, but are characterized by a slightly different analytic structure. In particular, we again seek functions with simple poles on the grid $x = -mb - nb^{-1}$, except now the non-negative integer pairs $(m,n)$ should be restricted to combinations of (even, even), (odd, odd), or (even, odd). This particular structure is captured by a product of two $\Gamma_b$'s. The two corresponding functions $\Gamma_b^{\mathrm{NS}}$ and $\Gamma_b^{\mathrm{R}}$ are defined as
\begin{equation}
    \Gamma_b^{\mathrm{NS}}(x)\equiv \Gamma_b\left(\frac{x}{2}\right)\Gamma_b\left(\frac{x+b+b^{-1}}{2}\right) \ , \ \ \
\Gamma_b^{\mathrm{R}}(x)\equiv\Gamma_b\left(\frac{x+b}{2}\right)\Gamma_b\left(\frac{x+b^{-1}}{2}\right)~.
\end{equation}
It is obvious that $\Gamma_b^{\mathrm{NS}(\mathrm{R})}=\Gamma_{b^{-1}}^{\mathrm{NS}(\mathrm{R})}$. 
As a function of $x$, $\Gamma_b^{\mathrm{NS}}$ and $\Gamma_b^{\mathrm{R}}$ have simple poles at
\begin{align}
   \Gamma_b^{\mathrm{NS}}(x)^{-1}&=0 \  \Leftrightarrow\  x= -kb-lb^{-1}~, ~~~~  \ (k,l)\in (2\mathbb{Z}_{\geq 0},2\mathbb{Z}_{\geq 0}) \text{ or }(2\mathbb{Z}_{\geq 0}+1,2\mathbb{Z}_{\geq 0}+1)\nonumber \\ 
\Gamma_b^{\mathrm{R}}(x)^{-1}&=0 \  \Leftrightarrow\  x=  -k'b-l'b^{-1} ~,~ \ (k',l')\in (2\mathbb{Z}_{\geq 0},2\mathbb{Z}_{\geq 0}+1) \text{ or }(2\mathbb{Z}_{\geq 0}+1,2\mathbb{Z}_{\geq 0}) . 
\end{align}
Their \textit{shift relations} are deduced from (\ref{eq:shiftsG}) and read
\begin{equation}\label{eq:shiftsG01}
\frac{\Gamma_b^{\mathrm{NS}}(x+b^{\pm1})}{\Gamma_b^{\mathrm{R}}(x)}=\frac{\sqrt{2\pi} \ b^{\pm \frac{xb^{\pm1}}{2}}}{\Gamma\left(\frac{1+b^{\pm1}x}{2}\right)} \ , \ \ \ \ \frac{\Gamma_b^{\mathrm{R}}(x+b^{\pm 1})}{\Gamma_b^{\mathrm{NS}}(x)}=\frac{\sqrt{2\pi} \ b^{\mp\frac{1}{2}(1-b^{\pm1}x)}}{\Gamma\left(\frac{xb^{\pm1}}{2}\right)} .
\end{equation}
Notice that shifts in $b^{\pm1}$ necessarily ``couple'' the two different functions $\Gamma_b^{\mathrm{R}},\Gamma_b^{\mathrm{NS}}$ with each other. We would need to have shifts by either $2b$, $2b^{-1}$, $Q=b+b^{-1}$, or $\hat{Q}=b^{-1}-b$ to get shift relations between the same functions. For example, for shifts by $Q$ we get
\begin{equation}
    \frac{\Gamma_b^{\mathrm{NS}}(x+Q)}{\Gamma_b^{\mathrm{NS}}(x)}= \frac{2 \pi \  b^{\frac{1}{2} \left(b x-b^{-1}x+2\right)}}{\Gamma \left(\frac{xb^{-1}}{2}\right) \Gamma \left(1+\frac{b x}{2}\right)}, \quad \frac{\Gamma_b^{\mathrm{R}}(x+Q)}{\Gamma_b^{\mathrm{R}}(x)}= \frac{2 \pi \ b^{\frac{1}{2}(bx-b^{-1}x)}}{\Gamma \left(\frac{1+b^{-1}x}{2}\right) \Gamma \left(\frac{1 + b x}{2} \right)} .
\end{equation}

\paragraph{$S^{\mathrm{NS}}$ and $S^{\mathrm{R}}$.}
In analogy with the $\mathcal{N}=0$ case, we further introduce two analogs of the $S_b$ function as follows:
\begin{equation}
    S_b^{\mathrm{NS}}(x):=\frac{\Gamma_b^{\mathrm{NS}}(x)}{\Gamma_b^{\mathrm{NS}}(Q-x)} \ , \ \ \ S_b^{\mathrm{R}}(x):=\frac{\Gamma_b^{\mathrm{R}}(x)}{\Gamma_b^{\mathrm{R}}(Q-x)}~ .
\end{equation}
They have \textit{poles} at
\begin{align}
   S_b^{\mathrm{NS}}(x)^{-1}&=0 \  \Longrightarrow \  x= -kb-lb^{-1} \ \ \ \ \ \ \ \text{for} \ \ k,l\in\mathbb{N} \ | \ (k,l)\in (2\mathbb{N},2\mathbb{N}) \text{ or }(2\mathbb{N}+1,2\mathbb{N}+1)\nonumber \\
S_b^{\mathrm{R}}(x)^{-1}&=0 \  \Longrightarrow \  x= -k'b-l'b^{-1} \ \ \ \ \ \text{for} \ k',l'\in\mathbb{N} \ | \ (k',l')\in (2\mathbb{N},2\mathbb{N}+1) \text{ or }(2\mathbb{N}+1,2\mathbb{N}) . 
\end{align}
They have \textit{zeroes} at
\begin{align}
S_b^{\mathrm{NS}}(x)&=0 \  \Longrightarrow \  x= kb+lb^{-1} \ \ \ \ \ \ \ \text{for} \ \ k,l\in\mathbb{N} \ | \ (k,l)\in (2\mathbb{N}+1,2\mathbb{N}+1) \text{ or }(2\mathbb{N}+2,2\mathbb{N}+2)\nonumber \\
S_b^{\mathrm{R}}(x)&=0 \  \Longrightarrow \  x= k'b+l'b^{-1} \ \ \ \ \ \text{for} \ k',l'\in\mathbb{N} \ | \ (k',l')\in (2\mathbb{N}+2,2\mathbb{N}+1) \text{ or }(2\mathbb{N}+1,2\mathbb{N}+2) .
\end{align}
They satisfy the \textit{shift relations}
\begin{equation}
\frac{S_b^{\mathrm{NS}}(x+b^{\pm1})}{S_b^{\mathrm{R}}(x)}=2\cos{\left(\frac{\pi b^{\pm1}x}{2}\right)} \ , \ \ \ \ \frac{S_b^{\mathrm{R}}(x+b^{\pm 1})}{S_b^{\mathrm{NS}}(x)}=2\sin{\left(\frac{\pi b^{\pm1}x}{2}\right)} .
\end{equation}
Shifts by $Q$ yield
\begin{equation}
\frac{S_b^{\mathrm{NS}}(x+Q)}{S_b^{\mathrm{NS}}(x)}=-4 \sin \left(\frac{\pi  b x}{2}\right)\sin \left(\frac{\pi  b^{-1} x}{2}\right)  \ \ , \ \ \ \ \ \frac{S_b^{\mathrm{R}}(x+Q)}{S_b^{\mathrm{R}}(x)}= 4 \cos \left(\frac{\pi  b x}{2}\right)\cos \left(\frac{\pi  b^{-1} x}{2}\right)  .
\end{equation}

\paragraph{$\Upsilon^{\mathrm{NS}}$ and $\Upsilon^{\mathrm{R}}$.}
Finally we introduce two analogs of the $\Upsilon_b$ function as follows:
\begin{equation}
    \Upsilon_b^{\mathrm{NS}}(x)\equiv\frac{1}{\Gamma^{\mathrm{NS}}_b(x)\Gamma^{\mathrm{NS}}_b(Q-x)}~ , \ \ \ \Upsilon_b^{\mathrm{R}}(x)\equiv\frac{1}{\Gamma_b^{R}(x)\Gamma_b^{\mathrm{R}}(Q-x)}~.
\end{equation}
These are entire functions with \textit{zeroes} at
\begin{align}
&\Upsilon_b^{\mathrm{NS}}(x)=0 \  \Longrightarrow \  x=kb+lb^{-1}, \ \  \text{for} \ \ k,l\in\mathbb{Z} \ | \ \text{sgn}(k\cdot l)=1 | \ (k,l)\in (2\mathbb{Z},2\mathbb{Z}) \text{ or }(2\mathbb{Z}+1,2\mathbb{Z}+1), \nonumber \\
&\Upsilon_b^{\mathrm{R}}(x)=0 \  \Longrightarrow \  x=k'b+l'b^{-1}, \  \text{for} \  k',l'\in\mathbb{Z} \ | \ \text{sgn}(k'\cdot l')=1 | \ (k',l')\in (2\mathbb{Z},2\mathbb{Z}+1) \text{ or }(2\mathbb{Z},2\mathbb{Z}+1).
\end{align}
They satisfy the \textit{shift relations}
\begin{equation}
    \frac{\Upsilon_b^{\mathrm{NS}}(x+b^{\pm1})}{\Upsilon_b^{\mathrm{R}}(x)}=b^{\mp b^{\pm1}x}\gamma\left(\frac{1+xb^{\pm1}}{2}\right) \ , \ \ \ \ \frac{\Upsilon_b^{\mathrm{R}}(x+b^{\pm1})}{\Upsilon_b^{\mathrm{NS}}(x)}=b^{\pm(1-b^{\pm1}x)}\gamma\left(\frac{xb^{\pm1}}{2}\right),
\end{equation}
    where $\gamma(x)\equiv \frac{\Gamma(x)}{\Gamma(1-x)}$.


\section{Review of the analytic bootstrap in $\mathcal{N}=0$ Liouville CFT}\label{app:N0Liouville}

In this appendix we give a pedagogical and self-contained review of the derivation of the structure constants for ordinary bosonic Liouville CFT with central charge $c\geq 25$ (spacelike) and $c\leq 1$ (timelike). In particular, we follow the chain of reasoning that we presented in the main body of the paper for the $\mathcal{N}=1$ case, implementing Teschner's trick and emphasizing the importance of shift relations for the normalization-independent bootstrap data as well as their symmetries under Virasoro-Wick Rotation. 
\par For the purposes of this appendix we use the following notations for the central charge and conformal dimensions:
\begin{equation}\label{eq:hcN0Liouville}
    c=1+6Q^2, ~~~~~ Q=b+b^{-1} , ~~~~~ h=\frac{Q^2}{4}-p^2 \ .
\end{equation}
The physical spectrum of bosonic Liouville consists of scalar primary operators with $p\in i\mathbb{R}$.
Degenerate representations of the Virasoro algebra occur at (real) momenta $\pm p_{\langle r,s\rangle}$, where
\begin{equation}
p_{\langle r,s\rangle}=\frac{1}{2}\left(rb+sb^{-1}\right) \ , \ \ r,s\in\mathbb{Z}_{\geq 1} \ .
\end{equation}

\subsection*{Derivation of the shift relations}

We consider the analytically continued (i.e. outside Liouville theory's spectrum) four-point function on the sphere
\begin{equation}\label{eq:N04pt}
\langle V_{p_{\langle2,1\rangle}}(z_0)V_{p_1}(z_1)V_{p_2}(z_2)V_{p_3}(z_3)\rangle \ ,
\end{equation}
with a single insertion of a degenerate operator $V_{p_{\langle2,1\rangle}}$  and three physical operators with $p_1,p_2,p_3\in i\mathbb{R}$. The degenerate operator gives rise to a null vector at level 2
\begin{equation}\label{eq:N0nullv}
\left[\frac{1}{b^2}L_{-1}^2+L_{-2}\right]V_{p_{\langle2,1\rangle}}=0. 
\end{equation}
The usual way that this null state equation leads to the BPZ equation\cite{Belavin:1984vu} is as follows.

Since $T(z)=O(z^{-4}$) as $z\rightarrow \infty$, we have $\oint_{\infty}\d z \ \varepsilon(z)T(z)=0$ for any holomorphic function $\varepsilon(z)$ with the property $\varepsilon(z)=O(z^n)$, and $n\leq 2$, as $z\rightarrow \infty$. In the case of the four-point function (\ref{eq:N04pt}), and since the OPE of primary fields with $T$ behaves as
\begin{equation}
    T(z)V_{p}(z_i)= O((z-z_i)^{-2}),
\end{equation}
it is convenient to choose $\varepsilon(z)$ to be a monomial in $(z-z_i)$ for $i=1,\cdots,3$ and, to ensure the behaviour at infinity, we divide with the monomial in the remaining point. In other words, we choose $\varepsilon(z)=\frac{(z-z_1)(z-z_2)(z-z_3)}{z-z_0}$. We then get the following Ward identity
\begin{equation}\label{eq: N0 null vector0}
     \oint_{\infty}\d z \ \varepsilon(z)\langle T(z)V_{p_{\langle2,1\rangle}}(z_0)V_{p_1}(z_1)V_{p_2}(z_2)V_{p_3}(z_3)\rangle=0 .
\end{equation}

We next deform the contour to wrap all the circles around the points $z_0,z_1,z_2,z_3$ and use the corresponding OPE. In other words,
\begin{align}\label{eq: N0 null vector1}
    &\oint_{z_0}\d z \ \varepsilon(z)\langle \left[T(z)V_{p_{\langle2,1\rangle}}(z_0)\right]_{\text{OPE}}V_{p_1}(z_1)V_{p_2}(z_2)V_{p_3}(z_3)\rangle \nonumber \\
&+ \sum_{i=1}^3\oint_{z_i}\d z \  \varepsilon(z) \langle V_{p_{\langle 2,1\rangle}}(z_0)\bigg[T(z)V_{p_i}(z_i)\bigg]_{\text{OPE}}\prod_{j\neq i}V_{p_j}(z_j)\rangle = 0 \ .
\end{align}
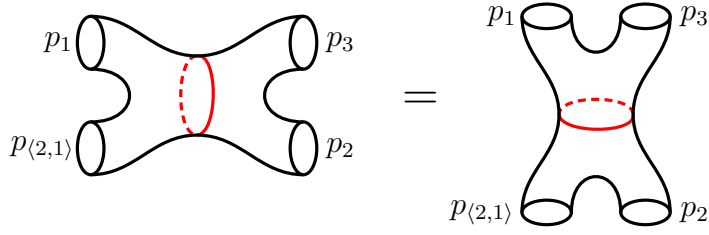
\begin{figure}
\centering 
\begin{tikzpicture}
\begin{scope}[shift={(1.1,0)}, scale=.7]
\draw[very thick, red, looseness=.7] (2,1.75) to[out=10,in=5] (2,.25);
\draw[very thick, densely dashed, red, looseness=.7] (2,1.75) to[out=-180,in=-180] (2,.25);    
\draw[very thick] (0,0) ellipse (0.25 and .5);
\draw[very thick] (0,2) ellipse (0.25 and .5);
\draw[very thick] (4,0) ellipse (0.25 and .5);
\draw[very thick] (4,2) ellipse (0.25 and .5);
\node[scale=1., thick] at (-0.9,0) {$p_{\langle2,1\rangle}$};
\node[scale=1., thick] at (-0.6,2) {$p_1$};  
\node[scale=1., thick] at (4.7,2) {$p_3$};
\node[scale=1., thick] at (4.7,0) {$p_2$};
\draw[very thick] (0,.5) to[out=0, in=0, looseness=2.5] (0,1.5);
\draw[very thick] (4,.5) to[out=180, in=180, looseness=2.5] (4,1.5);
\draw[very thick] (0,2.5) to[out=0, in=180] (2,1.75) to[out=0, in=180] (4,2.5);
\draw[very thick] (0,-.5) to[out=0, in=180] (2,.25) to[out=0, in=180] (4,-.5); 
\node[scale=1.5] at (6.2,.9) {$=$};
\end{scope}
\begin{scope}[shift={(7.1,1.75)}, scale=.65, rotate=270]
\draw[very thick, red, looseness=.7] (2,1.75) to[out=10,in=5] (2,.25);
\draw[very thick, densely dashed, red, looseness=.7] (2,1.75) to[out=-180,in=-180] (2,.25);    
\draw[very thick] (0,0) ellipse (0.25 and .5);
\draw[very thick] (0,2) ellipse (0.25 and .5);
\draw[very thick] (4,0) ellipse (0.25 and .5);
\draw[very thick] (4,2) ellipse (0.25 and .5);
\node[scale=1., thick] at (0,-0.9) {$p_1$};
\node[scale=1., thick] at (0,3) {$p_3$};  
\node[scale=1., thick] at (4,3) {$p_2$};
\node[scale=1., thick] at (4,-1.3) {$p_{\langle2,1\rangle}$};
\draw[very thick] (0,.5) to[out=0, in=0, looseness=2.5] (0,1.5);
\draw[very thick] (4,.5) to[out=180, in=180, looseness=2.5] (4,1.5);
\draw[very thick] (0,2.5) to[out=0, in=180] (2,1.75) to[out=0, in=180] (4,2.5);
\draw[very thick] (0,-.5) to[out=0, in=180] (2,.25) to[out=0, in=180] (4,-.5);    
\end{scope}
\end{tikzpicture}
\caption{Teschner's trick in usual bosonic Liouville theory: the analytic bootstrap problem involving crossing of the sphere four-point function between a Virasoro degenerate field $V_{p_{\langle2,1\rangle}}$ and three basic fields $V_{p_1},V_{p_2},V_{p_3}$. The analysis leads to the shift relations (\ref{eq:N0cond2}).
}
\label{fig:N1N0bootstrap}
\end{figure}
\noindent
\par For the second term (i.e. the sum over the three remaining points) it is obvious what the residues are. For the first term involving the degenerate field, we will have contributions up to, and including, order $(z-z_0)^0$ in the OPE, since this term is $L_{-2}V_{p_{\langle2,1\rangle}}$ and should be replaced by $-\frac{1}{b^2}L_{-1}^2V_{p_{\langle2,1\rangle}}=-\frac{1}{b^2}\partial_{z_0}^2V_{p_{\langle2,1\rangle}}(z_0)$ from (\ref{eq:N0nullv}). Doing all that, we obtain
\begin{align}
    \bigg\{\prod_{i=1}^3(z_0-z_i)&\left(-\frac{1}{b^2}\partial_{z_0}^2+\sum_{i=1}^3\frac{\partial_{z_0}}{z_0-z_i}\right)+(3z_0-z_1-z_2-z_3)h_{\langle2,1\rangle}\nonumber \\
&+\frac{z_{12}z_{13}}{z_1-z_0}h_1+\frac{z_{21}z_{23}}{z_2-z_0}h_2+\frac{z_{31}z_{32}}{z_3-z_0}h_3\bigg\}\langle V_{p_{\langle2,1\rangle}}(z_0)V_{p_1}(z_1)V_{p_2}(z_2)V_{p_3}(z_3)\rangle=0 \ .
\end{align}
We next write the correlation function in terms of the conformally invariant cross-ratio $z\equiv \frac{z_{01}z_{23}}{z_{03}z_{21}}$ (with $z_{ij}=z_i-z_j$)
\begin{equation}
   \langle V_{p_{\langle2,1\rangle}}(z_0)V_{p_1}(z_1)V_{p_2}(z_2)V_{p_3}(z_3)\rangle = \left[\prod_{i<j}(z_i-z_j)^{\mu_{ij}}\right] \mathcal{F}(z) 
\end{equation}
with $\mu_{ij}\equiv \frac{1}{3}\sum_{k=0}^3 h_{k} - h_i-h_j$.
Taking the limits $(z_1,z_2,z_3)\rightarrow (0,1,\infty)$ and rescaling
\begin{equation}
    \mathcal{F}(z) = z^{-\mu_{01}}(1-z)^{-\mu_{02}}\mathcal{F}(z)
\end{equation}
we finally get the desired BPZ differential equation
\begin{equation}\label{eq:NOBPZ}
\frac{z(1-z)}{b^2}\partial_z^2\mathcal{F}(z)+(2z-1)\partial_z\mathcal{F}(z)+\left(h_{\langle2,1\rangle}+\frac{h_1}{z}-h_3+\frac{h_2}{1-z}\right)\mathcal{F}(z)=0.
\end{equation}

There are \textit{two} linearly independent solutions of (\ref{eq:NOBPZ}) given by
\begin{align}\label{eq:NOschannblocks}
\mathcal{F}_{s}^{(\varepsilon=\pm)}(z)&=z^{\frac{bQ}{2}-\varepsilon b p_1}(1-z)^{\frac{bQ}{2}-bp_2}\nonumber \\
&\times {}_2F_1\left(\frac{1}{2}-bp_2+\varepsilon b(p_3-p_1),\frac{1}{2}-bp_2+\varepsilon b (-p_3-p_1);1-2\varepsilon b p_1 ; z\right) .
\end{align}
These are the two conformal blocks propagating in the $s-$channel OPE (between $V_{p_{\langle2,1\rangle}}V_{p_1}$ and $V_{p_{3}}V_{p_2}$, see left of Fig.\ref{fig:N0crossing}). The internal Liouville momenta are $p_s=p_1+\frac{\epsilon b}{2}$ and can be read from the exponent of the leading $z\rightarrow 0$ power
\begin{equation}
    \frac{bQ}{2} - \varepsilon b p_1 = - \left(\frac{Q^2}{4}- p_1^2\right)- \left(\frac{Q^2}{4}- p_{\langle 2,1\rangle}^2\right) +\left(\frac{Q^2}{4} - \left(p_1+ \frac{\varepsilon b}{2}\right)^2\right).
\end{equation}
The $t-$channel blocks $\mathcal{F}_{t}^{(\eta)}(z)$ have internal momenta $p_t=p_2+\frac{\eta b}{2} \ (\eta=\pm$) (see right of Fig.\ref{fig:N0crossing}) and are given by the same expression as in (\ref{eq:NOschannblocks}) after exchanging $p_1 \leftrightarrow p_2$ and $z\leftrightarrow 1-z$, namely
\begin{align}\label{eq:NOtchannblocks}
\mathcal{F}_{t}^{(\eta=\pm)}(z)&=z^{\frac{bQ}{2}- b p_1}(1-z)^{\frac{bQ}{2}-\eta bp_2}\nonumber \\
&\times {}_2F_1\left(\frac{1}{2}-bp_1+\eta b(p_3-p_2),\frac{1}{2}-bp_1+\eta b (-p_3-p_2);1-2\eta b p_2 ; 1-z\right) .
\end{align}
As before, one reads the internal momenta from the exponent in the leading $z\rightarrow1$ power
\begin{equation}
\frac{bQ}{2} - \eta bp_2=  -\left(\frac{Q^2}{4}- p_2^2\right) -\left(\frac{Q^2}{4}- p_{\langle 2,1\rangle}^2\right) +\left(\frac{Q^2}{4} - \left(p_2+ \frac{\eta b}{2}\right)^2\right).
\end{equation}
\begin{figure}
\begin{tikzpicture}
\begin{scope}[shift={(-5,0)}, scale=.7]
\draw[very thick, red, looseness=.7] (2,1.75) to[out=10,in=5] (2,.25);
\draw[very thick, densely dashed, red, looseness=.7] (2,1.75) to[out=-180,in=-180] (2,.25);    
\draw[fill=white, draw=gray, opacity=.1] (0,2.5) to[out=0, in=180] (2,1.75) to[out=0, in=180] (4,2.5) 
    to[out=200, in=160, looseness=.95] (4,1.5) to[out=180, in=180, looseness=2.5] (4,.5) 
    to[out=200, in=160, looseness=.95] (4,-.5) to[out=180, in=0] (2,.25) 
    to[out=180, in=0] (0,-.5) to[out=20, in=-20, looseness=.95] (0,.5) 
    to[out=0, in=0, looseness=2.5] (0,1.5) to[out=20, in=-20, looseness=.95] (0,2.5);
\draw[very thick] (0,0) ellipse (0.25 and .5);
\draw[very thick] (0,2) ellipse (0.25 and .5);
\draw[very thick] (4,0) ellipse (0.25 and .5);
\draw[very thick] (4,2) ellipse (0.25 and .5);
\node[scale=1., thick] at (-0.9,0) {$p_{\langle2,1\rangle}$};
\node[scale=1., thick] at (-0.6,2) {$p_1$};  
\node[scale=1., thick] at (4.7,2) {$p_3$};
\node[scale=1., thick] at (4.7,0) {$p_2$};
\draw[very thick] (0,.5) to[out=0, in=0, looseness=2.5] (0,1.5);
\draw[very thick] (4,.5) to[out=180, in=180, looseness=2.5] (4,1.5);
\draw[very thick] (0,2.5) to[out=0, in=180] (2,1.75) to[out=0, in=180] (4,2.5);
\draw[very thick] (0,-.5) to[out=0, in=180] (2,.25) to[out=0, in=180] (4,-.5); 
\node[scale=2.] at (7.3,1.) {$~~~~~~~~= \  \sum_{\eta=\pm}\mathbb{F}_{\epsilon,\eta}$};
\node[scale=1., red] at (2,-.5) {$p_1+ \frac{\varepsilon b}{2}$};
\end{scope}
\begin{scope}[shift={(4.5,1.75)}, scale=.65, rotate=270]
\draw[very thick, red, looseness=.7] (2,1.75) to[out=10,in=5] (2,.25);
\draw[very thick, densely dashed, red, looseness=.7] (2,1.75) to[out=-180,in=-180] (2,.25);    
\draw[fill=white, draw=gray, opacity=.1] (0,2.5) to[out=0, in=180] (2,1.75) to[out=0, in=180] (4,2.5) 
    to[out=200, in=160, looseness=.95] (4,1.5) to[out=180, in=180, looseness=2.5] (4,.5) 
    to[out=200, in=160, looseness=.95] (4,-.5) to[out=180, in=0] (2,.25) 
    to[out=180, in=0] (0,-.5) to[out=20, in=-20, looseness=.95] (0,.5) 
    to[out=0, in=0, looseness=2.5] (0,1.5) to[out=20, in=-20, looseness=.95] (0,2.5);
\draw[very thick] (0,0) ellipse (0.25 and .5);
\draw[very thick] (0,2) ellipse (0.25 and .5);
\draw[very thick] (4,0) ellipse (0.25 and .5);
\draw[very thick] (4,2) ellipse (0.25 and .5);
\node[scale=1., thick] at (0,-0.9) {$p_1$};
\node[scale=1., thick] at (0,3) {$p_3$};  
\node[scale=1., thick] at (4,3) {$p_2$};
\node[scale=1., thick] at (4,-1.3) {$p_{\langle2,1\rangle}$};
\draw[very thick] (0,.5) to[out=0, in=0, looseness=2.5] (0,1.5);
\draw[very thick] (4,.5) to[out=180, in=180, looseness=2.5] (4,1.5);
\draw[very thick] (0,2.5) to[out=0, in=180] (2,1.75) to[out=0, in=180] (4,2.5);
\draw[very thick] (0,-.5) to[out=0, in=180] (2,.25) to[out=0, in=180] (4,-.5);    
\node[scale=1., red] at (2,3) {$p_2+ \frac{\eta b}{2}$};
\end{scope}
\end{tikzpicture}
\caption{Crossing transformation of sphere four-point conformal blocks with one insertion of a $\langle2,1\rangle$ degenerate operator. The fusion kernel is a $2\times 2$ matrix given in (\ref{eq:N0crossing}). }
\label{fig:N0crossing}
\end{figure}
\par The correlation function then admits as usual the $s,t$ channel expansions (involving holomorphic and anti-holomorphic conformal blocks)
\begin{equation}\label{eq:NOcrossingst}
    \mathcal{F}(z)= \sum_{\varepsilon= \pm 1} c_\varepsilon^{(s)} \mathcal{F}_s^{(\varepsilon)}(z)\overline{\mathcal{F}_s^{(\varepsilon)}(z)} =\sum_{\eta = \pm 1} c_\eta^{(t)} \mathcal{F}_t^{(\eta)}(z)\overline{\mathcal{F}_t^{(\eta)}(z)}
\end{equation}
where
\begin{align}
c_\varepsilon^{(s)}&:=\frac{C^{(b)}(p_1,p_{\langle2,1\rangle},p_1+\frac{\varepsilon b}{2})C^{(b)}(p_1+\frac{\varepsilon b}{2},p_2,p_3)}{B^{(b)}\left(p_1+\frac{\varepsilon b}{2}\right)} \ , \nonumber\\ c_\eta^{(t)}&:=\frac{C^{(b)}(p_2,p_{\langle2,1\rangle},p_2+\frac{\eta b}{2})C^{(b)}(p_2+\frac{\eta b}{2},p_1,p_3)}{B^{(b)}\left(p_2+\frac{\eta b}{2}\right)}.
\end{align}
Here $C^{(b)}(p_i,p_j,p_k)$ are the structure constants of the theory (that we are eventually interested in computing), and $B^{(b)}(p_i)$ is an arbitrary two-point function normalization.

\par Using the properties of the hypergeometric function\footnote{In particular, the identity
\begin{align}
    _2 F_1 \left(a, f, c, z\right) &= \frac{\Gamma(c)\Gamma(c-a-f)}{\Gamma(c-a) \Gamma(c-f)} \, _2F_1(a, f, a+f-c+1, 1-z)  \cr  
    &\quad + \frac{\Gamma (c)\Gamma(a+f-c)}{\Gamma(a)\Gamma(f)} (1-z)^{c-a-f} \, _2F_1 (c-a,c-f,c-a-f+1,1-z).\nonumber
\end{align}
} it is straightforward to show that a $s-$channel conformal block can be written as a linear combination of $t-$channel blocks as
\begin{equation}\label{eq:N0crossing}
\mathcal{F}_{s}^{(\epsilon)}(z) = \sum_{\eta=\pm
} \mathbb{F}_{\epsilon,\eta} \ \mathcal{F}_{t}^{(\eta)}(z) \ , ~~~~~~~ \mathbb{F}_{\epsilon,\eta} := \frac{\Gamma\left(1-2b\epsilon p_1\right)\Gamma\left(2b\eta p_2\right)}{\Gamma\left(\frac{1}{2}+b\left(-\epsilon p_1+\eta p_2\pm p_3\right)\right)} \ , \ \  \epsilon,\eta=\pm.
\end{equation}
The coefficients $\mathbb{F}_{\epsilon,\eta}$ define a $2\times 2$ \textit{fusion kernel}. The form of this fusion kernel severely constraints the CFT data entering in the expansions (\ref{eq:NOcrossingst}). Indeed, crossing symmetry (\ref{eq:NOcrossingst}) combined with (\ref{eq:N0crossing}) imply the constraint
\begin{equation}\label{eq:N0cond1}
\sum_{\epsilon=\pm}\frac{C^{(b)}(p_1,p_{\langle 2,1\rangle},p_1+\frac{\epsilon b}{2})C^{(b)}(p_1+\frac{\epsilon b}{2},p_2,p_3)}{B^{(b)}(p_1+\frac{\epsilon b}{2})}\mathbb{F}_{\epsilon,+}\mathbb{F}_{\epsilon,-} = 0.
\end{equation}
At this point it is desirable to eliminate the structure constants involving the degenerate field $p_{\langle2,1\rangle}$. We can do this easily by setting $p_2=p_{\langle2,1\rangle}$ and $p_3=p_1$ in (\ref{eq:N0cond1}), which then yields an equation that involves only (the squares of) $C^{(b)}(p_1,p_{\langle 2,1\rangle},p_1\pm\frac{\epsilon b}{2})$. Using the latter back in (\ref{eq:N0cond1}) and after some rearrangement we get
\begin{equation}\label{eq:N0cond2}
\frac{C^{(b)}(p_1+\frac{b}{2},p_2,p_3)^2/B^{(b)}(p_1+\frac{b}{2})}{C^{(b)}(p_1-\frac{b}{2},p_2,p_3)^2/B^{(b)}(p_1-\frac{b}{2})}=\kappa_b(p_1|p_2,p_3),
\end{equation}
where
\begin{align}
\kappa_b(p_1|p_2,p_3):= -&\frac{\Gamma\left(\frac{1}{2}\pm(b^2+\frac{1}{2})+2bp_1\right)\Gamma\left(1+2bp_1\right)^2}{\Gamma\left(\frac{1}{2}\pm(b^2+\frac{1}{2})-2bp_1\right)\Gamma\left(1-2bp_1\right)^2} \frac{\Gamma\left(\frac{1}{2}+b(-p_1\pm p_2 \pm p_3)\right)^2}{\Gamma\left(\frac{1}{2}+b(p_1\pm p_2 \pm p_3)\right)^2}  .
\end{align}
As in the main text, the $\pm$ indicates a product over all possible combinations. 
Equation (\ref{eq:N0cond2}) is a shift relation for the normalization-independent bootstrap quantity $(C^{(b)}_{p_1,p_2,p_3})^2/B^{(b)}(p_1)$, which we now interpret as the expansion coefficient of an `auxiliary' four-point correlation function with pairwise identical external operators $p_2,p_3$ ($p_1$ being the exchanged internal one). In other words, even though we started with the correlation function (\ref{eq:N04pt}) involving the degenerate field $p_{\langle2,1\rangle}$, we ended up with an equation that involves only the unspecified momenta $p_1,p_2,p_3$\footnote{The information about the degenerate field is basically encoded in the function $\kappa_b$.}.
\par We emphasize that even though (\ref{eq:N0cond2}) is a shift relation in $p_1$, we could have equivalently derived identical equations involving shifts in either $p_2,p_3$. At this point we claim that it is natural to search for a solution of the structure constants that is permutation-invariant under all exchanges of momenta. There is furthermore an identical shift relation after replacing $b\rightarrow b^{-1}$, since all the quantities are ultimately functions of the central charge $c$. The $b$ and $b^{-1}$ shift relations are incommensurable only in the case where $b^{2}\notin \mathbb{Q}$. In what follows, we will mainly focus on the shift relation (\ref{eq:N0cond2}), while keeping in mind that there is a counterpart involving shifts by $b^{-1}$.
\par We pause to track some properties of $\kappa_b$. It is permutation symmetric under exchanging $p_2\leftrightarrow p_3$, but not with $p_1$. It is further invariant under reflections of either $p_2$ or $p_3$, but \textit{not} under reflections of $p_1$. It is only invariant under reflections of $p_1$ when we perform a simultaneous reflection on $b$. In summary,
\begin{equation}
\kappa_{b}(p_1|p_3,p_2)=\kappa_b(p_1|p_2,p_3) \ , \ \ \kappa_{-b}(-p_1|p_2,p_3)=\kappa_b(p_1|p_2,p_3)~.
\end{equation}
Furthermore, under the VWR map $b\rightarrow ib$, $p_j\rightarrow i p_j$ it satisfies an almost inverse relation with the original quantity: 
\begin{equation}\label{eq:N0vwrcond}
\kappa_{b}(p_1|p_2,p_3)= \frac{\left(\frac{p_1+\frac{b}{2}}{p_1-\frac{b}{2}}\right)^2}{\kappa_{-ib}(-ip_1|-ip_2,-ip_3)} \ .
\end{equation}
Just like in the main text for the $\mathcal{N}=1$ case, this property will turn out to be crucial shortly when we discuss solutions of the shift relations in the timelike central charge regime $c\leq 1$.

\subsection*{Spacelike structure constants}

 Our main objective is to construct a solution of (\ref{eq:N0cond2}) for the structure constants $C^{(b)}$ that is permutation symmetric under the exchange of $p_1,p_2,p_3$, reflection symmetric in each $p_i\rightarrow -p_i$, and invariant under $b\leftrightarrow b^{-1}$. With the additional assumption of meromorphicity in $p_i$'s and continuity in $b^2\in\mathbb{R}$, such a solution is essentially unique \cite{Teschner:1995yf}\footnote{When $b^2\in\mathbb{C}$ the uniqueness argument fails and two given solutions can differ by a doubly-periodic function of the momenta with periods $b$ and $b^{-1}$ \cite{Zamolodchikov:2005fy}. }. The so-called `spacelike' structure constants refer to solutions of (\ref{eq:N0cond2}) for $c\geq 25$. Therefore we want to solve 
 \begin{equation}\label{eq:N0breal}
\frac{C^{(b)}(p_1+\frac{b}{2},p_2,p_3)^2/B^{(b)}(p_1+\frac{b}{2})}{C^{(b)}(p_1-\frac{b}{2},p_2,p_3)^2/B^{(b)}(p_1-\frac{b}{2})}=\kappa_b(p_1|p_2,p_3)\ , \ \ \ \ \ \ \ \ \text{for} \ \ b\in\mathbb{R}_{(0,1]} \ .
\end{equation}
 Starting from (\ref{eq:N0breal}) we can make several convenient choices for the two-point structure constant $B^{(b)}$ which will then lead to different solutions for the structure constants with the aforementioned properties. We record two such choices that are of particular interest in the literature.
\paragraph{DOZZ normalization.} The most common normalization of the two- and three-point structure constants is inspired by the action of spacelike Liouville theory (see e.g. eqns (2.2),(2.6) in \cite{Zamolodchikov:2005fy}). It reads
\begin{equation}
B^{(b)}_{\text{DOZZ}}(p)= \lambda^{2b^{-1}p}\frac{\gamma(1-2bp)}{b^2\gamma(1+2b^{-1}p)}
\end{equation}
where $\lambda\equiv\pi \mu \gamma(b^2)$ and $\mu$ the usual cosmological constant that enters in the Lagrangian of the theory. With this two-point function, the structure constants that solve (\ref{eq:N0breal}) read
\begin{equation}
C^{(b)}_{\text{DOZZ}}(p_1,p_2,p_3)=(\lambda b^{2-2b^2})^{b^{-1}(-\frac{Q}{2}+p_1+p_2+p_3)}\Upsilon_0\frac{\Gamma_b(\frac{Q}{2}\pm p_1\pm p_2 \pm p_3)}{\prod_{j=1}^3\Gamma_b(Q-2p_j)\Gamma_b(2p_j)}
\end{equation}
where $\Upsilon_0\equiv \text{Res}_{x=0}[\partial_x\Upsilon_b(x)]$ is a convenient constant (which doesn't affect the shift relations).
\paragraph{Natural normalization.} The following normalization was recently used in \cite{Collier:2023cyw} (see also \cite{Collier:2019weq,Eberhardt:2023mrq}). In this normalization we identify the two-point function with the inverse of the modular kernel of the identity torus character in 2d CFTs. Indeed, it is known that
\begin{equation}
\chi_{\mathds{1}}(-1/\tau)=\int_{i\mathbb{R}}\frac{\d p}{2}  \rho^{(b)}_0(p) \chi_{p}(\tau) \ ,
\end{equation}
with
\begin{equation}
 \rho_0^{(b)}(p)=-4\sqrt{2}\sin{(2\pi b p)}\sin{(2\pi b^{-1}p)}~,\quad \chi_p(\tau) = \frac{\e^{-2\pi i \tau p^2}}{\eta(\tau)}~.
\end{equation}
We then make the choice
\begin{equation}\label{eq:n02ptfn}
B_{\text{VMS}}^{(b)}(p)=\frac{1}{\rho_0^{(b)}(p)}~.
\end{equation}
This causes an essential simplification in (\ref{eq:N0breal}) since known quantities form a total square. It is straightforward to see that the equation takes the form 
\begin{equation}\label{eq:N0VMSsimpler}
\frac{C_{\text{VMS}}^{(b)}(p_1-\frac{b}{2},p_2,p_3)}{C_{\text{VMS}}^{(b)}(p_1+\frac{b}{2},p_2,p_3)}= \frac{\Gamma\left(1-2bp_1\right)\Gamma\left(1+b^2-2bp_1\right)\Gamma\left(\frac{1}{2}+b(p_1\pm p_2 \pm p_3)\right)}{\Gamma\left(1+2bp_1\right)\Gamma\left(1+b^2+2bp_1\right)\Gamma\left(\frac{1}{2}+b(-p_1\pm p_2 \pm p_3)\right)} \ .
\end{equation}
A unique solution to (\ref{eq:N0VMSsimpler}) that is meromorphic in the momenta, continuous in $b\in(0,1]$, symmetric under reflections $p_i\rightarrow-p_i$, symmetric under any permutation of $p_{1},p_{2},p_{3}$, invariant under $b\leftrightarrow b^{-1}$, and reproduces the two-point structure constant normalization as
\begin{equation}
\lim_{p_3\rightarrow Q/2} C^{(b)}_{\text{VMS}}(p_1,p_2,p_3) = \frac{1}{\rho^{(b)}_0(p_1)}(\delta(p_1-p_2)  + \delta(p_1+p_2))\ ,
\end{equation}
is given by 
\begin{equation}\label{eq:stractconstVMS}
C^{(b)}_{\text{VMS}}(p_1,p_2,p_3)=\frac{\Gamma_b(2Q)\Gamma_b(\frac{Q}{2}\pm p_1\pm p_2 \pm p_3)}{\sqrt{2}\Gamma_b(Q)^3\prod_{j=1}^3\Gamma_b(Q\pm 2p_j)}.
\end{equation}

\subsection*{Timelike structure constants}\label{sec:VWRN0}
\par Just as we did for the $\mathcal{N}=1$ case in the main text, we will describe the derivation of the timelike structure constants using the symmetry of the shift relations (\ref{eq:N0cond2}) under VWR. 
\par For $b\in(0,1]$ (or $c\geq 25$) we have already found the solutions (\ref{eq:n02ptfn}), (\ref{eq:stractconstVMS}), i.e.
\begin{equation}\label{eq:N0bspacelike}
\frac{C^{(b)}(p_1+\frac{b}{2},p_2,p_3)^2/B^{(b)}(p_1+\frac{b}{2})}{C^{(b)}(p_1-\frac{b}{2},p_2,p_3)^2/B^{(b)}(p_1-\frac{b}{2})}=\kappa_b(p_1|p_2,p_3)\ , \ \ \ \ \ \ \ \ \text{for} \ \ c\geq 25~.
\end{equation}
For $b=-i\hat{b}$, with $\hat{b}\in\mathbb{R}_{(0,1]}$ (or $c\leq1$), let us denote the corresponding (and still unknown) solutions as $\widehat{B}^{(\hat{b})}, \widehat{C}^{(\hat{b})}$. They obviously satisfy the same shift relation in the corresponding central charge regime
\begin{equation}\label{eq:N0btimelike}
\frac{\widehat{C}^{(\hat{b})}(p_1+\frac{(-i\hat{b})}{2},p_2,p_3)^2/\widehat{B}^{(\hat{b})}(p_1+\frac{(-i\hat{b})}{2})}{\widehat{C}^{(\hat{b})}(p_1-\frac{(-i\hat{b})}{2},p_2,p_3)^2/\widehat{B}^{(\hat{b})}(p_1-\frac{(-i\hat{b})}{2})}=\kappa_{-i\hat{b}}(p_1|p_2,p_3)\ , \ \ \ \ \ \ \ \ \text{for} \ \ \ \  c\leq 1~.
\end{equation}
We can now pin down the new structure constants $\widehat{B}^{(\hat{b})}, \widehat{C}^{(\hat{b})}$ by implementing the VWR of the shift equations. Indeed, our strategy is to start from (\ref{eq:N0bspacelike}) and, by leveraging the analytic properties of 
$\kappa_b$ (in particular using (\ref{eq:N0vwrcond})), gradually construct a solution to (\ref{eq:N0btimelike}). 
\par Relabelling $b\equiv \hat{b}\in\mathbb{R}$ in (\ref{eq:N0bspacelike}) and using (\ref{eq:N0vwrcond}) yields
\begin{equation}
\frac{C^{(\hat{b})}(p_1+\frac{\hat{b}}{2},p_2,p_3)^2/B^{(\hat{b})}(p_1+\frac{\hat{b}}{2})}{C^{(\hat{b})}(p_1-\frac{\hat{b}}{2},p_2,p_3)^2/B^{(\hat{b})}(p_1-\frac{\hat{b}}{2})}=\kappa_{\hat{b}}(p_1|p_2,p_3)=\frac{\left(\frac{p_1+\frac{\hat{b}}{2}}{p_1-\frac{\hat{b}}{2}}\right)^2}{\kappa_{-i\hat{b}}(-ip_1|-ip_2,-ip_3)}~. 
\end{equation}
Next, let's re-name $p_k\rightarrow  ip_k$. This gives 
\begin{align}\label{eq:VWRshiftrel}
&\frac{C^{(\hat{b})}(i(p_1+\frac{(-i\hat{b})}{2}),ip_2,ip_3)^2/B^{(\hat{b})}(i(p_1+\frac{(-i\hat{b})}{2}))}{C^{(\hat{b})}(i(p_1-\frac{(-i\hat{b})}{2}),ip_2,ip_3)^2/B^{(\hat{b})}(i(p_1-\frac{(-i\hat{b})}{2}))}=\frac{\left(\frac{p_1-\frac{i\hat{b}}{2}}{p_1+\frac{i\hat{b}}{2}}\right)^2}{\kappa_{-i\hat{b}}(p_1|p_2,p_3)} \nonumber \\
&\Leftrightarrow \ \ \left(\frac{p_1-\frac{i\hat{b}}{2}}{p_1+\frac{i\hat{b}}{2}}\right)^2 \frac{C^{(\hat{b})}(i(p_1-\frac{(-i\hat{b})}{2}),ip_2,ip_3)^2/B^{(\hat{b})}(i(p_1-\frac{(-i\hat{b})}{2}))}{C^{(\hat{b})}(i(p_1+\frac{(-i\hat{b})}{2}),ip_2,ip_3)^2/B^{(\hat{b})}(i(p_1+\frac{(-i\hat{b})}{2}))} =\kappa_{-i\hat{b}}(p_1|p_2,p_3)
\end{align}
The last equation has exactly the form (\ref{eq:N0btimelike}) which we wanted to solve. 
\par There is of course some freedom on how to identify $\widehat{B}^{(\hat{b})}, \widehat{C}^{(\hat{b})}$ in (\ref{eq:VWRshiftrel}). The most natural choice is the one where the three-point structure constant is just the \textit{inverse} of the VWR-ed structure constant (\ref{eq:stractconstVMS}). We denote this particular choice as $\widehat{C}^{(\hat{b})}_{\text{VMS}}$. We then get  
\begin{equation}\label{eq:structconsttimelikeVMS}
\widehat{C}^{(\hat{b})}_{\text{VMS}}(p_1,p_2,p_3) \equiv \frac{1}{C_{\text{VMS}}^{(\hat{b})}(ip_1,ip_2,ip_3)} =\frac{\sqrt{2}\Gamma_{\hat{b}}(\hat{b}+\hat{b}^{-1})^3\prod_{j=1}^3\Gamma_{\hat{b}}(\hat{b}+\hat{b}^{-1}\pm 2ip_j)}{\Gamma_{\hat{b}}(2\hat{b}+2\hat{b}^{-1})\Gamma_{\hat{b}}(\frac{\hat{b}+\hat{b}^{-1}}{2}\pm ip_1\pm ip_2 \pm ip_3)}.
\end{equation}
This then leaves no room for the choice of two-point function (modulo momentum-independent factors), namely
\begin{equation}\label{eq:2ptfnVMStimelike}
\widehat{B}^{(\hat{b})}_{\text{VMS}}(p) \equiv \frac{1}{p^2 B^{(\hat{b})}_{\text{VMS}}(ip)} = \frac{\rho^{{(\hat{b})}}_0(ip)}{p^2}.
\end{equation}

\paragraph{Zamolodchikov's normalization.}

\par The timelike structure constants of $\mathcal{N}=0$ Liouville theory were originally derived in \cite{Zamolodchikov:2005fy}, where the following normalization was adopted\footnote{To connect with e.g. (5.1) of the paper we should take $\alpha_k=\frac{\hat{b}^{-1}-\hat{b}}{2}-ip_k$.}
\begin{equation}
\begin{aligned}
\widehat{B}^{(\hat{b})}_{\text{Z}}(p)&=1 \ , \\
\widehat{C}^{(\hat{b})}_{\text{Z}}(p_1,p_2,p_3)&= A \frac{\left[\prod_{j=1}^3\Gamma_{\hat{b}}\left(\pm2ip_j+\hat{b}\right)\Gamma_{\hat{b}}\left(\pm2ip_j+\hat{b}^{-1}\right)\right]^{1/2}}{\Gamma_{\hat{b}}\left(\frac{\hat{b}+\hat{b}^{-1}}{2}\pm i p_1 \pm ip_2 \pm ip_3\right)}
\end{aligned}.
\end{equation}
Here $A\equiv \frac{\hat{b}^{\hat{b}^{-2}-\hat{b}^2-1}[\gamma(\hat{b}^2)\gamma(\hat{b}^{-2}-1)]^{1/2}}{\Upsilon_{\hat{b}}(\hat{b})}$ and is chosen such that $\widehat{C}^{(\hat{b})}_{\text{Z}}(p,p,i(\hat{b}^{-1}-\hat{b})/2)=1$. One can verify that these expressions satisfy (\ref{eq:N0btimelike}). Despite the bizarre-looking square-root, this particular normalization of the three-point structure constants can be shown to match the structure constants of the Virasoro minimal models\cite{Zamolodchikov:2005fy} when evaluated at the appropriate values of the central charge and conformal dimensions. 
\par Another important and different normalization of the timelike structure constants was discussed in \cite{Harlow:2011ny}, where their particular choice originated from the interpretation of the timelike Liouville
path integral as being a different integration cycle of ordinary Liouville theory.

\section{Derivation of the shift relations in the NS-sector}\label{app:NS}
In this appendix we revisit the derivation of the shift relations for the NS-sector structure constants. Our approach relies on the NS-sector null vector (\ref{eq:nullvectorNS}) and is conceptually analogous to the non-supersymmetric case which we revisited in appendix \ref{app:N0Liouville}, i.e. it involves no superspace formalism. The only difference will be that instead of a second order differential equation, inserting the degenerate field $V_{p_{\langle 1,3\rangle}}^{\mathrm{NS}}$ into a four point function leads to a third order differential equation. For notational convenience we omit the NS superscript. All the fields in this appendix are in the NS-sector. 
\par Our approach toward the null vector differential equations follows the logic in \cite{Belavin:2007eq, Belavin:2007gz} (see also \cite{Ivanova:2025pkv}). 
We will derive two differential equations for the four-point functions
\begin{equation}\label{eq:N04ptNS}
\langle V_{p_{\langle 1,3\rangle}}(z_0)V_{p_1}(z_1)V_{p_2}(z_2)V_{p_3}(z_3)\rangle \ ,\quad \langle \Lambda_{p_{\langle 1,3\rangle}}(z_0)V_{p_1}(z_1)\Lambda_{p_2}(z_2)V_{p_3}(z_3)\rangle~
\end{equation}
with the degenerate fields $V_{p_{\langle 1,3\rangle}}$ and $\Lambda_{p_{\langle 1,3\rangle}} = G_{-1/2}V_{p_{\langle 1,3\rangle}}$, with $p_{\langle 1,3 \rangle } = \frac{1}{2b}+ \frac{3b}{2}$, $p_1,p_2,p_3 \in i \mathbb{R}$.
We then combine these two equations to obtain a third order differential equation for $\langle V_{p_{\langle 1,3\rangle}}(z_0)V_{p_1}(z_1)V_{p_2}(z_2)V_{p_3}(z_3)\rangle$. 
\paragraph{1st equation.}
We start by acting on (\ref{eq:nullvectorNS}) with $G_{-1/2}$ from the left and using the algebra (\ref{eq:N1algebra}) we obtain
\begin{equation}\label{eq: modified NS null vector}
    \left[\frac{1}{b^2}L_{-1}^2+2L_{-2}\right]V_{p_{\langle1,3\rangle}}=G_{-3/2}\Lambda_{p_{\langle1,3\rangle}}~.
\end{equation}
The left hand side now looks similar to the non-supersymmetric null vector equation (\ref{eq:N0nullv}). Borrowing the result (\ref{eq: N0 null vector1}) we obtain 
\begin{align}
    &\bigg\{\prod_{i=1}^3(z_0-z_i)\left(-\frac{1}{2b^2}\partial_{z_0}^2+\sum_{i=1}^3\frac{\partial_{z_0}}{z_0-z_i}\right)+(3z_0-z_1-z_2-z_3)h_{\langle1,3\rangle}\nonumber \\
&+\frac{z_{12}z_{13}}{z_1-z_0}h_1+\frac{z_{21}z_{23}}{z_2-z_0}h_2+\frac{z_{31}z_{32}}{z_3-z_0}h_3\bigg\}\langle V_{p_{\langle 1,3\rangle}}(z_0)V_{p_1}(z_1)V_{p_2}(z_2)V_{p_3}(z_3)\rangle\cr
&= -\frac{1}{2}\prod_{i=1}^3(z_0-z_i) \langle G_{-3/2}\Lambda_{p_{\langle1,3\rangle}}(z_0) V_1(z_1)V_2(z_2)V_{3}(z_3)\rangle\ .
\end{align}
For the right hand side we use
\begin{equation}\label{eq: N0 null vector0}
     \oint_{\infty}\d z \ \varepsilon(z)\langle G(z)\Lambda_{p_{\langle3,1\rangle}}(z_0)V_{p_1}(z_1)V_{p_2}(z_2)V_{p_3}(z_3)\rangle=0 ~.
\end{equation}
We now choose $\varepsilon(z)=(z-z_0)^{-1}$ and also use the fact that $G(z) = \mathcal{O}(z^{-3})$ as $z\rightarrow \infty$\footnote{\label{fn:Gbhviour}Under the conformal transformation $z\rightarrow \omega = 1/z$ the supercurrent transforms as
\begin{equation}
    G'(\omega) = \left(\frac{\d\omega}{\d z}\right)^{-\frac{3}{2}} G(z) = z^3 G(z)~,
\end{equation}
where we used $h= \tilde{h} = \frac{3}{2}$.
Regularity then implies that $G(z) \sim \mathcal{O}(z^{-3})$ as $z\rightarrow \infty$.
}. Deforming the contour to include the points $z_i$, $i= \{0,1,2,3\}$ (which we denote collectively by $\boldsymbol{z}$) we obtain 
\begin{equation}
  \langle G_{-3/2}\Lambda_{p_{\langle1,3\rangle}} (z_0)V_1(z_1)V_2(z_2)V_{3}(z_3)\rangle  = -\frac{f_{01}(\boldsymbol{z})}{(z_0-z_1)}-\frac{f_{02}(\boldsymbol{z})}{(z_0-z_2)} -\frac{f_{03}(\boldsymbol{z})}{(z_0-z_3)}~,
\end{equation}
where the subscript indicates the $z_i$ locations of $\Lambda_p$ operators, e.g.
\begin{align}
    f_{01}(\boldsymbol{z}) \equiv \langle \Lambda_{p_{\langle1,3\rangle}}(z_0)\Lambda_{p_1}(z_1)V_{p_2}(z_2)V_{p_3}(z_3)\rangle~,
\end{align}
Finally we consider 
\begin{equation}\label{eq: N0 null vectora}
     \oint_{\infty}\d z \ \varepsilon(z)\langle G(z)\Lambda_{p_{\langle3,1\rangle}}(z_0)V_{p_1}(z_1)V_{p_2}(z_2)V_{p_3}(z_3)\rangle=0 ~,
\end{equation}
with $\varepsilon(z)=1$. Performing the by now standard contour deformation we obtain
\begin{align}
    0=&\partial_{z_0}  \langle V_{p_{\langle3,1\rangle}}(z_0)V_{p_1}(z_1)V_{p_2}(z_2)V_{p_3}(z_3)\rangle - f_{01}(\boldsymbol{z}) - f_{02}(\boldsymbol{z}) - f_{03}(\boldsymbol{z}) ~.
\end{align}
We can eliminate for example $f_{01}$ in favour of $f_{02}$ and $f_{03}$, leading to
\begin{align}
    &\bigg\{\prod_{i=1}^3(z_0-z_i)\left(-\frac{1}{2b^2}\partial_{z_0}^2+\sum_{i=1}^3\frac{(1-\frac{1}{2}\delta_{i,1})\partial_{z_0}}{z_0-z_i}\right)+(3z_0-z_1-z_2-z_3)h_{\langle1,3\rangle}\nonumber \\
&\quad +\frac{z_{12}z_{13}}{z_1-z_0}h_1+\frac{z_{21}z_{23}}{z_2-z_0}h_2+\frac{z_{31}z_{32}}{z_3-z_0}h_3\bigg\}\langle V_{p_{\langle 1,3\rangle}}(z_0)V_{p_1}(z_1)V_{p_2}(z_2)V_{p_3}(z_3)\rangle\cr
&= -\frac{1}{2}(z_0-z_3)(z_1-z_2)(f_{02}(\boldsymbol{z})+f_{03}(\boldsymbol{z}))~.
\end{align}
Expressing the four point functions in terms of the cross ratio $z$ we get
\begin{align}
   &\langle V_{p_{\langle 1,3\rangle}}(z_0)V_{p_1}(z_1)V_{p_2}(z_2)V_{p_3}(z_3)\rangle = \left[\prod_{i<j}(z_i-z_j)^{\mu_{ij}}\right] \mathcal{F}(z)~,\cr
   &f_{02}(\boldsymbol{z})= \left[\prod_{i<j}(z_i-z_j)^{\nu_{ij}}\right] \mathcal{F}_{02}(z)~,\quad 
   f_{03}(\boldsymbol{z})= \left[\prod_{i<j}(z_i-z_j)^{\hat{\nu}_{ij}}\right] \mathcal{F}_{03}(z)~,\label{eq:appNSstepa}
\end{align}   
where $\mu_{ij}$ is again given by
\begin{equation}
    \mu_{ij} = \frac{1}{3}\sum_{k=0}^3h_k -h_i-h_j, 
\end{equation}
and the $\nu_{ij}$'s are defined completely analogously with the only difference that in $h_0$ and $h_2$ we account for the $1/2$ shift (or $h_0$ and $h_3$ respectively) since these are the conformal dimensions of the superdescendant fields $\Lambda_p = G_{-1/2}V_p$.
The derivatives with respect to $z_0$ can furthermore be expressed in terms of the cross ratio $z$ as
\begin{align}
    \partial_{z_0} &= \frac{\mu_{01}}{z}+\frac{\mu_{02}}{(z-1)} +\partial_z~,\nonumber \\
\partial_{z_0}^2 &= \frac{\mu_{01}(\mu_{01}-1)}{z^2} + \frac{\mu_{02}(\mu_{02}-1)}{(z-1)^2} +\frac{2\mu_{01}\mu_{02}}{z(z-1)} + 2\left(\frac{\mu_{01}}{z} +\frac{\mu_{02}}{(z-1)}\right)\partial_z + \partial_z^2~.
\end{align}
Taking the limits
\begin{equation}\label{eq:appNSlim}
    z_0 \rightarrow z~,\quad z_1 \rightarrow 0~,\quad z_2 \rightarrow 1~,\quad z_3 \rightarrow \infty
\end{equation}
we finally obtain
\begin{align}\nonumber
 &(-1)^{5/6}z^{1/6} (1 - z)^{2/3} \bigg[\frac{1}{b^{2}}\partial_z^2 + \left(2b^{-2} \left(\frac{\mu_{01}}{z} +\frac{\mu_{02}}{(z-1)}\right) +2\frac{1-2z}{z(z-1)} +\frac{1}{z}\right)\partial_z \\ \nonumber
 &+ \frac{1}{b^{2}}\left(\frac{\mu_{01}(\mu_{01}-1)}{z^2} +\frac{\mu_{02}(\mu_{02}-1)}{(z-1)^2} +\frac{2\mu_{01}\mu_{02}}{z(z-1)}\right) + 2\frac{h_1-\mu_{01}}{z^2} \\
 &+2\frac{h_2-\mu_{02}}{(z-1)^2} +2\frac{\mu_{12}}{z(z-1)} +\frac{\mu_{01}}{z^2} +\frac{\mu_{02}}{z(z-1)}\bigg] \mathcal{F}(z) = -\frac{1}{z(z-1)}\mathcal{F}_{02}(z)~.
\end{align}
Performing the shift
\begin{equation}
    \mathcal{F}(z) \rightarrow z^{-\mu_{01}}(1-z)^{-\mu_{02}}\mathcal{F}(z)~,\quad \mathcal{F}_{02}(z) \rightarrow z^{-\nu_{01}}(1-z)^{-\nu_{02}}\mathcal{F}_{02}(z)~,
\end{equation}
this leads to our first desired equation
\begin{multline}\label{eq: first equation NS}
\left(\frac{1}{b^{2}}\partial_z^2+\frac{1-3z}{z(z-1)}\partial_z +\frac{2h_1}{z^2}+ \frac{2h_2}{(z-1)^2} -2\frac{h_0+h_1+h_2 -h_3}{z(z-1)}\right)\mathcal{F}(z) \cr
=- \frac{1}{z(z-1)}\mathcal{F}_{02}(z)~.
\end{multline}
\paragraph{2nd equation.}
To obtain a second equation we take $\varepsilon(z)= (z-z_0)^{-1}$ and consider
\begin{equation}
    0=\oint_\infty \d z\, \varepsilon(z) \langle G(z) V_{p_{\langle 1,3\rangle}}(z_0)V_1(z_1)\Lambda_2(z_2)V_3(z_3)\rangle ~.
\end{equation}
Deforming the contour we obtain
\begin{align}
    0 &= \oint_{z_0} \d z\, \varepsilon(z) \langle [G(z) V_{p_{\langle 1,3\rangle}}(z_0)]_{\mathrm{OPE}}V_1(z_1)\Lambda_2(z_2)V_3(z_3)\rangle \cr
    &\quad + \oint_{z_1} \d z\, \varepsilon(z) \langle V_{p_{\langle 1,3\rangle}}(z_0)[G(z)V_1(z_1)]_{\mathrm{OPE}}\Lambda_2(z_2)V_3(z_3)\rangle \cr
    &\quad +  \oint_{z_2} \d z\, \varepsilon(z) \langle V_{p_{\langle 1,3\rangle}}(z_0)[G(z)V_1(z_1)[G(z)\Lambda_2(z_2)]_{\mathrm{OPE}}V_3(z_3)\rangle~\cr
    &\quad - \oint_{z_3} \d z\, \varepsilon(z) \langle V_{p_{\langle 1,3\rangle}}(z_0)V_1(z_1)\Lambda_2(z_2)[G(z)V_3(z_3)]_{\mathrm{OPE}}\rangle~. 
\end{align}
In the first contour integral around $z_0$ we expand $G(z)$ to order $(z-z_0)^0$, picking up a term $G_{-3/2} V_{p_{\langle 1,3\rangle}}$ which, using the null vector (\ref{eq:nullvectorNS}), we replace by 
\begin{equation}
G_{-3/2}V_{p_{\langle 1,3\rangle}} = -\frac{1}{b^2}G_{-1/2}^3 V_{p_{\langle 1,3\rangle}} =-\frac{1}{b^2} L_{-1}G_{-1/2} V_{p_{\langle 1,3\rangle}}= -\frac{1}{b^2}\partial_{z_0}\Lambda_{p_{\langle 1,3\rangle}}(z_0)~.
\end{equation}
We thus obtain
\begin{align}
    0=&-\frac{1}{b^2}\partial_{z_0} \langle  \Lambda_{p_{\langle 1,3\rangle}}(z_0)V_1(z_1)\Lambda_2(z_2)V_3(z_3)\rangle - \frac{1}{z_0-z_1}\langle  V_{p_{\langle 1,3\rangle}}(z_0)\Lambda_1(z_1)\Lambda_2(z_2)V_3(z_3)\rangle\cr
    &-\left(\frac{2h_2}{(z_0-z_2)^2} + \frac{1}{z_0-z_2}\partial_{z_2}\right) \langle  V_{p_{\langle 1,3\rangle}}(z_0)V_1(z_1)V_2(z_2)V_3(z_3)\rangle 
    \cr
    &\quad +\frac{1}{z_0-z_3}\langle  V_{p_{\langle 1,3\rangle}}(z_0)V_1(z_1)\Lambda_2(z_2)\Lambda_3(z_3)\rangle.\label{eq:appNSstep}
\end{align}
We can use
\begin{equation}
   0= \oint_\infty \d z  \langle G(z) V_{p_{\langle 1,3\rangle}}(z_0) V_1(z_1) \Lambda_2(z_2) V_3(z_3) \rangle ~,
\end{equation}
which implies
\begin{align}
    0 &= \langle \Lambda_{p_{\langle 1,3\rangle}}(z_0) V_1(z_1) \Lambda_2(z_2) V_3(z_3) \rangle + \langle V_{p_{\langle 1,3\rangle}}(z_0) \Lambda_1(z_1) \Lambda_2(z_2) V_3(z_3)  \rangle\cr
    &\quad +\partial_{z_2}\langle V_{p_{\langle 1,3\rangle}}(z_0) V_1(z_1) V_2(z_2) V_3(z_3) \rangle - \langle V_{p_{\langle 1,3\rangle}}(z_0) V_1(z_1) \Lambda_2(z_2) \Lambda_3(z_3) \rangle,
\end{align}
in order to eliminate $\langle V_{p_{\langle 1,3\rangle}}(z_0) \Lambda_1(z_1) \Lambda_2(z_2) V_3(z_3)  \rangle$ in (\ref{eq:appNSstep}). We then get
\begin{align}
    0&=\frac{1}{b^2}\partial_{z_0}f_{02}(\boldsymbol{z}) -\frac{1}{(z_0-z_1)}f_{02}(\boldsymbol{z}) -\left(\frac{1}{z_0-z_1}-\frac{1}{z_0-z_2}\right)\partial_{z_2}\langle V_{p_{\langle 1,3\rangle}}(z_0) V_1(z_1) \Lambda_2(z_2) V_3(z_3) \rangle\cr
&+\frac{2h_2}{(z_0-z_2)^2} \langle V_{p_{\langle 1,3\rangle}}(z_0) V_1(z_1) \Lambda_2(z_2) V_3(z_3) \rangle + \left(\frac{1}{z_0-z_1} - \frac{1}{z_0-z_3}\right)f_{03}(\boldsymbol{z})~.
\end{align}
Taking the limits (\ref{eq:appNSlim}) and using (\ref{eq:appNSstepa}), we obtain 
\begin{align}
   0&=(-1)^{5/6}z^{1/6} (1 - z)^{2/3}\left(\frac{1}{z(z-1)}\left(-\frac{\mu_{02}}{z-1} +\mu_{12} -z\partial_z\right)\mathcal{F}(z) + \frac{2h_2}{(z-1)^2}\mathcal{F}(z)\right)\cr
    &+\left(\frac{1}{b^2}\left(\frac{\nu_{01}}{z} + \frac{\nu_{02}}{z-1} +\partial_z\right)\mathcal{F}_{02}(z) -\frac{1}{z}\mathcal{F}_{02}(z)\right)~.
\end{align}
Finally, performing the shift
\begin{equation}
    \mathcal{F}(z) \rightarrow z^{-\mu_{01}}(1-z)^{-\mu_{02}}\mathcal{F}(z)~,\quad \mathcal{F}_{02}(z) \rightarrow z^{-\nu_{01}}(1-z)^{-\nu_{02}}\mathcal{F}_{02}(z)~,
\end{equation}
we land on our second equation
\begin{align}\label{eq:second equation NS}
&\left(\frac{1}{b^{2}}\partial_z-\frac{1}{z}\right)\mathcal{F}_{02}(z)+\left(\frac{1}{1-z}\partial_z +\frac{2h_2}{(z-1)^2} -\frac{h_0+h_1+h_2 -h_3}{z(z-1)} \right)\mathcal{F}(z)=0~.
\end{align}
\paragraph{Combined differential equation.}Combining (\ref{eq: first equation NS}) and (\ref{eq:second equation NS})  and using that the operator $V_{\langle 1,3\rangle }$ has conformal dimension $h_{\langle 1,3\rangle } = -1/2-b^2$ we obtain the differential equation \cite{Belavin:2007gz, Belavin:2007eq}
\begin{align}\label{eq:N1diffeqNS}
    &\frac{1}{b^2}\mathcal{F}{'''}+\frac{1-2b^2}{b^2}\frac{1-2z}{z(1-z)}\mathcal{F}{''}+\left(\frac{b^2+2h_1}{z^2}+\frac{b^2+2h_2}{(1-z)^2}+\frac{2-3b^2+2h_{1+2-3}}{z(1-z)}\right)\mathcal{F}{'} \nonumber \\
    +&\left(\frac{2h_2(1+b^2)}{(1-z)^3}-\frac{2h_1(1+b^2)}{z^3}+\frac{h_{2-1}+(1-2z)(b^4+b^2(1/2-h_{1+2-3})-h_{1+2})}{z^2(1-z)^2}\right)\mathcal{F}=0.
\end{align}
Similar equation holds for $\bar{z}$. 
\paragraph{The shift relations.}
There are now \textit{three} linearly independent solutions of (\ref{eq:N1diffeqNS}) given by\footnote{Note that, compared to \cite{Belavin:2007gz}, we use $\alpha_k=\frac{Q}{2}-p_k$.}
\begin{align}\label{eq:NSschanblocks}
\mathcal{F}_s^{(+)}(z) &=z^{\frac{bQ}{2}- bp_1}(1-z)^{\frac{bQ}{2}-bp_2} \times \mathcal{L}_{(+)}^2 \times \mathcal{I}_{1}\left(z\right),\nonumber \\
\mathcal{F}_s^{(0)}(z)
&= z^{\frac{bQ}{2}- bp_1}(1-z)^{\frac{bQ}{2}-bp_2}\times \mathcal{L}_{(0)}^2 \times \mathcal{I}_{2}\left(z\right), \nonumber \\
\mathcal{F}_s^{(-)}(z) &=z^{\frac{bQ}{2}- bp_1}(1-z)^{\frac{bQ}{2}-bp_2} \times \mathcal{L}_{(-)}^2\times \mathcal{I}_{3}\left(z\right),
\end{align}
where the momentum-dependent leg factors read
\begin{align}\label{eq:legfactors}
\textstyle
\mathcal{L}_{(\varepsilon=\pm)}\equiv
\frac{\Gamma\left(\frac{-1-b^2}{2}\right)\Gamma\left(1-\varepsilon b p_1\right)\Gamma\left(\frac{1-b^2}{2}-\varepsilon b p_1\right)\Gamma\left(\frac{3+b^2}{4}-\frac{b}{2}(p_2+\varepsilon(p_1-p_3))\right)^{-1}\Gamma\left(\frac{1-b^2}{4}-\frac{b}{2}(p_2+\varepsilon(p_1-p_3))\right)^{-1}}{\Gamma\left(-1-b^2\right)\Gamma\left(\frac{3+b^2}{4}+\frac{b}{2}(p_2-\varepsilon  (p_1+p_3))\right)\Gamma\left(\frac{1-b^2}{4}+\frac{b}{2}(p_2-\varepsilon  (p_1+p_3))\right)},\\
\textstyle
\mathcal{L}_{(0)}
\equiv \frac{b^2\Gamma\left(\frac{1+b^2}{2}+ bp_1\right)\Gamma\left(\frac{1+b^2}{2}- bp_1\right)}{\Gamma\left(\frac{3+b^2}{4}+\frac{b}{2}(p_1+p_2+p_3)\right)\Gamma\left(\frac{3+b^2}{4}+\frac{b}{2}(p_1-p_2-p_3)\right)\Gamma\left(\frac{3+b^2}{4}+\frac{b}{2}(-p_1+p_2-p_3)\right)\Gamma\left(\frac{3+b^2}{4}+\frac{b}{2}(-p_1-p_2+p_3)\right)}~.
\end{align}
The special functions $\mathcal{I}(z)$ depend additionally on the momenta $p_1,p_2,p_3$ as well as the central charge $c$, and admit integral representations of Dotsenko-Fateev type. Their properties are explained in detail in \cite{Belavin:2007gz} and we will not repeat them here\footnote{To avoid confusion, we actually adopt the same notation as in eqn. (C.7) in Appendix C of \cite{Belavin:2007gz}.}. What is important for us is their small $z$ expansions
\begin{align}
    \mathcal{I}_{1}(z) &= \frac{1}{\mathcal{L}_{(+)}}(1+\cdots), \nonumber \\
    \mathcal{I}_{2}(z) &= \frac{1}{\mathcal{L}_{(0)}}z^{bp_1+\frac{1+b^2}{2}}(1+\cdots), \nonumber \\
    \mathcal{I}_{3}(z) &= \frac{1}{\mathcal{L}_{(-)}}z^{2bp_1}(1+\cdots).
\end{align}
The functions (\ref{eq:NSschanblocks}) are the three $s-$channel conformal blocks in the OPE between $V_{p_{\langle1,3\rangle}}V_{p_1}$ and $V_{p_{3}}V_{p_2}$ (see left of Fig.\ref{fig:N1NScrossing}). The internal Liouville momenta are now $p_s=p_1+\varepsilon b$ with $\varepsilon=\pm$,  and $h_s = h_1+1/2 $. They can be read consistently from the corresponding exponents of the leading $z\rightarrow 0$ powers of those functions, namely from $\mathcal{F}_s^{(\pm)}$ we get
\begin{align}
    \frac{bQ}{2} + \varepsilon b p_1 &= - \frac{1}{2}\left(\frac{Q^2}{4}- p_1^2\right)- \frac{1}{2}\left(\frac{Q^2}{4}- p_{\langle 1,3\rangle}^2\right) +\frac{1}{2}\left(\frac{Q^2}{4} - \left(p_1- \varepsilon b\right)^2\right),
\end{align}
and from $\mathcal{F}_s^{(0)}$,
\begin{equation}
    1+b^2 = - \frac{1}{2}\left(\frac{Q^2}{4}- p_1^2\right)- \frac{1}{2}\left(\frac{Q^2}{4}- p_{\langle 1,3\rangle}^2\right) +\frac{1}{2}\left(\frac{Q^2}{4} - p_1^2+1\right)~.
\end{equation}
The $t-$channel blocks $\mathcal{F}_t^{(\eta=\pm,0)}(z)$ are obtained by the same expressions as in (\ref{eq:NSschanblocks}) after exchanging $p_1 \leftrightarrow p_2$ and $z \leftrightarrow 1-z$, namely
\begin{align}\label{eq:NStchanblocks}
\mathcal{F}_t^{(+)}(z) &=z^{\frac{bQ}{2}- bp_1}(1-z)^{\frac{bQ}{2}-bp_2} \times (\mathcal{L}'_{(+)})^2 \times \mathcal{J}_{1}\left(z\right),\nonumber \\
\mathcal{F}_t^{(0)}(z)
&= z^{\frac{bQ}{2}- bp_1}(1-z)^{\frac{bQ}{2}-bp_2}\times (\mathcal{L}'_{(0)})^2 \times \mathcal{J}_{2}\left(z\right), \nonumber \\
\mathcal{F}_t^{(-)}(z) &=z^{\frac{bQ}{2}- bp_1}(1-z)^{\frac{bQ}{2}-bp_2} \times (\mathcal{L}'_{(-)})^2\times \mathcal{J}_{3}\left(z\right),
\end{align}
where we denoted as $\mathcal{L}'$ the leg factors given in (\ref{eq:legfactors}) after exchanging $p_1\leftrightarrow p_2$, and $\mathcal{J}_i(z)$ are the special functions that result from $\mathcal{I}_i(z)$ accordingly. Their expansions near $z\rightarrow 1$ are
\begin{align}
    \mathcal{J}_{1}(z) &= \frac{1}{\mathcal{L}'_{(+)}}(1+\cdots), \nonumber \\
    \mathcal{J}_{2}(z) &= \frac{1}{\mathcal{L}'_{(0)}}(1-z)^{bp_2+\frac{1+b^2}{2}}(1+\cdots), \nonumber \\
    \mathcal{J}_{3}(z) &= \frac{1}{\mathcal{L}'_{(-)}}(1-z)^{2bp_2}(1+\cdots).
\end{align}
\par Combining the holomorphic and the anti-holomorphic sectors, the four-point function admits the $s,t$ channel expansions
\begin{align}\label{eq:N1nscrossingst}
    z^{bp_1-\frac{bQ}{2}}(1-z)^{bp_2-\frac{bQ}{2}}&\mathcal{F}(z)\nonumber \\
    &=\sum_{i=1}^3 c_i^{(s)} \mathcal{I}_i(z)\overline{\mathcal{I}_i}(z) =\sum_{i=1}^3 c_i^{(t)} \mathcal{J}_i(z)\overline{\mathcal{J}_i}(z)
\end{align}
where
\begin{align}
&c_{1}^{(s)}:=\mathcal{L}_{(+)}^2\frac{C_{\text{NS}}^{(b)}(p_1,p_{\langle1,3\rangle},p_1+ b)C_{\text{NS}}^{(b)}(p_1+ b,p_2,p_3)}{B_{\text{NS}}^{(b)}\left(p_1+ b\right)} , \nonumber \\
&c_{2}^{(s)}:=-\mathcal{L}_{(0)}^2\frac{\tilde{C}_{\text{NS}}^{(b)}(p_1,p_{\langle1,3\rangle},p_1)\tilde{C}_{\text{NS}}^{(b)}(p_1,p_2,p_3)}{B_{\text{NS}}^{(b)}\left(p_1\right)} , \nonumber \\
&c_{3}^{(s)}:=\mathcal{L}_{(-)}^2\frac{C_{\text{NS}}^{(b)}(p_1,p_{\langle1,3\rangle},p_1- b)C_{\text{NS}}^{(b)}(p_1- b,p_2,p_3)}{B_{\text{NS}}^{(b)}\left(p_1- b\right)} , \nonumber \\
\end{align}
and analogously for $c^{(t)}_i$. Here $C^{(b)}_{\text{NS}},\tilde{C}_{\text{NS}}^{(b)}$ are the NS structure constants of the theory that we are eventually interested in computing, and $B_{\text{NS}}^{(b)}$ is an arbitrary two-point function normalization.
\par There is now a non-trivial \textit{fusion kernel} that expresses a given special function $\mathcal{I}_i$ as a linear combination of $\mathcal{J}_i$'s.
In appendix C of \cite{Belavin:2007gz}, the authors wrote down explicitly the matrix that implements this non-trivial relation. We get
\begin{equation}\label{eq:N1crossing}
    \mathcal{I}_k(z)= \sum_{j=1
}^3 \mathbb{F}^{(\text{NS})}_{k,j} \ \mathcal{J}_j(z).
\end{equation}
where the entries of this $3\times3$ fusion matrix read
\begin{align}\label{eq: NS fusion matrix entries}
\mathbb{F}^{(\text{NS})}_{11}&= \frac{s\left(b^2-1+2bp_{1-2-3}\right)s\left(-3-b^2+2bp_{1-2-3}\right)}{s\left(-2+2b^2+4bp_2\right)s\left(4(bp_2-1)\right)} \ , \nonumber \\
   \mathbb{F}^{(\text{NS})}_{12}&=\frac{s\left(b^2-1+2bp_{1-2-3}\right)s\left(b^2-1+2bp_{1+2+3}\right)}{s\left(-2+2b^2+4bp_2\right)s\left(2b^2-4bp_2+6\right)} \ , \nonumber \\
   \mathbb{F}^{(\text{NS})}_{13}&=\frac{s\left(b^2-1+2bp_{1+2+3}\right)s\left(-3-b^2+2bp_{1+2+3}\right)}{s\left(4(bp_2-1)\right)s\left(-2b^2+4bp_2-6\right)} \ , \nonumber \\
   \mathbb{F}^{(\text{NS})}_{21}&=\frac{s\left(-3-b^2+2bp_{1-2-3}\right)s\left(b^2-5+2bp_{1+2-3}\right)}{(2s\left(2b^2\right))^{-1}s\left(-2+2b^2+4bp_2\right)s\left(4(bp_2-1)\right)} \ , \nonumber \\
   \mathbb{F}^{(\text{NS})}_{22}&=\frac{2 \left(\cos (\pi  b p_1) \cos (\pi  b p_2)-s \left(2b^2\right) \cos (\pi  b p_3)\right)}{\cos \left(\pi  b^2\right)+\cos (2 \pi  b p_2)} \ , \nonumber \\
   \mathbb{F}^{(\text{NS})}_{23}&=\frac{s\left(-3-b^2+2bp_{1+2+3}\right)s\left(3+b^2+2bp_{1-2+3}\right)}{(2s\left(2b^2\right))^{-1}s\left(4(bp_2-1)\right)s\left(-2b^2+4bp_2-6\right)} \ , \nonumber \\
   \mathbb{F}^{(\text{NS})}_{31}&=\frac{s\left(b^2-5+2bp_{1+2-3}\right)s\left(-7-b^2+2bp_{1+2-3}\right)}{s\left(-2+2b^2+4bp_2\right)s\left(4(bp_2-1)\right)} \ , \nonumber \\
   \mathbb{F}^{(\text{NS})}_{32}&=\frac{s\left(b^2-1+2bp_{2-1-3}\right)s\left(-7-b^2+2bp_{1+2-3}\right)}{s\left(-2+2b^2+4bp_2\right)s\left(-2b^2+4bp_2-6\right)} \ , \nonumber \\
   \mathbb{F}^{(\text{NS})}_{33}&=\frac{s\left(b^2-1+2bp_{2-1-3}\right)s\left(-3-b^2+2bp_{2-1-3}\right)}{s\left(4(bp_2-1)\right)s\left(-2b^2+4bp_2-6\right)} ~.
\end{align}
Here $s(x)\equiv \sin{(\pi x/4)}$, and we also use the notation $p_{1\pm2\pm3}=p_1\pm p_2\pm p_3$ for brevity.
\begin{figure}
\begin{tikzpicture}
\begin{scope}[shift={(-5,0)}, scale=.7]
\draw[very thick, red, looseness=.7] (2,1.75) to[out=10,in=5] (2,.25);
\draw[very thick, densely dashed, red, looseness=.7] (2,1.75) to[out=-180,in=-180] (2,.25);    
\draw[fill=white, draw=gray, opacity=.1] (0,2.5) to[out=0, in=180] (2,1.75) to[out=0, in=180] (4,2.5) 
    to[out=200, in=160, looseness=.95] (4,1.5) to[out=180, in=180, looseness=2.5] (4,.5) 
    to[out=200, in=160, looseness=.95] (4,-.5) to[out=180, in=0] (2,.25) 
    to[out=180, in=0] (0,-.5) to[out=20, in=-20, looseness=.95] (0,.5) 
    to[out=0, in=0, looseness=2.5] (0,1.5) to[out=20, in=-20, looseness=.95] (0,2.5);
\draw[very thick] (0,0) ellipse (0.25 and .5);
\draw[very thick] (0,2) ellipse (0.25 and .5);
\draw[very thick] (4,0) ellipse (0.25 and .5);
\draw[very thick] (4,2) ellipse (0.25 and .5);
\node[scale=1., thick] at (-0.9,0) {$p_{\langle1,3\rangle}$};
\node[scale=1., thick] at (-0.6,2) {$p_1$};  
\node[scale=1., thick] at (4.7,2) {$p_3$};
\node[scale=1., thick] at (4.7,0) {$p_2$};
\draw[very thick] (0,.5) to[out=0, in=0, looseness=2.5] (0,1.5);
\draw[very thick] (4,.5) to[out=180, in=180, looseness=2.5] (4,1.5);
\draw[very thick] (0,2.5) to[out=0, in=180] (2,1.75) to[out=0, in=180] (4,2.5);
\draw[very thick] (0,-.5) to[out=0, in=180] (2,.25) to[out=0, in=180] (4,-.5); 
\node[scale=1.5] at (7.3,.5) {$~~~~~~~~= \  \sum\limits_{j\in \{+,0,-\}}\mathbb{F}^{(\text{NS})}_{k,j}$};
\node[scale=1., red] at (2,-.5) {$p_1+ kb$};
\end{scope}
\begin{scope}[shift={(4.5,1.75)}, scale=.65, rotate=270]
\draw[very thick, red, looseness=.7] (2,1.75) to[out=10,in=5] (2,.25);
\draw[very thick, densely dashed, red, looseness=.7] (2,1.75) to[out=-180,in=-180] (2,.25);    
\draw[fill=white, draw=gray, opacity=.1] (0,2.5) to[out=0, in=180] (2,1.75) to[out=0, in=180] (4,2.5) 
    to[out=200, in=160, looseness=.95] (4,1.5) to[out=180, in=180, looseness=2.5] (4,.5) 
    to[out=200, in=160, looseness=.95] (4,-.5) to[out=180, in=0] (2,.25) 
    to[out=180, in=0] (0,-.5) to[out=20, in=-20, looseness=.95] (0,.5) 
    to[out=0, in=0, looseness=2.5] (0,1.5) to[out=20, in=-20, looseness=.95] (0,2.5);
\draw[very thick] (0,0) ellipse (0.25 and .5);
\draw[very thick] (0,2) ellipse (0.25 and .5);
\draw[very thick] (4,0) ellipse (0.25 and .5);
\draw[very thick] (4,2) ellipse (0.25 and .5);
\node[scale=1., thick] at (0,-0.9) {$p_1$};
\node[scale=1., thick] at (0,3) {$p_3$};  
\node[scale=1., thick] at (4,3) {$p_2$};
\node[scale=1., thick] at (4,-1.3) {$p_{\langle1,3\rangle}$};
\draw[very thick] (0,.5) to[out=0, in=0, looseness=2.5] (0,1.5);
\draw[very thick] (4,.5) to[out=180, in=180, looseness=2.5] (4,1.5);
\draw[very thick] (0,2.5) to[out=0, in=180] (2,1.75) to[out=0, in=180] (4,2.5);
\draw[very thick] (0,-.5) to[out=0, in=180] (2,.25) to[out=0, in=180] (4,-.5);    
\node[scale=1., red] at (2,3) {$p_2+ jb$};
\end{scope}
\end{tikzpicture}
\caption{Crossing transformation of sphere four-point conformal blocks with one insertion of a $\langle1,3\rangle$ degenerate NS-sector operator. The fusion kernel is a $3\times 3$ matrix given in (\ref{eq: NS fusion matrix entries}). }
\label{fig:N1NScrossing}
\end{figure}
\noindent
\par The form of this fusion kernel severely constraints the CFT data in the NS-sector of $\mathcal{N}=1$ Liouville. Indeed, crossing symmetry (\ref{eq:N1nscrossingst}) combined with (\ref{eq:N1crossing}) imply the following constraints which we express in matrix form:
\begin{equation}\label{eq:N1NSconstrmatrix}
    M\cdot \vec{c}=0
\end{equation}
where 
\begin{equation}
 M= 
    \renewcommand{\arraystretch}{2.3}
    \begin{pmatrix} 
    \mathbb{F}^{(\text{NS})}_{11}\mathbb{F}^{(\text{NS})}_{12} & ~~ \mathbb{F}^{(\text{NS})}_{21}\mathbb{F}^{(\text{NS})}_{22} & ~~\mathbb{F}^{(\text{NS})}_{31}\mathbb{F}^{(\text{NS})}_{32} \\
    \mathbb{F}^{(\text{NS})}_{11}\mathbb{F}^{(\text{NS})}_{13} & ~~\mathbb{F}^{(\text{NS})}_{21}\mathbb{F}^{(\text{NS})}_{23} & ~~\mathbb{F}^{(\text{NS})}_{31}\mathbb{F}^{(\text{NS})}_{33} \\ 
    \mathbb{F}^{(\text{NS})}_{12}\mathbb{F}^{(\text{NS})}_{13} & ~~\mathbb{F}^{(\text{NS})}_{22}\mathbb{F}^{(\text{NS})}_{23} & ~~\mathbb{F}^{(\text{NS})}_{32}\mathbb{F}^{(\text{NS})}_{33}
    \end{pmatrix}~~,~~
    \vec{c}= \begin{pmatrix}
    c^{(s)}_1 \\
    c^{(s)}_2 \\
    c^{(s)}_3 
\end{pmatrix} . 
\end{equation}
We will now establish that there are \textit{two independent} constraints for the structure constants $C_{\text{NS}}(p_1,p_2,p_3),\tilde{C}_{\text{NS}}(p_1,p_2,p_3)$ -- i.e. without involving the degenerate momentum $p_{\langle1,3\rangle}$ -- implied by (\ref{eq:N1NSconstrmatrix}). We will proceed analogously to the case of bosonic Liouville in Appendix \ref{app:N0Liouville}. In manipulating the constraints (\ref{eq:N1NSconstrmatrix}) our strategy is again to eliminate the structure constants that involve the degenerate momentum $p_{\langle1,3\rangle}$. 
\par In preparation for the above, let us first obtain relations that involve only $c_1^{(s)},c_{3}^{(s)}$. We can do that very simply by adding or subtracting appropriate linear combinations of two out of the three equations implied by (\ref{eq:N1NSconstrmatrix}). Let us, for example, subtract the following combinations of the two constraints from rows 1 and 2 of $M$:
\begin{align}
    c^{(s)}_1M_{11}+c^{(s)}_2M_{12}+c^{(s)}_3M_{13}&=0 \\
 \frac{\mathbb{F}^{(\text{NS})}_{22}}{\mathbb{F}^{(\text{NS})}_{23}}\times \left(c^{(s)}_1M_{21}+c^{(s)}_2M_{22}+c^{(s)}_3M_{23}\right)  &=0
\end{align}
resulting, after squaring both sides, to
\begin{equation}\label{eq:NSratiofirst1}
   \frac{C_{\text{NS}}^{(b)}(p_1+ b,p_2,p_3)^2/B^{(b)}_{\text{NS}}(p_1+b)}{C_{\text{NS}}^{(b)}(p_1- b,p_2,p_3)^2/B^{(b)}_{\text{NS}}(p_1-b)}=\frac{C_{\text{NS}}^{(b)}(p_1,p_{\langle1,3\rangle},p_1- b)^2/B^{(b)}_{\text{NS}}(p_1-b)}{C_{\text{NS}}^{(b)}(p_1,p_{\langle1,3\rangle},p_1+ b)^2/B^{(b)}_{\text{NS}}(p_1+b)}\times \mathsf{M}^2~.
\end{equation}
Here
\begin{equation}
    \mathsf{M}\equiv \left(\frac{\mathcal{L}_{(-)}}{\mathcal{L}_{(+)}}\right)^2\frac{\mathbb{F}^{(\text{NS})}_{31}\mathbb{F}^{(\text{NS})}_{33}\mathbb{F}^{(\text{NS})}_{22}-\mathbb{F}^{(\text{NS})}_{31}\mathbb{F}^{(\text{NS})}_{23}\mathbb{F}^{(\text{NS})}_{32}}{\mathbb{F}^{(\text{NS})}_{11}\mathbb{F}^{(\text{NS})}_{12}\mathbb{F}^{(\text{NS})}_{23}-\mathbb{F}^{(\text{NS})}_{11}\mathbb{F}^{(\text{NS})}_{13}\mathbb{F}^{(\text{NS})}_{22}}
\end{equation}
is a purely kinematic and known quantity. The ratio on the LHS of (\ref{eq:NSratiofirst1}) is the first unambiguous bootstrap datum that we are interested in computing eventually. However, there is still an unknown ratio on the RHS that we need to determine and we will discuss that shortly. It is obvious at this stage that we could have obtained relations involving only $c_1^{(s)},c_{3}^{(s)}$ by combining instead constraints coming from rows 2 and 3, or, 1 and 3 of $M$ in a similar manner. However, one quickly realizes that the two latter cases boil down back to the same equation (\ref{eq:NSratiofirst1})\footnote{This is due to some non-trivial relations between the matrix elements of $M$, which we leave as homework exercise to the interested reader.}.
\par The second thing we can do is to obtain relations between $c_2^{(s)}$ and $c_3^{(s)}$. Once we have those, and together with the previous relations between $c_1^{(s)},c_3^{(s)}$, it is obvious that we can straightforwardly generate the remaining relations between $c_1^{(s)},c_2^{(s)}$. Working similarly as before, we can subtract the following combinations of the two constraints from rows 1 and 2 of $M$:
\begin{align}
    c^{(s)}_1M_{11}+c^{(s)}_2M_{12}+c^{(s)}_3M_{13}&=0 \\
 \frac{\mathbb{F}^{(\text{NS})}_{12}}{\mathbb{F}^{(\text{NS})}_{13}}\times \left(c^{(s)}_1M_{21}+c^{(s)}_2M_{22}+c^{(s)}_3M_{23}\right)  &=0
\end{align}
resulting, after squaring both sides, to
\begin{equation}\label{eq:NSratiosecond1}
   \frac{\tilde{C}_{\text{NS}}^{(b)}(p_1,p_2,p_3)^2/B^{(b)}_{\text{NS}}(p_1)}{C_{\text{NS}}^{(b)}(p_1- b,p_2,p_3)^2/B^{(b)}_{\text{NS}}(p_1-b)}=\frac{C_{\text{NS}}^{(b)}(p_1,p_{\langle1,3\rangle},p_1- b)^2/B^{(b)}_{\text{NS}}(p_1-b)}{\tilde{C}_{\text{NS}}^{(b)}(p_1,p_{\langle1,3\rangle},p_1)^2/B^{(b)}_{\text{NS}}(p_1)}\times \widetilde{\mathsf{M}}^2~,
\end{equation}
where now
\begin{equation}
    \widetilde{\mathsf{M}}\equiv \left(\frac{\mathcal{L}_{(-)}}{\mathcal{L}_{(0)}}\right)^2\frac{\mathbb{F}^{(\text{NS})}_{31}\mathbb{F}^{(\text{NS})}_{32}\mathbb{F}^{(\text{NS})}_{13}-\mathbb{F}^{(\text{NS})}_{31}\mathbb{F}^{(\text{NS})}_{33}\mathbb{F}^{(\text{NS})}_{12}}{\mathbb{F}^{(\text{NS})}_{21}\mathbb{F}^{(\text{NS})}_{22}\mathbb{F}^{(\text{NS})}_{13}-\mathbb{F}^{(\text{NS})}_{21}\mathbb{F}^{(\text{NS})}_{23}\mathbb{F}^{(\text{NS})}_{12}} .
\end{equation}
The ratio on the LHS of (\ref{eq:NSratiosecond1}) is the second unambiguous bootstrap datum that we are interested in computing, barring the undetermined ratio involving the degenerate momentum $p_{\langle1,3\rangle}$ on the RHS. It is again easy to verify that the relations between $c^{(s)}_2,c^{(s)}_3$ coming from rows 2 and 3, or, 1 and 3 of $M$ under the same manipulations end up being the same as (\ref{eq:NSratiosecond1}). 
\par Let us now determine the unknown ratios on the RHS of (\ref{eq:NSratiofirst1}), (\ref{eq:NSratiosecond1}) and write down the final form of the shift relations in the NS-sector. We can do this easily by setting $p_2=p_{\langle1,3\rangle}$ and $p_3=p_1$ in (\ref{eq:N1NSconstrmatrix}). We first find
\begin{equation}
\left.\text{det}\ M\right|_{p_2\rightarrow p_{\langle1,3\rangle},p_3\rightarrow p_1}=0
\end{equation}
    which means that this particular matrix is not full rank. One can actually see that it is of rank 2. The vector $\left.\vec{c}\right|_{p_2\rightarrow p_{\langle1,3\rangle},p_3\rightarrow p_1}$ involves the \textit{squares} of the structures constants $C_{\text{NS}}^{(b)}(p_1,p_{\langle1,3\rangle},p_1\pm b), \tilde{C}_{\text{NS}}^{(b)}(p_1,p_{\langle1,3\rangle},p_1)$ (as well as the corresponding two-point functions) which are exactly the missing pieces in (\ref{eq:NSratiofirst1}), (\ref{eq:NSratiosecond1}). By performing the same manipulations as before, namely taking the subtraction of
\begin{align}
\left.c^{(s)}_1M_{11}+c^{(s)}_2M_{12}+c^{(s)}_3M_{13}\right|_{p_2\rightarrow p_{\langle1,3\rangle},p_3\rightarrow p_1}&=0~, \\
\left. \frac{\mathbb{F}^{(\text{NS})}_{22}}{\mathbb{F}^{(\text{NS})}_{23}}\times \left(c^{(s)}_1M_{21}+c^{(s)}_2M_{22}+c^{(s)}_3M_{23}\right)  \right|_{p_2\rightarrow p_{\langle1,3\rangle},p_3\rightarrow p_1}&=0~,
\end{align}
as well as the subtraction of
\begin{align}
\left.c^{(s)}_1M_{11}+c^{(s)}_2M_{12}+c^{(s)}_3M_{13}\right|_{p_2\rightarrow p_{\langle1,3\rangle},p_3\rightarrow p_1}&=0~, \\
\left. \frac{\mathbb{F}^{(\text{NS})}_{12}}{\mathbb{F}^{(\text{NS})}_{13}}\times \left(c^{(s)}_1M_{21}+c^{(s)}_2M_{22}+c^{(s)}_3M_{23}\right)  \right|_{p_2\rightarrow p_{\langle1,3\rangle},p_3\rightarrow p_1}&=0 ~,
\end{align}
we obtain respectively
\begin{align}\label{eq:NSratiosec1}
    &\frac{C_{\text{NS}}^{(b)}(p_1,p_{\langle1,3\rangle},p_1- b)^2/B^{(b)}_{\text{NS}}(p_1-b)}{C_{\text{NS}}^{(b)}(p_1,p_{\langle1,3\rangle},p_1+ b)^2/B^{(b)}_{\text{NS}}(p_1+b)}=\left(\left.\mathsf{M}\right|_{p_2\rightarrow p_{\langle1,3\rangle},p_3\rightarrow p_1}\right)^{-1} ~ , \nonumber \\ &\frac{C_{\text{NS}}^{(b)}(p_1,p_{\langle1,3\rangle},p_1- b)^2/B^{(b)}_{\text{NS}}(p_1-b)}{\tilde{C}_{\text{NS}}^{(b)}(p_1,p_{\langle1,3\rangle},p_1)^2/B^{(b)}_{\text{NS}}(p_1)}=\left(\left.\widetilde{\mathsf{M}}\right|_{p_2\rightarrow p_{\langle1,3\rangle},p_3\rightarrow p_1}\right)^{-1}.
\end{align}
Combining (\ref{eq:NSratiofirst1}), (\ref{eq:NSratiosecond1}) with (\ref{eq:NSratiosec1}), we eventually get the two desired shift relations 
\begin{align}\label{eq: shift equations NS sector}
    &\frac{C_{\text{NS}}^{(b)}(p_1+ b,p_2,p_3)^2/B^{(b)}_{\text{NS}}(p_1+b)}{C_{\text{NS}}^{(b)}(p_1- b,p_2,p_3)^2/B^{(b)}_{\text{NS}}(p_1-b)}=\frac{\mathsf{M}^2}{\left.\mathsf{M}\right|_{p_2\rightarrow p_{\langle1,3\rangle},p_3\rightarrow p_1}}\equiv \kappa^{(\text{NS})}_{b}(p_1|p_2,p_3), \nonumber \\
    &\frac{\tilde{C}_{\text{NS}}^{(b)}(p_1,p_2,p_3)^2/B^{(b)}_{\text{NS}}(p_1)}{C_{\text{NS}}^{(b)}(p_1- b,p_2,p_3)^2/B^{(b)}_{\text{NS}}(p_1-b)}=\frac{\widetilde{\mathsf{M}}^2}{\left.\widetilde{\mathsf{M}}\right|_{p_2\rightarrow p_{\langle1,3\rangle},p_3\rightarrow p_1}}\equiv \lambda^{(\text{NS})}_{b}(p_1|p_2,p_3),
\end{align}
where the explicit form of $\kappa^{(\text{NS})},\lambda^{(\text{NS})}$ read
\begin{align}
&\kappa^{(\text{NS})}_{b}(p_1|p_2,p_3)=-\frac{\gamma \left(-b^2+ b p_1\right)}{\gamma \left(-b^2- b p_1\right) } \left[ \frac{\Gamma ( b p_1)}{\Gamma (- b p_1)} \frac{\gamma \left(\frac{1-b^2}{2}+ b p_1\right)}{\gamma \left(\frac{1-b^2}{2}- b p_1\right)}\frac{\gamma \left(\frac{1}{4} \left(1-b^2-2  b p_{1-2-3}\right)\right)}{\gamma \left(\frac{1}{4} (1-b^2+2  b p_{1-2-3})\right)}\right]^2 \nonumber \\
& ~~~~~~~\times  \left[\frac{ \gamma \left(\frac{1}{4} \left(1-b^2-2  b p_{1+2-3}\right)\right) \gamma \left(\frac{1}{4} \left(1-b^2-2  b p_{1-2+3}\right)\right) \gamma \left(\frac{1}{4} \left(1-b^2-2  b p_{1+2+3}\right)\right)}{ \gamma \left(\frac{1}{4} (1-b^2+2  b p_{1+2-3})\right) \gamma \left(\frac{1}{4} (1-b^2+2  b p_{1-2+3})\right) \gamma \left(\frac{1}{4} (1-b^2+2  b p_{1+2+3})\right)}\right]^2~, \label{eq:kappaNS} \\
\nonumber \\
&\lambda^{(\text{NS})}_{b}(p_1|p_2,p_3)= -\frac{4\Gamma\left(1+bp_1\right)}{b^4\Gamma\left(-bp_1\right)}\frac{\gamma\left(-b^2+bp_1\right)}{\gamma\left(\frac{1+b^2}{2}-bp_1\right)^2}\nonumber \\
&\quad \quad \quad \quad \quad \quad \quad \quad \quad \times \left[\frac{\gamma\left(\frac{1}{4}\left(3+b^2-2bp_{1-2-3}\right)\right)\gamma\left(\frac{1}{4}\left(3+b^2-2bp_{1+2-3}\right)\right)}{\gamma\left(\frac{1}{4}\left(1-b^2+2bp_{1-2+3}\right)\right)\gamma\left(\frac{1}{4}\left(1-b^2+2bp_{1+2+3}\right)\right)}\right]^2 \label{eq:lambdaNS} ~.
\end{align}
with $\gamma(x)= \Gamma(x)/\Gamma(1-x)$ and e.g. $p_{1+2-3}=p_1+p_2-p_3$ etc. This concludes the derivation of the shift relations in the NS-sector.

\section{Derivation of the shift relations in the R-sector}\label{app:R}
We will next discuss the shift relations for the Ramond-sector structure constants\footnote{For a slightly different approach to the same problem see \cite{Belavin:2025rtg}.}. We start as in the non-supersymmetric and NS-sector cases with the null vector equation. In the R-sector the level 1 null vector is (\ref{eq:nullvectorR}). In the following we derive, as in the NS-sector, two coupled differential equations which we will use to obtain a single second-order (hypergeometric-type) differential equation for
\begin{equation}
    \langle  V_{p_{\langle 1,2 \rangle}}^{\mathrm{R},\epsilon_0}(z_0) V_{p_1}^{\mathrm{R},\epsilon_1}(z_1)V_{p_2}^{\mathrm{NS}}(z_2)V_{p_3}^{\mathrm{NS}}(z_3) \rangle ~.
\end{equation}
The reader not interested in all the detailed calculations leading up to the main differential equation can proceed to the discussion after (\ref{eq: gfinal}).
\paragraph{1st equation.}
We start with the following expression\footnote{We recall the behaviour of $G(z)$ as $z\rightarrow\infty$ from footnote \ref{fn:Gbhviour}.}
\begin{align}
    \oint_\infty \frac{\d z}{2\pi i} \varepsilon(z)\,\langle G(z)G_0 V_{p_{\langle 1,2 \rangle}}^{\mathrm{R},\epsilon_0}(z_0) V_{p_1}^{\mathrm{R},\epsilon_1}(z_1)V_{p_2}^{\mathrm{NS}}(z_2)V_{p_3}^{\mathrm{NS}}(z_3) \rangle =0~,
\end{align}
where
\begin{equation}\label{eq: epsilonRsector}
    \varepsilon(z) \equiv \sqrt{\frac{(z-z_1)}{(z-z_0)(z_0-z_1)}}~.
\end{equation}
The square root $\sqrt{\frac{z-z_1}{z-z_0}}$ avoids branch cuts in the OPE of the supercurrent $G(z)$ with the Ramond fields, while the exta $\sqrt{z_0-z_1}$ is merely a cosmetical contribution. 
Deforming the contour and using 
\begin{align}
    G(\omega)V_{p}^{\mathrm{R},\epsilon} &= \epsilon V_{p}^{\mathrm{R},\epsilon} G(\omega)~,\quad  G(\omega)G_0V_{p}^{\mathrm{R},\epsilon} = -\epsilon G_0V_{p}^{\mathrm{R},\epsilon} G(\omega)~,\cr
    G(\omega)G_0^2V_{p}^{\mathrm{R},\epsilon} &= \epsilon G_0^2V_{p}^{\mathrm{R},\epsilon}G(\omega)
\end{align}
we obtain 
\begin{align}
    0&=\oint_{z_0}\frac{\d z}{2\pi i}\varepsilon(z) \,\langle \big[G(z)G_0 V_{p_{\langle 1,2 \rangle}}^{\mathrm{R},\epsilon_0}(z_0)\big]_{\mathrm{OPE}} V_{p_1}^{\mathrm{R},\epsilon_1}(z_1)V^{\mathrm{NS}}_{p_2}(z_2)V^{\mathrm{NS}}_{p_3}(z_3) \rangle\cr
    &\quad -\epsilon_0 \oint_{z_1}\frac{\d z}{2\pi i}\varepsilon(z) \,\langle G_0 V_{p_{\langle 1,2 \rangle}}^{\mathrm{R},\epsilon_0}(z_0) \big[G(z)V_{p_1}^{\mathrm{R},\epsilon_1}(z_1)\big]_{\mathrm{OPE}}V^{\mathrm{NS}}_{p_2}(z_2)V^{\mathrm{NS}}_{p_3}(z_3) \rangle\cr
    &\quad -\epsilon_0 \epsilon_1\oint_{z_2}\frac{\d z}{2\pi i}\varepsilon(z) \,\langle G_0 V_{p_{\langle 1,2 \rangle}}^{\mathrm{R},\epsilon_0}(z_0) V_{p_1}^{\mathrm{R},\epsilon_1}(z_1)\big[G(z)V^{\mathrm{NS}}_{p_2}(z_2)\big]_{\mathrm{OPE}}V^{\mathrm{NS}}_{p_3}(z_3) \rangle\cr
    &\quad  -\epsilon_0 \epsilon_1\oint_{z_3}\frac{\d z}{2\pi i}\varepsilon(z) \,\langle G_0 V_{p_{\langle 1,2 \rangle}}^{\mathrm{R},\epsilon_0}(z_0) V_{p_1}^{\mathrm{R},\epsilon_1}(z_1)V^{\mathrm{NS}}_{p_2}(z_2)\big[G(z)V^{\mathrm{NS}}_{p_3}(z_3)\big]_{\mathrm{OPE}} \rangle~.
\end{align}
Using the OPE between the supercurrent and the NS- and R-fields respectively, in particular
\begin{equation}
G(z)V_{p}^{\mathrm{R},\epsilon} \sim \frac{1}{z^{3/2} } G_0 V_{p}^{\mathrm{R},\epsilon} +\frac{1}{z^{1/2} } G_{-1} V_{p}^{\mathrm{R},\epsilon}~,
\end{equation}
the above leads to 
\begin{align}
   0&=\oint_{z_0}\frac{\d z}{2\pi i}\varepsilon(z) \,\bigg\langle \frac{G_0^2 V_{p_{\langle 1,2 \rangle}}^{\mathrm{R},\epsilon_0}}{(z-z_0)^{\frac{3}{2}}} V_{p_1}^{\mathrm{R},\epsilon_1}(z_1)V_{p_2}^{\mathrm{NS}}(z_2)V_{p_3}^{\mathrm{NS}}(z_3) \bigg\rangle \cr
   &\quad +\oint_{z_0}\frac{\d z}{2\pi i}\varepsilon(z) \,\bigg\langle \frac{G_{-1} G_{0} V_{p_{\langle 1,2 \rangle}}^{\mathrm{R},\epsilon_0}(z_0)}{(z-z_0)^{\frac{1}{2}}} V_{p_1}^{\mathrm{R},\epsilon_1}(z_1)V_{p_2}^{\mathrm{NS}}(z_2)V_{p_3}^{\mathrm{NS}}(z_3) \bigg\rangle\cr 
   &\quad -\epsilon\oint_{z_1}\frac{\d z}{2\pi i}\varepsilon(z) \,\bigg\langle G_0 V_{p_{\langle 1,2 \rangle}}^{\mathrm{R},\epsilon_0}(z_0) \frac{G_0V_{p_1}^{\mathrm{R},\epsilon_1}(z_1)}{(z-z_1)^{\frac{3}{2}}}V^{\mathrm{NS}}_{p_2}(z_2)V^{\mathrm{NS}}_{p_3}(z_3) \bigg\rangle\cr
   &\quad -\epsilon \epsilon_1\oint_{z_2}\frac{\d z}{2\pi i}\varepsilon(z) \,\left\langle G_0 V_{p_{\langle 1,2 \rangle}}^{\mathrm{R},\epsilon_0}(z_0) V_{p_1}^{\mathrm{R},\epsilon_1}(z_1)\frac{\Lambda_{p_2}(z_2)}{(z-z_2)}V^{\mathrm{NS}}_{p_3}(z_3) \right\rangle\cr  
   &\quad -\epsilon \epsilon_1\oint_{z_3}\frac{\d z}{2\pi i}\varepsilon(z) \,\left\langle G_0 V_{p_{\langle 1,2 \rangle}}^{\mathrm{R},\epsilon_0}(z_0) V_{p_1}^{\mathrm{R},\epsilon_1}(z_1)V_{p_2}^{\mathrm{NS}}(z_2)\frac{\Lambda^{\mathrm{NS}}_{p_3}(z_3)}{(z-z_3)} \right\rangle~.
\end{align}
Now using $\varepsilon(z)$ given in (\ref{eq: epsilonRsector}) we obtain 
\begin{align}\label{eq:Gm1G0 Rsector}
    0 &= -\frac{1}{2z_{01}} \langle G_0^2 V_{p_{\langle 1,2 \rangle}}^{\mathrm{R},\epsilon_0} (z_0) V_{p_1}^{\mathrm{R},\epsilon_1}(z_1)V_{p_2}^{\mathrm{NS}}(z_2)V_{p_3}^{\mathrm{NS}}(z_3) \rangle - \langle G_{-1} G_{0} V_{p_{\langle 1,2 \rangle}}^{\mathrm{R},\epsilon_0}(z_0)V_{p_1}^{\mathrm{R},\epsilon_1} V_{p_2}^{\mathrm{NS}}(z_2)V_{p_3}^{\mathrm{NS}}(z_3)\rangle  \cr
    &\quad +\epsilon \epsilon_1\sqrt{\frac{z_{21}}{z_{01}z_{20}}}\langle G_0 V_{p_{\langle 1,2 \rangle}}^{\mathrm{R},\epsilon_0} (z_0) V_{p_1}^{\mathrm{R},\epsilon_1}(z_1)\Lambda_{p_2}^{\mathrm{NS}}(z_2)V_{p_3}^{\mathrm{NS}}(z_3) \rangle \cr
    &\quad +\epsilon \epsilon_1\sqrt{\frac{z_{31}}{z_{01}z_{30}}}\langle G_0 V_{p_{\langle 1,2 \rangle}}^{\mathrm{R},\epsilon_0} (z_0) V_{p_1}^{\mathrm{R},\epsilon_1}(z_1)V_{p_2}^{\mathrm{NS}}(z_2)\Lambda_{p_3}^{\mathrm{NS}}(z_3) \rangle~\cr
     &\quad +\epsilon\frac{\sqrt{-1}}{z_{01}} \langle G_0 V_{p_{\langle 1,2 \rangle}}^{\mathrm{R},\epsilon_0} (z_0) G_0V_{p_1}^{\mathrm{R},\epsilon_1}(z_1)V_{p_2}^{\mathrm{NS}}(z_2)V_{p_3}^{\mathrm{NS}}(z_3) \rangle~.
\end{align}
Note that we have
\begin{equation}\label{eq:G0squared on R}
    G_0 V_{p}^{\mathrm{R},\epsilon} = i \beta \e^{-\frac{\pi i\epsilon}{4}}V_{p}^{\mathrm{R},\epsilon}~\Rightarrow G_0^2 R_{a}^\epsilon = - \beta^2V_{p}^{\mathrm{R},\epsilon}~,\quad \beta \equiv \frac{p}{\sqrt{2}}~.
\end{equation}
Using the R-sector null vector (\ref{eq:nullvectorR}) we have
\begin{align}
    \kappa \partial_{z_0} \langle V_{p_{\langle 1,2 \rangle}}^{\mathrm{R},\epsilon_0}(z_0)V_{p_1}^{\mathrm{R},\epsilon_1}(z_1)V_{p_2}^{\mathrm{NS}}(z_2)V_{p_3}^{\mathrm{NS}}(z_3)\rangle  = \langle G_{-1}G_0 V_{p_{\langle 1,2 \rangle}}^{\mathrm{R},\epsilon_0}(z_0)V_{p_1}^{\mathrm{R},\epsilon_1}(z_1)V_{p_2}^{\mathrm{NS}}(z_2)V_{p_3}^{\mathrm{NS}}(z_3) \rangle ~,
\end{align}
with $\kappa \equiv \frac{2b^2+1}{2b^2}$. We then replace the right hand side with (\ref{eq:Gm1G0 Rsector}) which leads to 
\begin{align}
      &\kappa \partial_{z_0} \langle V_{p_{\langle 1,2 \rangle}}^{\mathrm{R},\epsilon_0}(z_0)V_{p_1}^{\mathrm{R},\epsilon_1}(z_1)V_{p_2}^{\mathrm{NS}}(z_2)V_{p_3}^{\mathrm{NS}}(z_3)\rangle = -\frac{1}{2z_{01}} \langle G_0^2 V_{p_{\langle 1,2 \rangle}}^{\mathrm{R},\epsilon_0} (z_0) V_{p_1}^{\mathrm{R},\epsilon_1}(z_1)V_{p_2}^{\mathrm{NS}}(z_2)V_{p_3}^{\mathrm{NS}}(z_3) \rangle \cr
      &\quad +\epsilon \epsilon_1\sqrt{\frac{z_{21}}{z_{01}z_{20}}}\langle G_0 V_{p_{\langle 1,2 \rangle}}^{\mathrm{R},\epsilon_0}(z_0) V_{p_1}^{\mathrm{R},\epsilon_1}(z_1)\Lambda_{p_2}^{\mathrm{NS}}(z_2)V_{p_3}^{\mathrm{NS}}(z_3) \rangle\cr
      &\quad +\epsilon \epsilon_1\sqrt{\frac{z_{31}}{z_{01}z_{30}}}\langle G_0 V_{p_{\langle 1,2 \rangle}}^{\mathrm{R},\epsilon_0} (z_0) V_{p_1}^{\mathrm{R},\epsilon_1}(z_1)V_{p_2}^{\mathrm{NS}}(z_2)\Lambda_{p_3}^{\mathrm{NS}}(z_3) \rangle\cr
      &\quad +\epsilon\frac{\sqrt{-1}}{z_{01}} \langle G_0 V_{p_{\langle 1,2 \rangle}}^{\mathrm{R},\epsilon_0}(z_0) G_0V_{p_1}^{\mathrm{R},\epsilon_1}(z_1)V_{p_2}^{\mathrm{NS}}(z_2)V_{p_3}^{\mathrm{NS}}(z_3) \rangle~.
\end{align}
Next we look at 
\begin{equation}
    \oint_\infty \frac{\d z}{2\pi i}\varepsilon(z) \langle G(z) G_0 V_{p_0}^{\mathrm{R},\epsilon_0}(z_0)V_{p_1}^{\mathrm{R},\epsilon_1}(z_1) V_{p_2}^{\mathrm{NS}}(z_2)V_{p_3}^{\mathrm{NS}}(z_3) \rangle =0~,\quad \varepsilon(z) \equiv \sqrt{\frac{(z-z_0)(z-z_1)}{(z_3-z_0)(z_3-z_1)}}~.
\end{equation}
This choice leads to 
\begin{align}
    0 &= \oint_{z_0}\frac{\d z}{2\pi i}\varepsilon(z) \left\langle\frac{G_0^2 V_{p_0}^{\mathrm{R},\epsilon_0}(z_0)}{(z-z_0)^{\frac{3}{2}}} V_{p_1}^{\mathrm{R},\epsilon_1}(z_1) V_{p_2}^{\mathrm{NS}}(z_2)V_{p_3}^{\mathrm{NS}}(z_3) \right\rangle \cr
    &\quad -\epsilon_0 \oint_{z_1}\frac{\d z}{2\pi i}\varepsilon(z) \left\langle V_{p_0}^{\mathrm{R},\epsilon_0}(z_0)\frac{G_0 V_{p_1}^{\mathrm{R},\epsilon_1}(z_1)}{(z-z_0)^{\frac{3}{2}}} V_{p_2}^{\mathrm{NS}}(z_2)V_{p_3}^{\mathrm{NS}}(z_3) \right\rangle \cr
    &\quad -\epsilon_0\epsilon_1\oint_{z_2}\frac{\d z}{2\pi i}\varepsilon(z) \left\langle G_0 V_{p_0}^{\mathrm{R},\epsilon_0}(z_0) V_{p_1}^{\mathrm{R},\epsilon_1}(z_1) \frac{\Lambda_{p_2}^{\mathrm{NS}}(z_2)}{(z-z_2)}V_{p_3}^{\mathrm{NS}}(z_3) \right\rangle\cr
    &\quad -\epsilon_0\epsilon_1\oint_{z_3}\frac{\d z}{2\pi i}\varepsilon(z) \left\langle G_0 V_{p_0}^{\mathrm{R},\epsilon_0}(z_0) V_{p_1}^{\mathrm{R},\epsilon_1}(z_1)V_{p_2}^{\mathrm{NS}}(z_2) \frac{\Lambda_{p_3}^{\mathrm{NS}}(z_3)}{(z-z_3)} \right\rangle~,
\end{align}
or, after evaluating the residues, 
\begin{align}
    0=&\sqrt{\frac{z_{01}}{z_{30}z_{31}}} \langle G_0^2 V_{p_0}^{\mathrm{R},\epsilon_0}(z_0)V_{p_1}^{\mathrm{R},\epsilon_1}(z_1) V_{p_2}^{\mathrm{NS}}(z_2)V_{p_3}^{\mathrm{NS}}(z_3) \rangle \cr
    &-  \epsilon_0 \sqrt{-1}\sqrt{\frac{z_{01}}{z_{30}z_{31}}} \langle G_0 V_{p_0}^{\mathrm{R},\epsilon_0}(z_0)G_0V_{p_1}^{\mathrm{R},\epsilon_1}(z_1) V_{p_2}^{\mathrm{NS}}(z_2)V_{p_3}^{\mathrm{NS}}(z_3) \rangle\cr
    &-\epsilon_0\epsilon_1 \sqrt{\frac{z_{20}z_{21}}{z_{30}z_{31}}} \langle G_0 V_{p_0}^{\mathrm{R},\epsilon_0}(z_0)V_{p_1}^{\mathrm{R},\epsilon_1}(z_1)\Lambda_{p_2}^{\mathrm{NS}}(z_2)V_{p_3}^{\mathrm{NS}}(z_3) \rangle \cr
    &-\epsilon_0\epsilon_1 \langle G_0 V_{p_0}^{\mathrm{R},\epsilon_0}(z_0)V_{p_1}^{\mathrm{R},\epsilon_1}(z_1)V_{p_2}^{\mathrm{NS}}(z_2)\Lambda_{p_3}^{\mathrm{NS}}(z_3) \rangle~.
\end{align}
We therefore obtain
\begin{align}\label{eq: R sector equation 1}
      &\kappa \partial_{z_0} g^{\epsilon_0,\epsilon_1}(\mathbf{z}) = \frac{\beta_{p_{0}}^2}{2z_{01}} g^{\epsilon_0,\epsilon_1}(\mathbf{z}) +\epsilon_0 \epsilon_1 \sqrt{\frac{z_{21}}{z_{01}z_{20}}} h^{\epsilon_0,\epsilon_1}(\mathbf{z}) - \epsilon_0 \beta_{p_{0}}\beta_{p_1}\frac{\sqrt{-1}}{z_{01}} \e^{-\frac{\pi i}{4}(\epsilon_0+\epsilon_1)} g^{-\epsilon_0,\epsilon_1}(\mathbf{z})\cr
      &\quad -\epsilon \epsilon_1\sqrt{\frac{z_{20}z_{21}}{z_{01}z_{30}^2}} h^{\epsilon_0,\epsilon_1}(\mathbf{z}) +\epsilon_1\beta_{p_0}\beta_{p_1}\frac{\sqrt{-1} }{z_{30}} \e^{-\frac{\pi i}{4}(\epsilon_0+\epsilon_1)} g^{-\epsilon_0,\epsilon_1}(\mathbf{z})- \epsilon_0\epsilon_1\frac{\beta_{p_0}^2}{z_{30}}  g^{\epsilon_0,\epsilon_1}(\mathbf{z})~,
\end{align}
where $p_0 = p_{\langle 1,2\rangle}$ and 
\begin{align}\label{eq: Gepsilon Hepsilon}
    g^{\epsilon_0,\epsilon_1}(\mathbf{z}) &\equiv \langle V_{p_0}^{\mathrm{R},\epsilon_0}(z_0)V_{p_1}^{\mathrm{R},\epsilon_1}(z_1)V_{p_2}^{\mathrm{NS}}(z_2)V_{p_3}^{\mathrm{NS}}(z_3)\rangle~,\cr h^{\epsilon_0,\epsilon_1}(\mathbf{z}) &\equiv \langle G_0V_{p_0}^{\mathrm{R},\epsilon_0}(z_0)V_{p_1}^{\mathrm{R},\epsilon_1}(z_1)\Lambda_{p_2}^{\mathrm{NS}}(z_2)V_{p_3}^{\mathrm{NS}}(z_3)\rangle~.
\end{align}
We will denote (\ref{eq: R sector equation 1}) as the first equation in the R-sector. 
\paragraph{2nd equation.}
For the second equation we start with 
\begin{align}
    \oint_\infty \frac{\d z}{2\pi i} \varepsilon(z)\,\langle G(z)G_0^2 V_{p_0}^{\mathrm{R},\epsilon_0}(z_0) V_{p_1}^{\mathrm{R},\epsilon_1}(z_1)\Lambda_{p_2}^{\mathrm{NS}}(z_2)V_{p_3}^{\mathrm{NS}}(z_3) \rangle =0~.
\end{align}
This leads to 
\begin{align}
    0&=-\oint_{z_0}\frac{\d z}{2\pi i}\varepsilon(z) \,\langle \big[G(z)G_0^2 V_{p_0}^{\mathrm{R},\epsilon_0}(z_0)\big]_{\mathrm{OPE}} V_{p_1}^{\mathrm{R},\epsilon_1}(z_1)\Lambda_2(z_2)V_3(z_3) \rangle\cr
    &\quad -\epsilon_0\oint_{z_1}\frac{\d z}{2\pi i}\varepsilon(z) \,\langle G_0^2 V_{p_0}^{\mathrm{R},\epsilon_0}(z_0) \big[G(z)V_{p_1}^{\mathrm{R},\epsilon_1}(z_1)\big]_{\mathrm{OPE}}\Lambda_{p_2}^{\mathrm{NS}}(z_2)V_{p_3}^{\mathrm{NS}}(z_3) \rangle\cr
    &\quad -\epsilon_0\epsilon_1\oint_{z_2}\frac{\d z}{2\pi i}\varepsilon(z) \,\langle G_0^2 V_{p_0}^{\mathrm{R},\epsilon_0}(z_0) V_{p_1}^{\mathrm{R},\epsilon_1}(z_1)\big[G(z)\Lambda_{p_2}^{\mathrm{NS}}(z_2)\big]_{\mathrm{OPE}}V_{p_3}^{\mathrm{NS}}(z_3) \rangle\cr
    &\quad  + \epsilon_0\epsilon_1\oint_{z_3}\frac{\d z}{2\pi i}\varepsilon(z) \,\langle G_0^2 V_{p_0}^{\mathrm{R},\epsilon_0}(z_0) V_{p_1}^{\mathrm{R},\epsilon_1}(z_1)\Lambda_{p_2}^{\mathrm{NS}}(z_2)\big[G(z)V_{p_3}^{\mathrm{NS}}(z_3)\big]_{\mathrm{OPE}} \rangle~.
\end{align}
Using the OPE between the supercurrent and the NS- and R-fields respectively we obtain 
\begin{align}
   0&=-\oint_{z_0}\frac{\d z}{2\pi i}\varepsilon(z) \,\bigg\langle \frac{G_0 G_0^2V_{p_0}^{\mathrm{R},\epsilon_0}(z_0)}{(z-z_0)^{\frac{3}{2}}} V_{p_1}^{\mathrm{R},\epsilon_1}(z_1)\Lambda_{p_2}^{\mathrm{NS}}(z_2)V_{p_3}^{\mathrm{NS}}(z_3) \bigg\rangle \cr
   &\quad -\oint_{z_0}\frac{\d z}{2\pi i}\varepsilon(z) \,\bigg\langle \frac{G_{-1} G_0^2V_{p_0}^{\mathrm{R},\epsilon_0}(z_0)}{(z-z_0)^{\frac{1}{2}}} V_{p_1}^{\mathrm{R},\epsilon_1}(z_1)\Lambda_{p_2}^{\mathrm{NS}}(z_2)V_{p_3}^{\mathrm{NS}}(z_3) \bigg\rangle\cr  
   &\quad -\epsilon_0\oint_{z_1}\frac{\d z}{2\pi i}\varepsilon(z) \,\bigg\langle G_0^2 V_{p_0}^{\mathrm{R},\epsilon_0}(z_0) \frac{G_0V_{p_1}^{\mathrm{R},\epsilon_1}(z_1)}{(z-z_1)^{\frac{3}{2}}}\Lambda_{p_2}^{\mathrm{NS}}(z_2)V_{p_3}^{\mathrm{NS}}(z_3) \bigg\rangle\cr
   &\quad -\epsilon_0\epsilon_1\oint_{z_2}\frac{\d z}{2\pi i}\varepsilon(z) \,\left\langle G_0^2 V_{p_0}^{\mathrm{R},\epsilon_0}(z_0) V_{p_1}^{\mathrm{R},\epsilon_1}(z_1)\left(\frac{2h_{p_2}^{\mathrm{NS}}}{(z-z_2)^2} + \frac{1}{(z-z_2)}\partial_{z_2}\right)V_{p_2}^{\mathrm{NS}}(z_2)V_{p_3}^{\mathrm{NS}}(z_3) \right\rangle\cr  
   &\quad + \epsilon_0\epsilon_1\oint_{z_3}\frac{\d z}{2\pi i}\varepsilon(z) \,\left\langle G_0^2 V_{p_0}^{\mathrm{R},\epsilon_0}(z_0) V_{p_1}^{\mathrm{R},\epsilon_1}(z_1)\Lambda_{p_2}^{\mathrm{NS}}(z_2)\frac{\Lambda_{p_3}^{\mathrm{NS}}(z_3)}{(z-z_3)} \right\rangle~.
\end{align}
Now using $\varepsilon(z)$ from (\ref{eq: epsilonRsector}), we obtain 
\begin{align}
    0 &= \frac{\beta_0^2}{2z_{01}} \langle G_0  V_{p_0}^{\mathrm{R},\epsilon_0} (z_0) V_{p_1}^{\mathrm{R},\epsilon_1}(z_1)\Lambda_{p_2}^{\mathrm{NS}}(z_2)V_{p_3}^{\mathrm{NS}}(z_3) \rangle  \cr
    &\quad -\langle G_{-1}G_0^2 V_{p_0}^{\mathrm{R},\epsilon_0} (z_0) V_{p_1}^{\mathrm{R},\epsilon_1}(z_1)\Lambda_{p_2}^{\mathrm{NS}}(z_2)V_{p_3}^{\mathrm{NS}}(z_3) \rangle\cr
    &\quad +\epsilon_0\beta_0^2\frac{\sqrt{-1}}{z_{01}} \langle  V_{p_0}^{\mathrm{R},\epsilon_0} (z_0) G_0V_{p_1}^{\mathrm{R},\epsilon_1}(z_1)\Lambda_{p_2}^{\mathrm{NS}}(z_2)V_{p_3}^{\mathrm{NS}}(z_3) \rangle \cr
    &\quad -\epsilon_0\epsilon_1\beta_0^2 h_{p_2}^{\mathrm{NS}}\sqrt{\frac{z_{01}}{z_{20}^3z_{21}}}\langle  V_{p_0}^{\mathrm{R},\epsilon_0} (z_0) V_{p_1}^{\mathrm{R},\epsilon_1}(z_1)V_{p_2}^{\mathrm{NS}}(z_2)V_{p_3}^{\mathrm{NS}}(z_3) \rangle \cr
&\quad+\epsilon_0\epsilon_1\beta_0^2\sqrt{\frac{z_{21}}{z_{20}z_{01}}}\partial_{z_2}\langle  V_{p_0}^{\mathrm{R},\epsilon_0} (z_0) V_{p_1}^{\mathrm{R},\epsilon_1}(z_1)V_{p_2}^{\mathrm{NS}}(z_2)V_{p_3}^{\mathrm{NS}}(z_3) \rangle\cr
&\quad -\epsilon_0\epsilon_1\beta_0^2\sqrt{\frac{z_{31}}{z_{01}z_{30}}}\langle  V_{p_0}^{\mathrm{R},\epsilon_0} (z_0) V_{p_1}^{\mathrm{R},\epsilon_1}(z_1)\Lambda_{p_2}^{\mathrm{NS}}(z_2)\Lambda_{p_3}^{\mathrm{NS}}(z_3) \rangle~. 
\end{align}
To eliminate the last term with two $\Lambda$ fields we use
\begin{equation}
    \oint_\infty \frac{\d z}{2\pi i}\varepsilon(z) \langle G(z) G_0 V_{p_0}^{\mathrm{R},\epsilon_0}(z_0)V_{p_1}^{\mathrm{R},\epsilon_1}(z_1) \Lambda_{p_2}^{\mathrm{NS}}(z_2)V_{p_3}^{\mathrm{NS}}(z_3) \rangle =0~,\quad \varepsilon(z) \equiv \sqrt{\frac{(z-z_0)(z-z_1)}{(z_3-z_0)(z_3-z_1)}}~,
\end{equation}
which leads to 
\begin{align}
 &\langle V_{p_0}^{\mathrm{R},\epsilon_0} (z_0) V_{p_1}^{\mathrm{R},\epsilon_1}(z_1)\Lambda_{p_2}^{\mathrm{NS}}(z_2)\Lambda_{p_3}^{\mathrm{NS}}(z_3) \rangle = - h_{p_2}^{\mathrm{NS}} \frac{z_{02}+z_{12}}{\sqrt{z_{30}z_{20}z_{31}z_{21}}} \langle  V_{p_0}^{\mathrm{R},\epsilon_0} (z_0) V_{p_1}^{\mathrm{R},\epsilon_1}(z_1)V_{p_2}^{\mathrm{NS}}(z_2)V_{p_3}^{\mathrm{NS}}(z_3) \rangle  \cr
 &\quad + \sqrt{\frac{z_{20}z_{21}}{z_{30}z_{31}}} \partial_{z_2}\langle  V_{p_0}^{\mathrm{R},\epsilon_0} (z_0) V_{p_1}^{\mathrm{R},\epsilon_1}(z_1)V_2(z_2)V_3(z_3) \rangle \cr
 &\quad + \epsilon_0 \epsilon_1\sqrt{\frac{z_{01}}{z_{30}z_{31}}} \langle  G_0V_{p_0}^{\mathrm{R},\epsilon_0} (z_0) V_{p_1}^{\mathrm{R},\epsilon_1}(z_1)\Lambda_{p_2}^{\mathrm{NS}}(z_2)V_{p_3}^{\mathrm{NS}}(z_3) \rangle\cr
 &\quad +\epsilon_1 \sqrt{-1} \sqrt{\frac{z_{01}}{z_{31}z_{30}}} \langle  V_{p_0}^{\mathrm{R},\epsilon_0}(z_0) G_0V_{p_1}^{\mathrm{R},\epsilon_1}(z_1)\Lambda_{p_2}^{\mathrm{NS}}(z_2)V_{p_3}^{\mathrm{NS}}(z_3) \rangle~.
\end{align}
Combining the above equation with the R-sector null-vector (\ref{eq:nullvectorR}) we obtain
\begin{align}\label{eq: R sector equation 2}
    \kappa\partial_{z_0}h^{\epsilon_0, \epsilon_1}(\mathbf{z})=&\langle G_{-1}G_0^2 V_{p_0}^{\mathrm{R},\epsilon_0} (z_0) V_{p_1}^{\mathrm{R},\epsilon_1}(z_1)\Lambda_{p_2}^{\mathrm{NS}}(z_2)V_{p_3}^{\mathrm{NS}}(z_3) \rangle = \frac{\beta_0^2}{2z_{01}} h^{\epsilon_0, \epsilon_1}(\mathbf{z})  \cr
    &\quad +\epsilon_0\beta\beta_1\frac{\sqrt{-1}}{z_{01}} \e^{-\frac{\pi i}{4}(\epsilon_0+\epsilon_1)}  h^{-\epsilon_0, -\epsilon_1}(\mathbf{z})
    -\epsilon_0 \epsilon_1\beta_0^2 h_{p_2}^{\mathrm{NS}}\sqrt{\frac{z_{01}}{z_{20}^3z_{21}}} g^{\epsilon_0,\epsilon_1}(\mathbf{z})\cr
     &\quad+\epsilon_0 \epsilon_1\beta_0^2 \sqrt{\frac{z_{21}}{z_{20}z_{01}}}\partial_{z_2}g^{\epsilon_0,\epsilon_1}(\mathbf{z})-\epsilon_0 \frac{\sqrt{-1}}{z_{30}}\beta_0\beta_1\e^{-\frac{\pi i}{4}(\epsilon_0+\epsilon_1)}  h^{-\epsilon_0, -\epsilon_1}(\mathbf{z})\cr 
      &\quad -\frac{\beta_0^2}{z_{30}} h^{\epsilon_0,\epsilon_1}(\mathbf{z}) -\epsilon_0 \epsilon_1\beta_0^2\sqrt{\frac{z_{20}z_{21}}{z_{01}z_{30}^2}} \partial_{z_2}g^{\epsilon_0,\epsilon_1}(\mathbf{z})\cr
      &\quad -\epsilon_0 \epsilon_1\beta_0^2h_{p_2}^{\mathrm{NS}}\frac{z_{20}+z_{21}}{\sqrt{{z_{01}z_{30}^2 z_{20}z_{21}}}}g^{\epsilon_0,\epsilon_1}(\mathbf{z})~.
\end{align}
Note that here we used
\begin{align}
    &\beta_0\langle  V_{p_0}^{\mathrm{R},\epsilon_0} (z_0) G_0V_{p_1}^{\mathrm{R},\epsilon_1}(z_1)\Lambda_{p_2}^{\mathrm{NS}}(z_2)V_{p_3}^{\mathrm{NS}}(z_3) \rangle \cr
    &= -i \e^{-\frac{\pi i}{4}\epsilon_0}\langle  G_0V_{p_0}^{\mathrm{R},-\epsilon_0} (z_0) G_0V_{p_1}^{\mathrm{R},\epsilon_1}(z_1)\Lambda_{p_2}^{\mathrm{NS}}(z_2)V_{p_3}^{\mathrm{NS}}(z_3) \rangle\cr
    &= -i \e^{-\frac{\pi i}{4}\epsilon_0}(i\e^{-\frac{\pi i}{4}\epsilon_1 })\beta_1\langle  G_0V_{p_0}^{\mathrm{R},-\epsilon_0}  (z_0) V_{p_1}^{\mathrm{R},-\epsilon_1} (z_1)\Lambda_{p_2}^{\mathrm{NS}}(z_2)V_{p_3}^{\mathrm{NS}}(z_3) \rangle \cr
    &= \e^{-\frac{\pi i}{4}(\epsilon_0+\epsilon_1)}\beta_1 h^{-\epsilon_0,-\epsilon_1}(\mathbf{z})~.
\end{align}
This gives us the second equation (\ref{eq: R sector equation 2}). 
\par We hence encounter the following different cases:
\paragraph{$\epsilon_0 = \epsilon_1$.}
If $\epsilon_0 = \epsilon_1$ the two equations (\ref{eq: R sector equation 1}) and (\ref{eq: R sector equation 2}) simplify to
\begin{subequations}
\begin{align}
      \kappa \partial_{z_0} g^{\epsilon_0}(\mathbf{z}) &= \frac{\beta_0^2}{2z_{01}} g^{\epsilon_0}(\mathbf{z}) +\sqrt{\frac{z_{21}}{z_{01}z_{20}}} h^{\epsilon_0}(\mathbf{z}) -\epsilon_0 \beta_0 \beta_1\frac{\sqrt{-1}}{z_{01}} \e^{- \frac{\pi i}{2}\epsilon_0} g^{-\epsilon_0}(\mathbf{z}) \cr
      &\quad -\sqrt{\frac{z_{20}z_{21}}{z_{01}z_{30}^2}} h^{\epsilon_0}(\mathbf{z}) +\beta_0 \beta_1\epsilon_0\frac{\sqrt{-1} }{z_{30}}\e^{-\frac{\pi i}{2}\epsilon_0} g^{-\epsilon_0}(\mathbf{z}) -\frac{\beta^2}{z_{30}} g^{\epsilon_0}(\mathbf{z})~,\\
    \kappa\partial_{z_0}h^{\epsilon_0}(\mathbf{z})&= \frac{\beta_0^2}{2z_{01}} h^{\epsilon_0}(\mathbf{z})  +\epsilon_0\beta_0\beta_1\frac{\sqrt{-1}}{z_{01}} \e^{-\frac{\pi i}{2}\epsilon_0}  h^{-\epsilon_0}(\mathbf{z})
     -\beta_0^2 h_{p_2}^{\mathrm{NS}}\sqrt{\frac{z_{01}}{z_{20}^3z_{21}}} g^{\epsilon_0}(\mathbf{z})\cr
     &\quad +\beta_0^2 \sqrt{\frac{z_{21}}{z_{20}z_{01}}}\partial_{z_2}g^{\epsilon_0}(\mathbf{z})-\epsilon_0 \frac{\sqrt{-1}}{z_{30}}\beta_0\beta_1\e^{-\frac{\pi i}{2}\epsilon_0}  h^{-\epsilon_0}(\mathbf{z})-\frac{\beta_0^2}{z_{30}} h^{\epsilon_0}(\mathbf{z}) \cr
     &\quad -\beta_0^2\sqrt{\frac{z_{20}z_{21}}{z_{01}z_{30}^2}} \partial_{z_2}g^{\epsilon_0}(\mathbf{z}) -\beta_0^2h_{p_2}^{\mathrm{NS}}\frac{z_{20}+z_{21}}{\sqrt{{z_{01}z_{30}^2 z_{20}z_{21}}}}g^{\epsilon_0}(\mathbf{z})~.
\end{align}      
\end{subequations}

\paragraph{$\epsilon_0 = -\epsilon_1$.} In this case,

\begin{subequations}
\begin{align}\label{}
      \kappa \partial_{z_0} g^{\epsilon_0,-\epsilon_0}(\mathbf{z}) &= \frac{\beta_{p_{0}}^2}{2z_{01}} g^{\epsilon_0,-\epsilon_0}(\mathbf{z}) - \sqrt{\frac{z_{21}}{z_{01}z_{20}}} h^{\epsilon_0,-\epsilon_0}(\mathbf{z}) - \epsilon_0 \beta_{p_{0}}\beta_{p_1}\frac{\sqrt{-1}}{z_{01}} g^{-\epsilon_0,-\epsilon_0}(\mathbf{z})\cr
      &\quad +\sqrt{\frac{z_{20}z_{21}}{z_{01}z_{30}^2}} h^{\epsilon_0,-\epsilon_0}(\mathbf{z}) -\epsilon_0\beta_{p_0}\beta_{p_1}\frac{\sqrt{-1} }{z_{30}}  g^{-\epsilon_0,-\epsilon_0}(\mathbf{z})+\frac{\beta_{p_0}^2}{z_{30}}  g^{\epsilon_0,-\epsilon_0}(\mathbf{z})~,\\
    \kappa\partial_{z_0}h^{\epsilon_0,- \epsilon_0}(\mathbf{z})&= \frac{\beta_0^2}{2z_{01}} h^{\epsilon_0, -\epsilon_0}(\mathbf{z})  +\epsilon_0\beta_0\beta_1\frac{\sqrt{-1}}{z_{01}}   h^{-\epsilon_0, \epsilon_0}(\mathbf{z})+\beta_0^2 h_{p_2}^{\mathrm{NS}}\sqrt{\frac{z_{01}}{z_{20}^3z_{21}}} g^{\epsilon_0,-\epsilon_0}(\mathbf{z})
    \cr
    &\quad -\beta_0^2 \sqrt{\frac{z_{21}}{z_{20}z_{01}}}\partial_{z_2}g^{\epsilon_0,-\epsilon_0}(\mathbf{z})-\epsilon_0 \frac{\sqrt{-1}}{z_{30}}\beta_0\beta_1  h^{-\epsilon_0, \epsilon_0}(\mathbf{z})-\frac{\beta_0^2}{z_{30}} h^{\epsilon_0,\epsilon_1}(\mathbf{z}) \cr
    &\quad+\beta_0^2\sqrt{\frac{z_{20}z_{21}}{z_{01}z_{30}^2}} \partial_{z_2}g^{\epsilon_0,-\epsilon_0}(\mathbf{z})
       +\beta_0^2h_{p_2}^{\mathrm{NS}}\frac{z_{20}+z_{21}}{\sqrt{{z_{01}z_{30}^2 z_{20}z_{21}}}}g^{\epsilon_0,-\epsilon_0}(\mathbf{z})~.
\end{align}
\end{subequations}
As a reminder,
\begin{align}\label{eq: gepsilon hepsilon}
    g^{\epsilon_0,\epsilon_1}(\mathbf{z}) &\equiv \langle V_{p_0}^{\mathrm{R},\epsilon_0}(z_0)V_{p_1}^{\mathrm{R},\epsilon_1}(z_1)V_{p_2}^{\mathrm{NS}}(z_2)V_{p_3}^{\mathrm{NS}}(z_3)\rangle~,\cr h^{\epsilon_0,\epsilon_1}(\mathbf{z}) &\equiv \langle G_0V_{p_0}^{\mathrm{R},\epsilon_0}(z_0)V_{p_1}^{\mathrm{R},\epsilon_1}(z_1)\Lambda_{p_2}^{\mathrm{NS}}(z_2)V_{p_3}^{\mathrm{NS}}(z_3)\rangle~.
\end{align}
We now make the ansatz
\begin{equation}
   g^{\epsilon_0,\epsilon_1}(\mathbf{z}) = \prod_{i<j}z_{ij}^{\mu_{ij}} g_0^{\epsilon_0,\epsilon_1}(z)~,\quad h^{\epsilon_0,\epsilon_1}(\mathbf{z}) = \prod_{i<j}z_{ij}^{\nu_{ij}} h_0^{\epsilon_0,\epsilon_1}(z)~,
\end{equation}
where $z \equiv \frac{z_{01}z_{23}}{z_{03}z_{21}}$ is the cross-ratio. In the limit
\begin{equation}
    z_0 \rightarrow z~,\quad z_1 \rightarrow 0~,\quad z_2 \rightarrow 1~,\quad z_3 \rightarrow \infty
\end{equation}
the derivatives transform as 
\begin{equation}
    \partial_{z_0} = \frac{\mu_{01}}{z} + \frac{\mu_{02}}{z-1} +\partial_z~,\quad \partial_{z_2} = -\frac{\mu_{02}}{z-1} +\mu_{12} - z\partial_z
\end{equation}
when acting on $g^{\epsilon_0,\epsilon_1}(\mathbf{z})$, and with $\mu \leftrightarrow \nu$ when we act on $h^{\epsilon_0,\epsilon_1}(\mathbf{z})$. Finally performing the by now standard replacement 
\begin{equation}
 g_0^{\epsilon_0,\epsilon_1}(z) \rightarrow z^{-\mu_{01}}(1-z)^{-\mu_{02}}g_0^{\epsilon_0,\epsilon_1}(z) ~,\quad h_0^{\epsilon_0,\epsilon_1}(z) \rightarrow z^{-\nu_{01}}(1-z)^{-\nu_{02}}h_0^{\epsilon_0,\epsilon_1}(z)   
\end{equation}
we find for $\epsilon_1 = \epsilon_0$:
\begin{subequations}\label{eq: R sector final epsilon0 = epsilon1}
\begin{align}
     \kappa \partial_{z} g_0^{\epsilon_0}(z) &= \frac{\beta_0^2}{2z} g_0^{\epsilon_0}(z) +\frac{1}{\sqrt{z(1-z)}}\, h_0^{\epsilon_0}(z) -\epsilon_0 \beta_0 \beta_1\frac{\sqrt{-1}}{z} \e^{- \frac{\pi i}{2}\epsilon_0} g_0^{-\epsilon_0}(z)~, \\
     \kappa\partial_{z}h_0^{\epsilon_0}(z)&= 
      \frac{\beta_0^2}{2z }h_0^{\epsilon_0}(z)  +
     \epsilon_0\beta_0\beta_1\frac{\sqrt{-1}}{z} \e^{-\frac{\pi i}{2}\epsilon_0}  h_0^{-\epsilon_0}(z)
       -\beta_0^2 h_{p_2}^{\mathrm{NS}}\sqrt{\frac{z}{(1-z)^3}}\, g_0^{\epsilon_0}(z)
      \cr
     &\quad 
     +\frac{\beta_0^2 }{\sqrt{z(1-z)}}(\mu_{01}+\mu_{02}+\mu_{12} -z \partial_z)g_0^{\epsilon_0}(z)~.
\end{align}
\end{subequations}
In principle we also have an equation for $\epsilon_1 = -\epsilon_0$, however it can be shown that none of the solutions of that differential equation satisfy crossing \cite{Fukuda:2002bv} and hence we will not discuss this situation further.
\par Focusing on the case $\epsilon_0 = \epsilon_1$ we now fix the branch of the square root. For this we look at the first equation in (\ref{eq: R sector final epsilon0 = epsilon1})  in the limit where $V_{p_2}^{\mathrm{NS}}$ and $V_{p_3}^{\mathrm{NS}}$ are the identity. We also use that $\epsilon_0 \,\e^{-\frac{\pi i}{2}\epsilon_0} = -i$. This then leads to two options:
\begin{equation}
  \kappa \partial_{z} \langle V_{p_{\langle 1,2 \rangle }}^{\mathrm{R},\epsilon_0}(z)  V_{p_{1}}^{\mathrm{R},\epsilon_1}(0)\rangle = \frac{\beta^2}{2z} \langle V_{p_{\langle 1,2 \rangle }}^{\mathrm{R},\epsilon_0}(z)  V_{p_{1}}^{\mathrm{R},\epsilon_1}(0)\rangle\pm \frac{ \beta^2}{z} \langle V_{p_{\langle 1,2 \rangle }}^{\mathrm{R},-\epsilon_0}(z)  V_{p_{1}}^{\mathrm{R},-\epsilon_1}(0)\rangle~,  
\end{equation}
Using that the two point function is of the form
\begin{equation}
    \langle V_{p_{0}}^{\mathrm{R},\epsilon_0}(z)  V_{p_{1}}^{\mathrm{R},\epsilon_1}(0)\rangle = \frac{C}{z^{2h_{[p_0]}}} \delta(p_0-p_1)~,
\end{equation}
where $C$ is independent of the coordinates we see that only the $+$ sign makes sense. Therefore, in summary, we obtain 
\begin{subequations}\label{eq: R sector final epsilon0 = epsilon1v2s}
\begin{align}
     \mathrm{I})\quad \kappa \partial_{z} g_0^{\epsilon_0}(z) &= \frac{\beta_0^2}{2z} g_0^{\epsilon_0}(z) +\frac{1}{\sqrt{z(1-z)}}\, h_{0}^{\epsilon_0}(z) + \frac{\beta_0 \beta_1}{z}  g_0^{-\epsilon_0}(z)~, \\
     \mathrm{II})\quad \kappa\partial_{z}h_0^{\epsilon_0}(z)&= 
      \frac{\beta_0^2}{2z }h_0^{\epsilon_0}(z) -\frac{\beta_0\beta_1}{z}   h_0^{-\epsilon_0}(z)
       -\beta_0^2 h_{p_2}^{\mathrm{NS}}\sqrt{\frac{z}{(1-z)^3}}\, g_0^{\epsilon_0}(z)
      \cr
     &\quad 
     +\frac{\beta_0^2 }{\sqrt{z(1-z)}}(\mu_{01}+\mu_{02}+\mu_{12} -z \partial_z)g_0^{\epsilon_0}(z)    ~,
\end{align}
\end{subequations}
where
\begin{equation}
    \mu_{01}+\mu_{02}+\mu_{12}= -h^{\mathrm{R}}_{[p_{0}]}-h^{\mathrm{R}}_{[p_1]}- h^{\mathrm{NS}}_{p_2}+h^{\mathrm{NS}}_{p_3} ~.
\end{equation}
The sign in the second equation is fixed by taking $V_{p_3}^{\mathrm{NS}}$ to the identity and using the general form of the OPE
\begin{align}
\left[V^{\text{R},\epsilon}\right]\left[V^{\text{R},\epsilon}\right]\sim \left[V\right] +\left[W\right]\ , \ \ \left[V^{\text{R},\epsilon}\right]\left[V^{\text{R},-\epsilon}\right]\sim \left[\Lambda\right] +\left[\widetilde{\Lambda}\right] \ , ~~~~~~ \epsilon=\pm~.
\end{align}
Adding $\mathrm{I})$ and $\mathrm{II})$ to themselves with $\epsilon_0 \rightarrow - \epsilon_0$, and with 
\begin{equation}
    \mathsf{g}^{\epsilon_0} = \frac{1}{2}(\epsilon_0 g_0^{\epsilon_0} +g_0^{-\epsilon_0})~,\quad \mathsf{h}^{\epsilon_0} = \frac{1}{2}(\epsilon_0 h_0^{\epsilon_0} +h_0^{-\epsilon_0}),
\end{equation}
we obtain
\begin{subequations}\label{eq:Rfinalform}
    \begin{align}
        \kappa \partial_z \mathsf{g}^{\epsilon}(z) &= \frac{\beta_0^2}{2z}\mathsf{g}^{\epsilon}(z) + \frac{1}{\sqrt{z(1-z)}}\mathsf{h}^\epsilon(z) + \epsilon \frac{\beta_0 \beta_1}{z}\mathsf{g}^\epsilon(z)~,\\
        \kappa \partial_z \mathsf{h}^\epsilon(z) &= \frac{\beta_0^2}{2z }\mathsf{h}^{\epsilon}(z) -\epsilon\frac{\beta_0\beta_1}{z}   \mathsf{h}^{\epsilon}(z)
       -\beta_0^2 h_{p_2}^{\mathrm{NS}}\sqrt{\frac{z}{(1-z)^3}}\, \mathsf{g}^{\epsilon}(z)
      \cr
     &\quad 
     +\frac{\beta_0^2 }{\sqrt{z(1-z)}}(\mu_{01}+\mu_{02}+\mu_{12} -z \partial_z)\mathsf{g}^{\epsilon}(z) ~,
    \end{align}
\end{subequations}
where we replaced $\epsilon_0 = \epsilon$ and $\beta \equiv \frac{p}{\sqrt{2}}$ with $p_0 = p_{\langle 1,2\rangle }$ and $\kappa \equiv \frac{2b^2+1}{2b^2}$. This is the final form of the equations that we will now study in order to obtain the R-sector shift relations.
\paragraph{Shift-equation.}
We can eliminate $\mathsf{h}^{\epsilon}$ in (\ref{eq:Rfinalform}) by solving for it in the first equation and plugging it into the second. Making the ansatz 
\begin{equation}\label{eq: gfinal}
  \mathsf{g}^{\epsilon}(z) = z^{\frac{1}{8} +\frac{b^2}{4} +\epsilon \frac{b p_1}{2}}(1-z)^{\frac{b}{2}\left(\frac{Q}{2} -p_2\right)} F(z)~,  
\end{equation}
we finally obtain the main differential equation 
\begin{equation}
    z(1-z) F''(z) + (c_\epsilon-(a_\epsilon+b_\epsilon+1)z)F'(z) - a_\epsilon b_\epsilon F(z)=0~,
\end{equation}
where 
\begin{equation}
        a_\epsilon= \frac{1}{4} - \frac{b}{2}(-\epsilon p_1 + p_2+p_3)~,\quad b_\epsilon = \frac{1}{4} - \frac{b}{2}(-\epsilon p_1 + p_2 -p_3)~, \quad c_\epsilon = \frac{1}{2} +\epsilon bp_1~.
\end{equation}
We have two linearly independent solutions for each $\epsilon$. In the s-channel these are given by
\begin{multline}
\mathsf{g}_s^{\epsilon}(z) = z^{\frac{1}{8} +\frac{b^2}{4} +\epsilon \frac{b p_1}{2}}(1-z)^{\frac{b}{2}(\frac{Q}{2}-p_2)}\,_2 F_1 \left(a_\epsilon, b_\epsilon, c_\epsilon; z\right)\cr
+z^{\frac{5}{8} +\frac{b^2}{4} -\epsilon \frac{b p_1}{2}}(1-z)^{\frac{b}{2}(\frac{Q}{2}-p_2)}\, _2 F_1 \left(a_\epsilon - c_\epsilon+1, b_\epsilon - c_\epsilon+1,2-c_\epsilon; z\right)~.
\end{multline}
Similarly we have two linearly independent solutions (for each $\epsilon$) in the $t$-channel
\begin{multline}
    \mathsf{g}_t^{\epsilon}(z) = z^{\frac{1}{8} +\frac{b^2}{4} +\epsilon \frac{b p_1}{2}}(1-z)^{\frac{b}{2}(\frac{Q}{2}-p_2)}\,_2 F_1 \left(a_\epsilon, b_\epsilon, a_\epsilon+ b_\epsilon +1 -c_\epsilon; 1-z\right)\cr
    +  z^{\frac{1}{8} +\frac{b^2}{4} +\epsilon \frac{b p_1}{2}}(1-z)^{\frac{b}{2}(\frac{Q}{2}+p_2)}\,_2 F_1 \left(c_\epsilon - a_\epsilon, c_\epsilon -b_\epsilon, c_\epsilon-a_\epsilon- b_\epsilon +1; 1-z\right)~.
\end{multline}
We can read the internal momenta in the $s$-channel from the $z\rightarrow 0$ limit. We have
\begin{subequations}\label{eq:internal momenta R sector s channel}
\begin{align}
   \frac{1}{8} + \frac{b^2}{4} + \epsilon \frac{b p_1}{2} &= - \left(\frac{Q^2}{8} - \frac{p^2_{\langle 1,2\rangle}}{2} + \frac{1}{16}\right) - \left(\frac{Q^2}{8} - \frac{p^2_1}{2} + \frac{1}{16}\right)
   + \left(\frac{Q^2}{8} - \frac{\left(p_1 - \frac{\epsilon b}{2}\right)^2}{2}\right)~\\
   \frac{5}{8} + \frac{b^2}{4} -\epsilon \frac{b p_1}{2} &= - \left(\frac{Q^2}{8} - \frac{p^2_{\langle 1,2\rangle}}{2} + \frac{1}{16}\right) - \left(\frac{Q^2}{8} - \frac{p^2_1}{2} + \frac{1}{16}\right)
   + \left(\frac{Q^2}{8} - \frac{\left(p_1 + \frac{\epsilon b}{2}\right)^2}{2} + \frac{1}{2}\right)~.
\end{align}   
\end{subequations}
Similarly, in the $t$-channel we can read off the internal momenta in the $z\rightarrow 1$ limit
\begin{equation}\label{eq: internal t channel R sector}
\frac{b}{2}\left(\frac{Q}{2}\mp p_2\right) = -\left(\frac{Q^2}{8} - \frac{p_{\langle 2,1\rangle}^2}{2} + \frac{1}{16}\right) -\left(\frac{Q^2}{8} - \frac{p_2^2}{2}  \right) +\left(\frac{Q^2}{8} - \frac{(p_2 \pm\frac{b}{2})^2}{2} +\frac{1}{16}\right)~.
\end{equation}
Since the exponent of $(1-z)$ in $\mathsf{g}_t^{\epsilon}(z)$ is independent of $\epsilon$,  (\ref{eq: internal t channel R sector}) is valid for either $\epsilon = \pm 1$. 
\par Note that, since $V_p^{\mathrm{R},+}$ is a combination of the two operators $\Theta^{\pm\pm}$, $\mathsf{g}^{+}(z)$ encodes the four-point functions
\begin{equation}\label{eq:plusfptfns}
  \langle\Theta^{\pm\pm}_{p_{\langle 1,2\rangle}} (z)\Theta^{\pm\pm}_{p_1} (0)V_{p_2}^{\mathrm{NS}} (1)V_{p_3}^{\mathrm{NS}}(\infty)\rangle~,\quad  \langle\Theta^{\pm\pm}_{p_{\langle 1,2\rangle}}(z) \Theta^{\mp\mp}_{p_1} (0)V_{p_2}^{\mathrm{NS}} (1)V_{p_3}^{\mathrm{NS}} (\infty)\rangle  ~. 
\end{equation}
Similarly, since $V_p^{\mathrm{R},-}$ is a combination of the two operators $\Theta^{\pm\mp}$, $\mathsf{g}^{-}(z)$ encodes the four-point functions
\begin{equation}\label{eq: OPE appendix}
  \langle\Theta^{\pm\mp}_{p_{\langle 1,2\rangle}} (z)\Theta^{\pm\mp}_{p_1} (0)V_{p_2}^{\mathrm{NS}} (1)V_{p_3}^{\mathrm{NS}}(\infty)\rangle~,\quad  \langle\Theta^{\pm\mp}_{p_{\langle 1,2\rangle}}(z) \Theta^{\mp\pm}_{p_1} (0)V_{p_2}^{\mathrm{NS}} (1)V_{p_3}^{\mathrm{NS}} (\infty)\rangle  ~. 
\end{equation}
We now discuss the case $\epsilon=+1$ (i.e. (\ref{eq:plusfptfns})). The case $\epsilon=-1$ follows analogously. We denote as
\begin{equation}
    C_{\mathrm{odd}}^{(b)}(p_1,p_2;p_3) \equiv \langle \Theta^{\pm\pm}_{p_1}\Theta^{\pm \pm}_{p_2}V_{p_3}^{\mathrm{NS}} \rangle~,\quad  C_{\mathrm{even}}^{(b)}(p_1,p_2;p_3) \equiv  \langle \Theta^{\pm\pm}_{p_1}\Theta^{\mp \mp}_{p_2}V_{p_3}^{\mathrm{NS}} \rangle~,
\end{equation}
and from the OPEs \cite{Fukuda:2002bv}
\begin{align} \left[\Theta_{p_{\langle 1,2 \rangle}}^{\pm\pm}\right]\left[\Theta_p^{\pm\pm}\right]  &\sim  \left[\Theta_{p_{\langle 1,2 \rangle}}^{\pm\mp}\right]\left[\Theta_p^{\pm\mp}\right] = [V^{\mathrm{NS}}_{p+\frac{b}{2}}] +[W^{\mathrm{NS}}_{p-\frac{b}{2}}] ~,\cr
\left[\Theta_{p_{\langle 1,2 \rangle}}^{\pm\pm}\right]\left[\Theta_p^{\mp\mp}\right]  &\sim  \left[\Theta_{p_{\langle 1,2 \rangle}}^{\pm\mp}\right]\left[\Theta_p^{\mp\pm}\right] =  [V^{\mathrm{NS}}_{p-\frac{b}{2}}] +[W^{\mathrm{NS}}_{p+\frac{b}{2}}] ~,\cr
    [\Theta^{\pm\pm}_{p_{\langle 1,2\rangle}}V_{p}^{\mathrm{NS}}] &\sim  [\Theta_{p+\frac{b}{2}}^{\pm\pm}] + [\Theta_{p -\frac{b}{2}}^{\mp \mp}]~, \label{eq: OPE appendix2}
\end{align}
it is clear that $C_{\mathrm{odd}}^{(b)}(p_1,p_2;p_3) \propto \langle \Theta^{\pm\mp}_{p_1}\Theta^{\pm \mp}_{p_2}V_{p_3}^{\mathrm{NS}} \rangle$ and $C_{\mathrm{even}}^{(b)}(p_1,p_2;p_3) \propto \langle \Theta^{\pm\mp}_{p_1}\Theta^{\mp \pm}_{p_2}V_{p_3}^{\mathrm{NS}} \rangle$. Consequently, we can expand
\begin{align}\label{eq: gepsilon crossing0}
     \langle \Theta_{p_{\langle 1,2\rangle}}^{++}\Theta_{p_1}^{--} V^{\mathrm{NS}}_{p_2}V^{\mathrm{NS}}_{p_3}\rangle &= c_1^{(s)} f_{1,s}(z)\overline{f_{1,s}(z)} + c_{2}^{(s)} f_{2,s}(z)\overline{f_{2,s}(z)}~\cr
    &= c_1^{(t)} f_{1,t}(z)\overline{f_{1,t}(z)} + c_{2}^{(t)} f_{2,t}(z)\overline{f_{2,t}(z)}~,
\end{align}
where 
\begin{align}\label{eq: c1t}
    c_1^{(t)} &= \frac{C^{(b)}_{\mathrm{odd}}\left(p_{\langle 1,2 \rangle}, p_2 +\frac{b}{2}; p_2\right)C^{(b)}_{\mathrm{even}}\left( p_2 +\frac{b}{2}, p_1; p_3\right)}{B_{\mathrm{R}}^{(b)}(p_2+\frac{b}{2})}\cr
    c_2^{(t)} &= \frac{C^{(b)}_{\mathrm{even}}\left(p_{\langle 1,2 \rangle}, p_2 -\frac{b}{2}; p_2\right)C^{(b)}_{\mathrm{odd}}\left( p_2 -\frac{b}{2}, p_1; p_3\right)}{B_{\mathrm{R}}^{(b)}(p_2-\frac{b}{2})}~.
\end{align} 
We note in passing that from the OPE (\ref{eq: OPE appendix2}) one can infer the property
\begin{equation}
    C_{\mathrm{even}}^{(b)}(p_1,-p_2;p_3) = C_{\mathrm{odd}}^{(b)}(p_1,p_2;p_3) = C_{\mathrm{even}}^{(b)}(-p_1,p_2;p_3)~. 
\end{equation}
 \begin{figure}[H]
\begin{tikzpicture}
\begin{scope}[shift={(-5,.4)}, scale=.65, rotate=270]
\draw[very thick, red, looseness=.7] (2,1.75) to[out=10,in=5] (2,.25);
\draw[very thick, densely dashed, red, looseness=.7] (2,1.75) to[out=-180,in=-180] (2,.25);    
\draw[fill=white, draw=gray, opacity=.1] (0,2.5) to[out=0, in=180] (2,1.75) to[out=0, in=180] (4,2.5) 
    to[out=200, in=160, looseness=.95] (4,1.5) to[out=180, in=180, looseness=2.5] (4,.5) 
    to[out=200, in=160, looseness=.95] (4,-.5) to[out=180, in=0] (2,.25) 
    to[out=180, in=0] (0,-.5) to[out=20, in=-20, looseness=.95] (0,.5) 
    to[out=0, in=0, looseness=2.5] (0,1.5) to[out=20, in=-20, looseness=.95] (0,2.5);
\draw[very thick] (0,0) ellipse (0.25 and .5);
\draw[very thick] (0,2) ellipse (0.25 and .5);
\draw[very thick] (4,0) ellipse (0.25 and .5);
\draw[very thick] (4,2) ellipse (0.25 and .5);
\node[scale=1., thick] at (-.9,0) {$\Theta^{--}_{p_{1}}$};
\node[scale=1., thick] at (-.8,2.6) {$V^{\mathrm{NS}}_{p_3}$};  
\node[scale=1., thick] at (4.8,2.8) {$V^{\mathrm{NS}}_{p_2}$};
\node[scale=1., thick] at (5,0) {$\Theta^{++}_{p_{\langle 1,2\rangle}}$};
\draw[very thick] (0,.5) to[out=0, in=0, looseness=2.5] (0,1.5);
\draw[very thick] (4,.5) to[out=180, in=180, looseness=2.5] (4,1.5);
\draw[very thick] (0,2.5) to[out=0, in=180] (2,1.75) to[out=0, in=180] (4,2.5);
\draw[very thick] (0,-.5) to[out=0, in=180] (2,.25) to[out=0, in=180] (4,-.5);    
\node[scale=1., red] at (2,3) {$h_{[p_2+ \frac{\eta b}{2}]}$};
\node[scale=1., red] at (2,-1) {$\Theta^{\eta\eta}_{p_{\langle 1,2\rangle}}$};
\end{scope}
\begin{scope}[shift={(2.8,-1.5)}, scale=.7]
\draw[very thick, red, looseness=.7] (2,1.75) to[out=10,in=5] (2,.25);
\draw[very thick, densely dashed, red, looseness=.7] (2,1.75) to[out=-180,in=-180] (2,.25);    
\draw[fill=white, draw=gray, opacity=.1] (0,2.5) to[out=0, in=180] (2,1.75) to[out=0, in=180] (4,2.5) 
    to[out=200, in=160, looseness=.95] (4,1.5) to[out=180, in=180, looseness=2.5] (4,.5) 
    to[out=200, in=160, looseness=.95] (4,-.5) to[out=180, in=0] (2,.25) 
    to[out=180, in=0] (0,-.5) to[out=20, in=-20, looseness=.95] (0,.5) 
    to[out=0, in=0, looseness=2.5] (0,1.5) to[out=20, in=-20, looseness=.95] (0,2.5);
\draw[very thick] (0,0) ellipse (0.25 and .5);
\draw[very thick] (0,2) ellipse (0.25 and .5);
\draw[very thick] (4,0) ellipse (0.25 and .5);
\draw[very thick] (4,2) ellipse (0.25 and .5);
\node[scale=1., thick] at (-1.1,-.05) {$\Theta_{p_{\langle 1,2\rangle}}^{++}$};
\node[scale=1., thick] at (-0.9,2) {$\Theta_{p_{1}}^{--}$};
\node[scale=1., thick] at (4.9,2) {$V^{\mathrm{NS}}_{p_3}$};
\node[scale=1., thick] at (4.9,-.05) {$V^{\mathrm{NS}}_{p_2}$};
\draw[very thick] (0,.5) to[out=0, in=0, looseness=2.5] (0,1.5);
\draw[very thick] (4,.5) to[out=180, in=180, looseness=2.5] (4,1.5);
\draw[very thick] (0,2.5) to[out=0, in=180] (2,1.75) to[out=0, in=180] (4,2.5);
\draw[very thick] (0,-.5) to[out=0, in=180] (2,.25) to[out=0, in=180] (4,-.5); 
\node[scale=1.2] at (-5.3,1.) {$~~~~~~~~= \  \sum_{\eta=\pm}{\mathbb{F}}_{\delta,\eta}$};
\node[scale=1., red] at (2.1,-1) {$h_{p_1-\delta \frac{p_1}{2}}$};
\end{scope}

\end{tikzpicture}
\caption{Crossing transformation of sphere four-point conformal blocks with one insertion of a $\langle1,2\rangle$ degenerate R-sector operator. The fusion kernel is a $2\times 2$ matrix given in (\ref{eq:FmatrixRamond}). }
\label{fig:N1crossing_Rsector}
\end{figure}
\noindent
We now impose crossing in (\ref{eq: gepsilon crossing0}), which implies $c_1^{(t)} {\mathbb{F}}_{11}{\mathbb{F}}_{12} + c_2^{(t)} {\mathbb{F}}_{21}{\mathbb{F}}_{22} =0$. The matrix entries ${\mathbb{F}}_{ij}$ follow again from the basic properties of the hypergeometric function, i.e.
\begin{multline}
    _2 F_1 \left(a, f, c, 1-z\right) = \frac{\Gamma(c)\Gamma(c-a-f)}{\Gamma(c-a) \Gamma(c-f)} \, _2F_1(a, f, a+f-c+1, z)  \cr  
     + \frac{\Gamma (c)\Gamma(a+f-c)}{\Gamma(a)\Gamma(f)} z^{c-a-f} \, _2F_1 (c-a,c-f,c-a-f+1,z)~.
\end{multline}
These are explicitly given by
 \begin{align}
     \mathbb{F}_{11} &=\frac{\Gamma(\frac{1}{2} - bp_1)\Gamma(1-bp_2)}{\Gamma(\frac{3}{4} - \frac{b}{2}p_{1+2-3})\Gamma(\frac{3}{4}-\frac{b}{2}p_{1+2+3})} ~,\quad \mathbb{F}_{12} =\frac{\Gamma(-\frac{1}{2}+ bp_1)\Gamma(1-bp_2)}{\Gamma(\frac{1}{4}+\frac{b}{2}p_{1-2+3})\Gamma(\frac{1}{4}-\frac{b}{2}p_{-1+2+3})}~,\cr
     \mathbb{F}_{21} &= \frac{\Gamma (1+bp_2)\Gamma(\frac{1}{2}-bp_1)}{\Gamma(\frac{3}{4} -\frac{b}{2}p_{1-2+3}) \Gamma(\frac{3}{4} + \frac{b}{2}p_{-1+2+3})}~,\quad \mathbb{F}_{22} =\frac{\Gamma(1+bp_2)\Gamma(-\frac{1}{2} + bp_1)}{\Gamma(\frac{1}{4}+\frac{b}{2}p_{1+2-3})\Gamma(\frac{1}{4} +\frac{b}{2}p_{1+2+3})} ~.\label{eq:FmatrixRamond}
 \end{align}
 We thus find 
 \begin{align}\label{eq: shiftequation R sector oddeven}
    \frac{C_{\mathrm{even}}^{(b)}(p_2+\frac{b}{2},p_1; p_3)^2/B_{\mathrm{R}}^{(b)}(p_2+\frac{b}{2})}{C_{\mathrm{odd}}^{(b)}(p_2-\frac{b}{2}, p_1; p_3)^2/B_{\mathrm{R}}^{(b)}(p_2-\frac{b}{2})} &= -\left(\frac{{\mathbb{F}}_{21}{\mathbb{F}}_{22}}{{\mathbb{F}}_{11}{\mathbb{F}}_{12}}\right)^{\!\!2}\times \left(\frac{{\mathbb{F}}_{11}{\mathbb{F}}_{12}}{{\mathbb{F}}_{22}{\mathbb{F}}_{21}}\bigg|_{\substack{p_1 = p_{\langle 1,2\rangle }\\ p_3 = p_2}}\right)~\cr
    &\equiv \kappa_{\mathrm{R}}^{(b)}(p_2|p_1;p_3)~.
\end{align}
Similarly, crossing symmetry of the following four-point function
\begin{align}\label{eq: gepsilon crossing}
     \langle \Theta_{p_{\langle 1,2\rangle}}^{++}\Theta_{-p_1}^{++} V^{\mathrm{NS}}_{p_2}V^{\mathrm{NS}}_{p_3}\rangle &= c_1^{(s)} f_{1,s}(z)\overline{f_{1,s}(z)} + c_{2}^{(s)} f_{2,s}(z)\overline{f_{2,s}(z)}~\cr
    &= c_1^{(t)} f_{1,t}(z)\overline{f_{1,t}(z)} + c_{2}^{(t)} f_{2,t}(z)\overline{f_{2,t}(z)}~,
\end{align}
where now 
\begin{align}\label{eq: c1t}
    c_1^{(t)} &= \frac{C^{(b)}_{\mathrm{odd}}\left(p_{\langle 1,2 \rangle}, p_2 +\frac{b}{2}; p_2\right)C^{(b)}_{\mathrm{odd}}\left( p_2 +\frac{b}{2}, p_1; p_3\right)}{B_{\mathrm{R}}^{(b)}(p_2+\frac{b}{2})}\cr
    c_2^{(t)} &= \frac{C^{(b)}_{\mathrm{even}}\left(p_{\langle 1,2 \rangle}, p_2 -\frac{b}{2}; p_2\right)C^{(b)}_{\mathrm{even}}\left( p_2 -\frac{b}{2}, p_1; p_3\right)}{B_{\mathrm{R}}^{(b)}(p_2-\frac{b}{2})}~,
\end{align}
yields
\begin{equation}\label{eq: shiftequation R sector evenodd}
\frac{C_{\mathrm{odd}}^{(b)}(p_2+\frac{b}{2},p_1; p_3)^2/B_{\mathrm{R}}^{(b)}(p_2+\frac{b}{2})}{C_{\mathrm{even}}^{(b)}(p_2-\frac{b}{2}, p_1; p_3)^2/B_{\mathrm{R}}^{(b)}(p_2-\frac{b}{2})} =\kappa^{(b)}_{\mathrm{R}}\left(p_2|-p_1;p_3\right)~.
\end{equation}
In summary (and after relabelling $p_1\leftrightarrow p_2$ appropriately) we find 
\begin{align}\label{eq: shiftequations R sector appendix}
\frac{C_{\mathrm{even}}^{(b)}(p_1+\frac{b}{2},p_2; p_3)^2/B_{\mathrm{R}}^{(b)}(p_1+\frac{b}{2})}{C_{\mathrm{odd}}^{(b)}(p_1-\frac{b}{2}, p_2; p_3)^2/B_{\mathrm{R}}^{(b)}(p_1-\frac{b}{2})} &=\kappa^{(b)}_{\mathrm{R}}\left(p_1|p_2;p_3\right) ~ ,  \cr
\frac{C_{\mathrm{odd}}^{(b)}(p_1+\frac{b}{2},p_2; p_3)^2/B_{\mathrm{R}}^{(b)}(p_1+\frac{b}{2})}{C_{\mathrm{even}}^{(b)}(p_1-\frac{b}{2}, p_2; p_3)^2/B_{\mathrm{R}}^{(b)}(p_1-\frac{b}{2})} &=\kappa^{(b)}_{\mathrm{R}}\left(p_1|-p_2;p_3\right)~,
\end{align}
where $\kappa_{\mathrm{R}}^{(b)}(p_1|p_2;p_3)$ is given in (\ref{eq:kappaR}) in the main text. These are exactly the two shift relations for the normalization-independent bootstrap data on the sphere in the R-sector.

\section{Relation to the $\mathcal{N}=1$ fusion kernel}\label{app:fusionkernel}
In this last appendix we discuss the fusion kernel of (spacelike) $\mathcal{N}=1$ Liouville theory in the NS-sector as obtained in \cite{Hadasz:2007wi,Chorazkiewicz:2008es,Hadasz:2013bwa,Pawelkiewicz:2013wga,Poghosyan:2016kvd}. Our motivation is to highlight the fact that the NS Liouville structure constants that we discussed in the main text are realized as particular instances of this fusion kernel, which is in a sense a more general (representation-theoretic) object. This parallels the story in bosonic Liouville where the DOZZ structure constants (in the appropriate normalization) can also be written in terms of the analogous fusion kernel for the usual Virasoro conformal blocks \cite{Teschner:2001rv,Collier:2019weq}.
\par The crossing transformation of an even (e) or odd (o) NS four-point conformal block with four external momenta $p_{1,2,3,4}$\footnote{See e.g. \cite{Belavin:2007gz,Belavin:2007eq} for more details on the construction of the NS conformal blocks. Here the even conformal blocks are normalized as $\mathcal{F}_{p_s}^{\mathrm{e}}\left[\begin{smallmatrix} p_3 & p_2 \\ p_4 & p_1 \end{smallmatrix}\right](z)=z^{h_{s}-h_1-h_2}\left(1+O(z)\right)$, while the odd conformal blocks as $\mathcal{F}_{p_s}^{\mathrm{o}}\left[\begin{smallmatrix} p_3 & p_2 \\ p_4 & p_1 \end{smallmatrix}\right](z)=z^{h_s+1/2-h_{1}-h_2}\left(\frac{1}{2h_s}+O(z)\right)$.} defines the $\mathcal{N}=1$ NS fusion kernel via
\begin{equation}
    \mathcal{F}_{p_s}^{\eta}\left[\begin{smallmatrix} p_3 & p_2 \\ p_4 & p_1 \end{smallmatrix}\right](z)=\int_{i\mathbb{R}_+}\frac{\d p_t}{i}\sum_{\rho=e,o}\mathbb{F}_{p_sp_t}\left[\begin{smallmatrix} p_3 & p_2 \\ p_4 & p_1 \end{smallmatrix}\right]^{\eta}\text{}_{\rho} \ \mathcal{F}_{p_t}^{\rho}\left[\begin{smallmatrix} p_1 & p_2 \\ p_4 & p_3 \end{smallmatrix}\right](1-z)~, \quad ~ \eta=\mathrm{e}, \mathrm{o}.
   \end{equation}
 The components of the fusion kernel were first written down in \cite{Hadasz:2007wi,Chorazkiewicz:2008es},
 and in our notation read explicitly
\begin{align}
    &\mathbb{F}_{p_sp_t}\left[\begin{smallmatrix} p_3 & p_2 \\ p_4 & p_1 \end{smallmatrix}\right]^{\mathrm{e}}\text{}_{\mathrm{e}}= \nonumber \\
    &\frac{\Gamma_b^{\text{NS}}\left(\frac{Q}{2}+p_t+p_2-p_3\right)\Gamma_b^{\text{NS}}\left(\frac{Q}{2}+p_t+p_2+p_3\right)\Gamma_b^{\text{NS}}\left(\frac{Q}{2}+p_t-p_1-p_4\right)\Gamma_b^{\text{NS}}\left(\frac{Q}{2}+p_t+p_1-p_4\right)}{\Gamma_b^{\text{NS}}\left(\frac{Q}{2}+p_s-p_3-p_4\right)\Gamma_b^{\text{NS}}\left(\frac{Q}{2}+p_s+p_3-p_4\right)\Gamma_b^{\text{NS}}\left(\frac{Q}{2}+p_s-p_1+p_2\right)\Gamma_b^{\text{NS}}\left(\frac{Q}{2}+p_s+p_1+p_2\right)} \nonumber \\
    &\times \frac{\Gamma_b^{\text{NS}}\left(\frac{Q}{2}-p_t+p_2-p_3\right)\Gamma_b^{\text{NS}}\left(\frac{Q}{2}-p_t+p_2+p_3\right)\Gamma_b^{\text{NS}}\left(\frac{Q}{2}-p_t-p_1-p_4\right)\Gamma_b^{\text{NS}}\left(\frac{Q}{2}-p_t+p_1-p_4\right)}{\Gamma_b^{\text{NS}}\left(\frac{Q}{2}-p_s-p_3-p_4\right)\Gamma_b^{\text{NS}}\left(\frac{Q}{2}-p_s+p_3-p_4\right)\Gamma_b^{\text{NS}}\left(\frac{Q}{2}-p_s-p_1+p_2\right)\Gamma_b^{\text{NS}}\left(\frac{Q}{2}-p_s+p_1+p_2\right)} \nonumber \\
    &\times \frac{1}{2}\frac{\Gamma_b^{\text{NS}}\left(Q+2p_s\right)\Gamma_b^{\text{NS}}\left(Q-2p_s\right)}{\Gamma_b^{\text{NS}}\left(2p_t\right)\Gamma_b^{\text{NS}}\left(-2p_t\right)}\int_{i\mathbb{R}}\frac{\d t}{i}J^{(\mathrm{ee})}_{p_sp_t}\left[\begin{smallmatrix} p_3 & p_2 \\ p_4 & p_1 \end{smallmatrix}\right] \nonumber , \\
\label{eq:eveneven} \\
 &\mathbb{F}_{p_sp_t}\left[\begin{smallmatrix} p_3 & p_2 \\ p_4 & p_1 \end{smallmatrix}\right]^{\mathrm{e}}\text{}_{\mathrm{o}}= \nonumber \\
    &\frac{\Gamma_b^{\text{R}}\left(\frac{Q}{2}+p_t+p_2-p_3\right)\Gamma_b^{\text{R}}\left(\frac{Q}{2}+p_t+p_2+p_3\right)\Gamma_b^{\text{R}}\left(\frac{Q}{2}+p_t-p_1-p_4\right)\Gamma_b^{\text{R}}\left(\frac{Q}{2}+p_t+p_1-p_4\right)}{\Gamma_b^{\text{NS}}\left(\frac{Q}{2}+p_s-p_3-p_4\right)\Gamma_b^{\text{NS}}\left(\frac{Q}{2}+p_s+p_3-p_4\right)\Gamma_b^{\text{NS}}\left(\frac{Q}{2}+p_s-p_1+p_2\right)\Gamma_b^{\text{NS}}\left(\frac{Q}{2}+p_s+p_1+p_2\right)} \nonumber \\
    &\times \frac{\Gamma_b^{\text{R}}\left(\frac{Q}{2}-p_t+p_2-p_3\right)\Gamma_b^{\text{R}}\left(\frac{Q}{2}-p_t+p_2+p_3\right)\Gamma_b^{\text{R}}\left(\frac{Q}{2}-p_t-p_1-p_4\right)\Gamma_b^{\text{R}}\left(\frac{Q}{2}-p_t+p_1-p_4\right)}{\Gamma_b^{\text{NS}}\left(\frac{Q}{2}-p_s-p_3-p_4\right)\Gamma_b^{\text{NS}}\left(\frac{Q}{2}-p_s+p_3-p_4\right)\Gamma_b^{\text{NS}}\left(\frac{Q}{2}-p_s-p_1+p_2\right)\Gamma_b^{\text{NS}}\left(\frac{Q}{2}-p_s+p_1+p_2\right)} \nonumber \\
    &\times \frac{1}{2}\frac{\Gamma_b^{\text{NS}}\left(Q+2p_s\right)\Gamma_b^{\text{NS}}\left(Q-2p_s\right)}{\Gamma_b^{\text{NS}}\left(2p_t\right)\Gamma_b^{\text{NS}}\left(-2p_t\right)}\int_{i\mathbb{R}}\frac{\d t}{i}J^{(\mathrm{eo})}_{p_sp_t}\left[\begin{smallmatrix} p_3 & p_2 \\ p_4 & p_1 \end{smallmatrix}\right]~.\label{eq:evenodd}   
\end{align}
The integrands $J$ are given respectively by\footnote{As a point of reference, note that compared to the conventions of \cite{Hadasz:2013bwa} we have shifted the contour by $t\rightarrow t-Q/2+p_s$.}
\begin{align}
J^{(\mathrm{ee})}_{p_sp_t}\left[\begin{smallmatrix} p_3 & p_2 \\ p_4 & p_1 \end{smallmatrix}\right]&=\sum_{\nu=\text{NS,R}}\frac{S_b^{\nu}\left(t+U_1\right)S_b^{\nu}\left(t+U_2\right)S_b^{\nu}\left(t+U_3\right)S_b^{\nu}\left(t+U_4\right)}{S_b^{\nu}\left(t+V_1\right)S_b^{\nu}\left(t+V_2\right)S_b^{\nu}\left(t+V_3\right)S_b^{\nu}\left(t+V_4\right)} \nonumber , \\
J^{(\mathrm{eo})}_{p_sp_t}\left[\begin{smallmatrix} p_3 & p_2 \\ p_4 & p_1 \end{smallmatrix}\right]&=\frac{S_b^{\text{NS}}\left(t+U_1\right)S_b^{\text{NS}}\left(t+U_2\right)S_b^{\text{NS}}\left(t+U_3\right)S_b^{\text{NS}}\left(t+U_4\right)}{S_b^{\text{R}}\left(t+V_1\right)S_b^{\text{R}}\left(t+V_2\right)S_b^{\text{NS}}\left(t+V_3\right)S_b^{\text{NS}}\left(t+V_4\right)}
\nonumber \\
&~~~~-~\frac{S_b^{\text{R}}\left(t+U_1\right)S_b^{\text{R}}\left(t+U_2\right)S_b^{\text{R}}\left(t+U_3\right)S_b^{\text{R}}\left(t+U_4\right)}{S_b^{\text{NS}}\left(t+V_1\right)S_b^{\text{NS}}\left(t+V_2\right)S_b^{\text{R}}\left(t+V_3\right)S_b^{\text{R}}\left(t+V_4\right)},
\end{align}
where
\begin{align}
    U_1&=p_3-p_4 ~, \quad ~~~ V_1=\frac{Q}{2}+p_t+p_2-p_4 ~, \nonumber \\
    U_2&=-p_3-p_4 ~, ~~~~  V_2=\frac{Q}{2}-p_t+p_2-p_4 ~, \nonumber \\
    U_3&=p_1+p_2 ~, ~~~ ~~~~ V_3=\frac{Q}{2}+p_s ~, \nonumber \\
    U_4&=-p_1+p_2 ~, \quad ~~ V_4=\frac{Q}{2}-p_s ~. \nonumber
\end{align}
The expressions for the remaining two components $\mathbb{F}_{p_sp_t}\left[\begin{smallmatrix} p_3 & p_2 \\ p_4 & p_1 \end{smallmatrix}\right]^{\mathrm{o}}\text{}_{\mathrm{e}},\mathbb{F}_{p_sp_t}\left[\begin{smallmatrix} p_3 & p_2 \\ p_4 & p_1 \end{smallmatrix}\right]^{\mathrm{o}}\text{}_{\mathrm{o}}$ can be found e.g. in \cite{Poghosyan:2016kvd} and we will not need them here.
\par In particular, we will be interested in the fusion of the identity block in the s-channel of pairwise identical external operators  (i.e. when $p_1=p_2\equiv p_1$ and $p_3=p_4\equiv p_2$) which is an even block \cite{Poghosyan:2016kvd}. We then have
\begin{equation}
\mathcal{F}_{\mathds{1}}^{\mathrm{e}}\left[\begin{smallmatrix} p_2 & p_1 \\ p_2 & p_1 \end{smallmatrix}\right](z)=\int_{i\mathbb{R}_+}\frac{\d p_t}{i}\sum_{\rho=\mathrm{e},\mathrm{o}}\mathbb{F}_{\mathds{1},p_t}\left[\begin{smallmatrix} p_2 & p_1 \\ p_2 & p_1 \end{smallmatrix}\right]^{\mathrm{e}}\text{}_{\rho} \ \mathcal{F}_{p_t}^{\rho}\left[\begin{smallmatrix} p_1 & p_1 \\ p_2 & p_2 \end{smallmatrix}\right](1-z)~.
   \end{equation}
 It is a small calculation to start from the expressions (\ref{eq:eveneven}),(\ref{eq:evenodd}) and compute the identity limit: 
  \begin{equation}\label{eq:idlimit}
      \ p_2\rightarrow p_1+\delta, \  p_3\rightarrow p_2-\delta \ , \ p_4\equiv p_2, \ \ \ \ \ \ \ \text{as} \ \ p_s\rightarrow \frac{Q}{2}-\delta , \ \delta\rightarrow0.
 \end{equation}
We find
 \begin{align}
     \mathbb{F}_{\mathds{1},p_t}\left[\begin{smallmatrix} p_2 & p_1 \\ p_2 & p_1 \end{smallmatrix}\right]^{\mathrm{e}}\text{}_{\mathrm{e}}&=\rho^{(b)}_{\text{NS}}(p_t)C_{\mathrm{NS}}^{(b)}(p_t,p_1,p_2) \ , \nonumber \\
     \mathbb{F}_{\mathds{1},p_t}\left[\begin{smallmatrix} p_2 & p_1 \\ p_2 & p_1 \end{smallmatrix}\right]^{\mathrm{e}}\text{}_{\mathrm{o}}&=\frac{1}{2i}\rho^{(b)}_{\text{NS}}(p_t)\widetilde{C}_{\mathrm{NS}}^{(b)}(p_t,p_1,p_2)~.\label{eq:idlimitsCvms}
 \end{align}
 The quantities $\rho_{\text{NS}}$, $C_{\mathrm{NS}},\widetilde{C}_{\mathrm{NS}}$ are exactly the NS structure constants of Liouville theory given in (\ref{eq: NS spacelike 1 and 2 summary}). This relation establishes the generalization of the result given in \cite{Teschner:2001rv,Collier:2019weq}, for the case of bosonic Liouville's structure constant and the Virasoro fusion kernel.
 \par It is worth mentioning how the calculation leading up to this result actually works, since a careful analysis can give us slightly more general formulae than (\ref{eq:idlimitsCvms}) which, to the extent of our knowledge, have not appeared before in the literature and are interesting in their own rights. In other words, we will see that there is a simplification of the fusion kernel in the limit (\ref{eq:idlimit}) even for finite $\delta\neq0$. 
 \par The logic is as follows. In the limit (\ref{eq:idlimit}), and \textit{before} taking $p_s\rightarrow \frac{Q}{2}-\delta$ or $\delta\rightarrow0$, we find for the `e-e' component
 \begin{align}
     &\mathbb{F}_{p_s,p_t}\left[\begin{smallmatrix} p_2-\delta & p_1+\delta \\ p_2 & p_1 \end{smallmatrix}\right]^{\mathrm{e}}\text{}_{\mathrm{e}}=\rho^{(b)}_{\text{NS}}(p_t)C_{\mathrm{NS}}^{(b)}(p_t,p_1,p_2)\times \frac{f(p_s;p_i)}{\Gamma_b^{\text{NS}}\left(Q/2-\delta-p_s\right)}\int_{i\mathbb{R}}\frac{\d t}{i}J^{(ee)}_{p_sp_t}\left[\begin{smallmatrix} p_2-\delta & p_1+\delta \\ p_2 & p_1 \end{smallmatrix}\right], \label{eq:feedelta}
 \end{align}
 where $f$ is a meromorphic function of the momenta that is regular as $p_s\rightarrow p_s^*\equiv Q/2-\delta$. It specifically reads
\begin{align}
&f(p_s;p_i):=\frac{\Gamma_b^{\text{NS}}\left(Q\right)^3\Gamma_b^{\text{NS}}\left(Q\pm2p_1\right)\Gamma_b^{\text{NS}}\left(Q\pm2p_2\right)\Gamma_b^{\text{NS}}\left(\frac{Q}{2}\pm p_t+p_1-p_2+2\delta\right)}{\Gamma_b^{\text{NS}}\left(2Q\right)\Gamma_b^{\text{NS}}\left(\frac{Q}{2}\pm p_t+p_2-p_1\right)}\nonumber\\
     &~~~~~\times \frac{\Gamma_b^{\text{NS}}\left(Q\pm 2p_s\right)}{\Gamma_b^{\text{NS}}\left(\frac{Q}{2}\pm p_s+2p_1+\delta\right)\Gamma_b^{\text{NS}}\left(\frac{Q}{2}\pm p_s-2p_2+\delta\right)\Gamma_b^{\text{NS}}\left(\frac{Q}{2}+\delta\pm p_s\right)\Gamma_b^{\text{NS}}\left(\frac{Q}{2}-\delta+p_s\right)}~.
 \end{align}
 We therefore see that the prefactor in (\ref{eq:feedelta}) has a \textit{simple zero} when $p_s\rightarrow p_s^*$, even for finite $\delta\neq0$. We hold on to that observation.
 \par Next, for the integrand we get
 \begin{align}
 \frac{1}{i}J^{(\mathrm{ee})}_{p_sp_t}\left[\begin{smallmatrix} p_2-\delta & p_1+\delta \\ p_2 & p_1 \end{smallmatrix}\right]&= \sum_{\nu=\text{NS,R}}\frac{S_b^{\mathrm{\nu}}\left(t\pm\delta\right)S_b^{\mathrm{\nu}}\left(t+\delta+2p_1\right)S_b^{\mathrm{\nu}}\left(t+\delta-2p_2\right)}{i\times S_b^{\mathrm{\nu}}\left(\frac{Q}{2}+t\pm p_s\right)S_b^{\mathrm{\nu}}\left(\frac{Q}{2}+t+\delta \pm p_t+p_1-p_2 \right)}. \label{eq:sumintegrandlim}
 \end{align}
 In particular, the first summand in this expression only includes the functions $S_b^{\mathrm{NS}}$, while the second only $S_b^{\mathrm{R}}$'s. We recall (cf. Appendix \ref{app:SpecialFunctions}) that $S_b^{\mathrm{NS}}(x)$ has a simple pole when $x=0$, whereas for $x=Q$ has a simple zero. At these two points, on the other hand, the function $S_b^{\mathrm{R}}$ is regular. Therefore, when $p_s\rightarrow p_s^*$ the full integral develops a \textit{pinching singularity} since the contour must pass between two colliding poles originating from the first summand in (\ref{eq:sumintegrandlim}): a single pole at $t=\delta$ coming from the factor $S_b^{\mathrm{NS}}\left(t-\delta\right)$ and a single pole at $t=Q/2-p_s$ coming from the factor $S_b^{\mathrm{NS}}\left(\frac{Q}{2}+t+p_s\right)$. The singularity can be evaluated by the residue of the latter. Using the fact that $\text{Res}_{x\rightarrow0}\Gamma_b(x)=\frac{\Gamma_b(Q)}{2\pi}$, we deform appropriately and obtain for the full integral 
 \begin{align}
&\frac{S_b^{\mathrm{NS}}\left(Q/2-p_s\pm\delta\right)S_b^{\mathrm{NS}}\left(Q/2-p_s+\delta+2p_1\right)S_b^{\mathrm{NS}}\left(Q/2-p_s+\delta-2p_2\right)}{ S_b^{\mathrm{NS}}\left(Q-2p_s\right)S_b^{\mathrm{NS}}\left(Q-p_s+\delta \pm p_t+p_1-p_2 \right)}\nonumber\\
&\quad \quad \quad \quad \quad \quad \quad \quad \quad \quad \quad \quad \quad \quad \quad \quad \quad   +\int_{i\mathbb{R}} \d t \ \left(\text{reg. at $t\rightarrow Q/2-p_s$}\right).
\end{align} 
Crucially, the first factor above has a \textit{simple pole} at $p_s=p_s^*$ reflecting exactly the pinching singularity that we mentioned before. With these expressions at hand, we now take the limit $p_s\rightarrow p_s^*$ in the full kernel (prefactor times integral), in which case the simple zero of the prefactor that we encountered before will \textit{cancel} with the simple pole from the pinching singularity, and this will end up giving us the following general simple expression for the `e-e' component of the full kernel:
\begin{equation}
\mathbb{F}_{Q/2-\delta,p_t}\left[\begin{smallmatrix} p_2-\delta & p_1+\delta \\ p_2 & p_1 \end{smallmatrix}\right]^{\mathrm{e}}\text{}_{\mathrm{e}}=\zeta^{(e)}_\delta(p_1,p_2,p_t)\times \mathbb{F}_{\mathds{1},p_t}\left[\begin{smallmatrix} p_2 & p_1 \\ p_2 & p_1 \end{smallmatrix}\right]^{\mathrm{e}}\text{}_{\mathrm{e}} \ , \ \ \ \ \ \delta\neq0\label{eq:generaleecomp}
\end{equation}
where
\begin{align}
&\zeta^{(\mathrm{e})}_\delta(p_1,p_2,p_t):=\\
&\frac{\Gamma_b^{\mathrm{NS}}\left(2Q-2\delta\right)\Gamma_b^{\mathrm{NS}}\left(Q\right)\Gamma_b^{\mathrm{NS}}\left(Q-2p_1\right)\Gamma_b^{\mathrm{NS}}\left(Q+2p_2\right)\Gamma_b^{\mathrm{NS}}\left(\frac{Q}{2}\pm p_t+p_2-p_1-2\delta\right)}{\Gamma_b^{\mathrm{NS}}\left(2Q\right)\Gamma_b^{\mathrm{NS}}\left(Q-2\delta\right)\Gamma_b^{\mathrm{NS}}\left(Q-2p_1-2\delta\right)\Gamma_b^{\mathrm{NS}}\left(Q+2p_2-2\delta\right)\Gamma_b^{\mathrm{NS}}\left(\frac{Q}{2}\pm p_t+p_2-p_1\right)}~,
\end{align}
and
\begin{equation}
\mathbb{F}_{\mathds{1},p_t}\left[\begin{smallmatrix} p_2 & p_1 \\ p_2 & p_1 \end{smallmatrix}\right]^{\mathrm{e}}\text{}_{\mathrm{e}}:=\lim_{\delta\rightarrow0}\mathbb{F}_{Q/2-\delta,p_t}\left[\begin{smallmatrix} p_2-\delta & p_1+\delta \\ p_2 & p_1 \end{smallmatrix}\right]^{\mathrm{e}}\text{}_{\mathrm{e}}=\rho^{(b)}_{\text{NS}}(p_t)C_{\mathrm{NS}}^{(b)}(p_t,p_1,p_2).
\end{equation}
The last equation is exactly what we wanted to establish, but equation (\ref{eq:generaleecomp}) is more interesting since it highlights a relation of $\rho_{\mathrm{NS}},C_{\mathrm{NS}}$ with a more general fusion kernel entry with $\delta\neq0$.
\par Doing similar manipulations for the `e-o' component, we find
\begin{equation}
\mathbb{F}_{Q/2-\delta,p_t}\left[\begin{smallmatrix} p_2-\delta & p_1+\delta \\ p_2 & p_1 \end{smallmatrix}\right]^{\mathrm{e}}\text{}_{\mathrm{o}}=\zeta^{(\mathrm{o})}_\delta(p_1,p_2,p_t)\times \mathbb{F}_{\mathds{1},p_t}\left[\begin{smallmatrix} p_2 & p_1 \\ p_2 & p_1 \end{smallmatrix}\right]^{\mathrm{e}}\text{}_{\mathrm{o}} \ , \ \ \ \ \ \delta\neq0\label{eq:generaleocomp}
\end{equation}
where
\begin{align}
&\zeta^{(\mathrm{o})}_\delta(p_1,p_2,p_t):=\\
&\frac{\Gamma_b^{\mathrm{NS}}\left(2Q-2\delta\right)\Gamma_b^{\mathrm{NS}}\left(Q\right)\Gamma_b^{\mathrm{NS}}\left(Q-2p_1\right)\Gamma_b^{\mathrm{NS}}\left(Q+2p_2\right)\Gamma_b^{\mathrm{R}}\left(\frac{Q}{2}\pm p_t+p_2-p_1-2\delta\right)}{\Gamma_b^{\mathrm{NS}}\left(2Q\right)\Gamma_b^{\mathrm{NS}}\left(Q-2\delta\right)\Gamma_b^{\mathrm{NS}}\left(Q-2p_1-2\delta\right)\Gamma_b^{\mathrm{NS}}\left(Q+2p_2-2\delta\right)\Gamma_b^{\mathrm{R}}\left(\frac{Q}{2}\pm p_t+p_2-p_1\right)},
\end{align}
and
\begin{equation}
\mathbb{F}_{\mathds{1},p_t}\left[\begin{smallmatrix} p_2 & p_1 \\ p_2 & p_1 \end{smallmatrix}\right]^{\mathrm{e}}\text{}_{\mathrm{o}}:=\lim_{\delta\rightarrow0}\mathbb{F}_{Q/2-\delta,p_t}\left[\begin{smallmatrix} p_2-\delta & p_1+\delta \\ p_2 & p_1 \end{smallmatrix}\right]^{\mathrm{e}}\text{}_{\mathrm{o}}=\frac{1}{2i}\rho^{(b)}_{\text{NS}}(p_t)\widetilde{C}_{\mathrm{NS}}^{(b)}(p_t,p_1,p_2).
\end{equation}
The last equation is again what we wanted to show, although (\ref{eq:generaleocomp}) holds more generally for $\delta\neq0$.
\bibliographystyle{JHEP}
\bibliography{bib}
\end{document}